\documentclass[runningheads,openany]{svmult}

\usepackage{palatino}
\usepackage{euler}
\usepackage{helvet}

\usepackage{graphicx}  
\usepackage{physprbb}  

\usepackage{proc}  
\usepackage{epsfig}

\usepackage{amssymb}
\usepackage{psfrag}
\usepackage{curves}
\usepackage{epic}
\usepackage{eepic}
\usepackage{rotating}

\makeindex             

\usepackage{amsmath}   
\DeclareMathOperator{\tr}{Tr}

\let\tocauthor\authorrunning

\begin{document}

\frontmatter

\thispagestyle{empty}
\parindent=0pt

{\Large\sc Blejske delavnice iz fizike \hfill Letnik~4, \v{s}t. 2--3}

\smallskip

{\large\sc Bled Workshops in Physics \hfill Vol.~4, No.~2--3}

\smallskip

\hrule

\hrule

\hrule

\vspace{0.5mm}

\hrule

\medskip
{\sc ISSN 1580--4992}

\vfill

\bigskip\bigskip
\begin{center}

{\bfseries 
{\Huge  Proceedings to the 
\smallskip
Euroconference on Symmetries 
\smallskip
Beyond the Standard Model}

\vspace{5mm}
{\Large  Portoro\v z, July 12 -- 17, 2003}

\medskip
{\Huge (Part 1 of 2)} 
\vspace{10mm}}

\vfill

{\bfseries\large
Edited by

\vspace{5mm}
Norma Manko\v c Bor\v stnik\rlap{$^{1,2}$}

\smallskip

Holger Bech Nielsen\rlap{$^{3}$}

\smallskip

Colin D. Froggatt\rlap{$^{4}$}

\smallskip

Dragan Lukman\rlap{$^2$}

\bigskip

{\em\normalsize $^1$University of Ljubljana, $^2$PINT, %
$^3$ Niels Bohr Institute, $^4$ Glasgow University}

\vspace{12pt}

\vspace{3mm}

\vrule height 1pt depth 0pt width 54 mm}

\vspace*{3cm}

{\large {\sc  DMFA -- zalo\v{z}ni\v{s}tvo} \\[6pt]
{\sc Ljubljana, december 2003}}
\end{center}
\newpage

\thispagestyle{empty}
\parindent=0pt
\begin{flushright}
{\parskip 6pt
{\bfseries\large
                  The \textit{Euroconference on Symmetries Beyond  
                  the Standard Model},\\
                  12.--17. July 2003, Portoro\v z, Slovenia}

\bigskip\bigskip

{\bf was organized by}

{\parindent8pt
\textit{European Science Foundation, EURESCO office, Strassbourg}
}

\bigskip

{\bf and sponsored by}

{\parindent8pt
\textit{Ministry of Education, Science and Sport of Slovenia}

\textit{Luka Koper d.d., Koper, Slovenia}

\textit{Department of Physics, Faculty of Mathematics and Physics,
University of Ljubljana}

\textit{Primorska Institute of Natural Sciences and Technology, Koper}}

\bigskip
\medskip

{\bf Scientific Organizing Committee}

\medskip

{\parindent9pt
\textit{Norma Manko\v c Bor\v stnik (Chairperson), Ljubljana and Koper, Slovenia}

\textit{Holger Bech Nielsen (Vice Chairperson), Copenhagen, Denmark}

\textit{Colin D. Froggatt, Glasgow, United Kingdom}

\textit{Loriano Bonora, Trieste, Italy}

\textit{Roman Jackiw, Cambridge, Massachussetts, USA}

\textit{Kumar S. Narain, Trieste, Italy}

\textit{Julius Wess, Munich, Germany}

\textit{Donald L. Bennett, Copenhagen, Denmark}

\textit{Pavle Saksida, Ljubljana, Slovenia}

\textit{Astri Kleppe, Oslo, Norway}

\textit{Dragan Lukman, Koper, Slovenia} }}

\end{flushright}

\setcounter{tocdepth}{0}
\tableofcontents
\cleardoublepage
\chapter*{Preface}
\addcontentsline{toc}{chapter}{Preface}

This was the first EURESCO conference in the series of EURESCO
conferences entitled ''What comes beyond the Standard Models''. It has
started to confront in open-minded and friendly discussions new ideas,
models, approaches and theories in elementary particle physics,
cosmology and those parts of mathematics, which are essential in these
two (and other) fields of physics. As an Euresco conference, the
conference was expected to offer a lot of possibilities for all the
participants, and in particular for young scientists, to put questions
and comments to senior scientists.

The invited speakers presented their own work in an intelligible way
for nonexperts, discussing also their own work in light of other
approaches. All the participants were very active in discussions,
following the talks, and in several round tables,

There is as yet no experimental evidence that points to a definite
approach to understanding the physics underlying the Standard Model of
the electroweak and colour interactions. For example, there is no
theoretical understanding of origin of the symmetries of the
Electroweak Standard Model or the values of about 20 free parameters
of the model. This being the case, it is very important to
open-mindedly examine a broad range of possible approaches including
of course the most popular ones (strings and branes, noncommutative
geometry, conformal field theory, etc.) as well as some perhaps less
well known approaches. Because the chance a priori of guessing the
correct theory behind the electroweak Standard Model is very small, it
is important to effectively use the hints that can be gleaned from
careful analysis of the known physics that is not explained by the
Electroweak Standard Model (e.g., the observed symmetry groups and
values of free parameters, number of generations, etc.). It is also
important to use the input from the Standard Cosmological Model. By
carefully using the hints from these models we have the best chance of
building a correct model for the physics behind the Standard Model
that unifies all interactions including gravity. In pursuing this
goal, it is also important to have potentially promising mathematical
concepts at our disposal (e.g., noncommutative algebras, Hopf
algebras, Moyal products, etc.).

Among the questions. that were pertinently discussed, are:
\begin{itemize}
\item Why, when and how has Nature in her evolution decided to
   demonstrate at low energies one time and three space coordinates?
\item How has the internal space of spins and charges dictated this
   decision?
\item Should not the whole internal space (of spins and charges) be
   unique?
\item Should not accordingly also all the interactions (interaction
   fields) be unique?
\item How have been in the evolution decisions about when and how to
break symmetries from the general one within the space-time of $q$
space and $d-q$ time coordinates, for $d$ may be very large and any
$q$, were made?
\item How topology and differential geometry are connected in
$d$-dimensional spaces and how this dependance changes with $d$?  How
topology and diferential geometry influenced in the evolution breaks
of symmetries?
\item When in the evolution has Nature made the decision of the Minkowski
signature and how the internal space has contributed to this decision?
\item Where does the observed symmetry between matter and anti-matter
   originate from?
\item Why do massless fields exist at all? Where and why does the Higgs
mass (the weak scale) come from? Where do the Yukawa couplings come
from?
\item Why do only the left handed spinors (fermions) carry the weak
charge? Why does the weak charge (and not the colour charge) break
parity?
\item Why are some spinor masses so small in comparison with the weak
    scale?
\item Where do the familes come from?
\item Is the anomaly cancellation a general feature or is a very special
    choice?
\item Do Majorana-like particles exist?
\item Why the observed representations of known charges are so small:
singlets, doublets and at most triplets?
\item What is our Universe made out of (besides the baryonic matter)?
\item What is the origin of fields which have caused the inflation?
\item Can all known elementary fermionic fields be understood as
different states of only one field, with the unique internal space of
spins and charges? Not only spinors (fermions) but also all the
fields?
\item How can all the bosonic fileds be unified and quantized?
\item What is the role of symmetries in Nature? And how symmetries and
    dimensions are connected?
\item \emph{and many others}
\end{itemize}

The conference gathered participants from several European countries,
United States, and European scientists working in the United
States. There were also participants from the Eastern European
countries and Japan.  Lively interaction between the participants
served to establish connections and links between the participants and
between their home institutions as was stated by several of them
already during the conference. Noted was the diversity of the European
countries from which the participants come, with no single country
dominating. Some younger participants later participated at the Annual
International Bled Workshop which is on the same topics, but dedicated
to longer in-depth discussion sessions and which took place
immediately after the conference in Portoroz. There was plenty of time
for discussion and it looked like that it was properly used, and that
nobody felt embarrassed to speak or under pressure. The participants
express a desire to meet in the second conference in the series.

Let us at the end thank all the participants for their contribution to
making this conference, in our view --- also shared by many participants,
a very successfull one. Special thanks goes to the contributors to this
volume. Your timely response, quality of your contributions and, last
but not least, polished form in which you submitted them, has made the
editors' work  much easier. We also thank the EURESCO office in
Strassbourg for organizing such a nice conference and the two sponsors
(Luka Koper d.d. and Ministry of Education, Science and Sports) for
their financial contributions.\\[2 cm]

\parbox[b]{40mm}{%
                 \textit{Norma Manko\v c Bor\v stnik}\\
                 \textit{Holger Bech Nielsen}\\
		 \textit{Colin Froggatt}\\
		 \textit{Dragan Lukman} }
\qquad\qquad\qquad\qquad\qquad\qquad\quad
\textit{Ljubljana, December 2003}

\cleardoublepage

\mainmatter
\parindent=20pt
\setcounter{page}{5}
\setcounter{page}{1}
\title*{Status of the Standard Model}
\author{Paul H. Frampton}
\institute{%
University of North Carolina at Chapel Hill\\
Department of Physics and Astronomy\\
Phillips Hall\\
Chapel Hill NC 27599-3255\\
United States}

\titlerunning{Status of the Standard Model}
\authorrunning{Paul H. Frampton}
\maketitle

\section{Introduction}
 
The standard model is healthy in all respects
except for the non-zero neutrino masses which
require an extension of the minimal version.

This talk was in three parts:

(I) Rise and Fall of the Zee Model 1998-2001.

(II) Classification of Two-Zero Textures.

(III) FGY Model relating Cosmological B to Neutrino CP violation.

For parts (I) and (II) references are provided. Part (III) is
included in this write-up.

\section{Rise and Fall of the Zee Model 1998-2001}

This first part was based on:

P.H. Frampton and S.L. Glashow, Phys. Lett.
{\bf 461,} 95 (1999). {\tt hep-ph/9906375}.

\section{Classification of Two-Zero Textures}

The second part was based on:

P.H. Frampton, S.L. Glashow and D. Marfatia,
Phys. Lett. {\bf B536,} 79 (2002).
{\tt hep-ph/0201008}

\section{FGY Model relating Cosmological B with Neutrino CP Violation}

One of the most profound ideas is\cite{Sakharov}
that baryon number asymmetry arises in the early universe
because of processes which violate CP symmetry and that
terrestrial experiments on CP violation
could therefore inform us of the details of such
cosmological baryogenesis.

The early discussions of baryogenesis focused on
the violation of baryon number and its possible relation to
proton decay. In the light of present evidence for neutrino masses
and oscillations
it is more fruitful to associate the baryon number
of the universe with violation of lepton number\cite{FY}.
In the present Letter
we shall show how, in one class of models, the sign of the
baryon number of the universe correlates with
the results of CP violation in neutrino oscillation
experiments which will be performed in the forseeable
future.

Present data on atmospheric and solar neutrinos suggest
that there are respective squared mass differences
$\Delta_a \simeq 3 \times 10^{-3} eV^2$
and
$\Delta_s \simeq 5 \times 10^{-5} eV^2$.
The corresponding mixing angles $\theta_1$
and $\theta_3$ satisfy
$tan^2 \theta_1 \simeq 1$
and $0.6 \leq sin^2 2\theta_3 \leq 0.96$
with $sin^2 \theta_3 = 0.8$ as the best fit.
The third mixing angle is much smaller than the other
two, since the data require $sin^2 2 \theta_2 \leq 0.1$.

A first requirement is that our model\cite{FGY} accommodate
these experimental facts at low energy.


\section{The Model}


In the minimal standard model, neutrinos are massless.
The most economical addition to the standard model
which accommodates both neutrino masses
and allows the violation
of lepton number to underly the cosmological baryon asymmetry
is two right-handed neutrinos $N_{1,2}$.

These lead to new terms in the lagrangian:

\begin{eqnarray}
{\cal L} & = & \frac{1}{2} (N_1, N_2) \left(
\begin{array}{cc} M_1 & 0 \\ 0 & M_2
\end{array} \right)
\left( \begin{array}{c} N_1 \\ N_2 \end{array} \right)
+  \nonumber \\
& + & (N_1, N_2)
\left( \begin{array}{ccc} a  &  a^{'}  &  0  \\
0  &  b  &  b^{'}  \end{array} \right)
\left( \begin{array}{c} l_1  \\  l_2 \\ l_3 \end{array}
\right)  H  + h.c.
\label{Lag}
\end{eqnarray}
where we shall denote the rectangular Dirac mass matrix
by $D_{ij}$. We have assumed a texture
for $D_{ij}$ in which the upper
right and lower left entries vanish.
The remaining parameters in our model
are both necessary and sufficient
to account for the data.

For the light neutrinos, the see-saw mechanism leads to
the mass matrix\cite{Y}
\begin{eqnarray}
\hat{L} & = & D^T M^{-1} D \nonumber \\
& = & \left( \begin{array}{ccc}
\frac{a^2}{M_1}  &  \frac{a a^{'}}{M_1}  &  0  \\
\frac{a a^{'}}{M_1}  &  \frac{(a^{'})^2}{M_1} + \frac{b^2}{M_2}
&  \frac{b b^{'}}{M_2} \\
0  &  \frac{b b^{'}}{M_2}  &  \frac{(b^{'})^2}{M_2}
\end{array}  \right)
\label{L}
\end{eqnarray}

We take a basis where $a, b, b^{'}$ are real and where
$a^{'}$ is complex $a^{'} \equiv |a^{'}|e^{i \delta}$.
To check consistency with low-energy
phenomenology we temporarily take the specific
values (these will be loosened later)
$b^{'} = b$ and $a^{'} = \sqrt{2} a$ and all parameters real.
In that case:
\begin{eqnarray}
\hat{L}
& = & \left( \begin{array}{ccc}
\frac{a^2}{M_1}  &  \frac{\sqrt{2}a^2}{M_1}  &  0  \\
\frac{\sqrt{2}a^2}{M_1}  &  \frac{2a^2}{M_1} + \frac{b^2}{M_2}
&  \frac{b^{2}}{M_2} \\
0  &  \frac{b^{2}}{M_2}  &  \frac{b^{^2}}{M_2}
\end{array}  \right)
\label{LL}
\end{eqnarray}
We now diagonalize to the mass basis by writing:
\begin{equation}
{\cal L} = \frac{1}{2} \nu^T \hat{L} \nu
= \frac{1}{2} \nu^{'T} U^T \hat{L} U \nu^{'}
\end{equation}
where
\begin{eqnarray}
U & = & \left( \begin{array}{ccc} 1/\sqrt{2}  &  1/\sqrt{2}  &  0  \\
- 1/2 & 1/2  &  1/\sqrt{2}  \\
1/2  &  -1/2 &  1/\sqrt{2}
\end{array} \right) \times \nonumber \\
& \times &
\left( \begin{array}{ccc}
1  &  0  &  0  \\
0  &  cos\theta &  sin\theta  \\
0  &  - sin\theta  &  cos\theta
\end{array}
\right)
\end{eqnarray}
We deduce that the mass eigenvalues  and $\theta$
are given by

\begin{equation}
m(\nu_3^{'}) \simeq 2 b^2/M_2; ~~~ m(\nu_2^{'}) \simeq 2 a^2 /M_1; ~
~~
m(\nu_1^{'}) = 0
\end{equation}
and
\begin{equation}
\theta \simeq m(\nu_2^{'}) / (\sqrt{2} m(\nu_3^{'}))
\end{equation}
in which it was assumed that $a^2/M_1 \ll b^2/M_2$.

By examining the relation between the three mass eigenstates
and the corresponding flavor eigenstates
we find
that for the unitary matrix relevant to neutrino oscillations
that
\begin{equation}
U_{e3} \simeq sin\theta/\sqrt{2} \simeq m(\nu_2)/(2m(\nu_3))
\end{equation}

Thus the assumptions $a^{'} = \sqrt{2} a$, $
b^{'} = b$ adequately fit the
experimental data, but
$a^{'}$ and $b^{'}$ could
be varied around
$\sqrt{2}a$ and $b$ respectively
to achieve better fits.

But we may conclude that
\begin{eqnarray}
2b^2/M_2 & \simeq & 0.05 eV = \sqrt{\Delta_a} \nonumber \\
2a^2/M_1 & \simeq & 7 \times 10^{-3} eV  = \sqrt{\Delta_{s}}
\label{numass}
\end{eqnarray}

It follows from these values that $N_1$ decay satisfies the
out-of-equilibrium condition for leptogenesis (the
absolute requirement is $m < 10^{-2} eV$ \cite{buchpascos})
while $N_2$ decay does not.
This fact enables us to predict
the sign of CP violation in neutrino oscillations without ambiguity.


\section{Connecting Link}


Let us now come to the main result. Having
a model consistent with all low-energy data and with
adequate texture zeros\cite{FGM} in $\hat{L}$ and
equivalently $D$ we can
compute the sign both of the high-energy
CP violating parameter ($\xi_H$) appearing
in leptogenesis and of the CP violation parameter
which will be measured in low-energy
$\nu$ oscillations ($\xi_L$).

We find the baryon number $B$ of the universe
produced by $N_1$ decay
proportional to\cite{Buch}
\begin{eqnarray}
B & \propto & \xi_H =
(Im D D^{\dagger} )_{12}^2 = Im (a^{'} b)^2 \nonumber \\
& = & + Y^2a^2b^2 sin 2\delta
\label{highenergy}
\end{eqnarray}
in which $B$ is positive by observation of the universe.
Here we have loosened our assumption about $a^{'}$
to $a^{'} = Y a e^{i \delta}$.

At low energy the CP violation in neutrino oscillations is governed
by
the quantity\cite{branco}
\begin{equation}
\xi_L = Im (h_{12} h_{23} h_{31})
\end{equation}
where $h = \hat{L} \hat{L}^{\dagger}$.

Using Eq.(\ref{L}) we find:
\begin{eqnarray}
h_{12} & = & \left( \frac{a^3 a^{'*}}{M_1^2} + \frac{a |a^{'}|^2 a^{
'*}}{M_1^2}
\right) + \frac{a a^{'} b^2}{M_1M_2} \nonumber \\
h_{23} & = & \left( \frac{b b^{'} a^{'2}}{M_1 M_2} \right)
 + \left( \frac{b^3 b^{'}}{M_2^2} + \frac{b b^{'3}}{M_2^2} \right)
\nonumber \\
h_{31} & = & \left( \frac{a a^{'*} b b^{'}}{M_1 M_2}
\right) \nonumber \\
\end{eqnarray}
from which it follows that
\begin{equation}
\xi_L =  - \frac{a^6 b^6}{M_1^3 M_2^3} sin 2\delta [ Y^2 (2 + Y^2)]
\label{lowenergy}
\end{equation}
Here we have taken $b=b^{'}$ because
the mixing for the atmospheric neutrinos
is almost maximal.

Neutrinoless double beta decay $(\beta\beta)_{0\nu}$
is predicted at a rate corresponding to $\hat{L}_{ee} \simeq 3 \times 10^{-3}eV$.

The comparison between Eq.(\ref{highenergy})
and Eq.(\ref{lowenergy})
now gives a unique relation between the signs of $\xi_L$ and $\xi_H$
.

As a check of this assertion we consider
the equally viable alternative model

\begin{equation}
D = \left( \begin{array}{ccc} a & 0 & a^{'} \\
0 & b & b^{'}
\end{array} \right)
\label{alternative}
\end{equation}
in Eq.(\ref{Lag})
where $\xi_L$ reverses sign but the
signs of $\xi_H$ and $\xi_L$ are
still uniquely correlated once the $\hat{L}$
textures arising from the $D$ textures
of Eq.(\ref{Lag}) and Eq.(\ref{alternative})
are distinguished by low-energy phenomenology.
Note that such models have five parameters including
a phase
and that cases B1 and B2 in \cite{FGM}
can be regarded as (unphysical) limits of (\ref{Lag})
and (\ref{alternative}) respectively.

This fulfils in such a class of models the idea of \cite{Sakharov}
with only the small change that baryon
number violation is replaced by lepton number violation.


\section{Further Properties}


The model of \cite{FGY} has additional properties
which we allude to here briefly:


\noindent 1) It is important that the zeroes occurring
in Eq.(\ref{Lag}) can be associated with a global symmetry and hence
 are
not infinitely renormalized. This can be achieved.


\noindent 2) The model has four parameters in the texture
of Eq.(\ref{L})  and leads to a prediction of
$\theta_{13}$ in terms of the other four parameters
$\Delta_a, \Delta_S, \theta_{12},$ and $\theta_{23}$.
The result is that $\theta_{13}$ is predicted to be non-zero
with magnitude related to the smallness of
$\Delta_S/\Delta_a$.


\noindent Details of these properties are currently under
further investigation.


\section*{Acknowledgement}

I thank  Professor N. Mankoc-Borstnik of the University
of Ljubjana
for organizing this stimulating workshop in Portoroz,
Slovenia.
This work was supported in part
by the Department of Energy
under Grant Number
DE-FG02-97ER-410236.


\title*{Cosmological Constraints from Microwave Background Anisotropy
and Polarization}
\author{Alessandro Melchiorri}
\institute{%
 Universita' di Roma ``La Sapienza''\\ 
Ple Aldo Moro 2, 00185, Rome, Italy}

\titlerunning{Cosmological Constraints from Microwave Background Anisotropy}
\authorrunning{Alessandro Melchiorri}
\maketitle

\begin{abstract}
The recent high-quality measurements of the Cosmic Microwave Background 
anisotropies and polarization provided 
by ground-based, balloon-borne and satellite experiments have presented 
cosmologists with the possibility of studying the large scale properties of 
our universe with unprecedented precision.
Here I review the current status of observations and constraints
on theoretical models.
\end{abstract}

\section{Introduction}\label{AM:s1}

The nature of cosmology as a mature and testable science
lies in the realm of observations of Cosmic Microwave Background 
(CMB) anisotropy and polarization. The recent high-quality 
measurements of the CMB anisotropies provided by ground-based, 
balloon-borne  and satellite experiments have indeed presented cosmologists 
with the possibility of studying the large scale properties of our 
universe with unprecedented precision.\\ 
An increasingly complete cosmological image arises as the key parameters 
of the cosmological model  have now been constrained within a few percent 
accuracy. The impact of these results in different sectors than
cosmology has been extremely relevant since CMB studies can 
set stringent constraints on the early thermal history of the universe 
and its particle content.
For example, important constraints have been placed in fields 
related to particle physics or quantum gravity 
like neutrino physics, extra dimensions, and super-symmetry theories.\\
In the next couple of years, new and current on-going experiments
will provide datasets with even higher quality and information.
In particular, accurate measurements of the CMB polarization statistical 
properties represent a new research area. 
The CMB polarization has been detected by two experiments, but 
remains to be thoroughly investigated. In conjunction with our extensive 
knowledge about the CMB temperature anisotropies, new constraints on the 
physics of the early universe (gravity waves, isocurvature perturbations, 
variations in fundamental constants) as well as late universe phenomena 
(reionization, formation of the first objects, galactic foregrounds) will be 
investigated with implications for different fields ranging from particle 
physics to astronomy.\\
Moreover, new CMB observations at small (arcminute) angular scales will 
probe secondary fluctuations associated  with the first nonlinear objects.
This is where the first galaxies and the
first quasars may leave distinct imprints in the CMB and where an interface
between cosmology and the local universe can be established. 
In this proceedings I will briefly review the current status of CMB 
observations, I discuss the agreement with the current theoretical
scenario and I will finally draw some conclusions.

\section{The standard picture.}

The standard model of structure formation (described in great
detail in several reviews, see e.g. \cite{review}, 
\cite{review2}, \cite{review3}, \cite{review4}, \cite{review5}).
relies on the presence of a background of tiny 
(of order $10^{-6}$) primordial density 
perturbations on all scales (including those larger than
the causal horizon). \\
This primordial background of perturbations is assumed gaussian, adiabatic, 
and nearly scale-invariant as generally predicted by the inflationary 
paradigm. Once inflation is over, the evolution of all Fourier mode 
density perturbations is linear and passive (see \cite{review5}).\\
Moreover, prior to recombination, a given Fourier mode begins
oscillating as an acoustic wave once the horizon overtakes its wavelength.
Since all modes with a given wavelength begin evolving simultaneously
the resulting acoustic oscillations are phase-coherent, leading to
a structure of peaks in the temperature and polarization power spectra of
the Cosmic Microwave Background (\cite{Peeb1970}, \cite{SZ70}, 
\cite{wilson}).\\
The anisotropy with respect to the mean temperature $\Delta T=T-T_0$
 of the CMB sky in the direction ${\bf n}$ measured at
time $t$ and from the position $\vec x$ can be expanded in 
spherical harmonics:

\begin{equation}
{\Delta T\over T_0}({\bf n},t,\vec x) = \sum_{\ell=2}^\infty\sum_{m=-\ell}^{m=\ell}
	 a_{\ell m}(t, \vec x) Y_{\ell m}({\bf n})~,
\end{equation}

If the fluctuations are Gaussian all the statistical information is
contained in the $2$-point correlation function. In the case of isotropic
fluctuations, this can be written as:

\begin{equation}
\left \langle{\Delta T\over T_0}({{\bf n}_1}){\Delta T\over T_0}({{\bf n}_2})\right\rangle =
 {1\over 4\pi} \sum_\ell (2\ell+1)C_\ell P_\ell({\bf n}_1\cdot{\bf n}_2)~.
\end{equation}

where the average is an average over "all the possible universes"
i.e., by the ergodic theorem, over $\vec x$. The CMB power
spectrum $C_\ell$ are the ensemble average of the coefficients
$a_{\ell m}$,
\[ C_\ell = \langle|a_{\ell m}|^2\rangle ~. \]

A similar approach can be used for the cosmic microwave background
polarization and the cross temperature-polarization correlation
functions.
Since it is impossible to measure ${\Delta T\over T_0}$ in every position in
the universe, we cannot do an ensemble average.
This introduces a fundamental limitation for the precision of a 
measurement (the cosmic variance) which is 
important especially for low multipoles. 
If the temperature fluctuations are Gaussian, the $C_{\ell}$ have 
a chi-square distribution with $2\ell+1$ degrees of freedom and 
the observed mean deviates from the ensemble average by 

\begin{equation} 
{{\Delta C_\ell} \over C_\ell} = 
	\sqrt{2\over 2\ell + 1}~.  \label{2cv}
\end{equation}

Moreover, in a real experiment, one never obtain complete sky 
coverage because of the limited amount of observational time 
(ground based and balloon borne
experiments) or because of galaxy foreground contamination 
(satellite experiments). 
All the telescopes also have to deal with the noise of the detectors 
and are obviously not sensitive to scales smaller than the 
angular resolution.\\

\begin{figure}
\centerline{\includegraphics[width=4.0in]{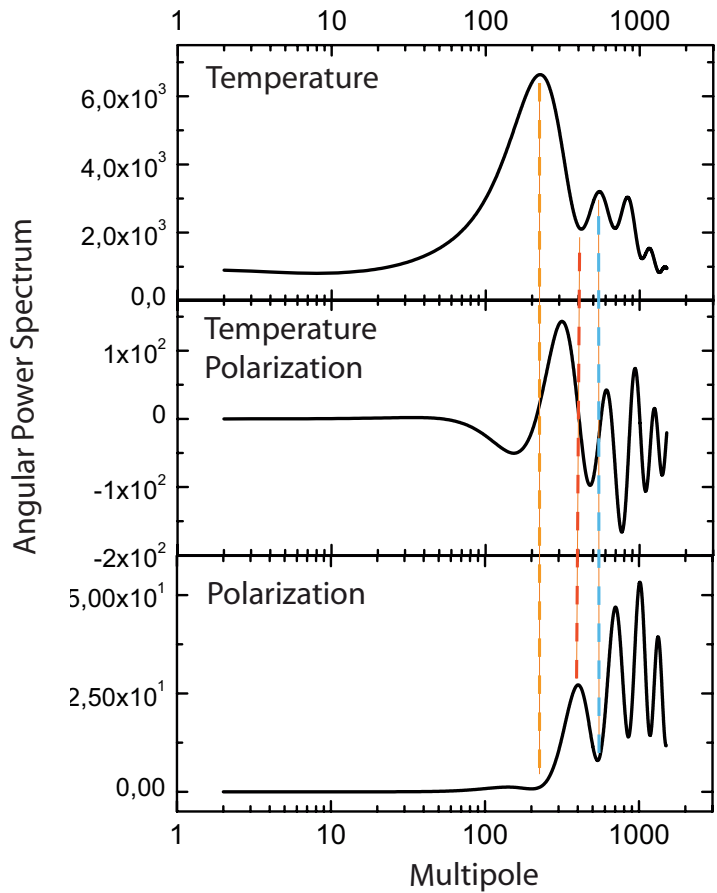}}
\caption{\label{fig1} Theoretical predictions for the CMB
temperature, polarization and cross temperature-polarization power
spectra in the case of the standard model of structure formation.
The peaks in the temperature and polarization spectra are alternate.}
\end{figure}

In Figure $1$ we plot the theoretical prediction for CMB temperature and 
polarization power spectra and the 
cross-correlation between temperature and polarization in the case of the 
so-called 'concordance' model. 
The major conclusions we can draw from these predictions are:

\begin{itemize}

\item The power spectra show an unique structure: 
The temperature power spectrum is flat on large scale
while shows oscillations on smaller scales. The polarization and
cross temperature-polarization spectra are also showing oscillations
on smaller scales, but the signal at large scale is expected to be 
negligible. This is a direct consequence of the fact that
different physical mechanisms control the growth of perturbations
on different scales.

\item On smaller scales, again, different mechanisms are responsible
for the oscillations. Gravity and density perturbations obey a cosine 
function, while velocity perturbations follow a sine function. 
On the scale of the thickness of the last scattering
surface, temperature anisotropies are more coupled to density and
gravity perturbations, while polarization is more coupled to velocity
perturbations. The most striking observational prediction of this is
the out-of-phase position of the peaks and dips in the temperature
and polarization power spectra. The peaks and dips of the
cross-correlation spectra fall in the middle.

\item The shape of the power spectra depends on the value
of the cosmological parameters assumed in the theoretical
computation. A measure of these spectra can therefore
produce indirect constraints on several parameters of the model.
The parameters that can be well identified are the overall curvature, 
the energy density in baryons $\omega_b$, the energy density in dark matter
$\omega_m$, the spectral tilt of primordial fluctuations $n_S$ and 
the value of the optical depth of the universe $\tau$. 
Presence of polarization at large angular scale, in particular, 
is evidence for a reionization of the intergalactic medium. 
Higher is the polarization at large scale, higher is the value of the 
optical of the universe at late redshifts. 
Since the overall picture must be consistent, the cosmological
parameters determined indirectly from CMB observations must agree
with the values inferred from independent observations (Big Bang
Nucleosynthesis, Galaxy Surveys, Ly-$\alpha$ forest clouds, simulations,
etc etc.).

\end{itemize}

\section{The latest measurements.}

The last years have been an exciting period for
the field of the CMB research. 
With the TOCO$-97/98$ (\cite{torbet},\cite{miller}) 
and BOOMERanG-$97$ (\cite{mauskopf}) experiments a firm detection of
a first peak in the CMB angular power spectrum 
on about degree scales has been obtained. 
In the framework of adiabatic Cold Dark Matter (CDM) models, the
position, amplitude and width of this peak provide strong supporting 
evidence for the inflationary predictions of
a low curvature (flat) universe and a scale-invariant primordial 
spectrum (\cite{knox}, \cite{melchiorri}, \cite{tegb97}).\\
The subsequent data from BOOMERanG LDB (\cite{netterfield}), 
DASI (\cite{halverson}), MAXIMA (\cite{lee}), 
VSAE (\cite{grainge}) and Archeops 
(\cite{benoit}) have provided further evidence for the presence of the 
first peak and refined the data at larger multipole hinting
towards the presence of multiple peaks in the spectrum.
Moreover, the very small scale observations made by the
CBI (\cite{pearson}) and ACBAR (\cite{acbar}) experiments 
have confirmed the presence of a damping tail, while the
new DASI results presented the first evidence for polarization
(\cite{dasipol}).\\
The combined data clearly confirmed the model
prediction of acoustic oscillations in the primeval plasma 
and shed new light on several cosmological and 
inflationary parameters ( see e.g. 
\cite{anze}, \cite{odman}, \cite{wang2}).\\
The recent results from the WMAP satellite experiment
(\cite{bennett}) have confirmed in a spectacular way all these 
previous results with a considerable reduction of the error
bars. In particular, the amplitude and position of the
first two peaks in the spectrum are now determined with a precision 
about $6$ times better than before (\cite{page}).\\
Furthermore, the WMAP team released the first high quality measurements
of the temperature-polarization spectrum \cite{kogut}.
The presence of polarization at intermediate angular scale helps
in discriminating inflationary models from causal scaling seed 
toy models. Moreover, the  position of the first anti-peak
and second peak in the spectrum are also in agreement with the
prediction of inflation (\cite{dode}).\\
As main intriguing discrepancy, the WMAP data shows (in agreement with the 
previous COBE data) a lower temperature quadrupole than expected.
The statistical significance of this discrepancy is still unclear
(see e.g. \cite{costa}, \cite{efstathiou2}, \cite{dore}).\\
The CMB anisotropies measured by WMAP are also in good agreement 
with the standard inflationary prediction of gaussianity 
(\cite{komatsu}).

\section{CMB constraints on the standard model.}

In principle, the standard scenario of structure formation based on adiabatic
primordial fluctuations can depend on more than $11$ parameters.\\
However for a first analysis and under the assumption of a flat universe
which is already well consistent with CMB data, it is 
possible to restrict ourselves to just $5$ parameters: the tilt of primordial 
spectrum of scalar perturbations $n_S$, the optical depth of the universe 
$\tau_c$, the physical energy densities in baryons and dark matter
$\omega_b=\Omega_bh^2$ and $\omega_{dm}=\Omega_{dm}h^2$ and 
the Hubble parameter $h$.

\begin{table}
\begin{tabular}{lllll}
\hline
& WMAP & WMAPext &  WMAPext+LSS  & Pre-WMAP+LSS \\
\hline
$\Omega_b h^2$ &\ensuremath{0.024 \pm 0.001} &
\ensuremath{0.023 \pm 0.001}&
\ensuremath{0.023 \pm 0.001}&
\ensuremath{0.021 \pm 0.003} \\
\hline
$\Omega_m h^2$ &\ensuremath{0.14 \pm 0.02} &
\ensuremath{0.13 \pm 0.01}&
\ensuremath{0.134 \pm 0.006}&
\ensuremath{0.14 \pm 0.02} \\
\hline
$h$ &\ensuremath{0.72 \pm 0.05} &
\ensuremath{0.73 \pm 0.05}&
\ensuremath{0.73 \pm 0.03}&
\ensuremath{0.69 \pm 0.07} \\
\hline
$n_s$ &\ensuremath{0.99 \pm 0.04} &
\ensuremath{0.97 \pm 0.03}&
\ensuremath{0.97 \pm 0.03}&
\ensuremath{0.97 \pm 0.04} \\
\hline
$\tau_c$ &\ensuremath{0.166^{+ 0.076}_{- 0.071}} &
\ensuremath{0.143^{+ 0.071}_{- 0.062}}&
\ensuremath{0.148^{+ 0.073}_{- 0.071}}&
\ensuremath{0.07^{+ 0.07}_{- 0.05}} \\
\hline
\end{tabular}
\caption{Current constraints on the $5$ parameters of the standard model
(flat universe). The WMAP results are taken from Spergel et al. 2003, 
the previous results are taken from Melchiorri and Odman 2003 (see also
Slosar et al. 2003 and Wang et al. 2003).}
\end{table}

In Table $1$ we report the constraints on these parameters 
obtained by the WMAP team (see \cite{spergel} and \cite{verde})
in $3$ cases: WMAP only,
WMAP+CBI+ACBAR and WMAP+CBI+ACBAR+LSS. Also, for comparison,
we present the CMB+LSS results previous to WMAP in the forth column.\\
As we can see a value for the baryon density $\omega_b =0.020\pm0.002$ 
as predicted by Standard Big Bang Nucleosynthesis (see e.g.\cite{cyburt})
is in very good agreement with the WMAP results.
The WMAP data is in agreement with the previous results 
and its inclusion reduces the error bar on this parameter by 
a factor $3$.\\
The amount of cold dark matter is also well constrained by the CMB data. 
The presence of power around the third peak is crucial in this sense, 
since it cannot be easily accommodated in models based on just baryonic 
matter (see e.g. \cite{melksilk} and references therein). 
As we can see, including the CMB data on those scales (not sampled
by WMAP) halves the error bars.
WMAP is again in agreement with the previous determination 
and its inclusion reduces the error bar on this parameter by 
of  factor $3$-$4$.\\

These values implies the existence of a cosmological constant
at high significance with $\Omega_{\Lambda}=0$, $\Omega_{M}=1$
excluded at $5$-$\sigma$ from the WMAP data alone and at
$\sim 15$-$\sigma$ when combined with supernovae data.
A cosmological constant is also suggested from the evidence of 
correlations of the WMAP data with large scale structure 
(\cite{scranton}, \cite{nolta}, \cite{boughn},
\cite{afshordi}).\\

Under the assumption of flatness, is possible to constrain 
the value of the Hubble parameter $h$. The constraints on
this parameter are in very well agreement with the
HST constraint (\cite{freedman}).\\

An increase in the optical depth $\tau_c$ after recombination 
by reionization damps the amplitude of the CMB peaks. 
This small scale damping can be somewhat compensated by an
increase in the spectral index $n_S$. This leads to a nearly perfect
degeneracy between $n_S$ and $\tau_c$ and, in practice, no 
significant upper bound on these parameters can be placed from temperature 
data. However, large scale polarization data, as measured by WMAP, and
LSS data can break this degeneracy. At the same time, inclusion of
galaxy clustering data can determine $n_S$ and further break the
degeneracy. As we can see, the current constraint on the spectral
index is close to scale invariance ($n_S \sim 1$) as predicted
by inflation. The best-fit value of the optical depth determined by WMAP
is slightly higher but consistent in between $1-\sigma$ with supercomputer 
simulations of reionization processes ($\tau_c \sim 0.10$,
see e.g. \cite{ciardi}).

\section{Constraints on possible extensions of the standard model.}

The standard model provides a reasonable fit to the data.
However it is possible to consider several modifications characterized
by the inclusion of new parameters. The data considered here 
doesn't show any definite evidence for those modifications, providing more
a set of useful constraints. The present constraints are as follows:

\begin{itemize}

\item Running of the Spectral Index

The possibility of a scale dependence of the
scalar spectral index, $n_S(k)$, has been considered
in various works (see e.g. \cite{kosowsky}, \cite{AMcopeland}, 
\cite{doste}). Even though this dependence is considered to 
have small effects on CMB scales in most of the slow-roll inflationary models, 
it is worthwhile to see if any useful constraint can be obtained.
The present CMB data is at the moment compatible with no
scale dependence (\cite{kkmr},\cite{bridle}), however, joint analyses with 
other datasets (like Lyman-$\alpha$) shows a $\sim 2-\sigma$ evidence for a 
negative running (\cite{peiris}). At the moment, the biggest case
against running comes from reionization models, which are unable
to reach the large optical depth observed in this case
by WMAP ($\tau \sim 0.17$, see \cite{spergel}) with the suppressed power 
on small scales (see e.g. \cite{cen}).

\item Gravitational Waves.

The metric perturbations created during inflation belong to two types:
{\it scalar} perturbations, which couple to the stress-energy of 
matter in the universe and form the ``seeds'' for structure formation 
and {\it tensor} perturbations, also known as 
gravitational wave perturbations.
A sizable background of gravity waves 
is expected in most of the inflationary scenarios and
a detection of the GW background 
can provide information on the second derivative
of the inflaton potential and shed light on the physics at
$\sim 10^{16} Gev$.

The amplitude of the GW background is weakly constrained
by the current CMB data. However, when information from BBN, local
cluster abundance and galaxy clustering are included, an upper limit
of about $r = C_2^T/C_2^S < 0.5$ (no running) is obtained
(see e.g. \cite{spergel}, \cite{kkmr}, \cite{barger}, \cite{leach}).

\item Isocurvature Perturbations
 
Another key assumption of the standard model is that the primordial 
fluctuations were adiabatic.
Adiabaticity is not a necessary consequence of inflation though and 
many inflationary models have been constructed where
isocurvature perturbations would have generically been concomitantly
produced (see e.g. \cite{langlois}, \cite{gordon}, \cite{bartolo}).
Pure isocurvature perturbations are highly excluded by present CMB
data. 
Due to degeneracies with other cosmological
parameters, the amount of isocurvature modes is still weakly
constrained by the present data (see e.g. \cite{peiris},
 \cite{jussi}, \cite{AMcrotty}).

\item Modified recombination.

The standard recombination process can be modified in several
ways. Extra sources of ionizing and resonance radiation at 
recombination or having a time-varying fine-structure constant, 
for example, can delay the recombination process and leave an
imprint on the CMB anisotropies. The present data is 
in agreement with the standard recombinations scheme. However,
non-standard recombination scenarios are still consistent with the 
current data and  may affect the current WMAP constraints on inflationary 
parameters like the spectral index, $n_{s}$, and its running
(see e.g. \cite{avelino}, \cite{martins}, \cite{bms}).

\item Neutrino Physics

The effective number of neutrinos and their effective 
mass can be both constrained by combining cosmological data.
The combination of present cosmological data under the assumption 
of several priors provide a constraint on the effective
neutrino mass of $m_{ee} <0.23 eV$ (\cite{spergel}). 
The data constraints the effective number of neutrino
species to about $N_{eff} < 7$ 
(see e.g. \cite{pierpaoli}, \cite{julien},\cite{hannestad}).

\item Dark Energy and its equation of state

The discovery that the universe's evolution may be dominated by
an effective cosmological constant is one of the most remarkable 
cosmological findings of recent years.
Observationally distinguishing a time variation in
the equation of state or finding it different from $-1$ is a 
powerful test for the cosmological constant.
The present constraints on $w$ obtained
combining the CMB data with several other cosmological 
datasets are consistent with $w=-1$, with models with
$w <-1$ slightly preferred (see e.g. \cite{mmot}, 
\cite{wlewis}). Other dark energy parameters, like
its sound speed, are weakly constrained (\cite{bo}, 
\cite{caldwell}).

\end{itemize}

\section{Conclusions}

The recent CMB data represent a beautiful success for the 
standard cosmological model. The acoustic oscillations in
the CMB temperature power spectrum, a major
prediction of the model, have now been detected 
with high statistical significance. The amplitude and
shape of the cross correlation temperature-polarization 
power spectrum is also in agreement with the expectations.

Furthermore, when constraints on cosmological parameters are 
derived under the assumption of adiabatic primordial perturbations
their values are in agreement with the predictions of the theory
and/or with independent observations.

The largest discrepancy between the standard predictions and
the data seems to come from the low value of the CMB quadrupole.
New physics has been proposed to explain this discrepancy
(see e.g. \cite{contaldi}, \cite{freese}, \cite{luminet}),
but the statistical significance is of difficult 
interpretation.

As we saw in the previous section modifications 
as isocurvature modes or topological defects, are still
compatible with current CMB observations, but are not necessary and
can be reasonably constrained when complementary datasets are included.\\

\section{Acknowledgements}

I wish to thank the organizers of the conference: 
Norma Mankoc Borstnik and  Holger Bech Nielsen.
Many thanks also to Rachel Bean, Will Kinney, 
Rocky Kolb, Carlos Martins, Laura Mersini, Roya Mohayaee, 
Carolina Oedman, Antonio Riotto, Graca Rocha, Joe Silk, and Mark Trodden 
for comments, discussions and help.

\title*{AdS/CFT Correspondence and Unification at About 4 TeV}
\author{Paul H. Frampton}
\institute{%
University of North Carolina at Chapel Hill\\
Department of Physics and Astronomy\\
Phillips Hall\\
Chapel Hill NC 27599-3255\\
United States}

\titlerunning{AdS/CFT Correspondence and Unification at About 4 TeV}
\authorrunning{Paul H. Frampton}
\maketitle

\section{Introduction}
 
I will give a brief summary of an approach to string phenomenology
which is inspired by AdS/CFT correspondence and which has been pursued
for the last five years. Finite-N non-SUSY theories as discussed here 
are not obtainable from AdS/CFT although a speculation, currently under study,
is that key UV properties of infinite-N theories which may be so obtained
can survive, at least in some(one?) cases for the finite-N case.
Future work will study the (non-)occurrence of quadratic divergences in
the resultant finite-N gauge theories. 

Independent of the outcome of that study, interesting possibilities emerge
for extending the standard model without low-energy supersymmetry.
These possibilities would become far more compelling if the quadratic 
divergences associated with fundamental scalars can indeed be eliminated.

\section{Quiver Gauge Theory}

The relationship of the Type IIB superstring to conformal gauge theory
in $d=4$ gives rise to an interesting class of gauge
theories.
Choosing the simplest compactification\cite{Maldacena}
on $AdS_5 \times S_5$ gives rise to an ${\cal N} = 4$ SU(N) gauge theory
which is known to be conformal due to
the extended global supersymmetry and non-renormalization theorems. All
of the RGE $\beta-$functions for this ${\cal N} = 4$
case are vanishing in perturbation theory. It is possible to break
the ${\cal N}=4$ to ${\cal N}=2,1,0$ by replacing
$S_5$ by an orbifold $S_5/\Gamma$
where $\Gamma$ is a discrete group with
$\Gamma \subset SU(2), \subset SU(3), \not\subset SU(3)$
respectively.

In building a conformal gauge theory model \cite{Frampton,FS,FV},
the steps are: (1) Choose the discrete group $\Gamma$; (2) Embed
$\Gamma \subset SU(4)$; (3) Choose the $N$ of $SU(N)$; and
(4) Embed the Standard Model $SU(3) \times SU(2) \times U(1)$
in the resultant gauge group $\bigotimes SU(N)^p$ (quiver
node identification). Here we shall look only
at abelian $\Gamma = Z_p$ and define
$\alpha = exp(2 \pi i/p)$. It is expected from the string-field
duality that the resultant field
theory is conformal in the $N\longrightarrow \infty$ limit,
and will have a fixed manifold, or at least a fixed point, for $N$ finite.

Before focusing on ${\cal N}=0$ non-supersymmetric cases, let
us first examine an ${\cal N}=1$ model first
put forward in the work of
Kachru and Silverstein\cite{KS}.
The choice is $\Gamma = Z_3$ and the {\bf 4} of $SU(4)$
is {\bf 4} = $(1, \alpha, \alpha, \alpha^2)$. Choosing N=3
this leads to the three chiral families under $SU(3)^3$
trinification\cite{DGG}
\begin{equation}
(3, \bar{3}, 1) + (1, 3, \bar{3}) + (\bar{3}, 1, 3)
\end{equation}

\section{Gauge Couplings.}

An alternative to conformality, grand unification with supersymmetry,
leads to an impressively accurate gauge coupling unification\cite{ADFFL}.
In particular it predicts an electroweak mixing angle
at the Z-pole, ${\tt sin}^2 \theta = 0.231$. This result
may, however, be fortuitous, but rather than
abandon gauge coupling unification, we can rederive ${\tt sin}^2 \theta = 0.231$
in a different way by embedding the electroweak $SU(2) \times U(1)$ in
$SU(N) \times SU(N) \times SU(N)$ to
find ${\tt sin}^2 \theta = 3/13 \simeq 0.231$\cite{FV,F2}.
This will be a common feature of the models in this paper.

\section{4 TeV Grand Unification}

Conformal invariance in two dimensions has had
great success in comparison to several condensed matter
systems. It is an
interesting question whether conformal symmetry
can have comparable success in a four-dimensional
description of high-energy physics.

Even before the standard model (SM)
$SU(2) \times U(1)$ electroweak theory
was firmly established by experimental
data, proposals were made
\cite{PS,GG} of models which would subsume it into
a grand unified theory (GUT) including also the dynamics\cite{GQW} of
QCD. Although the prediction of
SU(5) in its minimal form for the proton lifetime
has long ago been excluded, {\it ad hoc} variants thereof
\cite{FG} remain viable.
Low-energy supersymmetry improves the accuracy of
unification of the three 321 couplings\cite{ADF,ADFFL}
and such theories encompass a ``desert'' between the weak
scale $\sim 250$ GeV and the much-higher GUT scale
$\sim 2 \times 10^{16}$ GeV, although minimal supersymmetric
$SU(5)$ is by now ruled out\cite{Murayama}.

Recent developments in string theory are suggestive
of a different strategy for unification of electroweak
theory with QCD. Both the desert and low-energy
supersymmetry are abandoned. Instead, the
standard $SU(3)_C \times SU(2)_L \times U(1)_Y$
gauge group is embedded in a semi-simple gauge
group such as $SU(3)^N$ as suggested by gauge
theories arising from compactification of the IIB superstring
on an orbifold $AdS_5 \times S^5/\Gamma$ where
$\Gamma$ is the abelian finite group $Z_N$\cite{Frampton}.
In such nonsupersymmetric quiver gauge theories
the unification of couplings happens not by
logarithmic evolution\cite{GQW} over an
enormous desert covering, say, a dozen orders
of magnitude in energy scale. Instead the
unification occurs abruptly at $\mu = M$ through the
diagonal embeddings of 321 in $SU(3)^N$\cite{F2}.
The key prediction of such unification shifts from
proton decay to additional particle content,
in the present model
at $\simeq 4$ TeV.

Let me consider first the electroweak group
which in the standard model is still un-unified
as $SU(2) \times U(1)$. In the 331-model\cite{PP,PF}
where this is extended to $SU(3) \times U(1)$
there appears a Landau pole at $M \simeq 4$ TeV
because that is the scale at which ${\rm sin}^2
\theta (\mu)$ slides to the value
${\rm sin}^2 (M) = 1/4$.
It is also the scale at which the custodial gauged
$SU(3)$ is broken in the framework
of \cite{DK}.

There remains the question of embedding such unification
in an $SU(3)^N$ of the type described in \cite{Frampton,F2}.
Since the required embedding of $SU(2)_L \times U(1)_Y$
into an $SU(3)$ necessitates $3\alpha_Y=\alpha_H$
the ratios of couplings at $\simeq 4$ TeV
is: $\alpha_{3C} : \alpha_{3W} :  \alpha_{3H} :: 5 : 2 : 2$
and it is natural
to examine $N=12$ with diagonal embeddings of
Color (C), Weak (W) and Hypercharge (H)
in $SU(3)^2, SU(3)^5, SU(3)^5$ respectively.

To accomplish this I specify the embedding
of $\Gamma = Z_{12}$ in the global $SU(4)$ R-parity
of the ${\cal N} = 4$ supersymmetry of the underlying theory.
Defining $\alpha = {\rm exp} ( 2\pi i / 12)$ this specification
can be made by ${\bf 4} \equiv (\alpha^{A_1}, \alpha^{A_2},
\alpha^{A_3}, \alpha^{A_4})$ with $\Sigma A_{\mu} = 0 ({\rm mod} 12)$
and all $A_{\mu} \not= 0$ so that all four supersymmetries are
broken from ${\cal N} = 4$ to ${\cal N} = 0$.

Having specified $A_{\mu}$ I calculate the content
of complex scalars by investigating in $SU(4)$
the ${\bf 6} \equiv (\alpha^{a_1}, \alpha^{a_2}, \alpha^{a_3},
\alpha^{-a_3}, \alpha^{-a_2},\alpha^{-a_1})$ with
$a_1 = A_1 + A_2, a_2 = A_2 + A_3, a_3 = A_3 + A_1$ where
all quantities are defined (mod 12).

Finally I identify the nodes (as C, W or H)
on the dodecahedral quiver such that the complex scalars
\begin{equation}
\Sigma_{i=1}^{i=3} \Sigma_{\alpha=1}^{\alpha=12}
\left( N_{\alpha}, \bar{N}_{\alpha \pm a_i} \right)
\label{scalars2}
\end{equation}
are adequate to allow the required symmetry breaking to the
$SU(3)^3$ diagonal subgroup, and the chiral fermions
\begin{equation}
\Sigma_{\mu=1}^{\mu=4} \Sigma_{\alpha=1}^{\alpha=12}
\left( N_{\alpha}, \bar{N}_{\alpha + A_{\mu}} \right)
\label{fermions2}
\end{equation}
can accommodate the three generations of quarks and leptons.

It is not trivial to accomplish all of these requirements
so let me demonstrate by an explicit example.

For the embedding I take $A_{\mu} = (1, 2, 3, 6)$ and for
the quiver nodes take the ordering:
\begin{equation}
- C - W - H - C - W^4 - H^4 -
\label{quiver}
\end{equation}
with the two ends of (\ref{quiver}) identified.

The scalars follow from $a_i = (3, 4, 5)$
and the scalars in Eq.(\ref{scalars2})
\begin{equation}
\Sigma_{i=1}^{i=3} \Sigma_{\alpha=1}^{\alpha=12}
\left( 3_{\alpha}, \bar{3}_{\alpha \pm a_i} \right)
\label{modelscalars}
\end{equation}
are sufficient to break to all diagonal subgroups as
\begin{equation}
SU(3)_C \times SU(3)_{W} \times SU(3)_{H}
\label{gaugegroup}
\end{equation}

The fermions follow from $A_{\mu}$ in Eq.(\ref{fermions2}) as
\begin{equation}
\Sigma_{\mu=1}^{\mu=4} \Sigma_{\alpha=1}^{\alpha=12}
\left( 3_{\alpha}, \bar{3}_{\alpha + A_{\mu}} \right)
\label{modelfermions}
\end{equation}
and the particular dodecahedral quiver
in (\ref{quiver}) gives rise  to exactly {\it three}
chiral generations which transform under (\ref{gaugegroup})
as
\begin{equation}
3[ (3, \bar{3}, 1) + (\bar{3}, 1, 3) + (1, 3, \bar{3}) ]
\label{generations}
\end{equation}
I note that anomaly freedom of the underlying superstring
dictates that only the combination
of states in Eq.(\ref{generations})
can survive. Thus, it
is sufficient to examine one of the terms, say
$( 3, \bar{3}, 1)$. By drawing the quiver diagram
indicated by Eq.(\ref{quiver}) with the twelve nodes
on a ``clock-face'' and using
$A_{\mu} = (1, 2, 3, 6)$
I find
five $(3, \bar{3}, 1)$'s and two $(\bar{3}, 3, 1)$'s
implying three chiral families as stated in Eq.(\ref{generations}).

After further symmetry breaking at scale $M$ to
$SU(3)_C \times SU(2)_L \times U(1)_Y$ the
surviving chiral fermions are the quarks and leptons
of the SM. The appearance
of three families depends on both
the identification of modes in (\ref{quiver})
and on the embedding of $\Gamma \subset SU(4)$. The
embedding must simultaneously give adequate
scalars whose VEVs can break the symmetry
spontaneously to (\ref{gaugegroup}).
All of this is achieved successfully by the
choices made.
The three gauge couplings evolve
for $M_Z \leq \mu \leq M$. For $\mu \geq M$ the
(equal) gauge couplings of $SU(3)^{12}$
do not run if, as conjectured in \cite{Frampton,F2}
there is a conformal fixed point at $\mu = M$.

The basis of the conjecture in \cite{Frampton,F2}
is the proposed duality of Maldacena\cite{Maldacena}
which shows that in the $N \rightarrow \infty$
limit ${\cal N} = 4$ supersymmetric
$SU(N)$gauge  theory, as well as orbifolded versions with
${\cal N} = 2,1$ and $0$\cite{bershadsky1,bershadsky2}
become conformally invariant.
It was known long ago
that the
${\cal N} = 4$ theory is
conformally invariant for all finite ${\cal N} \geq 2$.
This led to the conjecture in \cite{Frampton}
that the ${\cal N} = 0$
theories might be conformally
invariant, at least in some case(s),
for finite $N$.
It should be emphasized that this
conjecture cannot be checked
purely
within a perturbative framework\cite{FMink}.
I assume that the local $U(1)$'s
which arise in this scenario
and which would lead to $U(N)$
gauge groups are non-dynamical,
as suggested by Witten\cite{Witten},
leaving $SU(N)$'s.

As for experimental tests of such
a TeV GUT, the situation at energies
below 4 TeV is predicted to be the standard model with
a Higgs boson still to be discovered at a mass
predicted by radiative corrections
\cite{PDG} to be below 267 GeV at 99\% confidence level.

There are many particles predicted
at $\simeq 4$ TeV beyond those of
the minimal standard model.
They include
as spin-0 scalars the states of Eq.(\ref{modelscalars}).
and
as spin-1/2 fermions the states
of Eq.(\ref{modelfermions}),
Also predicted are gauge bosons to fill out the gauge groups
of (\ref{gaugegroup}), and in the same energy region
the gauge bosons to fill out all of
$SU(3)^{12}$. All these extra particles are necessitated by
the conformality constraints of \cite{Frampton,F2} to lie
close to the conformal fixed point.

One important issue is whether this proliferation
of states at $\sim 4$ TeV is
compatible with precision
electroweak data in hand. This has
been studied in the related model of
\cite{DK} in a recent article\cite{Csaki}. Those results
are not easily translated to the present
model but it is possible that such an analysis
including limits on flavor-changing neutral currents
could rule out the entire framework.

\section{Predictivity}

The calculations have been done in the one-loop
approximation to the renormalization group equations
and threshold effects have been ignored.
These corrections are not expected to be large
since the couplings are weak in the entrire energy
range considered. There are possible further corrections
such a non-perturbative effects, and the effects of
large extra dimensions, if any.

In one sense the robustness of this TeV-scale
unification is almost self-evident, in that it follows from the weakness
of the coupling constants in the evolution from $M_Z$ to $M_U$.
That is, in order to define the theory at $M_U$,
one must combine the effects of
threshold corrections ( due to O($\alpha(M_U)$)
mass splittings )
and potential corrections from redefinitions
of the coupling constants and the unification scale.
We can then {\it impose} the coupling constant relations at $M_U$
as renormalization conditions and this is valid
to the extent that higher order corrections do
not destabilize the vacuum state.

We shall approach the comparison with data in two
different but almost equivalent ways. The first
is "bottom-up" where we use as input that the
values of $\alpha_3(\mu)/\alpha_2(\mu)$ and
$\sin^2 \theta (\mu)$ are expected to be $5/2$
and $1/4$ respectively at $\mu = M_U$.

Using the experimental ranges allowed for
$\sin^2 \theta (M_Z) = 0.23113 \pm 0.00015$,
$\alpha_3 (M_Z) = 0.1172 \pm 0.0020$ and
$\alpha_{em}^{-1} (M_Z) = 127.934 \pm 0.027$
\cite{PDG} we have calculated \cite{FRT}
the values of $\sin^2 \theta (M_U)$
and $\alpha_3 (M_U) / \alpha_2(M_U)$
for a range of $M_U$ between 1.5 TeV
and 8 TeV.
Allowing a maximum discrepancy of $\pm 1\%$ in
$\sin^2 \theta (M_U)$ and
$\pm 4\%$ in $\alpha_3 (M_U) / \alpha_2 (M_U)$
as reasonable estimates of corrections, we deduce that
the unification scale $M_U$ can lie anywhere
between 2.5 TeV and 5 TeV. Thus the theory is
robust in the sense that there is no singular
limit involved in choosing a particular value
of $M_U$.

\bigskip

\noindent Another test of predictivity of the same
model is to fix the unification values at $M_U$ of
$\sin^2 \theta(M_U) = 1/4$ and $\alpha_3 (M_U) /
\alpha_2 (M_U) = 5/2$. We then compute the
resultant predictions at the scale $\mu = M_Z$.

The results are shown for $\sin^2 \theta (M_Z)$
in \cite{FRT} with the allowed range\cite{PDG}
$\alpha_3 (M_Z) = 0.1172 \pm 0.0020$. The precise
data on $\sin^2 (M_Z)$ are indicated in \cite{FRT} and
the conclusion is that the model makes correct
predictions for $\sin^2 \theta (M_Z)$.
Similarly, in \cite{FRT}, there is a plot of the
prediction for $\alpha_3 (M_Z)$ versus
$M_U$ with $\sin^2 \theta(M_Z)$ held
with the allowed empirical range.
The two quantities plotted in \cite{FRT}
are consistent for similar ranges of $M_U$.
Both $\sin^2 \theta(M_Z)$ and $\alpha_3(M_Z)$ are within the empirical limits
if $M_U = 3.8 \pm 0.4$ TeV.

\bigskip

\noindent The model has many additional gauge bosons
at the unification scale, including neutral $Z^{'}$'s,
which could mediate flavor-changing processes
on which there are strong empirical upper limits.

A detailed analysis wll require specific identification
of the light families and quark flavors with
the chiral fermions appearing in the quiver diagram
for the model. We can make only the general
observation that the lower bound on a $Z^{'}$
which couples like the standard $Z$ boson
is quoted as $M(Z^{'}) < 1.5$ TeV \cite{PDG}
which is safely below the $M_U$ values considered here
and which we identify with the mass of the new gauge bosons.

This is encouraging to believe that flavor-changing
processes are under control in the model but
this issue will require more careful analysis when
a specific identification of the quark states
is attempted.

\bigskip

\noindent Since there are many new states predicted
at the unification scale $\sim 4$ TeV, there is a danger
of being ruled out by precision low energy data.
This issue is conveniently studied in terms
of the parameters $S$ and $T$ introduced in \cite{Peskin}
and designed to measure departure from the predictions
of the standard model.

Concerning $T$, if the new $SU(2)$ doublets are
mass-degenerate and hence do not violate a custodial
$SU(2)$ symmetry they contribute nothing to $T$.
This therefore provides a constraint on the spectrum
of new states.

\section{Discussion}

\bigskip

\noindent The plots we have presented clarify the accuracy
of the predictions of this TeV unification scheme for
the precision values accurately measured at the Z-pole.
The predictivity is as accurate for $\sin^2 \theta$ as
it is for supersymmetric GUT models\cite{ADFFL,ADF,DRW,DG}.
There is, in addition, an accurate prediction for $\alpha_3$
which is used merely as input in SusyGUT models.

At the same time, the accurate predictions are seen to be robust
under varying the unification scale around $\sim 4 TeV$
from about 2.5 TeV to 5 TeV.

One interesting question is concerning the accommodation
of neutrino masses in view of the popularity of the mechanisms
which require a higher mass scale than occurs in the present
type of model. For example, one would like to
know whether any of the recent studies in \cite{FGMY}
can be useful within this framework.

In conclusion, since this model ameliorates the GUT hierarchy
problem and naturally accommodates three families, it
provides a viable alternative to the widely-studied
GUT models which unify by logarithmic evolution
of couplings up to much higher GUT scales.

\section*{Acknowledgement}

I thank  Professor N. Mankoc-Borstnik of the University
of Ljubjana
for organizing this stimulating workshop in Portoroz,
Slovenia.
This work was supported in part
by the Department of Energy
under Grant Number
DE-FG02-97ER-410236.



\title*{New Solutions in String Field Theory}
\author{Loriano Bonora}
\institute{%
International School for Advanced Studies (SISSA/ISAS)\\
Via Beirut 2--4, 34014 Trieste, Italy, and INFN, Sezione di
Trieste\\e-mail:bonora@sissa.it}

\titlerunning{New Solutions in String Field Theory}
\authorrunning{Loriano Bonora}
\maketitle

\begin{abstract}
Solitonic solutions of new type are described in the framework of Vacuum String Field
Theory. They are deformations of the sliver solution and are characterized,
among other properties, by their norm and action being finite.
\end{abstract}

\section{Introduction}
Recently, as a consequence of the increasing interest in tachyon
condensation, String Field Theory (SFT) has received renewed attention.
There is no doubt that the most complete description of tachyon
condensation and related phenomena has been given so far in the framework
of Witten's Open String Field Theory, \cite{W1}. This is not surprising,
since the study of tachyon condensation involves off--shell
calculations, and SFT is the natural framework where off--shell
analysis can be carried out.

All these developments can be described in terms of
Sen's conjectures, \cite{Sen}. Sen's conjectures can be summarized as
follows. Bosonic open  string theory in D=26 dimensions is quantized
on an unstable vacuum, an instability which manifests itself through the
existence of the open string tachyon. The effective tachyonic potential
has, beside the local maximum where the theory is quantized, a local
minimum. Sen's conjectures concern the nature of the theory around
this local minimum. First of all, the energy density difference between
the maximum and the minimum should exactly compensate for the
D25--brane tension characterizing the unstable vacuum: this is a
condition for the stability of the theory at the minimum.
Therefore the theory around the minimum should not contain any quantum
fluctuation pertaining to the original (unstable) theory. The minimum
should therefore correspond to an entirely new theory, the bosonic
closed string theory. If so, in the new theory one should be able to
explicitly find in particular all the classical solutions
characteristic of closed string theory, specifically the D25--brane
as well as all the lower dimensional D--branes.

The evidence that has been collected so far for the above conjectures does 
not have a
uniform degree of accuracy and reliability, but it is enough to conclude
that they provide a correct description of tachyon condensation in
SFT. Especially elegant is the proof of the existence of solitonic
solutions in Vacuum String Field Theory (VSFT), the SFT version which is
believed to represent the theory near the minimum.

A time-dependent solution 
which describes the evolution from the maximum of the tachyon potential 
to such a minimum (a rolling tachyon), if it exists, would describe 
the decay of the D25--brane into closed string states. It has been 
argued in many ways that such a solution exists; in particular this has
led to the formulation of a new kind of duality between open and closed
strings. But all this has been possible so far only outside a SFT 
framework. It would make an important progress if we could describe
the rolling tachyon solution and the open--closed string duality in
the framework of SFT. 

In this regard VSFT could play an important role. VSFT is a simplified 
version of SFT, in which the BRST operator ${\cal Q}$ takes a very simple form
in terms of ghost oscillators alone. It is clearly simpler to work in
such a framework than in the original SFT. In fact many classical 
solutions have been shown to exist, which are candidates for representing
D--branes (the sliver,the butterfly,etc), and other classical
solutions have been found (lump solutions) which may represent
lower dimensional D--branes. In some cases the spectrum around such 
solutions have been analyzed and some aspects of the D--brane spectrum
have been reproduced. However the responses of VSFT are still far
from being satisfactory. There are a series of nontrivial problems left
behind. Let us consider for definitness the sliver solution.
To start with it has vanishing action for the matter part and
infinite action for the ghost part, but it is impossible to get a 
finite number out of them. Second, it is not at all clear whether
the solutions of the linear equations of motion around the sliver
can accommodate all the open string modes (as one would expect if the sliver
has to represent a D25--brane). Third, the other Virasoro constraints on
such modes are nowhere to be seen.

We believe that these drawbacks are due to the fact that the sliver is 
not the most suitable solution to represent a D25--brane.  
On the other hand the sliver has many interesting properties: 
it is simple and algebraically appealing (it is a squeezed state), 
its structure matrix $CS$ commutes with the twisted matrices of the 
three strings vertex coefficients and the calculations involving
the sliver are relatively simple. Therefore in order to define a new 
solution we choose to stay as close as possible to the sliver. In practice
we start from the sliver and `perturb it' by adding to $CS$ a suitable 
rank one 
projector. We can show not only that this is a solution of the VSFT
equations of motion, but that we can define infinite many independent 
such solutions. We call such solutions {\it dressed slivers}. 
They are characterized by finite norm and action. 

In this contribution I will limit myself to showing how one arrives
at the definition of the dressed sliver solution. The explicit calculations
as well as other properties of this solution are reported in two research 
papers, \cite{BMP1,BMP2}.

\section{A review of SFT and VSFT}

Let us start with a short review of SFT \`a la Witten.
The open string field theory action proposed by E.Witten, \cite{W1}, years
ago is
\begin{equation}
{\cal S}(\Psi)= - \frac 1{g_0^2} \int\left(\frac 12 \Psi *Q\Psi +
\frac 13 \Psi *\Psi *\Psi\right)\label{sftaction}
\end{equation}
In this expression $\Psi$ is the string field, which can be understood
either as a classical functional of the open string configurations or as
a vector in the Fock space of states of the open string. We will consider
in the following the second point of view. In the field theory limit
it makes sense to represent it as a superposition of Fock space states
with ghost number 1, with coefficient represented by local fields,
\begin{equation}
|{\Psi}\rangle = (\phi(x)+ A_\mu (x) a_1^{\mu\dagger}+
\ldots) c_1|{0}\rangle\label{stringfield}
\end{equation}
The BRST charge $Q$ has the same form as in the first quantized string
theory. The star product of two string fields $\Psi_1, \Psi_2$ represents
the process of identifying the right half of the first string
with the left half of the second string and integrating over the
overlapping degrees of freedom, to produce a third string which
corresponds to $\Psi_1 * \Psi_2$. This can be done in various ways, either
using the classical string functionals (as in the original formulation), or
using the three string vertex (see below), or the conformal field theory
language \cite{leclair1}.
Finally the integration in (\ref{sftaction}) corresponds to bending the left
half of the string over the right half and integrating over the
corresponding degrees of freedom in such a way as to produce a number.
The following rules are obeyed
\begin{eqnarray}
&&Q^2 =0 \nonumber\\
&& \int Q \Psi = 0\nonumber\\
&& (\Psi_1 *\Psi_2) *\Psi_3 = \Psi_1 * (\Psi_2 *\Psi_3)\nonumber\\
&& Q(\Psi_1 * \Psi_2) = (Q \Psi_1)*\Psi_2 +
(-1)^{|\Psi_1|} \Psi_1 * (Q\Psi_2)\label{rules}
\end{eqnarray}
where $|\Psi|$ is the Grassmannality of the string field $\Psi$,
which, for bosonic strings, coincides with the ghost number.
The action (\ref{sftaction}) is invariant under the BRST transformation
\begin{equation}
\delta \Psi = Q \Lambda + \Psi *\Lambda - \Lambda * \Psi\label{gaugetr}
\end{equation}
Finally, the ghost numbers of the various objects $Q, \Psi, \Lambda,
*, \int$ are $1,1,0,0,-3$, respectively.

\subsection{Vacuum string field theory}

The action (\ref{sftaction}) represents open string theory about the
trivial unstable vacuum $|\Psi_0\rangle=c_1|0\rangle$.
Vacuum string field theory (VSFT) is instead a version of Witten's open SFT
which is conjectured to correspond to the minimum of the tachyon potential.
As explained in the introduction, at the minimum of the tachyon
potential a dramatic change occurs in the theory, which, corresponding
to the new vacuum, is expected to represent closed string
theory rather that the open string theory we started with. In particular,
this theory should host tachyonic lumps representing unstable D--branes
of any dimension less than 25, beside the original D25--brane.
Unfortunately we have been so
far unable to find an exact classical solution, say $|\Phi_0\rangle$,
representing the new vacuum. One can nevertheless guess the form taken
by the theory at the new minimum, see \cite{RSZ1}. The VSFT action
has the same form as (\ref{sftaction}), where the new string field is
still denoted by $\Psi$, the $*$ product is the same as in the previous
theory, while the BRST operator
$Q$ is replaced by a new one, usually denoted ${\cal Q}$, which is
characterized by universality and vanishing cohomology.
Relying on such general arguments, one can even deduce
a precise form of $\cal Q$, \cite{HKw,GRSZ1}, 
\begin{equation}
{\cal {Q}} =  c_0 + \sum_{n>0} \,(-1)^n \,(c_{2n}+ c_{-2n})\label{calQ}
\end{equation}
Now, the equation of motion of VSFT is
\begin{equation}
{\cal Q} \Psi = - \Psi * \Psi\label{EOM}
\end{equation}
and nonperturbative solutions are looked for in the factorized form
\begin{equation}
\Psi= \Psi_m \otimes \Psi_g\label{ans}
\end{equation}
where $\Psi_g$ and $\Psi_m$ depend purely on ghost and matter
degrees of freedom, respectively. Then eq.(\ref{EOM}) splits into
\begin{eqnarray}
 {\cal Q} \Psi_g & = & - \Psi_g *_g \Psi_g\label{EOMg}\\
\Psi_m & = & \Psi_m *_m \Psi_m\label{EOMm}
\end{eqnarray}
The action for this type of solutions becomes
\begin{equation}
{\cal S}(\Psi)= - \frac 1{6 g_0^2} \langle \Psi_g |{\cal Q}|\Psi_g\rangle
\langle \Psi_m |\Psi_m\rangle \label{actionsliver}
\end{equation}
In the following, for simplicity, we will limit myself to the matter part.
$\langle \Psi_m |\Psi_m\rangle$ is the ordinary inner product, 
$\langle \Psi_m |$ being the $bpz$ conjugate of $|\Psi_m\rangle$ 
(see below).

Since here we are interested in the D25--brane, which is translational 
invariant, the $*_m$ product is simply defined as follows
\begin{equation}
\langle V_3|\Psi_1\rangle_1 \Psi_2\rangle_2 =_3\langle \Psi_1*_m\Psi_2|
\label{starm}
\end{equation}
where the reduced three strings vertex $V_3$ is defined by
\begin{equation}
|V_3\rangle= {\rm exp}(-E)\,|0\rangle_{123},\quad\quad
E= \frac 12 \sum_{a,b=1}^3\sum_{m,n\geq 1}\eta_{\mu\nu}
a_m^{(a)\mu\dagger}V_{mn}^{ab}
a_n^{(b)\nu\dagger} 
\label{V3}
\end{equation}
 
Summation over the Lorentz indices $\mu,\nu=0,\ldots,25$
is understood and $\eta$ denotes the flat Lorentz metric.
The operators $ a_m^{(a)\mu},a_m^{(a)\mu\dagger}$ denote the non--zero
modes matter oscillators of the $a$--th string, which satisfy
\begin{equation}
[a_m^{(a)\mu},a_n^{(b)\nu\dagger}]=
\eta^{\mu\nu}\delta_{mn}\delta^{ab},
\quad\quad m,n\geq 1 \label{CCR}
\end{equation}
Moreover $|0\rangle_{123}=|0\rangle_1\otimes|0\rangle_2\otimes |0\rangle_3$
is the tensor product of the Fock vacuum
states relative to the three strings.  
The symbols $V_{nm}^{ab},V_{0m}^{ab},V_{00}^{ab}$ will denote
the coefficients computed in \cite{GJ1}. We will use them
in the notation of Appendix A and B of \cite{RSZ2}. 

To complete the definition of the $*_m$ product we must specify the 
$bpz$ conjugation properties of the oscillators
\begin{equation}
bpz(a_n^{(a)\mu}) = (-1)^{n+1} a_{-n}^{(a)\mu}\nonumber
\end{equation}

\subsection{The sliver solution}

Let us now return to eq.(\ref{EOMm}).  Its solutions are projectors of
the $*_m$ algebra. We recall the simplest one, the sliver. It is
defined by
\begin{equation}
|\Xi\rangle = \cal{N} e^{-\frac 12 a^\dagger Sa^\dagger},\quad\quad
a^\dagger S a^\dagger = \sum_{n,m=1}^\infty a_n^{\mu\dagger} S_{nm}
 a_m^{\nu\dagger}\eta_{\mu\nu}\label{Xi}
\end{equation} 
This state satisfies eq.(\ref{EOMm}) provided the matrix $S$ satisfies
the equation
\begin{equation}
S= V^{11} +(V^{12},V^{21})({\bf I}-
\Sigma{\cal V})^{-1}\Sigma
\left(\begin{matrix}V^{21}\cr V^{12}\end{matrix}\right)\label{SS}
\end{equation}
where
\begin{equation}
\Sigma= \left(\begin{matrix}S&0\cr 0& S\end{matrix}\right),
\quad\quad\quad
{\cal V} = \left(\begin{matrix}V^{11}&V^{12}\cr V^{21}&V^{22}\end{matrix}\right),
\label{SigmaV}
\end{equation}
The proof of this fact is well--known, \cite{KP}. First one expresses 
eq.(\ref{SigmaV}) in terms of the twisted matrices $X=CV^{11},X_+=CV^{12}$
and $X_-=V^{21}$, together with $T=CS=SC$, where 
$C_{nm}= (-1)^n\delta_{nm}$. The matrices $X,X_+,X_-$ are mutually 
commuting. Then, requiring that $T$ commute with them as well, one can show 
that eq.(\ref{SigmaV}) reduces to the algebraic equation
\begin{equation}
XT^2-(1+X)T+X=0\label{algeq}
\end{equation}
The interesting solution is
\begin{equation}
T= \frac 1{2X} (1+X-\sqrt{(1+3X)(1-X)})\label{sliver}
\end{equation}

The normalization constant $\cal{N}$ is calculated to be
\begin{equation}
\cal{N}= (\det (1-\Sigma \cal{V}))^{13}\label{norm}
\end{equation}
The contribution of the sliver to the matter part of the action 
(see (\ref{actionsliver}) is given by
\begin{equation}
\langle \Xi|\Xi\rangle = \frac {\cal{N}^2}{(\det (1-S^2))^{13}}\label{ener}
\end{equation}
Both eq.(\ref{norm}) and (\ref{ener}) are ill--defined and need to be 
regularized, after which they both result to be vanishing (see below).  

\section{The dressed sliver solution}

Now we want to deform the sliver by adding some special matrix to $S$. 
To this end first we introduce the infinite vector $\xi=\{\xi_{n}\}$  
which are chosen to satisfy the condition
\begin{equation}
\rho_1 \xi =0,\quad\quad \rho_2 \xi =\xi, \label{xi}
\end{equation}
The operators $\rho_1,\rho_2$ are Fock space projectors into (left or
right) half string states. Next we require $\xi$ to be real and set
\begin{equation}
\xi^T \frac 1{1-T^2}\xi =1 ,\quad\quad
\xi^T \frac {T}{1-T^2}\xi= \kappa
\label{noncond}
\end{equation}
$\kappa$ turns out to be a negative real number.
We remark  that
the conditions (\ref{noncond}) are not very stringent. The only thing
one has to worry is that the LHS's are finite (this is the only true
condition). Once this is true the rest follows from suitably rescaling
$\xi$, so that the first equation is satisfied, and from the reality of 
$\xi$.

Our candidate for the {\it dressed sliver} solution is given by an
ansatz similar to (\ref{Xi}) 
\begin{equation}
|\hat\Xi\rangle = \hat{\cal N} e^{-\frac 12 a^\dagger \hat S a^\dagger},
\label{Xihat}
\end{equation}
with $S$ replaced by
\begin{equation}
\hat S = S +R,\quad\quad R_{nm}= \frac 1{\kappa +1}\left(\xi_n(-1)^m\xi_m
+\xi_m(-1)^n\xi_n\right)\label{Shat}
\end{equation}
As a consequence $\hat T$ is replaced by
\begin{equation}
\hat T_{\epsilon} = T +P,\quad\quad P_{nm}= 
\frac 1{\kappa +1}\left(\xi_n(-1)^{m+n}
\xi_m+\xi_m\xi_n\right)\label{That}
\end{equation}
The dressed sliver satisfies hermiticity. 

We claim that $|\hat\Xi\rangle$ is a projector.
The dressed sliver matrix $\hat T$ does not commute with $X,X_-,X_+$
(as T does), but we can nevertheless make use of the property
$C\hat T= \hat T C$, because $CP=PC$. 
Using this it is in fact possible to show, \cite{BMP1}, that  
\begin{equation}
V^{11} +(V^{12},V^{21})({1}-
\hat\Sigma{\cal V})^{-1}\hat\Sigma
\left(\begin{matrix}V^{21}\cr V^{12}\end{matrix}\right)=\hat S\label{hatShatS}
\end{equation}
where
\begin{equation}
\hat\Sigma= \left(\begin{matrix}\hat S&0\cr 0& \hat S\end{matrix}\right)\label{hatSigma}
\end{equation}
 
The defintion (\ref{Xihat}) is not yet satisfactory. The reason is that
its normalization and the corresponding action are still ill-defined.
We have to supplement the above definition with some specification.
To this end we introduce in (\ref{Xihat}) a deformation parameter
${\epsilon}$, which multiplies $R$, i.e. $\hat S \to \hat S_{\epsilon} = S +{\epsilon} R$. 
In this way we obtain a state
$\hat \Xi_{\epsilon}$ which interpolates between the sliver 
$\epsilon=0$ and the dressed sliver $\epsilon=1$.
Now, interpreting it as a sequence of states 
in the vicinity of $\epsilon=1$, we can give a precise 
definition of the dressed sliver, so that both its norm and its action 
can be made finite.

Let us see this in some detail.
As already mentioned above, the determinants in (\ref{norm}),
(\ref{ener}) relevant to the sliver are ill--defined. They are actually
well defined for any finite truncation of the matrix $X$ to level $L$ and 
need a regulator to account for its behavior when $L\to\infty$.
A regularization that fits particularly our needs here was introduced by
Okuyama,\cite{Oku2} and we will use it here. It consists in using an asymptotic
expression for the eigenvalue density $\rho(k)$ of $X$,
$\rho(k) \sim \frac 1{2\pi} \log L$, for large $L$. This lead to
asymptotic expressions for the various determinant we need. In particular
we get
\begin{eqnarray}
&&\det(1+T)=  L^{-\frac 13}+\ldots\nonumber\\
&&\det(1-T)=  L^{\frac 16}+\ldots\label{asymp}\\
&&\det(1-X)= L^{\frac 19}+\ldots\nonumber
\end{eqnarray}
where dots denote non--leading contribution when $L\to\infty$.
Our regularization scheme consists in tuning $L$ with $\epsilon$ in such a way as 
to obtain finite results.  

To define the norm of $\hat \Xi$ one would think that we have
to evaluate the limit of $ \langle \hat \Xi_{{\epsilon}} |\hat \Xi_{{\epsilon}} \rangle $
for ${\epsilon}\to 1$.
But this is not the right choice. We must instead use
\begin{equation}
\lim_{{\epsilon}_1\to 1}\left(\lim_{{\epsilon}_2\to 1} 
\langle \hat \Xi_{{\epsilon}_1} |\hat \Xi_{{\epsilon}_2} \rangle\right)\label{limlim}
\end{equation}
When ${\epsilon}_1$ and ${\epsilon}_2$ are in the vicinity of 1
we have
\begin{eqnarray}
&&\frac 1{\langle 0|0\rangle}\,\langle \hat \Xi_{{\epsilon}_1} |\hat \Xi_{{\epsilon}_2} 
\rangle\nonumber\\
&& = 
\left(\frac {\det(1-\Sigma\cal{V})}{\sqrt{\det(1-S^2)}}\right)^D
\left(\frac 1{4(\kappa+1)^2}\right)^{\frac D2}
\left(\frac{4}{(\kappa(1-{\epsilon}_1)(1-{\epsilon}_2)+1-{\epsilon}_1{\epsilon}_2)^2}\right)^{\frac D2}+
\ldots\nonumber
\end{eqnarray}
where dots denote non--leading terms. Taking the limit 
(\ref{limlim})
\begin{eqnarray}
&&\frac 1{\langle 0|0\rangle}\,\lim_{{\epsilon}_1\to 1}\left(\lim_{{\epsilon}_2\to 1} 
\langle \hat \Xi_{{\epsilon}_1} |\hat \Xi_{{\epsilon}_2}\rangle\right) \label{xi1xi2f}\\
&&= \lim_{{\epsilon}_1\to 1} \left(\frac {\det(1-\Sigma\cal{V})}
{\sqrt{\det(1-S^2)}}\right)^{\frac D2}
\left(\frac 1{4(\kappa+1)^2}\right)^{\frac D2}
\left(\frac{4}{(1-{\epsilon}_1)^2}\right)^{\frac D2}+\ldots\nonumber\\
&&= \lim_{{\epsilon}_1\to 1} 
\left(\frac{1}{(\kappa+1)^2 }\right)^{\!\frac D2}
\left(\frac {L^{-\frac 5{36}}}{1-{\epsilon}_1}\right)^{\!D}+\ldots = 
\left(\frac{1}{(\kappa+1)^2 s_1^2}\right)^{\!\frac D2}\nonumber
\end{eqnarray}
provided 
\begin{equation}
1-{\epsilon}_1 =s_1 L^{-\frac 5{36}}\label{regpres1}
\end{equation}
It is easy to see that if we reverse the order of the limits in
(\ref{limlim}) we obtain the same result.

The reason why we adopt this result as the norm of $\hat \Xi$ is because
it is consistent with the equations of motion, i.e.
that
\begin{equation}\label{boxlim}
\lim_{{\epsilon}_1\to 1}\left(\lim_{{\epsilon}_2\to 1} 
\langle \hat \Xi_{{\epsilon}_1} |\hat \Xi_{{\epsilon}_2}\rangle\right)=
\lim_{{\epsilon}_1\to 1}\left(\lim_{{\epsilon}_2\to 1}\left(\lim_{{\epsilon}_3\to 1}
\langle \hat \Xi_{{\epsilon}_1} |\hat \Xi_{{\epsilon}_2}*\hat \Xi_{{\epsilon}_3}\rangle\right)\right)
\end{equation}
Had we used $\lim_{{\epsilon}\to 1} 
\langle \hat \Xi_{{\epsilon}} |\hat \Xi_{{\epsilon}} \rangle$, we would have found a 
mismatch between the two members of the analogous equation.

Since something similar holds also for the ghost part of the solution
it is understandable that the corresponding action
may take any prescribed negative finite value, the negative of which,
divided by the volume factor, is identified with the brane tension.

It is not possible to define a state in the Hilbert space
to which $\hat \Xi_{\epsilon}$ tends in the limit ${\epsilon}\to 1$. The state
$\hat \Xi$, even though it is not a Hilbert space state,
due to its finite norm and action,
is a good candidate for the D25--brane. For this candidacy to be confirmed
one has to analyze the spectrum and show that it indeed accommodates 
all the states that appear in the spectrum of the open strings attached to
the brane. This is the task of ref.\cite{BMP2}

\section*{Acknowledgments}

This research was supported by the Italian MIUR
under the program ``Teoria dei Campi, Superstringhe e Gravit\`a''.

\title*{The Approach Unifying Spins and Charges in SO(1,13) and Its Predictions}
\author{Anamarija Bor\v stnik Bra\v ci\v c${}^1$ and Norma Manko\v c Bor\v stnik${}^2$}
\institute{%
${}^1$Educational Faculty, University of Ljubljana,
Kardeljeva plo\v s\v cad 17, Ljubljana 1000
and Primorska Institute for Natural Sciences and Technology, 
C. Mare\v zganskega upora 2, Koper 6000, Slovenia\\
${}^2$Department of Physics, University of
Ljubljana, Jadranska 19, Ljubljana 1111,
and Primorska Institute for Natural Sciences and Technology, 
C. Mare\v zganskega upora 2, Koper 6000, Slovenia}

\titlerunning{The Approach Unifying Spins and Charges in SO(1,13) and Its Predictions}
\authorrunning{Anamarija Bor\v stnik Bra\v ci\v c and Norma Manko\v c Bor\v stnik}
\maketitle

\begin{abstract} 
Ten years ago we have proposed the approach unifying all the internal degrees of freedom - that is the spin and 
all the charges\cite{norma92,norma93,norma97,norma01} within the group $SO(1,13)$. The approach is a kind of 
Kaluza-Klein-like theories. In this talk we present the advances of the approach and its success 
in answering the open questions of the Standard electroweak model. 
We demonstrate that (only!) one left handed Weyl multiplet of the group $SO(1,13)$ contains, if represented in a way 
to demonstrate the $SU(3), SU(2)$ and $U(1)$'s substructure, {\it the spinors - the quarks and the leptons -  
and the ``anti-spinors`` - the anti-quarks and the anti-leptons - of the Standard electroweak and colour
model}. We demonstrate why the weak charge breaks parity while the colour charge does not. 
We comment on a possible way of breaking the group $SO(1,13) $ leading to  
spins, charges and flavours of  leptons and quarks and antileptons and antiquarks. We comment on how
spinor representations of only one handedness might be chosen after each break of symmetry, although,
as Witten has commented\cite{witten81}, at each break of symmetry by the compactification, 
spinor representations of both 
handedness appear, which very likely ruins the mass protection mechanism. We comment on the 
appearance of spin connections and vielbeins as gauge fields 
connected with charges, and as  Yukawa couplings determining accordingly masses of families. 
We demonstrate the appearance of families, suggesting symmetries of mass matrices and argue for
the appearance of the fourth family, with all the properties (besides the masses) 
of the three known families (all in agreement with ref.\cite{okun}). 
We also comment on small charges of observed spinors (and ``antispinors``) 
and on anomaly cancellation.

\end{abstract}

\section{Introduction}
\label{SNMBintroduction}

There is no experimental data yet, which would not be in agreement with the Standard electroweak model. 
But the Standard electroweak model has more than 20 parameters  and assumptions, 
the origin of which is not at all understood. There are also no theoretical approaches yet which
would be able to explain all these 
assumptions and parameters. 

We expect a lot from experiments on new extremely sophisticated and expensive 
accelerators and spectrometers (in elementary particle physics and cosmology). 
But measurements will first of all corroborate or not with predictions for 
several events calculated with models and theories, none of which explains all the assumptions and parameters 
of the Standard model.
In the foreword of this proceedings (some of) the open questions of both standard models - electroweak and cosmological -
are presented. We repeat here only those of the open questions, to which our approach can contribute the answers.  

The Standard electroweak model assumes the left handed weak charged doublets which are either colour triplets 
(quarks) or colour singlets (leptons) and the right handed weak chargeless singlets which are again 
either colour triplets
or colour singlets. And the corresponding ``antispinors`` (antiquarks and antileptons). (Therefore, the Standard model
assumes that the spin and the weak charge are connected, as handedness is determined by the properties of the group
$SO(1,3)$.) It follows then that the weak charge breaks parity. The question arises, why does the weak charge break
parity while the
colour charge does not? The Standard model assumes three families of quarks and leptons and the corresponding
antiquarks and antileptons, without giving any explanation about the origin of families, or anifamilies 
(antiquarks and antileptons of three families).
It also assumes that the quarks and the leptons are massless - until gaining a (small) mass at low 
energies through the 
vacuum expectation value(s) of Higgs fields and Yukawa couplings, without giving any explanation, why is this so and
where does te weak scale comes from. The Standard model also assumes that the elementary fields are in ( the 
fundamental for spinors and the adjoint for
the corresponding gauge fields) representations of the group $U(1),SU(2), SU(3)$. But, why  
the observed representations of known 
charges are so small?.

The great advantage of the approach of (one of) us, unifying spins and 
charges \cite{norma92,norma93,norma95,norma97,norma01,holgernorma2002,pikanormaproceedings1,pikanormaproceedings2}, 
is, that it proposes possible answers to the above cited open
questions of the Standard electroweak model.
 We demonstrate that a left handed $SO(1,13)$ Weyl spinor multiplet includes, 
if the representation is interpreted
in terms of the subgroups $SO(1,3)$, $SU(2)$, $SU(3)$ and the sum of the two $U(1)$'s,  spinors and ``antispinors`` of
the Standard model - that is the left handed $SU(2)$ doublets and the right handed  $SU(2)$ singlets of with the group 
$SU(3)$ charged quarks and with the group $SU(3)$ chargeless leptons, while the ``antispinors`` are oppositely charged 
and have opposite handedness.
Right handed neutrinos and left handed antineutrinos - both weak chargeless - are also included, so that 
the multiplet
has 64 members, half with spin up and half with spin down. (Both representations of the spin are needed for 
the solution of the Weyl equations of motion).
We demonstrate that for the group $SO(1,13)$, that half of one left handed Weyl representation, 
which describes the spinors,
alone is anomaly free and the same is true for the ``antispinors`` half of one left handed Weyl alone.

We demonstrate that, when starting with a spinor of one handedness only, spinor representations of subgroups
always contain representations of both handedness - with respect to each of the subgroups. (This is what Witten
pointed out in his articles\cite{witten81} when analyzing possibilities that the Kaluza-Klein-like theories  lead 
at low energies to ''realistic'' world, that is to massless  - or almost massless - spinors.) But even then
each of the two representations might be distinguished
by the charges - a kind of the gauge Kaluza-KLein charges - of subgroups  and accordingly, the choice of the 
representation of a particular handedness is still possible. Therefore, 
if the compactification of a part of space is responsible for the break of symmetry, then one still might be able 
to choose the representation of only handedness\cite{holgernorma2004}, which assures then masslessness of spinors.

We demonstrate that the approach offers a possible 
explanation for families of spinors and their masses, since a part of the gravitational gauge fields
originating in higher than four dimensions appears as terms which simulate the Yukawa couplings, postulated by 
the Standard electroweak model\cite{pikanormaproceedings2}.

Our gauge group is $SO(1,13)$ - the smallest complex Lorentz group with a left handed Weyl spinor
containing the needed representations of the Standard model.  The gauge fields of this group are spin 
connections and vielbeins\cite{norma93,norma01}, determining the gravitational field in (d = 14)-dimensional space.
Then a gauge gravitational field manifests in four dimensional subspace as all the gauge fields of the known charges, 
and (as already written ) also as the Yukawa couplings.
 
We present an action for a Weyl (massless) spinor in a d-dimensional space 
in a gauge gravitational field, which demonstrates also the appearance of families of quarks and leptons.

We define the handedness of the group $SO(1,d-1)$ and  the subgroups  $SO(d')$ of this group (\ref{spinor}),
with $d=2n$ and $d'=2k$ in terms of appropriate products of the generators of the Lorentz transformations in 
internal 
space\cite{norma93,normasuper94}. We demonstrate that handedness of the group $SO(1,d-1)$ and of subgroups play an
essential role for spinors. 

We use the technique\cite{holgernorma2002}, which enables to follow explicitly the appearance of 
charged and chargeless states and antistates of an irreducible
(left handed Weyl) representation of the group $SO(1,13)$. It helps to understand the anomaly 
freedom of the representation, the
smallness of the representation, the appearance of the complex representation - needed to distinguish
between spinors and ``antispinors``.

We also introduce operators, transforming one family into another and pre\-sent
the mass matrices for the four families of quarks and leptons, suggested by our approach. We also comment on
more than four families, assuming the symmetry of mass matrices, proposed by the approach.

The approach was first formulated in the space of Grassmann coordinates of the same dimension as the ordinary space.
The formulation leads to two kinds of the Clifford algebra objects\cite{norma92,norma93,norma95,norma97,norma01}. 
Later the same approach was formulated by using the differential 
forms\cite{holgernorma00} and also with the Clifford algebra objects of two kinds  
alone\cite{holgernorma2003,pikanormaproceedings1,pikanormaproceedings2}. In this talk the formulation with the 
Clifford algebra objects is used.

\section{Properties of spinor representation of SO(1,13) and subgroups}
\label{spinor}

In this section we present properties of one Weyl spinor of the group $SO(1,13)$ in terms of properties of subgroups
$SO(1,7) \times SO(6)$, $SO(1,9) \times SO(4)$ and  of $SO(1,3) \times SU(3)\times SU(2) \times U(1)$. 

We formulate
spinor representations in terms of nilpotents and 
projectors \cite{normasuper94,norma01,holgernorma00,holgernorma2002,holgernorma2003,pikanormaproceedings1,pikanormaproceedings2}
which are Clifford algebra odd  and even binomials of $\gamma^a$'s, respectively.

We demonstrate, that one left handed $SO(1,13)$ multiplet contains the states with
the properties needed
to  describe all the quarks and the leptons as well the antiquarks and the antileptons of one family of the Standard model. 

We also demonstrate the appearance of families.

We comment on the properties of of the group $SO(1,13)$ and subgroups 
$SO(1,7) \times SO(6)$, $SO(1,9) \times SO(4)$ and  of $SO(1,3) \times SU(3)\times SU(2) \times U(1)$ from 
the point of view of the open question of the Standard model.

\subsection{Lorentz group  and Clifford algebra  }
\label{lorentz}

Let  operators $\gamma^a$ close the Clifford algebra
\begin{equation}
\{\gamma^a, \gamma^b \}_+ = 2\eta^{ab}, \quad {\rm for} \quad a,b \quad \in \{0,1,2,3,5,\cdots,d \},
\label{clif}
\end{equation}
for any $d$, even or odd, and let the Hermiticity property of $\gamma^a$'s be
\begin{eqnarray}
\gamma^{a+} = \eta^{aa} \gamma^a,
\label{cliffher}
\end{eqnarray}
in order that 
$\gamma^a$ be unitary as usual, i.e. ${\gamma^a}^{\dagger}\gamma^a=1$.

 The operators 
\begin{equation}
S^{ab} = \frac{i}{4} [\gamma^a, \gamma^b ] := \frac{i}{4} (\gamma^a \gamma^b - \gamma^b \gamma^a)
\label{sab}
\end{equation}
close the algebra of the Lorentz group 
\begin{equation}
\{S^{ab},S^{cd}\}_- = i (\eta^{ad} S^{bc} + \eta^{bc} S^{ad} - \eta^{ac} S^{bd} - \eta^{bd} S^{ac})
\label{loralg}
\end{equation}
and also fulfill the  spinor algebra 
$\{S^{ab},S^{ac}\}_+ = \frac{1}{2} \eta^{aa} \eta^{bc}.$
Recognizing from Eq.(\ref{loralg}) that two operators $S^{ab}, S^{cd}$ with all indices different 
commute, we readily select the Cartan subalgebra of the algebra of the Lorentz group with $m =  d/2$  
for $ d $ even commuting operators.

We define one of the Casimirs of the Lorentz group which determines the handedness of an
irreducible representation of the Lorentz group\footnote{To see the definition of the 
operator $\Gamma$ for any spin in even-dimensional spaces see references\cite{norma93,%
normasuper94,bojannorma2001,holgernorma00}.} for spinors 
\begin{eqnarray}
\Gamma^{(d)} :&=& 2^{d/2} \; \prod_a \sqrt{\eta^{aa}} \quad S^{03} S^{12} S^{56} \cdots S^{d-1\; d}= \nonumber\\
 &=& \;
(i)^{d/2}\;\prod_a \quad (\sqrt{\eta^{aa}} \gamma^a), \quad {\rm if } \quad d = 2n, \nonumber\\
\Gamma^{(d)} :&=& 
(i)^{(d-1)/2}\;\prod_a \quad (\sqrt{\eta^{aa}} \gamma^a), \quad {\rm if } \quad d = 2n +1,
\label{hand}
\end{eqnarray}
for any integer $n$. We understand the product of $\gamma^a$'s in the ascending order with respect to 
the index $a$: $\gamma^0 \gamma^1\cdots \gamma^d$. 
It follows for any choice of the signature $\eta^{aa}$  that 
$\Gamma^{(d)}$ is Hermitean and its square is equal to the unity operator
\begin{eqnarray}
\Gamma^{(d)\dagger}= \Gamma^{(d)},\quad
\Gamma^{(d)2} = I.
\label{SNMBprophand}
\end{eqnarray}
One also finds that in even-dimensional spaces $\Gamma^{(d)}$ anticommutes, while in odd-dimensional spaces
$\Gamma^{(d)}$ commutes with $\gamma^a$'s
($\{\Gamma^{(d)},\gamma^a\}_{+} = 0 \;{\rm for} \; d \; {\rm even}$ and
$\{\Gamma^{(d)},\gamma^a\}_{-} = 0, \; {\rm for} \; d \; {\rm odd}$.)
$\Gamma^{(d)}$ always commutes with the generators of the Lorentz algebra
\begin{eqnarray}
\{\Gamma^{(d)}, S^{ab}\}_- = 0.
\label{prophand}
\end{eqnarray}
%

We shall select operators belonging to the Cartan subalgebra of $SO(1,13)$ as follows
\begin{eqnarray}
S^{03}, S^{12}, S^{56}, \cdots, S^{13\; 14}.
\label{cartan}
\end{eqnarray}
We present the operators of handedness for the Lorentz group $SO(1,13)$ and the subgroups $SO(1,9), SO(1,7),
SO(6), SO(4)$ and $SO(1,3)$
\begin{eqnarray}
\Gamma^{(1,13)} &=& \; 2^{7}i \; S^{03} S^{12} S^{56} \cdots S^{13 \; 14},
\nonumber\\
\Gamma^{(1,9)}\; &=& \; 2^{5} i \; S^{03} S^{12} S^{9\;10} S^{11\;12} S^{13 \; 14},
\nonumber\\
\Gamma^{(1,7)} \;&=& \; - 2^{4} i \; S^{03} S^{12} S^{56} S^{78},
\nonumber\\
\Gamma^{(6)}\;\;\; &=& \; -8 \; S^{9 \;10} S^{11\;12} S^{13 \; 14},
\nonumber\\
\Gamma^{(1,3)} \; &=& \; - 4i \; S^{03} S^{12},
\nonumber\\
\Gamma^{(4)}\;\;\; &=&\; 4 \; S^{56} S^{78}.
\label{gammas}
\end{eqnarray}

\subsection{Subgroups of $SO(1,13)$}
\label{subgroups}


The group $SO(1,13)$ and the subgroups $SO(1,9)$ and $SO(6)$  
have complex representations as all  groups of the type $SO(2(2k+1)),$ for any $k$, 
have\cite{georgi}, offering the possibility to distinguish  
between spinors and ``antispinors``.
We shall later comment on the concept of spinors and ''antispinors'' in the context of charges and
''anticharges'', describ with respect to a complex group.

When analyzing properties of $SO(1,13)$ in terms of subgroups, $0$ will be the time index, in $SO(1,n)$ 
indices will run through $0,1,2,3, 5,..,n$, indices $5,6,7,8$ will be reserved for $SO(4)$, while 
the indices $9,10,11,12$ will be reserved for $SO(6)$. The generators of the subgroups $SO(1,9)$ and $SO(1,7)$
are the $S^{ab},$ with $a,b = \in\{0,1,2,3,5,..,n\}$ and $n$ equal to $10$ and $8$, respectively. 
The generators of the groups $SO(n)$ are $S^{ab}$ with the
indices $a,b = \in\{9,10,11,12,13,14\}$ and $a,b = \in\{5,6,7,8\}$ for $n =6$ and $n=4$, respectively.

The generators of the subgroups $SO(1,3)$, 
 $SU(2), SU(3)$ and $U(1)$'s, needed to determine
the spin, the weak charge, the colour charge and the hyper 
charges content of $SO(1,13)$  can be written in terms of the generators $S^{ab}$ as follows 
\begin{eqnarray}
\tau^{Ai} = \sum_{a,b} \;c^{ai}{ }_{ab} \; S^{ab},
\nonumber\\
\{\tau^{Ai}, \tau^{Bj}\}_- = i \delta^{AB} f^{Aijk} \tau^{Ak},
\label{tau}
\end{eqnarray}
with $A=1,2,3,4,5,6$ representing the corresponding subgroups and $f^{Aijk}$ the corresponding structure constants.
Coefficients $c^{Ai}{ }_{ab}$ have to be determined so that the commutation relations of Eq.(\ref{tau}) 
hold\cite{norma97}.

%
%
The $SU(2)\times SU(2)$ content of the compact group $SO(4)$ can then be demonstrated if expressing
\begin{eqnarray}
\tau^{11}: = \frac{1}{2} ( {\mathcal S}^{58} - {\mathcal S}^{67} ),\quad
\tau^{12}: = \frac{1}{2} ( {\mathcal S}^{57} + {\mathcal S}^{68} ),\quad
\tau^{13}: = \frac{1}{2} ( {\mathcal S}^{56} - {\mathcal S}^{78} )
\nonumber\\
\tau^{21}: = \frac{1}{2} ( {\mathcal S}^{58} + {\mathcal S}^{67} ), \quad
\tau^{22}: = \frac{1}{2} ( {\mathcal S}^{57} - {\mathcal S}^{68} ), \quad
\tau^{23}: = \frac{1}{2} ( {\mathcal S}^{56} + {\mathcal S}^{78} ).
\label{su12w}
\end{eqnarray}
 $\tau^{1i}, i=1,2,3$ will be used to describe the weak charge and $\tau^{23}$ 
to describe the  $U(1)$ content of $SO(4)$.
%
%
%

We express 
generators of  subgroups $SU(3)$ and $U(1)$ of the group
$SO(6)$ in terms of the generators ${\mathcal S}^{ab}$ as follows
\begin{eqnarray}
\tau^{31}: &=& \frac{1}{2} ( {\mathcal S}^{9\;12} - {\mathcal S}^{10\;11} ),\quad
\tau^{32}: = \frac{1}{2} ( {\mathcal S}^{9\;11} + {\mathcal S}^{10\;12} ),\quad
\nonumber\\
\tau^{33}: &=& \frac{1}{2} ( {\mathcal S}^{9\;10} - {\mathcal S}^{11\;12} ),\quad
\tau^{34}: = \frac{1}{2} ( {\mathcal S}^{9\;14} - {\mathcal S}^{10\;13} ),\quad \nonumber\\
\tau^{35}: &=& \frac{1}{2} ( {\mathcal S}^{9\;13} + {\mathcal S}^{10\;14} ),\quad
\tau^{36}: = \frac{1}{2} ( {\mathcal S}^{11\;14} - {\mathcal S}^{12\;13}
),\quad 
\nonumber\\
\tau^{37}: &=& \frac{1}{2} ( {\mathcal S}^{11\;13} + {\mathcal S}^{12\;14} ),\quad
\tau^{38}: = \frac{1}{2\sqrt{3}} ( {\mathcal S}^{9\;10} + {\mathcal
S}^{11\;12} - 2{\mathcal S}^{13\;14})
\nonumber\\
\tau^{41}: &=& -\frac{1}{3}( {\mathcal S}^{9\;10} + {\mathcal S}^{11\;12}
+ {\mathcal S}^{13\;14} ).
\label{su3u1so6}
\end{eqnarray}
Then one finds
$\; \{\tau^{3i}, \tau^{3j}\} = i f^{ijk} \tau^{3k}, \quad
\{\tau^{41}, \tau^{3i}\} = 0, {\mathrm \;\; for\;\; each \;\;i},\;$
with coefficients $f^{ijk}$ which are the structure constants of the group
$SU(3)$. 

We define two superpositions of the two $U(1)$'s generators as follows
\begin{eqnarray}
Y = \tau^{41} + \tau^{23}, \quad  Y' = \tau^{41} - \tau^{23}. 
\label{yyprime}
\end{eqnarray}

We made the above choice of subgroups of the group $SO(1,13)$ to be able to comment one Weyl spinor of the 
group $SO(1,13)$ in terms of the groups of the Standard model.

\subsection{Technique for generating spinor representations in terms of Clifford algebra objects}
\label{technique}

We breafly present a simple technique from refs.\cite{normasuper94,pikanormaproceedings1,%
holgernorma2002}, which makes  spinor representations and accordingly all their properties very transparent.
In this technique, members of spinor representations are polynomials  of the Clifford algebra objects 
$\gamma^a$'s, applied on a vacuum state. Each of basic vectors is chosen to be an eigenstate of all the
Cartan subalgebra members. $\gamma^a$'s and $S^{ab}$'s have to be applied on these vectors 
from the left hand side. The reader can find all the details, with the proofs included in the 
reference\cite{holgernorma2002}.

To make the technique simple, we introduce the graphic representation\cite{holgernorma2002}
as follows
\begin{eqnarray}
\stackrel{ab}{(k)}:&=& 
\frac{1}{2}(\gamma^a + \frac{\eta^{aa}}{ik} \gamma^b),\nonumber\\
\stackrel{ab}{[k]}:&=&
\frac{1}{2}(1+ \frac{i}{k} \gamma^a \gamma^b),
\label{signature}
\end{eqnarray}
where $k^2 = \eta^{aa} \eta^{bb}$.
One can easily check by taking into account the Clifford algebra relation (Eq.\ref{clif}) and the
definition of $S^{ab}$ (Eq.\ref{sab})
that if one multiplies from the left hand side by $S^{ab}$ the Clifford algebra objects $\stackrel{ab}{(k)}$
and $\stackrel{ab}{[k]}$,
it follows that
\begin{eqnarray}
S^{ab}\stackrel{ab}{(k)}=\frac{1}{2}k \stackrel{ab}{(k)},\nonumber\\
S^{ab}\stackrel{ab}{[k]}=\frac{1}{2}k \stackrel{ab}{[k]},
\label{grapheigen}
\end{eqnarray}
which  means that
$\stackrel{ab}{(k)}$ and $\stackrel{ab}{[k]}$ acting from the left hand side on anything (on a
vacuum state $|\psi_0\rangle$, for example ) are eigenvectors of $S^{ab}$ with the eigenvalues $k/2$.

We further find 
\begin{eqnarray}
\gamma^a \stackrel{ab}{(k)}&=&\eta^{aa}\stackrel{ab}{[-k]},\nonumber\\
\gamma^b \stackrel{ab}{(k)}&=& -ik \stackrel{ab}{[-k]}, \nonumber\\
\gamma^a \stackrel{ab}{[k]}&=& \stackrel{ab}{(-k)},\nonumber\\
\gamma^b \stackrel{ab}{[k]}&=& -ik \eta^{aa} \stackrel{ab}{(-k)}
\label{graphgammaaction}
\end{eqnarray}
It follows that
$
S^{ac}\stackrel{ab}{(k)}\stackrel{cd}{(k)} = -\frac{i}{2} \eta^{aa} \eta^{cc} 
\stackrel{ab}{[-k]}\stackrel{cd}{[-k]}$, 
$S^{ac}\stackrel{ab}{[k]}\stackrel{cd}{[k]} = \frac{i}{2}  
\stackrel{ab}{(-k)}\stackrel{cd}{(-k)}$, 
$S^{ac}\stackrel{ab}{(k)}\stackrel{cd}{[k]} = -\frac{i}{2} \eta^{aa}  
\stackrel{ab}{[-k]}\stackrel{cd}{(-k)}$, 
$S^{ac}\stackrel{ab}{[k]}\stackrel{cd}{(k)} = \frac{i}{2} \eta^{cc}  
\stackrel{ab}{(-k)}\stackrel{cd}{[-k]}$.
It is useful to deduce the following relations
\begin{eqnarray}
\stackrel{ab}{(k)}^{\dagger}=\eta^{aa}\stackrel{ab}{(-k)},\quad
\stackrel{ab}{[k]}^{\dagger}= \stackrel{ab}{[k]},
\label{graphher}
\end{eqnarray}
and
\begin{eqnarray}
\stackrel{ab}{(k)}\stackrel{ab}{(k)}& =& 0, \quad \quad \stackrel{ab}{(k)}\stackrel{ab}{(-k)}
= \eta^{aa}  \stackrel{ab}{[k]}, \quad \stackrel{ab}{(-k)}\stackrel{ab}{(k)}=
\eta^{aa}   \stackrel{ab}{[-k]},\quad
\stackrel{ab}{(-k)} \stackrel{ab}{(-k)} = 0 \nonumber\\
\stackrel{ab}{[k]}\stackrel{ab}{[k]}& =& \stackrel{ab}{[k]}, \quad \quad
\stackrel{ab}{[k]}\stackrel{ab}{[-k]}= 0, \;\;\quad \quad  \quad \stackrel{ab}{[-k]}\stackrel{ab}{[k]}=0,
 \;\;\quad \quad \quad \quad \stackrel{ab}{[-k]}\stackrel{ab}{[-k]} = \stackrel{ab}{[-k]}
 \nonumber\\
\stackrel{ab}{(k)}\stackrel{ab}{[k]}& =& 0,\quad \quad \quad \stackrel{ab}{[k]}\stackrel{ab}{(k)}
=  \stackrel{ab}{(k)}, \quad \quad \quad \stackrel{ab}{(-k)}\stackrel{ab}{[k]}=
 \stackrel{ab}{(-k)},\quad \quad \quad 
\stackrel{ab}{(-k)}\stackrel{ab}{[-k]} = 0
\nonumber\\
\stackrel{ab}{(k)}\stackrel{ab}{[-k]}& =&  \stackrel{ab}{(k)},
\quad \quad \stackrel{ab}{[k]}\stackrel{ab}{(-k)} =0,  \quad \quad 
\quad \stackrel{ab}{[-k]}\stackrel{ab}{(k)}= 0, \quad \quad \quad \quad
\stackrel{ab}{[-k]}\stackrel{ab}{(-k)} = \stackrel{ab}{(-k)}.
\label{graphbinoms}
\end{eqnarray}
We recognize in  the first equation of the first row and the first equation of the second row
the demonstration of the nilpotent and the projector character of the Clifford algebra objects $\stackrel{ab}{(k)}$ and 
$\stackrel{ab}{[k]}$, respectively. 

{\em The reader should note that whenever the Clifford algebra objects apply from the left hand side,
they always transform } $\stackrel{ab}{(k)}$ {\em to} $\stackrel{ab}{[-k]}$, {\em never to} $\stackrel{ab}{[k]}$,
{\em and similarly } $\stackrel{ab}{[k]}$ {\em to} $\stackrel{ab}{(-k)}$, {\em never to} $\stackrel{ab}{(k)}$.

According to ref.\cite{holgernorma2002},  we define a vacuum state $|\psi_0>$ so that one finds
\begin{eqnarray}
< \;\stackrel{ab}{(k)}^{\dagger}
 \stackrel{ab}{(k)}\; > = 1, \nonumber\\
< \;\stackrel{ab}{[k]}^{\dagger}
 \stackrel{ab}{[k]}\; > = 1.
\label{graphherscal}
\end{eqnarray}

Taking the above equations into account it is easy to find a Weyl spinor irreducible representation
for $d$-dimensional space, with $d$ even or odd. (We advise the reader to see the reference\cite{holgernorma2002}.) 

For $d$ even, we simply set the starting state as a product of $d/2$, let us say, only nilpotents 
$\stackrel{ab}{(k)}$, one for each $S^{ab}$ of the Cartan subalgebra  elements (Eq.(\ref{cartan})),  applying it 
on an (unimportant) vacuum state\cite{holgernorma2002}. 
Then the generators $S^{ab}$, which do not belong 
to the Cartan subalgebra, applied to the starting state from the left hand side, 
 generate all the members of one
Weyl spinor.  
\begin{eqnarray}
\stackrel{0d}{(k_{0d})} \stackrel{12}{(k_{12})} \stackrel{35}{(k_{35})}\cdots \stackrel{d-1\;d-2}{(k_{d-1\;d-2})}
\psi_0 \nonumber\\
\stackrel{0d}{[-k_{0d}]} \stackrel{12}{[-k_{12}]} \stackrel{35}{(k_{35})}\cdots \stackrel{d-1\;d-2}{(k_{d-1\;d-2})}
\psi_0 \nonumber\\
\stackrel{0d}{[-k_{0d}]} \stackrel{12}{(k_{12})} \stackrel{35}{[-k_{35}]}\cdots \stackrel{d-1\;d-2}{(k_{d-1\;d-2})}
\psi_0 \nonumber\\
\vdots \nonumber\\
\stackrel{0d}{[-k_{0d}]} \stackrel{12}{(k_{12})} \stackrel{35}{(k_{35})}\cdots \stackrel{d-1\;d-2}{[-k_{d-1\;d-2}]}
\psi_0 \nonumber\\
\stackrel{od}{(k_{0d})} \stackrel{12}{[-k_{12}]} \stackrel{35}{[-k_{35}]}\cdots \stackrel{d-1\;d-2}{(k_{d-1\;d-2})}
\psi_0 \nonumber\\
\vdots 
\label{graphicd}
\end{eqnarray}
All the states have the same handedness $\Gamma $, since $\{ \Gamma, S^{ab}\}_- = 0$.
States belonging to one multiplet  with respect to a group $SO(q,d-q)$, that is to one
irreducible representation of spinors (one Weyl spinor), can have any phase. We chose 
the simplest one, setting all  phases equal to one.

The above graphic representation demonstrated that for $d$ even 
all the states of one irreducible Weyl representation of a definite handedness follow from the starting state, 
which is, for example, a product of nilpotents $\stackrel{ab}{(k)}$, by transforming all possible pairs
of $\stackrel{ab}{(k)} \stackrel{mn}{(k)}$ into $\stackrel{ab}{[-k]} \stackrel{mn}{[-k]}$.
There are $S^{am}, S^{an}, S^{bm}, S^{bn}$, which do this.
The procedure gives $2^{(d/2-1)}$ states. A Clifford algebra object $\gamma^a$ applied from the left hand side
transforms  a 
Weyl spinor of one handedness into a Weyl spinor of the opposite handedness. Both Weyl spinors form a Dirac 
spinor. We call such a set of
states a ''family''. 


We shall speak about left handedness when $\Gamma= -1$ and right
handedness when $\Gamma =1$ for either $d$ even or odd. 



\subsection{Left handed representation of $SO(1,13)$ and subgroups}
\label{repso13}

For the group $SO(1,13)$, the starting state  of a left handed Weyl representation will be chosen  
as follows
\begin{eqnarray}
& &\stackrel{03}{(+i)}\stackrel{12}{(+)}|\stackrel{56}{(+)}\stackrel{78}{(+)}
||\stackrel{9 \;10}{(+)}\stackrel{11\;12}{(-)}\stackrel{13\;14}{(-)} |\psi \rangle =
\nonumber\\
& &(\gamma^0 -\gamma^3)(\gamma^1 +i \gamma^2)| (\gamma^5 + i\gamma^6)(\gamma^7 +i \gamma^8)|| \nonumber\\
& & (\gamma^9 +i\gamma^{10})(\gamma^{11} -i \gamma^{12})(\gamma^{13}-i\gamma^{14})|\psi \rangle.
\label{start}
\end{eqnarray}

The signs "$|$" and "$||$" are to point out the  $SO(1,3)$ (up to $|$), $SO(1,7)$ (up to $||$)
and $SO(6)$ (between $|$ and $||$) substructure of the starting state of the left handed multiplet of
$SO(1,13)$ which has $2^{14/2-1}= 64 $ vectors. Again $|\psi\rangle$ is any state, which is not transformed 
to zero. From now on we shall not write down $|\psi \rangle$ anylonger.
One easily finds that the eigenvalues of the chosen
Cartan subalgebra elements of Eq.(\ref{cartan}) are $+i/2, 1/2, 1/2,1/2,1/2,-1/2,-1/2$, 
respectively. This state is a right handed spinor with respect to $SO(1,3)$ ($\Gamma^{(1,3)} =1$, 
Eq.(\ref{gammas})), with spin up 
($S^{12} =1/2$), it is $SU(2)$  
singlet ($\tau^{33} = 0$, Eq.(\ref{su12w})), and it is the member 
of  the $SU(3)$ triplet (Eq.(\ref{su3u1so6})) with ($\tau^{53} =1/2, \tau^{58} = 1/(2 \sqrt{3})$),
it has $\tau^{43} = 1/2$ and $\tau^{6,1}= 1/2$. We further find
according to Eq.(\ref{gammas}) that $\Gamma^{(4)} =1, \Gamma^{(1,7)}= 1, \Gamma^{(6)} = -1$ and 
$\Gamma^{(1,9)} = -1$.

To obtain all the states of one Weyl spinor one only has to apply on the starting state of Eq.(\ref{start})
the generators $S^{ab}$. 

The generators $S^{01}, S^{02}, S^{31}, S^{32}$ transform spin up state (the 
$\stackrel{03}{(+i)}\stackrel{12}{(+)} $ part of the starting state (Eq.(\ref{start}) with
$S^{12}=1/2$ and $S^{03}=i/2$) into spin 
down state ($\stackrel{03}{[-i]}\stackrel{12}{[-]}$, which has
$S^{12}=-1/2$ and $S^{03} = -i/2$), leaving all the other parts of the state and accordingly also all the other properties of 
this state unchanged.  None of the generators $S^{mn},$ with $m,n= 0,1,2,3$, can change a right handed
$SO(1,3)$ Weyl spinor ($\Gamma^{(1,3)} =1$) into a left handed
$SO(1,3)$ Weyl spinor ($\Gamma^{(1,3)} =-1$).

The generators $S^{57}, S^{58}, S^{67}, S^{68}$ transform one $SU(2)$ singlet into another singlet
($\stackrel{56}{(+)}\stackrel{78}{(+)}$ into $\stackrel{56}{[-]}\stackrel{78}{[-]}$), changing
at the same time the value of 
$\tau^{43}$ from $1/2$ to $-1/2$. ($\tau^{3i}$ can not do that, of course.)

The generators $S^{mh}, m=0,1,2,3$, $\;h =5,6,7,8$  transform a right handed 
$SU(2)$ singlet  ($\Gamma^{(1,3)}=1$) with spin up into a member of the left handed ($\Gamma^{(1,3)}=-1$)
$SU(2)$ doublet, 
with spin up ($S^{05}$, for example, changes 
$\stackrel{03}{(+i)}\stackrel{12}{(+)} \stackrel{56}{(+)}\stackrel{78}{(+)}$ into $\stackrel{03}{[-i]}
\stackrel{12}{(+)}\stackrel{56}{[-]}\stackrel{78}{(+)}$) or spin down ($ S^{15}$, for example,
changes state $\stackrel{03}{(+i)}\stackrel{12}{(+)}\stackrel{56}{(+)}\stackrel{78}{(+)}$ into
$\stackrel{03}{(+i)}\stackrel{12}{[-]}\stackrel{56}{[-]}\stackrel{78}{(+)}$).
$S^{57}, S^{58}, S^{67}, S^{68}$
transform one state of a doublet into another state of the same doublet
($\stackrel{56}{(+)}\stackrel{78}{(+)}$ into $\stackrel{56}{[-]}\stackrel{78}{[-]}$), as also $\tau^{3i}$ do.
The $SU(3)$ quantum numbers $\tau^{53}$
and $\tau^{58}$ as well as $\tau^{61}$ stay unchanged.

The generators $S^{kl}$, with $k,l = 9,10,11,12,13,14$, transform one member of the triplet 
into the other
two members ($S^{9\;11}$, $ S^{9\;12}$, $ S^{10\;11}$, $ S^{10\;12}$, $ S^{9\;13}$, $ S^{9\;14}$, $ S^{10\;13}$, $ S^{10\;14}$ 
transform $\stackrel{9 \;10}{(+)}\stackrel{11\;12}{(-)}\stackrel{13\;14}{(-)}$ either into 
$\stackrel{9 \;10}{[-]} \; \stackrel{11\;12}{[+]} \; \stackrel{13\;14}{(-)}$ with $\tau^{53} = -1/2$, 
$\tau^{58}= 1/(2\sqrt{3})$ and $\tau^{61} = 1/6$ or into 
$\stackrel{9 \;10}{[-]}\; \stackrel{11\;12}{(-)}\; \stackrel{13\;14}{[+]}$ with $\tau^{53} = 0$,
$\tau^{58}= -1/\sqrt{3}$ and $\tau^{61}= 1/6$) or 
they transform a triplet into a singlet ($S^{11\;13}, S^{11\;,14}, S^{12\;13}, S^{12\;14}$
transform $\stackrel{9 \;10}{(+)}\; \stackrel{11\;12}{(-)} \; \stackrel{13\;14}{(-)}$ 
into $\stackrel{9 \;10}{(+)} \; \stackrel{11\;12}{[+]} \; \stackrel{13\;14}{[+]}$ with  $\tau^{53}=0=\tau^{58}$ and
$\tau^{61}=-1/2$).

The generators $S^{h,k}$, with $h=0,1,2,3,5,6,7,8$ and $k=9,10,11,12,13,14$, transform a triplet of 
Eq.(\ref{start}) into an antitriplet. $S^{7\;13}$ for example, applied on the starting state, transforms it into
the right handed member of the $SU(2)$ (anti)doublet with $\tau^{33}= 1/2, \tau^{43}=0, \tau^{53}=1/2,
\tau^{58}= -1/(2\sqrt{3})$ and $\tau^{61}= -1/6$. Both $\Gamma^{(4)}$ and $\Gamma^{(6)}$ change sign.

We present the discussed left handed Weyl (spinor) representation of $SO(1,13)$ with 64 states
in the next subsection in Table I.

\subsection{Complex representations and charges}
\label{complex}

Let us look at one left handed Weyl spinor representation of the complex group $SO(1,9)$ with $2^{10/2-1}= 16$ members. 
Each of the member of the representation follows from the starting state $\stackrel{03}{(+i)}\stackrel{12}{(+)}
||\stackrel{9 \;10}{(+)}\stackrel{11\;12}{(=-)}\stackrel{13\;14}{(-)}$ by the application of $S^{ab}$ from the 
left hand side. We drop the indices $5,6,7,8$ in the notation, and use the indices $9,10,11,12,13,14$ instead. 
For the later use but also in order to use the Standard model nomination of states. We present this sixteenplet  
as if the group $SO(1,9)$ would be embedded into the group $SO(1,13)$ and make a choice of the  
Cartan subalgebra elements for the two dropped generators of $SO(1,13)$ as follows: $S^{5,6}=1/2$, $S^{78}=1/2$ 
which corresponds to $\Gamma^{(4)}=1$ of the dropped subgroup, or accordingly with 
 $\tau^{13}= 0$ and $\tau^{23}=1/2$. This means that the representation of the group $SO(1,9)$, embedded into
 $SO(1,13)$ corresponds from the point of view of the Standard model nomination to weak chargeless particles and antiparticles. 
 The sixteenplet is presented on Table I.

\begin{table}
\begin{center}
\begin{tabular}{|r|c||c||c|c|c||c|c|c|c|}
\hline
i&$$&$|^a\psi_i>$&$\Gamma^{(1,3)}$&$ S^{12}$&$S^{03}$&$\tau^{33}$&$\tau^{38}$&$\tau^{41}$&$\Gamma^{(6)}$\\
\hline\hline
&& ${\rm Sixteenplet\; },\;\Gamma^{(1,9)} =-1$&&&&&&& \\
\hline\hline
1&$u_{R}^{c1}$&$\stackrel{03}{(+i)}\stackrel{12}{(+)}
||\stackrel{9 \;10}{(+)}\stackrel{11\;12}{(=-)}\stackrel{13\;14}{(-)}$
&1&1/2&+i/2&1/2&$1/(2\sqrt{3})$&1/6&-1\\
\hline 
2&$u_{R}^{c1}$&$\stackrel{03}{[-i]}\stackrel{12}{[-]}
||\stackrel{9 \;10}{(+)}\stackrel{11\;12}{(-)}\stackrel{13\;14}{(-)}$
&1&-1/2&-i/2&1/2&$1/(2\sqrt{3})$&1/6&-1\\
\hline
3&$u_{R}^{c2}$&$\stackrel{03}{(+i)}\stackrel{12}{(+)}
||\stackrel{9 \;10}{[-]}\stackrel{11\;12}{[+]}\stackrel{13\;14}{(-)}$
&1&1/2&+i/2&-1/2&$1/(2\sqrt{3})$&1/6&-1\\
\hline 
4&$u_{R}^{c2}$&$\stackrel{03}{[-i]}\stackrel{12}{[-]}
||\stackrel{9 \;10}{[-]}\stackrel{11\;12}{[+]}\stackrel{13\;14}{(-)}$
&1&-1/2&-i/2&-1/2&$1/(2\sqrt{3})$&1/6&-1\\
\hline
5&$u_{R}^{c3}$&$\stackrel{03}{(+i)}\stackrel{12}{(+)}
||\stackrel{9 \;10}{[-]}\stackrel{11\;12}{(-)}\stackrel{13\;14}{[+]}$
&1&1/2&+i/2&0&$-1/\sqrt{3}$&1/6&-1\\
\hline 
6&$u_{R}^{c3}$&$\stackrel{03}{[-i]}\stackrel{12}{[-]}
||\stackrel{9 \;10}{[-]}\stackrel{11\;12}{(-)}\stackrel{13\;14}{[+]}$
&1&-1/2&-i/2&0&$-1/\sqrt{3}$&1/6&-1\\
\hline
7&$\nu_{R}$&$\stackrel{03}{(+i)}\stackrel{12}{(+)}
||\stackrel{9 \;10}{(+)}\stackrel{11\;12}{[+]}\stackrel{13\;14}{[+]}$
&1&1/2&+i/2&0&0&-1/2&-1\\
\hline 
8&$\nu_{R}$&$\stackrel{03}{[-i]}\stackrel{12}{[-]}
||\stackrel{9 \;10}{(+)}\stackrel{11\;12}{[+]}\stackrel{13\;14}{[+]}$
&1&-1/2&-i/2&0&0&-1/2&-1\\
\hline
9&$\bar{d}_{L}^{\bar{c1}}$&$\stackrel{03}{[-i]}\stackrel{12}{(+)}
||\stackrel{9 \;10}{[-]}\stackrel{11\;12}{[+]}\stackrel{13\;14}{[+]}$
&-1&1/2&-i/2&-1/2&$-1/(2\sqrt{3})$&-1/6&1\\
\hline 
10&$\bar{d}_{L}^{\bar{c1}}$&$\stackrel{03}{(+i)}\stackrel{12}{[-]}
||\stackrel{9 \;10}{[-]}\stackrel{11\;12}{[+]}\stackrel{13\;14}{[+]}$
&-1&-1/2&+i/2&-1/2&$-1/(2\sqrt{3})$&-1/6&1\\
\hline
11&$\bar{d}_{L}^{\bar{c2}}$&$\stackrel{03}{[-i]}\stackrel{12}{(+)}
||\stackrel{9 \;10}{(+)}\stackrel{11\;12}{(-)}\stackrel{13\;14}{[+]}$
&-1&1/2&-i/2&1/2&$-1/(2\sqrt{3})$&-1/6&1\\
\hline 
12&$\bar{d}_{L}^{\bar{c2}}$&$\stackrel{03}{(+i)}\stackrel{12}{[-]}
||\stackrel{9 \;10}{(+)}\stackrel{11\;12}{(-)}\stackrel{13\;14}{[+]}$
&-1&-1/2&+i/2&1/2&$-1/(2\sqrt{3})$&-1/6&1\\
\hline
13&$\bar{d}_{L}^{\bar{c3}}$&$\stackrel{03}{[-i]}\stackrel{12}{(+)}
||\stackrel{9 \;10}{(+)}\stackrel{11\;12}{[+]}\stackrel{13\;14}{(-)}$
&-1&1/2&-i/2&0&$1/\sqrt{3}$&-1/6&1\\
\hline 
14&$\bar{d}_{L}^{\bar{c3}}$&$\stackrel{03}{(+i)}\stackrel{12}{[-]}
||\stackrel{9 \;10}{(+)}\stackrel{11\;12}{[+]}\stackrel{13\;14}{(-)}$
&-1&-1/2&+i/2&0&$1/\sqrt{3}$&-1/6&1\\
\hline
15&$\bar{e}_{L}$&$\stackrel{03}{[-i]}\stackrel{12}{(+)}
||\stackrel{9 \;10}{[-]}\stackrel{11\;12}{(-)}\stackrel{13\;14}{(-)}$
&-1&1/2&-i/2&0&$0$&1/2&1\\
\hline 
16&$\bar{e}_{L}$&$\stackrel{03}{(+i)}\stackrel{12}{[-]}
||\stackrel{9 \;10}{[-]}\stackrel{11\;12}{(-)}\stackrel{13\;14}{(-)}$
&-1&-1/2&+i/2&0&$0$&1/2&1\\
\hline\hline 
\end{tabular}
\end{center}
\caption{
The sixteenplet of the group $SO(1,9)$, presented as the subgroup of the group $SO(1,13)$, 
with fixed values of $S^{56}= +1/2$ and $S^{78}=+1/2$. It accordingly contains, interpreted in terms of
the Standard model charge groups, the right handed weak chargeless ($\tau^{13}=0\;\tau^{23} =1/2$) u-quarks of 
three colours, the right handed weak chargeless colourless neutrino, the left handed weak chargeless anti-d-quarks
of three anti-colours and the left handed weak chargeless positron.}
\end{table}

One easily sees from Table I that the interpretation,
in which to each right handed particle of a particular colour charge the left handed anti-particle of the anti-colour 
charge (all
the charge values appear with the opposite sign) corresponds, is only possible, since the subgroup 
used to describe the charge
is the complex one ($SO(2(2k+1))$, with $k=1$).
If we would take $SO(4)$ instead (with the Cartan subalgebra elements $S^{11 \;12}$ and $S^{13\;14}$, for example) to
describe the  charge part, and $SO(1,5)$ to describe the spin part, it would not be possible to say that a
right handed particle of a particular 
charge and a left handed anti-particle of the opposite charge appear in the same representation of the group.

\subsection{One Weyl spinor contains all the quarks and the leptons and all the anti-quarks and the
anti-leptons of one family of the Standard model}
\label{left}

All the $2^{14/2-1}= 64$ basic states of one Weyl left handed ($\Gamma^{1,13}=-1$) spinor can be obtained
from the starting state, presented in Eq.(\ref{start}),
which is a member of a triplet state with respect to the group $SU(3)$  and it is
a right handed ($\Gamma^{(1,3)} =1$) singlet with respect to the  $ SU(2)$ subgroup of the group $SO(4)$. According to the
Standard model notation, we shall call it $u^{c1}{}_{R}$. By changing simultaneously the types and the signs
of any pair of brackets, the $64$-plet, 
presented in Table II, follows.  States on Table II
are dressed by the Standard model names and  the two 
hyper charges $Y$ and $Y'$ of Eq.(\ref{yyprime}) - following from the group $SO(4)$and $SO(6)$, respectively - 
are also presented. To point out the Standard model content
of the multiplet (with the right handed neutrino and left handed antineutrino included), 
we first generate the octet of $SO(1,7)$ by
applying on a starting state the generators $S^{ab},$ with $a,b = 0,1,2,3,5,6,7,8.$ 
This particular octet, with $\Gamma^{(1,7)}=1$, has
the $SU(3)$ charge equal to ($\tau^{33}=1/2$, $\tau^{38}=1/(2\sqrt{3})$) and the $U(1)$ charge equal to 
$\tau^{41}= 1/6$, while  $\Gamma^{(6)}= -1$. It contains  {\em two right handed ($\Gamma^{(1,3)}=1$)
$SU(2)$ singlets} ($ \Gamma^{(4)}=1$), each with spin up and spin down, and {\em one left handed 
($\Gamma^{(1,3)}=-1$) $SU(2)$ doublet ($\Gamma^{(4)}=-1$)}, 
each state of the doublet again appears with spin up and spin down. This part of the $64$ multiplet is presented
in Octet I of Table II. 

We want to point out here that a solution of the Weyl equation in $d = (1+13)$ is a generic  
superposition of all the states of one Weyl representation. Accordingly, a solution of the
Weyl equation in terms of the spin and the charges in $d=(1+3)$ part of the space is a
superposition of both - spin up and spin down - members of a representation, for either spinors or ,,anti-spinors''.
Therefore, {\bf the number of members of one left handed $SO(1,13)$ spinor includes sixteen quarks and leptons 
as well as sixteen anti-quarks and anti-leptons of the Standard model}.

The generators $\tau^{5i}$ (or $ S^{hk}$, $h = 9,10$ and $k = 11,12,13,14$) do not change the handedness
$\Gamma^{(6)}= -1$, but they do change the charge within the  
$SU(3)$ triplet - from $c^1$ to $c^i, i=2,3$. Accordingly, the two octets -
 Octets II and III on Table II - follow, all together representing the triplet of {\em right handed
weak chargeless quarks and left handed weak charged quarks} of one family.

By applying the generators $S^{hk}$, with $h,k = 11,12,13,14$, $\Gamma^{(6)} =-1$ does not change,
but the $SU(3)$ triplet changes into a $SU(3)$
singlet with $\tau^{53}=0=\tau^{58}$. $\; \tau^{61}$ changes to $-1/2$. The octet describes now
the {\em weak chargeless right handed
leptons and the weak charged left handed leptons} presented as Octet IV on Table II.

If we apply on the starting state generators $S^{lh},$ with $l=0,1,2,3,5,6,7,8$ and
$h=11,12,13,14$, the member of a $SU(3)$ triplet will transform into the corresponding member of an anti-triplet.
Applying, for example, $S^{0 \;11}$ on the starting state, an anti-triplet with $\tau^{53}=-1/2, 
\tau^{58}= -1/(2\sqrt{3})$, $\tau^{61}=-1/6$ and $\Gamma^{(6)}=-1$ follows. 
According to the Standard model notation 
we call it $\bar{d}^{\bar{c1}}{}_{L}$.  Following the above discussed procedure, one finds 
the whole octet, presented on Table II as Anti-octet V, which has
 {\em two left handed ($\Gamma^{(1,3)}=-1$)
$SU(2)$ singlets} ($ \Gamma^{(4)}=1$), each with spin up and spin down, and {\em one right handed 
($\Gamma^{(1,3)}=1$) $SU(2)$ doublet ($\Gamma^{(4)}=-1$)}, 
again with spin up and spin down.  
As in the case of the $SU(3)$ triplets
the application of $\tau^{5i}$ (or $S^{hk},$ with $h=11,12$ and $k = 9,10,13,14$, so that
only one out of three factors determining the colour charge is of the ``-`` type, that is 
either $(-)$ or $[-]$)  generates 
the other two anti-triplets (Anti-octets 
VI and VII on Table II). 

When $S^{hk}; h,k = 9,10,11,12 $ changes two``+`` into ``-``, the ,,anti-singlet'' states
are generated, with the same octet content all the time, 
 $\Gamma^{(6)}-1$, while 
$\tau^{53}=0=\tau^{58}$, while $\tau^{61} = 1/2$. One finds  {\em weak charged right handed
leptons and weak chargeless left handed leptons} ( Anti-octet VIII of Table II).

\begin{sidewaystable}
\centering
\begin{tabular}{|r|c||c||c|c||c|c|c||c|c|c||r|r|}
\hline
i&$$&$|^a\psi_i>$&$\Gamma^{(1,3)}$&$ S^{12}$&$\Gamma^{(4)}$&
$\tau^{33}$&$\tau^{43}$&$\tau^{53}$&$\tau^{58}$&$\tau^{61}$&$Y$&$Y'$\\
\hline\hline
&& ${\rm Octet\; I},\;\Gamma^{(1,7)} =1,\;\Gamma^{(6)} = -1,$&&&&&&&&&& \\
&& ${\rm of \; quarks}$&&&&&&&&&&\\
\hline\hline
1&$u_{R}^{c1}$&$\stackrel{03}{(+i)}\stackrel{12}{(+)}|\stackrel{56}{(+)}\stackrel{78}{(+)}
||\stackrel{9 \;10}{(+)}\stackrel{11\;12}{(-)}\stackrel{13\;14}{(-)}$
&1&1/2&1&0&1/2&1/2&$1/(2\sqrt{3})$&1/6&2/3&-1/3\\
\hline 
2&$u_{R}^{c1}$&$\stackrel{03}{[-i]}\stackrel{12}{[-]}|\stackrel{56}{(+)}\stackrel{78}{(+)}
||\stackrel{9 \;10}{(+)}\stackrel{11\;12}{(-)}\stackrel{13\;14}{(-)}$
&1&-1/2&1&0&1/2&1/2&$1/(2\sqrt{3})$&1/6&2/3&-1/3\\
\hline
3&$d_{R}^{c1}$&$\stackrel{03}{(+i)}\stackrel{12}{(+)}|\stackrel{56}{[-]}\stackrel{78}{[-]}
||\stackrel{9 \;10}{(+)}\stackrel{11\;12}{(-)}\stackrel{13\;14}{(-)}$
&1&1/2&1&0&-1/2&1/2&$1/(2\sqrt{3})$&1/6&-1/3&2/3\\
\hline
4&$d_{R}^{c1}$&$\stackrel{03}{[-i]}\stackrel{12}{[-]}|\stackrel{56}{[-]}\stackrel{78}{[-]}
||\stackrel{9 \;10}{(+)}\stackrel{11\;12}{(-)}\stackrel{13\;14}{(-)}$
&1&-1/2&1&0&-1/2&1/2&$1/(2\sqrt{3})$&1/6&-1/3&2/3\\
\hline
5&$d_{L}^{c1}$&$\stackrel{03}{[-i]}\stackrel{12}{(+)}|\stackrel{56}{[-]}\stackrel{78}{(+)}
||\stackrel{9 \;10}{(+)}\stackrel{11\;12}{(-)}\stackrel{13\;14}{(-)}$
&-1&1/2&-1&-1/2&0&1/2&$1/(2\sqrt{3})$&1/6&1/6&1/6\\
\hline
6&$d_{L}^{c1}$&$\stackrel{03}{(+i)}\stackrel{12}{[-]}|\stackrel{56}{[-]}\stackrel{78}{(+)}
||\stackrel{9 \;10}{(+)}\stackrel{11\;12}{(-)}\stackrel{13\;14}{(-)}$
&-1&-1/2&-1&-1/2&0&1/2&$1/(2\sqrt{3})$&1/6&1/6&1/6\\
\hline
7&$u_{L}^{c1}$&$\stackrel{03}{[-i]}\stackrel{12}{(+)}|\stackrel{56}{(+)}\stackrel{78}{[-]}
||\stackrel{9 \;10}{(+)}\stackrel{11\;12}{(-)}\stackrel{13\;14}{(-)}$
&-1&1/2&-1&1/2&0&1/2&$1/(2\sqrt{3})$&1/6&1/6&1/6\\
\hline
8&$u_{L}^{c1}$&$\stackrel{03}{(+i)}\stackrel{12}{[-]}|\stackrel{56}{(+)}\stackrel{78}{[-]}
||\stackrel{9 \;10}{(+)}\stackrel{11\;12}{(-)}\stackrel{13\;14}{(-)}$
&-1&-1/2&-1&1/2&0&1/2&$1/(2\sqrt{3})$&1/6&1/6&1/6\\
\hline\hline 
\end{tabular}\\[1mm]
\begin{tabular}{|r|c||c||c|c||c|c|c||c|c|c||r|r|}
\hline
i&$$&$|^a\psi_i>$&$\Gamma^{(1,3)}$&$ S^{12}$&$\Gamma^{(4)}$&
$\tau^{33}$&$\tau^{43}$&$\tau^{53}$&$\tau^{58}$&$\tau^{61}$&$Y$&$Y'$\\
\hline\hline
&& ${\rm Octet\; II},\;\Gamma^{(1,7)} =1,\;\Gamma^{(6)} = -1,$&&&&&&&&&& \\
&& ${\rm of \; quarks}$&&&&&&&&&&\\
\hline\hline
9&$u_{R}^{c2}$&$\stackrel{03}{(+i)}\stackrel{12}{(+)}|\stackrel{56}{(+)}\stackrel{78}{(+)}
||\stackrel{9 \;10}{[-]}\stackrel{11\;12}{[+]}\stackrel{13\;14}{(-)}$
&1&1/2&1&0&1/2&-1/2&$1/(2\sqrt{3})$&1/6&2/3&-1/3\\
\hline
10&$u_{R}^{c2}$&$\stackrel{03}{[-i]}\stackrel{12}{[-]}|\stackrel{56}{(+)}\stackrel{78}{(+)}
||\stackrel{9 \;10}{[-]}\stackrel{11\;12}{[+]}\stackrel{13\;14}{(-)}$
&1&-1/2&1&0&1/2&-1/2&$1/(2\sqrt{3})$&1/6&2/3&-1/3\\
\hline
11&$d_{R}^{c2}$&$\stackrel{03}{(+i)}\stackrel{12}{(+)}|\stackrel{56}{[-]}\stackrel{78}{[-]}
||\stackrel{9 \;10}{[-]}\stackrel{11\;12}{[+]}\stackrel{13\;14}{(-)}$
&1&1/2&1&0&-1/2&-1/2&$1/(2\sqrt{3})$&1/6&-1/3&2/3\\
\hline
12&$d_{R}^{c2}$&$\stackrel{03}{[-i]}\stackrel{12}{[-]}|\stackrel{56}{[-]}\stackrel{78}{[-]}
||\stackrel{9 \;10}{[-]}\stackrel{11\;12}{[+]}\stackrel{13\;14}{(-)}$
&1&-1/2&1&0&-1/2&-1/2&$1/(2\sqrt{3})$&1/6&-1/3&2/3\\
\hline
13&$d_{L}^{c2}$&$\stackrel{03}{[-i]}\stackrel{12}{(+)}|\stackrel{56}{[-]}\stackrel{78}{(+)}
||\stackrel{9 \;10}{[-]}\stackrel{11\;12}{[+]}\stackrel{13\;14}{(-)}$
&-1&1/2&-1&-1/2&0&-1/2&$1/(2\sqrt{3})$&1/6&1/6&1/6\\
\hline
14&$d_{L}^{c2}$&$\stackrel{03}{(+i)}\stackrel{12}{[-]}|\stackrel{56}{[-]}\stackrel{78}{(+)}
||\stackrel{9 \;10}{[-]}\stackrel{11\;12}{[+]}\stackrel{13\;14}{(-)}$
&-1&-1/2&-1&-1/2&0&-1/2&$1/(2\sqrt{3})$&1/6&1/6&1/6\\
\hline
15&$u_{L}^{c2}$&$\stackrel{03}{[-i]}\stackrel{12}{(+)}|\stackrel{56}{(+)}\stackrel{78}{[-]}
||\stackrel{9 \;10}{[-]}\stackrel{11\;12}{[+]}\stackrel{13\;14}{(-)}$
&-1&1/2&-1&1/2&0&-1/2&$1/(2\sqrt{3})$&1/6&1/6&1/6\\
\hline
16&$u_{L}^{c2}$&$\stackrel{03}{(+i)}\stackrel{12}{[-]}|\stackrel{56}{(+)}\stackrel{78}{[-]}
||\stackrel{9 \;10}{[-]}\stackrel{11\;12}{[+]}\stackrel{13\;14}{(-)}$
&-1&-1/2&-1&1/2&0&-1/2&$1/(2\sqrt{3})$&1/6&1/6&1/6\\
\hline\hline
\end{tabular}
\caption{Parts I and II of Table II.}
\end{sidewaystable}
\begin{sidewaystable}
\centering
\begin{tabular}{|r|c||c||c|c||c|c|c||c|c|c||r|r|}
\hline
i&$$&$|^a\psi_i>$&$\Gamma^{(1,3)}$&$ S^{12}$&$\Gamma^{(4)}$&
$\tau^{33}$&$\tau^{43}$&$\tau^{53}$&$\tau^{58}$&$\tau^{61}$&$Y$&$Y'$\\
\hline\hline
&& ${\rm Octet\; III},\;\Gamma^{(1,7)} =1,\;\Gamma^{(6)} = -1,$&&&&&&&&&& \\
&& ${\rm of \; quarks}$&&&&&&&&&&\\
\hline\hline
17&$u_{R}^{c3}$&$\stackrel{03}{(+i)}\stackrel{12}{(+)}|\stackrel{56}{(+)}\stackrel{78}{(+)}
||\stackrel{9 \;10}{[-]}\stackrel{11\;12}{(-)}\stackrel{13\;14}{[+]}$
&1&1/2&1&0&1/2&0&$-1/\sqrt{3}$&1/6&2/3&-1/3\\
\hline 
18&$u_{R}^{c3}$&$\stackrel{03}{[-i]}\stackrel{12}{[-]}|\stackrel{56}{(+)}\stackrel{78}{(+)}
||\stackrel{9 \;10}{[-]}\stackrel{11\;12}{(-)}\stackrel{13\;14}{[+]}$
&1&-1/2&1&0&1/2&0&$-1/\sqrt{3}$&1/6&2/3&-1/3\\
\hline
19&$d_{R}^{c3}$&$\stackrel{03}{(+i)}\stackrel{12}{(+)}|\stackrel{56}{[-]}\stackrel{78}{[-]}
||\stackrel{9 \;10}{[-]}\stackrel{11\;12}{(-)}\stackrel{13\;14}{[+]}$
&1&1/2&1&0&-1/2&0&$-1/\sqrt{3}$&1/6&-1/3&2/3\\
\hline 
20&$d_{R}^{c3}$&$\stackrel{03}{[-i]}\stackrel{12}{[-]}|\stackrel{56}{[-]}\stackrel{78}{[-]}
||\stackrel{9 \;10}{[-]}\stackrel{11\;12}{(-)}\stackrel{13\;14}{[+]}$
&1&-1/2&1&0&-1/2&0&$-1/\sqrt{3}$&1/6&-1/3&2/3\\
\hline
21&$d_{L}^{c3}$&$\stackrel{03}{[-i]}\stackrel{12}{(+)}|\stackrel{56}{[-]}\stackrel{78}{(+)}
||\stackrel{9 \;10}{[-]}\stackrel{11\;12}{(-)}\stackrel{13\;14}{[+]}$
&-1&1/2&-1&-1/2&0&0&$-1/\sqrt{3}$&1/6&1/6&1/6\\
\hline
22&$d_{L}^{c3}$&$\stackrel{03}{(+i)}\stackrel{12}{[-]}|\stackrel{56}{[-]}\stackrel{78}{(+)}
||\stackrel{9 \;10}{[-]}\stackrel{11\;12}{(-)}\stackrel{13\;14}{[+]}$
&-1&-1/2&-1&-1/2&0&0&$-1/\sqrt{3}$&1/6&1/6&1/6\\
\hline
23&$u_{L}^{c3}$&$\stackrel{03}{[-i]}\stackrel{12}{(+)}|\stackrel{56}{(+)}\stackrel{78}{[-]}
||\stackrel{9 \;10}{[-]}\stackrel{11\;12}{(-)}\stackrel{13\;14}{[+]}$
&-1&1/2&-1&1/2&0&0&$-1/\sqrt{3}$&1/6&1/6&1/6\\
\hline
24&$u_{L}^{c3}$&$\stackrel{03}{(+i)}\stackrel{12}{[-]}|\stackrel{56}{(+)}\stackrel{78}{[-]}
||\stackrel{9 \;10}{[-]}\stackrel{11\;12}{(-)}\stackrel{13\;14}{[+]}$
&-1&-1/2&-1&1/2&0&0&$-1/\sqrt{3}$&1/6&1/6&1/6\\
\hline\hline
\end{tabular}\\[1mm]
\begin{tabular}{|r|c||c||c|c||c|c|c||c|c|c||r|r|}
\hline
i&$$&$|^a\psi_i>$&$\Gamma^{(1,3)}$&$ S^{12}$&$\Gamma^{(4)}$&
$\tau^{33}$&$\tau^{43}$&$\tau^{53}$&$\tau^{58}$&$\tau^{61}$&$Y$&$Y'$\\
\hline\hline
&& ${\rm Octet\; IV},\;\Gamma^{(1,7)} =1,\;\Gamma^{(6)} = -1,$&&&&&&&&&& \\
&& ${\rm of \; leptons}$&&&&&&&&&&\\
\hline\hline
25&$\nu_{R}$&$\stackrel{03}{(+i)}\stackrel{12}{(+)}|\stackrel{56}{(+)}\stackrel{78}{(+)}
||\stackrel{9 \;10}{(+)}\stackrel{11\;12}{[+]}\stackrel{13\;14}{[+]}$
&1&1/2&1&0&1/2&0&$0$&-1/2&0&-1\\
\hline
26&$\nu_{R}$&$\stackrel{03}{[-i]}\stackrel{12}{[-]}|\stackrel{56}{(+)}\stackrel{78}{(+)}
||\stackrel{9 \;10}{(+)}\stackrel{11\;12}{[+]}\stackrel{13\;14}{[+]}$
&1&-1/2&1&0&1/2&0&$0$&-1/2&0&-1\\
\hline
27&$e_{R}$&$\stackrel{03}{(+i)}\stackrel{12}{(+)}|\stackrel{56}{[-]}\stackrel{78}{[-]}
||\stackrel{9 \;10}{(+)}\stackrel{11\;12}{[+]}\stackrel{13\;14}{[+]}$
&1&1/2&1&0&-1/2&0&$0$&-1/2&-1&0\\
\hline
28&$e_{R}$&$\stackrel{03}{[-i]}\stackrel{12}{[-]}|\stackrel{56}{[-]}\stackrel{78}{[-]}
||\stackrel{9 \;10}{(+)}\stackrel{11\;12}{[+]}\stackrel{13\;14}{[+]}$
&1&-1/2&1&0&-1/2&0&$0$&-1/2&-1&0\\
\hline
29&$e_{L}$&$\stackrel{03}{[-i]}\stackrel{12}{(+)}|\stackrel{56}{[-]}\stackrel{78}{(+)}
||\stackrel{9 \;10}{(+)}\stackrel{11\;12}{[+]}\stackrel{13\;14}{[+]}$
&-1&1/2&-1&-1/2&0&0&$0$&-1/2&-1/2&-1/2\\
\hline
30&$e_{L}$&$\stackrel{03}{(+i)}\stackrel{12}{[-]}|\stackrel{56}{[-]}\stackrel{78}{(+)}
||\stackrel{9 \;10}{(+)}\stackrel{11\;12}{[+]}\stackrel{13\;14}{[+]}$
&-1&-1/2&-1&-1/2&0&0&$0$&-1/2&-1/2&-1/2\\
\hline
31&$\nu_{L}$&$\stackrel{03}{[-i]}\stackrel{12}{(+)}|\stackrel{56}{(+)}\stackrel{78}{[-]}
||\stackrel{9 \;10}{(+)}\stackrel{11\;12}{[+]}\stackrel{13\;14}{[+]}$
&-1&1/2&-1&1/2&0&0&$0$&-1/2&-1/2&-1/2\\
\hline
32&$\nu_{L}$&$\stackrel{03}{(+i)}\stackrel{12}{[-]}|\stackrel{56}{(+)}\stackrel{78}{[-]}
||\stackrel{9 \;10}{(+)}\stackrel{11\;12}{[+]}\stackrel{13\;14}{[+]}$
&-1&-1/2&-1&1/2&0&0&$0$&-1/2&-1/2&-1/2\\
\hline\hline
\end{tabular}
\caption{Parts III and IV of Table II.}
\end{sidewaystable}
\begin{sidewaystable} 
\centering
\begin{tabular}{|r|c||c||c|c||c|c|c||c|c|c||r|r|}
\hline
i&$$&$|^a\psi_i>$&$\Gamma^{(1,3)}$&$ S^{12}$&$\Gamma^{(4)}$&
$\tau^{33}$&$\tau^{43}$&$\tau^{53}$&$\tau^{58}$&$\tau^{61}$&$Y$&$Y'$\\
\hline\hline
&& ${\rm Antioctet\; V},\;\Gamma^{(1,7)} =-1,\;\Gamma^{(6)} = 1,$&&&&&&&&&& \\
&& ${\rm of \; antiquarks}$&&&&&&&&&&\\
\hline\hline
33&$\bar{d}_{L}^{\bar{c1}}$&$\stackrel{03}{[-i]}\stackrel{12}{(+)}|\stackrel{56}{(+)}\stackrel{78}{(+)}
||\stackrel{9 \;10}{[-]}\stackrel{11\;12}{[+]}\stackrel{13\;14}{[+]}$
&-1&1/2&1&0&1/2&-1/2&$-1/(2\sqrt{3})$&-1/6&1/3&-2/3\\
\hline
34&$\bar{d}_{L}^{\bar{c1}}$&$\stackrel{03}{(+i)}\stackrel{12}{[-]}|\stackrel{56}{(+)}\stackrel{78}{(+)}
||\stackrel{9 \;10}{[-]}\stackrel{11\;12}{[+]}\stackrel{13\;14}{[+]}$
&-1&-1/2&1&0&1/2&-1/2&$-1/(2\sqrt{3})$&-1/6&1/3&-2/3\\
\hline
35&$\bar{u}_{L}^{\bar{c1}}$&$\stackrel{03}{[-i]}\stackrel{12}{(+)}|\stackrel{56}{[-]}\stackrel{78}{[-]}
||\stackrel{9 \;10}{[-]}\stackrel{11\;12}{[+]}\stackrel{13\;14}{[+]}$
&-1&1/2&1&0&-1/2&-1/2&$-1/(2\sqrt{3})$&-1/6&-2/3&1/3\\
\hline
36&$\bar{u}_{L}^{\bar{c1}}$&$\stackrel{03}{(+i)}\stackrel{12}{[-]}|\stackrel{56}{[-]}\stackrel{78}{[-]}
||\stackrel{9 \;10}{[-]}\stackrel{11\;12}{[+]}\stackrel{13\;14}{[+]}$
&-1&-1/2&1&0&-1/2&-1/2&$-1/(2\sqrt{3})$&-1/6&-2/3&1/3\\
\hline
37&$\bar{d}_{R}^{\bar{c1}}$&$\stackrel{03}{(+i)}\stackrel{12}{(+)}|\stackrel{56}{(+)}\stackrel{78}{[-]}
||\stackrel{9 \;10}{[-]}\stackrel{11\;12}{[+]}\stackrel{13\;14}{[+]}$
&1&1/2&-1&1/2&0&-1/2&$-1/(2\sqrt{3})$&-1/6&-1/6&-1/6\\
\hline
38&$\bar{d}_{R}^{\bar{c1}}$&$\stackrel{03}{[-i]}\stackrel{12}{[-]}|\stackrel{56}{(+)}\stackrel{78}{[-]}
||\stackrel{9 \;10}{[-]}\stackrel{11\;12}{[+]}\stackrel{13\;14}{[+]}$
&1&-1/2&-1&1/2&0&-1/2&$-1/(2\sqrt{3})$&-1/6&-1/6&-1/6\\
\hline
39&$\bar{u}_{R}^{\bar{c1}}$&$\stackrel{03}{(+i)}\stackrel{12}{(+)}|\stackrel{56}{[-]}\stackrel{78}{(+)}
||\stackrel{9 \;10}{[-]}\stackrel{11\;12}{[+]}\stackrel{13\;14}{[+]}$
&1&1/2&-1&-1/2&0&-1/2&$-1/(2\sqrt{3})$&-1/6&-1/6&-1/6\\
\hline
40&$\bar{u}_{R}^{\bar{c1}}$&$\stackrel{03}{[-i]}\stackrel{12}{[-]}|\stackrel{56}{[-]}\stackrel{78}{(+)}
||\stackrel{9 \;10}{[-]}\stackrel{11\;12}{[+]}\stackrel{13\;14}{[+]}$
&1&-1/2&-1&-1/2&0&-1/2&$-1/(2\sqrt{3})$&-1/6&-1/6&-1/6\\
\hline\hline
\end{tabular}\\[1mm]
\begin{tabular}{|r|c||c||c|c||c|c|c||c|c|c||r|r|}
\hline
i&$$&$|^a\psi_i>$&$\Gamma^{(1,3)}$&$ S^{12}$&$\Gamma^{(4)}$&
$\tau^{33}$&$\tau^{43}$&$\tau^{53}$&$\tau^{58}$&$\tau^{61}$&$Y$&$Y'$\\
\hline\hline
&& ${\rm Antioctet\; VI},\;\Gamma^{(1,7)} =-1,\;\Gamma^{(6)} = 1,$&&&&&&&&&& \\
&& ${\rm of \; antiquarks}$&&&&&&&&&&\\
\hline\hline
41&$\bar{d}_{L}^{\bar{c2}}$&$\stackrel{03}{[-i]}\stackrel{12}{(+)}|\stackrel{56}{(+)}\stackrel{78}{(+)}
||\stackrel{9 \;10}{(+)}\stackrel{11\;12}{(-)}\stackrel{13\;14}{[+]}$
&-1&1/2&1&0&1/2&1/2&$-1/(2\sqrt{3})$&-1/6&1/3&-2/3\\
\hline 
42&$\bar{d}_{L}^{\bar{c2}}$&$\stackrel{03}{(+i)}\stackrel{12}{[-]}|\stackrel{56}{(+)}\stackrel{78}{(+)}
||\stackrel{9 \;10}{(+)}\stackrel{11\;12}{(-)}\stackrel{13\;14}{[+]}$
&-1&-1/2&1&0&1/2&1/2&$-1/(2\sqrt{3})$&-1/6&1/3&-2/3\\
\hline
43&$\bar{u}_{L}^{\bar{c2}}$&$\stackrel{03}{[-i]}\stackrel{12}{(+)}|\stackrel{56}{[-]}\stackrel{78}{[-]}
||\stackrel{9 \;10}{(+)}\stackrel{11\;12}{(-)}\stackrel{13\;14}{[+]}$
&-1&1/2&1&0&-1/2&1/2&$-1/(2\sqrt{3})$&-1/6&-2/3&1/3\\
\hline 
44&$\bar{u}_{L}^{\bar{c2}}$&$\stackrel{03}{(+i)}\stackrel{12}{[-]}|\stackrel{56}{[-]}\stackrel{78}{[-]}
||\stackrel{9 \;10}{(+)}\stackrel{11\;12}{(-)}\stackrel{13\;14}{[+]}$
&-1&-1/2&1&0&-1/2&1/2&$-1/(2\sqrt{3})$&-1/6&-2/3&1/3\\
\hline
45&$\bar{d}_{R}^{\bar{c2}}$&$\stackrel{03}{(+i)}\stackrel{12}{(+)}|\stackrel{56}{(+)}\stackrel{78}{[-]}
||\stackrel{9 \;10}{(+)}\stackrel{11\;12}{(-)}\stackrel{13\;14}{[+]}$
&1&1/2&-1&1/2&0&1/2&$-1/(2\sqrt{3})$&-1/6&-1/6&-1/6\\
\hline
46&$\bar{d}_{R}^{\bar{c}}$&$\stackrel{03}{[-i]}\stackrel{12}{[-]}|\stackrel{56}{(+)}\stackrel{78}{[-]}
||\stackrel{9 \;10}{(+)}\stackrel{11\;12}{(-)}\stackrel{13\;14}{[+]}$
&1&-1/2&-1&1/2&0&1/2&$-1/(2\sqrt{3})$&-1/6&-1/6&-1/6\\
\hline
47&$\bar{u}_{R}^{\bar{c2}}$&$\stackrel{03}{(+i)}\stackrel{12}{(+)}|\stackrel{56}{[-]}\stackrel{78}{(+)}
||\stackrel{9 \;10}{(+)}\stackrel{11\;12}{(-)}\stackrel{13\;14}{[+]}$
&1&1/2&-1&-1/2&0&1/2&$-1/(2\sqrt{3})$&-1/6&-1/6&-1/6\\
\hline
48&$\bar{u}_{R}^{\bar{c2}}$&$\stackrel{03}{[-i]}\stackrel{12}{[-]}|\stackrel{56}{[-]}\stackrel{78}{(+)}
||\stackrel{9 \;10}{(+)}\stackrel{11\;12}{(-)}\stackrel{13\;14}{[+]}$
&1&-1/2&-1&-1/2&0&1/2&$-1/(2\sqrt{3})$&-1/6&-1/6&-1/6\\
\hline\hline 
\end{tabular}
\caption{Parts V and VI of Table II.}
\end{sidewaystable}
\begin{sidewaystable}
\centering
\begin{tabular}{|r|c||c||c|c||c|c|c||c|c|c||r|r|}
\hline
i&$$&$|^a\psi_i>$&$\Gamma^{(1,3)}$&$ S^{12}$&$\Gamma^{(4)}$&
$\tau^{33}$&$\tau^{43}$&$\tau^{53}$&$\tau^{58}$&$\tau^{61}$&$Y$&$Y'$\\
\hline\hline
&& ${\rm Antioctet\; VII},\;\Gamma^{(1,7)} =-1,\;\Gamma^{(6)} = 1,$&&&&&&&&&& \\
&& ${\rm of \; antiquarks}$&&&&&&&&&&\\
\hline\hline
49&$\bar{d}_{L}^{\bar{c3}}$&$\stackrel{03}{[-i]}\stackrel{12}{(+)}|\stackrel{56}{(+)}\stackrel{78}{(+)}
||\stackrel{9 \;10}{(+)}\stackrel{11\;12}{[+]}\stackrel{13\;14}{(-)}$
&-1&1/2&1&0&1/2&0&$1/\sqrt{3}$&-1/6&1/3&-2/3\\
\hline 
50&$\bar{d}_{L}^{\bar{c3}}$&$\stackrel{03}{(+i)}\stackrel{12}{[-]}|\stackrel{56}{(+)}\stackrel{78}{(+)}
||\stackrel{9 \;10}{(+)}\stackrel{11\;12}{[+]}\stackrel{13\;14}{(-)}$
&-1&-1/2&1&0&1/2&0&$1/\sqrt{3}$&-1/6&1/3&-2/3\\
\hline
51&$\bar{u}_{L}^{\bar{c}}$&$\stackrel{03}{[-i]}\stackrel{12}{(+)}|\stackrel{56}{[-]}\stackrel{78}{[-]}
||\stackrel{9 \;10}{(+)}\stackrel{11\;12}{[+]}\stackrel{13\;14}{(-)}$
&-1&1/2&1&0&-1/2&0&$1/\sqrt{3}$&-1/6&-2/3&1/3\\
\hline 
52&$\bar{u}_{L}^{\bar{c3}}$&$\stackrel{03}{(+i)}\stackrel{12}{[-]}|\stackrel{56}{[-]}\stackrel{78}{[-]}
||\stackrel{9 \;10}{(+)}\stackrel{11\;12}{[+]}\stackrel{13\;14}{(-)}$
&-1&-1/2&1&0&-1/2&0&$1/\sqrt{3}$&-1/6&-2/3&1/3\\
\hline
53&$\bar{d}_{R}^{\bar{c3}}$&$\stackrel{03}{(+i)}\stackrel{12}{(+)}|\stackrel{56}{(+)}\stackrel{78}{[-]}
||\stackrel{9 \;10}{(+)}\stackrel{11\;12}{[+]}\stackrel{13\;14}{(-)}$
&1&1/2&-1&1/2&0&0&$1/\sqrt{3}$&-1/6&-1/6&-1/6\\
\hline
54&$\bar{d}_{R}^{\bar{c3}}$&$\stackrel{03}{[-i]}\stackrel{12}{[-]}|\stackrel{56}{(+)}\stackrel{78}{[-]}
||\stackrel{9 \;10}{(+)}\stackrel{11\;12}{[+]}\stackrel{13\;14}{(-)}$
&1&-1/2&-1&1/2&0&0&$1/\sqrt{3}$&-1/6&-1/6&-1/6\\
\hline
55&$\bar{u}_{R}^{\bar{c3}}$&$\stackrel{03}{(+i)}\stackrel{12}{(+)}|\stackrel{56}{[-]}\stackrel{78}{(+)}
||\stackrel{9 \;10}{(+)}\stackrel{11\;12}{[+]}\stackrel{13\;14}{(-)}$
&1&1/2&-1&-1/2&0&0&$1/\sqrt{3}$&-1/6&-1/6&-1/6\\
\hline
56&$\bar{u}_{R}^{\bar{c3}}$&$\stackrel{03}{[-i]}\stackrel{12}{[-]}|\stackrel{56}{[-]}\stackrel{78}{(+)}
||\stackrel{9 \;10}{(+)}\stackrel{11\;12}{[+]}\stackrel{13\;14}{(-)}$
&1&-1/2&-1&-1/2&0&0&$1/\sqrt{3}$&-1/6&-1/6&-1/6\\
\hline\hline 
\end{tabular}\\[1mm]
\begin{tabular}{|r|c||c||c|c||c|c|c||c|c|c||r|r|}
\hline
i&$$&$|^a\psi_i>$&$\Gamma^{(1,3)}$&$ S^{12}$&$\Gamma^{(4)}$&
$\tau^{33}$&$\tau^{43}$&$\tau^{53}$&$\tau^{58}$&$\tau^{61}$&$Y$&$Y'$\\
\hline\hline
&& ${\rm Antioctet\; VIII},\;\Gamma^{(1,7)} =-1,\;\Gamma^{(6)} = 1,$&&&&&&&&&& \\
&& ${\rm of \; antileptons}$&&&&&&&&&&\\
\hline\hline
57&$\bar{e}_{L}$&$\stackrel{03}{[-i]}\stackrel{12}{(+)}|\stackrel{56}{(+)}\stackrel{78}{(+)}
||\stackrel{9 \;10}{[-]}\stackrel{11\;12}{(-)}\stackrel{13\;14}{(-)}$
&-1&1/2&1&0&1/2&0&$0$&1/2&1&0\\
\hline 
58&$\bar{e}_{L}$&$\stackrel{03}{(+i)}\stackrel{12}{[-]}|\stackrel{56}{(+)}\stackrel{78}{(+)}
||\stackrel{9 \;10}{[-]}\stackrel{11\;12}{(-)}\stackrel{13\;14}{(-)}$
&-1&-1/2&1&0&1/2&0&$0$&1/2&1&0\\
\hline
59&$\bar{\nu}_{L}$&$\stackrel{03}{[-i]}\stackrel{12}{(+)}|\stackrel{56}{[-]}\stackrel{78}{[-]}
||\stackrel{9 \;10}{[-]}\stackrel{11\;12}{(-)}\stackrel{13\;14}{(-)}$
&-1&1/2&1&0&-1/2&0&$0$&1/2&0&1\\
\hline
60&$\bar{\nu}_{L}$&$\stackrel{03}{(+i)}\stackrel{12}{[-]}|\stackrel{56}{[-]}\stackrel{78}{[-]}
||\stackrel{9 \;10}{[-]}\stackrel{11\;12}{(-)}\stackrel{13\;14}{(-)}$
&-1&-1/2&1&0&-1/2&0&$0$&1/2&0&1\\
\hline
61&$\bar{\nu}_{R}$&$\stackrel{03}{(+i)}\stackrel{12}{(+)}|\stackrel{56}{[-]}\stackrel{78}{(+)}
||\stackrel{9 \;10}{[-]}\stackrel{11\;12}{(-)}\stackrel{13\;14}{(-)}$
&1&1/2&-1&-1/2&0&0&$0$&1/2&1/2&1/2\\
\hline
62&$\bar{\nu}_{R}$&$\stackrel{03}{[-i]}\stackrel{12}{[-]}|\stackrel{56}{[-]}\stackrel{78}{(+)}
||\stackrel{9 \;10}{[-]}\stackrel{11\;12}{(-)}\stackrel{13\;14}{(-)}$
&1&-1/2&-1&-1/2&0&0&$0$&1/2&1/2&1/2\\
\hline
63&$\bar{e}_{R}$&$\stackrel{03}{(+i)}\stackrel{12}{(+)}|\stackrel{56}{(+)}\stackrel{78}{[-]}
||\stackrel{9 \;10}{[-]}\stackrel{11\;12}{(-)}\stackrel{13\;14}{(-)}$
&1&1/2&-1&1/2&0&0&$0$&1/2&1/2&1/2\\
\hline
64&$\bar{e}_{R}$&$\stackrel{03}{[-i]}\stackrel{12}{[-]}|\stackrel{56}{(+)}\stackrel{78}{[-]}
||\stackrel{9 \;10}{[-]}\stackrel{11\;12}{(-)}\stackrel{13\;14}{(-)}$
&1&-1/2&-1&1/2&0&0&$0$&1/2&1/2&1/2\\
\hline\hline
\end{tabular}
\caption{Parts VII and VIII of Table II.}
\end{sidewaystable}

In Table II (spread over the Tables 1--4) the 64-plet of one Weyl spinor of $SO(1,13)$ is presented. The multiplet contains  spinors - all the quarks 
(Octets I,II,III) and leptons (Octet IV)- and the corresponding ``anti-spinors''- anti-quarks 
(Anti-octets V, VI, VII) and anti-leptons 
(Anti-octet VIII) of the Standard model, that is left handed weak charged quarks and leptons and right handed 
weak chargeless anti-quarks and anti-leptons, as well as left handed weak chargeless anti-quarks and anti-leptons
and weak charged right handed anti-quarks and anti-leptons.  It also contains right handed weak chargeless 
neutrinos and left handed weak chargeless anti-neutrinos in addition. There are accordingly indeed 32 quarks 
and leptons and anti-quarks and anti-leptons in one Weyl representation, since a generic solution of the Weyl equation
in $d=(1+3)$ part of space is 
a superposition of spin up and spin down member of the representation.

Taking the first two states of each of the triplet (that is of Octet I, Octet II, Octet III), 
the anti-triplet (that is of Anti-octet V, Anti-octet VI,
Anti-octet VII), the singlet (that is of Octet IV) and the anti-singlet (that is of Anti-octet VIII) 
one recognizes the ($2^{10/2-1}$ =16)-plet of the group $SO(1,9)$, presented in Table I of the subsection \ref{complex}.
One notices that the $SO(4)$ part is now a ''spectator'', as it was the ''spectator'' the $SO(6)$ content of the 
group $SO(1,13)$, in representations of the subgroup $SO(1,7)$ octets on Table II.

\section{Mechanism generating families}
\label{mechanism}

We presented in subsection \ref{technique} the technique, which generates a left handed and a right handed Weyl
multiplet from the starting state expressed as products of projectors and nilpotents, which are the 
Clifford algebra objects. 

There are $2^d$ orthogonal polynomials, which are products of nilpotents and projectors. 
When all $2^d$ states are considered as a Hilbert space, we recognize that for $d$ even 
there are $2^{d/2}$ ''families'' and for $d$ odd $2^{(d+1)/2}$ ''families'' of spinors, just like the one presented 
in the previous section.

We present in this section  the operators, which transform the state of one ''family'' of spinors into the 
state of another ''family'' of spinors,
changing nothing but the ''family'' number. These operations can be used as  
a possible mechanism for generation families of quarks and leptons.

Let us define\cite{pikanorma2002}  the Clifford algebra objects
$\tilde{\gamma}^a$'s  as operations which operate formally from the left hand side (as $\gamma^a$'s do)
 on objects  $\stackrel{ab}{(k)}$  and $\stackrel{ab}{[k]}$,  transforming objects to
$\stackrel{ab}{[k]}$  and $\stackrel{ab}{(k)}$, respectively,  as $\gamma^a$'s  would 
if applied from the right hand side, up to a phase i
\begin{eqnarray}
\tilde{\gamma^a} \stackrel{ab}{(k)}: &=& -i\stackrel{ab}{(k)} \gamma^a = - i\eta^{aa}\stackrel{ab}{[k]},\noindent\\
\tilde{\gamma^b} \stackrel{ab}{(k)}: &=& -i\stackrel{ab}{(k)} \gamma^b = - k
\stackrel{ab}{[k]}.
\label{gammatilde}
\end{eqnarray}
One accordingly finds
\begin{eqnarray}
\tilde{\gamma^a} \stackrel{ab}{[k]}: &=& \;\; i \stackrel{ab}{[k]} \gamma^a = \;\;i\stackrel{ab}{(k)},\noindent\\
\tilde{\gamma^b} \stackrel{ab}{[k]}: &=& \;\; i \stackrel{ab}{[k]} \gamma^b = -k \eta^{aa} \stackrel{ab}{(k)}.
\label{gammatilde1}
\end{eqnarray}

We can prove\cite{holgernorma2003} that $\tilde{\gamma^a}$ obey the same Clifford algebra relation as $\gamma^a$,
and that $\tilde{\gamma^a}$ and $\gamma^a$ anti commute
\begin{eqnarray}
\{ \tilde{\gamma^a}, \gamma^b \}_+ &=& 0, \quad {\rm while} \quad
\{\tilde{\gamma^a}, \tilde{\gamma^b} \}_+ = 2 \eta^{ab}.
\label{gammatildegamma}
\end{eqnarray}
If we define
\begin{eqnarray}
\tilde{S}^{ab} = \frac{i}{4}\; [\tilde{\gamma}^a,\tilde{\gamma}^b] = \frac{1}{4} (\tilde{\gamma}^a
\tilde{\gamma}^b - \tilde{\gamma}^b\tilde{\gamma}^a),
\label{tildesab}
\end{eqnarray}
we can show that $\tilde{S}^{ab}$ fulfil the Lorentz algebra relation as $S^{ab}$ do.  
We further find
\begin{eqnarray}
\{\tilde{S}^{ab}, S^{ab}\}_- =0,\quad
\{\tilde{S}^{ab}, \gamma^c \}_-=0,\quad
\{S^{ab}, \tilde{\gamma}^c \}_-=0.
\label{sabtildesab}
\end{eqnarray}
One also finds 
\begin{eqnarray}
\{\tilde{S}^{ab}, \Gamma^{(d)} \}_- =0,\quad \{ \tilde{\gamma}^a, \Gamma^{(d)} \}_- = 0, \quad {\rm for \;\; d\;\; even,}\nonumber\\
\{\tilde{S}^{ab}, \Gamma^{(d)} \}_- =0,\quad \{ \tilde{\gamma}^a, \Gamma^{(d)} \}_+ = 0, \quad {\rm for \;\; d\;\; odd.}
\label{sabtildeGAMMA}
\end{eqnarray}

Since $\gamma^a$ transform $\stackrel{ab}{(k)}$ to $\stackrel{ab}{[-k]}$, never to $\stackrel{ab}{[k]}$, while 
$\tilde{\gamma}^a$ do transform $\stackrel{ab}{(k)}$ to $\stackrel{ab}{[k]}$, we declare the states formed as factors
of nilpotents and projectors of this second type as ''families'' of representations. 

The Eq.(\ref{sabtildeGAMMA}) means that transforming one ''family'' into another by operating on the starting family
with either $\tilde{S}^{ab}$ 
or $\tilde{\gamma}^a$ leaves in $d$ even the handedness $\Gamma$ unchanged, while in $d$ odd the transformation to another 
''family'' with $\tilde{\gamma}^a$ changes the handedness of states.

One notices that nilpotents $\stackrel{ab}{(k)}$ and projectors $\stackrel{ab}{[k]}$ are eigenvectors
not only of the Cartan subalgebra $S^{ab}$ but also of $\tilde{S}^{ab}$. Accordingly only $\tilde{S}^{ac}$, which
do not carry the Cartan subalgebra indices, cause the transition from one ''family'' to another ''family''.

We can conclude that the operators $\tilde{S}^{ab}$ (if  $S^{ab}$ do not belong to the Cartan subalgebra) transform
the starting state of one Weyl spinor of Eq.(\ref{start}) into the starting state of another ''family''.

\section{Lagrange function for spinors leading in d=4 to the Standard model one, with Yukawa couplings included}
\label{Yukawa}

We shall present in this section the Lagrange density for spinors, which in $d=(1+13)$ describes a Weyl
spinor in a gauge gravitational field (expressed by vielbeins and spin connections) and which manifests in 
the four-dimensional part of space as the Lagrange density of the Standard model, with the Yukawa couplings
included.

Refereeing to the work\cite{norma01,pikanormaproceedings1,pikanormaproceedings2} we write the Lagrange density 
function for a Weyl (massless)
spinor in $d(=1+13)$ - dimensional space as
\begin{eqnarray}
{\cal L} &=& \bar{\psi}\gamma^a p_{0a} \psi = \bar{\psi} \gamma^a f^{\mu}_a p_{0\mu}, 
\nonumber\\
\quad {\rm with}\quad
p_{0\mu} &=& p_{\mu} - \frac{1}{2}S^{ab} \omega_{ab\mu} - \frac{1}{2}\tilde{S}^{ab} \tilde{\omega}_{ab\mu}.
\label{lagrange}
\end{eqnarray}
Here $f^{\mu}_a$ are vielbeins, while 
 $\omega_{ab\mu}$ and $\tilde{\omega}_{ab\mu} $ are spin connections, the gauge fields of $S^{ab}$ and
$\tilde{S}^{ab}$, respectively.

According to what we have presented in section \ref{spinor},
one Weyl spinor in $d=(1+13)$ with the spin as the only internal degree of freedom, (might) manifests  in
four-dimensional part of space  as the ordinary ($SO(1,3)$) spinor with all the known charges 
of one family of the Standard model (as presented in Table II).
The gravitational field presented with spin connections and vielbeins might accordingly in four 
dimensions  manifest as all the known gauge fields as
well as the Yukawa couplings, if the break of symmetries occurs in an appropriate way. 

To see that let us first rewrite the Lagrange density of Eq.(\ref{lagrange}) in an appropriate way. 
According to section \ref{spinor} we can rewrite 
Eq.(\ref{lagrange}) as follows
\begin{eqnarray}
{\cal L} &=& \bar{\psi}\gamma^{\alpha} (p_{\alpha}- \sum_{A,i}\; g^{A}\tau^{Ai} A^{Ai}_{\alpha} \psi) 
\nonumber\\
&+& i\psi^+ S^{0h} S^{k k'} f^{\sigma}_h \omega_{k k' \sigma} \psi + \quad 
i \psi^+ S^{0h} \tilde{S}^{k k'} f^{\sigma}_h \tilde{\omega}_{kk' \sigma} \psi,
\label{lagrangein4}
\end{eqnarray}
with  $\psi$, which is (for low energy solution) assumed not to depend on coordinates $x^{\sigma}, \sigma
=\{5,6, \cdots ,14 \}$. We assume for simplicity in addition  that there is no gravitational field in 
four-dimensional subspace
($f^{\alpha}_m = \delta^{\alpha}_{m}, m=\{0,1,2,3 \}, \alpha =\{0,1,2,3 \},\; \omega_{mn\alpha} =0,
\; tilde{\omega}_{mn\alpha} =0$).

The second and the third term in Eq.(\ref{lagrangein4} ) look like a mass term, since 
$f^{\sigma}_h \tilde{\omega}_{kk' \sigma}$ behaves in $d(=1+3)-$
dimensional part of space like a scalar field, while the operator $S^{0h}, h=7,8$, for example, 
transforms a right handed
weak chargeless spinor into a left handed weak charged spinor, without changing the spin (just
what the Yukawa couplings with the Higgs doublet included, do in the Standard model formulation.
The reader should note, that no Higgs weak charge doublet is needed here, as $S^{0h}, h=7,8$ does his job.)

Since masses of quarks of one family differ from masses of leptons in the same family, it is meaningful to
rewrite the term $\psi^+ S^{0h} S^{k k'} f^{\sigma}_h \omega_{k k' \sigma} \psi $ in the form $- \gamma^0 \gamma^h
\tau^{Ai} A^{Ai}_{\sigma} f^{\sigma}_h$ to point out that hyper charges ($Y$ and $Y'$) are important for 
the appropriate Yukawa couplings. We also note that the term $i \psi^+ S^{0h} \tilde{S}^{k k'} 
f^{\sigma}_h \tilde{\omega}_{kk' \sigma} \psi$ contributes to the Yukawa couplings, which connect
different families.

One should of course ask oneself whether or not it is at all possible to choose 
spin connections and vielbeins $f^{\sigma}_h \omega_{k k' \sigma}$ and 
$f^{\sigma}_h \tilde{\omega}_{k k' \sigma}$ in a way to reproduce the masses of the three families of 
quarks and leptons and to predict, what are masses of a possible next (fourth) family in a way to be in agreement
with the experimental data, since we have seen that our approach unifying spins and 
charges does predict more than one family. 

The work on this topic is under consideration. To demonstrate that the approach used in this paper 
does suggest possible relations among Yukawa couplings
and consequently also the mass matrix, we briefly follow  in  subsection \ref{mass} the 
ref.\cite{pikanormaproceedings2} demonstrating
a possible choice  
of Yukawa couplings, suggested by Eq.(\ref{lagrangein4}), which leads to
four rather than to three generations of quarks and leptons, with the
values for the masses of the fourth generation, which do not contradict the experimental data\cite{okun}.

\subsection{Mass matrices for four generations of quarks and leptons, 
suggested by our approach unifying spins and charges}
\label{SNMBsec:mass}

If the break of symmetries leads (this is an assumption, suggested by our research on the break of symmetries
) to only terms like $\psi^+ S^{0h} \tilde{S}^{k k'} 
f^{\sigma}_h \tilde{\omega}_{kk' \sigma} \psi $
 in the Lagrange density, with $h $ and $\sigma \in 5,6,7,8$ and   $k,k'$ either both equal to 
$0,1,2,3$ or both equal to $5,6,7,8$, then there are only four families  measurable at low energies, with
the starting states in the $SO(1,7)$ segment as follows
namely
\begin{eqnarray}
\stackrel{03}{(+i)} \stackrel{12}{(+)} \stackrel{56}{(+)} \stackrel{78}{(+)} \nonumber\\
\stackrel{03}{[+i]} \stackrel{12}{[+]} \stackrel{56}{(+)} \stackrel{78}{(+)} \nonumber\\
\stackrel{03}{(+i)} \stackrel{12}{(+)} \stackrel{56}{[+]} \stackrel{78}{[+]} \nonumber\\
\stackrel{03}{[+i]} \stackrel{12}{[+]} \stackrel{56}{[+]} \stackrel{78}{[+]}, 
\label{threefamilies}
\end{eqnarray}
 while they all have the same $SO(6)$ segment, namely equal to
 $\stackrel{9 10}{(+)} \stackrel{11 12}{(-)} \stackrel{13 14}{(-)}$, the same one as on Table II.
 
 If we denote by $A_a$ the matrix element for the transition from  a right handed weak chargeless 
 spinor of type $a = u,d,e,\nu$ to the left handed weak charged spinor (these transitions occur within 
 one family and are caused by the second term $- \gamma^0 \gamma^h
\tau^{Ai} A^{Ai}_{\sigma} f^{\sigma}_h$) of Eq.(\ref{lagrangein4} ), by $B_a$ the matrix element 
causing the transition,
 in which $\stackrel{03}{(+i)} \stackrel{12}{(+)}$ changes to $\stackrel{03}{[+i]} \stackrel{12}{[+]}$
 or opposite (such are transitions between the first and the second family or transitions between the 
 third and the fourth family of Eq.(\ref{threefamilies}) caused by 
 $\tilde{S}^{mm'}f^{\sigma}_h \tilde{\omega}_{mm' \sigma}$ with $m,m'=0,1,2,3$ and $h = 5,6,7,8$), 
 by $C_a$  the matrix element causing the transition 
 in which $\stackrel{56}{(+)} \stackrel{78}{(+)}$ changes to $\stackrel{56}{[+]} \stackrel{78}{[+]}$
 or opposite (such are transitions between the first and the third family or transitions between the 
 second and the fourth family of Eq.(\ref{threefamilies}) caused by $ \tilde{S}^{kk'}f^{\sigma}_h
 \tilde{\omega}_{kk' \sigma}$ with $h,k,k'=5,6,7,8$) and by $D_a$  transitions in which all four 
 factors change, that is the transitions,
 in which $\stackrel{03}{(+i)} \stackrel{12}{(+)}$ changes to $\stackrel{03}{[+i]} \stackrel{12}{[+]}$
 or opposite and  $\stackrel{56}{(+)} \stackrel{78}{(+)}$ changes to $\stackrel{56}{[+]} \stackrel{78}{[+]}$ or
 opposite
 (such are transitions between the first and the fourth family or transitions between the 
 second  and the third family of Eq.(\ref{threefamilies}))  and if we further assume that 
 the elements are real numbers, we find 
 the following mass matrix
\begin{displaymath}
\left( \begin{array}{cccc}
A_a&B_a&C_a&D_a\\
B_a&A_a&D_a&C_a\\
C_a&D_a&A_a&B_a\\
D_a&C_a&B_a&A_a
\end{array} \right).
\label{mass}
\end{displaymath}

 To evaluate the matrix elements $A_a,B_a,C_a,D_a$ one should make a precise model, taking into account that
 matrix elements within one family depend on quantum numbers of the members
 of the family, like $Y,Y'$  and others, and accordingly also on the way how symmetries break. 
 All these need further study.
 
We study here only the symmetry of the mass matrix, namely
\begin{displaymath}
\left( \begin{array}{cc}
X&Y\\
Y&X
\end{array} \right),
\label{xy}
\end{displaymath}
which makes the diagonalization of the mass matrix of Eq.(\ref{mass})
simple. We find
\begin{eqnarray}
\lambda_{a_1} &=& A_a-B_a-C_a+D_a,\nonumber\\
\lambda_{a_2} &=& A_a-B_a+C_a-D_a,\nonumber\\
\lambda_{a_3} &=& A_a+B_a-C_a-D_a,\nonumber\\
\lambda_{a_4} &=& A_a+B_a+C_a+D_a. 
\label{formalabcd}
\end{eqnarray}
We immediately see that a ''democratic'' matrix with $A_a=B_b=C_c=D_d$  (ref.\cite{fritsch})
leads to $\lambda_{a_1}=\lambda_{a_2}=\lambda_{a_3}= 0, \lambda_{a_4} = 4 A_a$. The diagonal matrix leads to
four equal values $\lambda_{a_i} = A_a. $ We expect that break of symmetries of the group $SO(1,13)$ down
to $SO(1,3), SU(3)$ and $U(1)$ will lead to something in between.
If we fit $\lambda_{a_i}$ with the masses of families $m_{ai}$, with $a= u,d,\nu,e$ and $i$ is
the number of family,
we find
\begin{eqnarray}
A_a &=& \{(m_{a4}+ m_{a3}) + (m_{a2} + m_{a1})\}/4,\nonumber\\
B_a &=& \{(m_{a4}+ m_{a3}) - (m_{a2} + m_{a1})\}/4,\nonumber\\
C_a &=& \{(m_{a4}- m_{a3}) + (m_{a2} - m_{a1})\}/4,\nonumber\\
D_a &=& \{(m_{a4}- m_{a3}) - (m_{a2} - m_{a1})\}/4.
\label{formalabcdwithm}
\end{eqnarray}
For the masses of quarks and leptons to agree with the experimental masses  $m_{u_i}/GeV = 0.0004$, $1.4$, 
$180$, $285 (215)$ and $m_{d_i}/GeV = 0.009, 0.2, 6.3, 215 (285)$ for quarks, and
$m_{e_i}/GeV = 0.0005, 0.105, 1.78, 100$ and $m_{\nu_i}/GeV$ let say 
$1.10^{-11}$, $2.10^{-11}$, $6.10{-11}$
and $50$ for leptons, which would agree also with what Okun and coauthors\cite{okun} have found as possible values 
for masses of the fourth family, 
we find 
\begin{equation}
\begin{array}{cccc}
A_u = 116.601 & B_u = 115.899 & C_u = 26.599 & D_u = 25.901 \\
(A_u' = 99.101 & B_u' = 98.399 & C_u' = 9.099 & D_u' = 8.401)  \\
A_d = 55.377 & B_d = 55.2728 & C_d = 52.223 & D_d = 52.127 \\
(A_d' = 72.877 & B_d' = 72.773 & C_d' = 69.723 & D_d' = 69.627) \\
A_e = 25.471 & B_e = 25.419 & C_e = 24.581 & D_e = 24.529 \\
A_\nu = 12.5 & B_\nu = 12.5 & C_\nu = 12.5  & D_\nu = 12.5.  
\end{array}
\end{equation}
Values in  the parentheses correspond to the values in the parentheses for the masses of quarks.
The mass matrices are for leptons and even for $d$ quarks very close to a ''democratic''
one\cite{fritsch}. One could also notice that for 
quarks $A_a$ are roughly proportional to the charge $Y$.

Further studies are needed to 
comment more on mass matrices, suggested by our approach unifying spins and charges.
Some further discussions can be found in ref.\cite{astridragannorma}.

\section{Possible break of symmetries, suggested by our approach unifying spins and charges}
\label{break}

We want to comment on a possible break of symmetries which would lead to physics of the Standard model. 
Taking into account that mass protection mechanism occurs only in even dimensional 
spaces \cite{norma01,holgernorma2002,holgernormawhy}, that
spinors and ``anti-spinors`` should not transform into each others at low energies, that the colour charge 
should be a conserved quantity, as well as other above discussed phenomena, we find as a promising way of 
breaking symmetries\cite{holgernormaren} the following one\footnote{Two of the authors of this paper have 
shown\cite{holgernormaren} that the way of breaking the group $SO(1,13)$, presented also in this section, leads 
to unification of all the three  coupling constants at high enough energy and that the proton decay does not contradict
the experimental data.}
\[
\begin{array}{c}
\begin{array}{c}
\underbrace{%
\begin{array}{rrcll}
 & & SO(1,13) \\
 & & \downarrow \\
 & & SO(1,7) \otimes SO(6) \\
 & \swarrow & &  \searrow \\
 & SO(1,7) & & SU(4)\\
 \swarrow\qquad & & & \qquad\downarrow \\
 SO(1,3)\otimes SO(4) & & & SU(3)\otimes U(1) \\
 \downarrow\qquad & & & \qquad\downarrow \\
SO(1,3)\otimes SU(2)\otimes U(1) & & & SU(3)\otimes U(1)\\
& & & \\
\end{array}} \\
SO(1,3)\otimes SU(2)\otimes U(1)\otimes SU(3)\otimes U(1)\\
\end{array}\\
\downarrow \\
SO(1,3)\otimes SU(2)\otimes U(1)\otimes SU(3)\\
\end{array}
\]
%

In the first step of breaking $SO(1,13)$ the ${\cal S}^{ab}\omega_{ab \mu}$ term should break into
$S^{ab}\omega_{ab \mu} + $ $\tilde{S}^{ab}\tilde{\omega}_{ab \mu}$. Further breaks of symmetries then
should take care that particle-antiparticle transitions should not occur any longer, which means that
transitions caused by $S^{ab}$ or $\tilde{S}^{ab}$, with $a= 0,1,2,3,5,6,7,8$ and $b= 9,10,11,12 $
(or opposite, with $a$ and $b$ exchanged), should at lower energies appear with a negligible
probability. The same should true also for transitions transforming (coloured) quarks into (colourless)
leptons while  
quarks and leptons demonstrate a $SO(1,7)$ multiplet, with (as we have seen)
left handed weak charged
and right handed weak chargeless spinors. Further break should occur then close to the weak scale, leading
to all the Standard model spinors. This study is under considerations.

\section{Can Kaluza-Klein-like theories lead to massless spinors? }
\label{wittennogointogo}

We pay attention on the Witten's paper\cite{witten81}, who pointed out that when one 
starts with a Weyl spinor of only one handedness, and then a compactification of a part of space occurs,
in both parts of space - the compactified one and the noncompactified one - the representations of 
both handedness appear. Accordingly there is no mass protection mechanism, which would prevent
spinors from gaining masses of the scale which is inversely proportional to the radius of the
compactified part of space. 

Looking on Table II (or Table I) one immediately sees that Witten is obviously right. 
We started with a left handed
Weyl spinor. When analyzing the Weyl left handed representation in terms of representations
of subgroups, we allways find that the representations of both handedness appear for any of the subgroups.
If the sum of the Cartan subalgebra elements of all the subgroups is equal to the hole set 
of the Cartan subalgebra elements of the starting group, the product of the handedness of 
all the subgroups is equal to the handedness of the starting group.

But we also see that the representations of both handedness distinguish between themselves in properties in
the rest of space: The break of $SO(1,13)$ into $SO(1,7) \times SO(6)$ causes representations of both handednes
in the $SO(6)$ and also in $SO(1,7)$ part. But to  the left handed representation of $SO(1,7)$ ($\Gamma^{(1,7)}=-1$) 
represented in Table II as anti-quarks and anti-leptons the ''anti-charges'' and $\Gamma^{(6)} = 1$
correspond, while the right handed $SO(1,7)$ representation (with $\Gamma^{(1,7)} =1$) carries charges
in the $SO(6)$ part of space. 

{\em We  therefore conclude that although the representations of both handedness in the noncompactified 
part of space appear,
the two representations distinguish from each other in a kind of a Kaluza-Klein gauge charges. If having
a possibility to make a choice of a particular Kaluza-Klein charge, the handedness of only one type can 
be chosen as well, making the
mass protection mechanism working. This fact leaves a hope that Kaluza-Klein-like theories have a chance to 
lead to the ''realistic'' theory in the low (observable) enery limit.} 

But does a spontaneus compactification, caused by gravity itself, really lead at low energy 
limit to massless particles?
The paper\cite{holgernorma2004}, 
discussing the example when the compacification of a $SO(1,5)$ into $M^{4}$ and a flat torus with a
 torsion occurs, can be found in this book.

\section{Discussions and conclusions}
\label{conclusions}


We reviewed in this talk the success of the approach  of (one of) us unifying spins and charges in 
answering the open questions of the Standard electroweak
model. We used our  technique, suggested by the approach, to present spinor representations, since
it makes properties of spinor representations very transparent. The technique works for any dimension and any
signature.

In the approach the space is d-dimensional, with $d> 4$, with the Poincar\' e
gauge gravitational field as the only field in $d-$dimensional space, like in all the Kaluza-Klein 
theories. We pay attention
in particular on $d=1+13$, since  one Weyl left handed representation includes all the quarks and  the leptons 
and the anti-quarks and the anti-leptons of one family of the Standard model, with the right handed weak 
chargeless neutrino
and the left handed weak chargeless anti-neutrino included. We present a possible mechanism 
(and the technique) for generating families
of quarks and leptons.

We used the approach to look for answers to (some of) the open questions of the Standard model, 
like: Why does space-time
look four-dimensional\cite{holgernorma2002}? How can the spin and all the known  charges unify to only a spin and 
how do accordingly all the known interactions - with Yukawa couplings included - unify to only gravity?
Why are the charge representations so small? 
Where do masses of quarks and leptons come from? Where do families come from? Why does  the weak interaction
break parity, while the colour one does not?
Why do only left handed quarks and leptons and right handed anti-quarks and anti-leptons
carry the weak charge? Why the weak interaction breaks parity, while the colour does not?
Why do we have particles and anti-particles? How did in the evolution symmetries possibly break
down, leading to the observable symmetries and the observable coupling constants?

The approach is a type of Kaluza-KLein theories, for which Witten has shown, that they have a  very little
chance to lead to the ''realistic'' theory at the observable energies, due to the fact that whenever a
symmetry breaks due to a compactification of a part of space, representations of both handedness always appear,
and accordingly no mass protection mechanism works. We comment on this problem\cite{holgernorma2004}, showing
the way out.

We demonstrated that a left handed $SO(1,13)$ Weyl multiplet (we have accordingly no mirror 
symmetry in this approach!) 
contains, if represented in a way to demonstrate the
$SO(1,3)$, $SU(2)$ and the two $U(1)$'s substructure of the group $SO(1,13)$, just all the quarks and the leptons
and all the anti-quarks and the anti-leptons of the Standard model, with  
right handed (charged only with
$Y'$) neutrinos and left handed (again charged only with $Y'$) anti-neutrinos in addition. 

The group $SO(1,13)$ (with the rank $7$) has  complex 
representations. 
Its two regular subgroups  $SO(1,7)$ (with the rank $ 4$ and the possible regular subgroups  
$(SU(2)\times SU(2))^2$) and
$SO(6)$ (with the rank $3$ and possible regular subgroups $SU(3)$ and $U(1)$ - 
neither $SO(1,13)$ nor $SO(6)$ have as regular
subgroups groups $(SU(2)\times SU(2))^k$) have real and complex representations,
respectively. Complexity of the representations of $SO(1,13)$ and accordingly of $SO(6)$ enables the concept
of spi\-nors and ``anti-spinors``.

We further see that due to real representations of the group $SO(1,7)$, the
left handed weak charged quarks and leptons together with right handed weak chargeless quarks and leptons 
appear in the Weyl multiplet of $SO(1,13)$, causing that the weak charge violets the parity. 

A spinor multiplet of the  subgroup $SO(1,9)$  instead contains, 
due to the complex character of its representations, 
the colour charged and chargeless spinors of left handedness and the colour ''anti-charged'' and ''anti-chargeless'' 
``anti-spinors`` of right handedness and could accordingly not
break the parity. (All spinors of one multiplet of $SO(1,9)$ have namely the same handedness 
while ``anti-spinors`` of the same multiplet have the opposite handedness.) 
It is the real and the complex nature of representations of the two subgroups $SO(1,7)$ and
$SO(1,9)$, respectively, which is responsible for the fact that weak charge breaks parity 
while the colour charge can not, if the spinor-anti-spinor concept should stay. Accordingly, also the
colour charge can be conserved.

We pay attention to the invariant,
which  is the operator of 
handedness\cite{normasuper94,norma01,bojannorma2001,holgernorma0203}.
For an even $d$ it can be defined for any spin as 
$$\Gamma = \alpha \varepsilon_{a_1 a_2 \dots a_d}
S^{a_1 a_2} S^{a_3 a_4} \cdots S^{a_{d-1} a_d} ,$$ 
with a constant $\alpha $, which can be chosen in a way
that for any spin $\Gamma = \pm 1$ and $\Gamma^2 =1, \Gamma^+ = \Gamma$. 
According to Table II, the value of $\Gamma^{(1,13)} = -1$ and  stays  (of course) unchanged also if only a
part of the group is concerned. We also see that $\Gamma^{(1,7)}$ is for all the spinors equal to 1 and
for all the ``anti-spinors'` equal to -1. One also sees that when the subgroup $SU(2)$ (the weak charge group) is broken,
that is when $Q= (Y + \tau^{33})$ (the electromagnetic charge) only is the well defined quantity besides the 
colour charge ($\tau^{53}, \tau^{58}$), the handedness 
$\Gamma^{(1,3)}$  is broken, leading to the superposition of states $u_L$ and $u_R$, for example, 
since both states have
the same colour properties ($\tau^{53}$, $\tau^{58}$), and the electromagnetic charge $Q.$

We saw that the trace of all the Cartan subalgebra operators within half a 
Weyl representation of $SO(1,13)$, namely the particle part alone (as well as the anti-particle part alone),
is equal
to zero. This is true also for all the other generators of $SO(1,13)$ and is accordingly true for all the linear
superpositions of $S^{ab}$. Due to this fact the anomaly cancellation for spinors is guaranteed.

A break of a symmetry leads to smaller subgroups, of smaller ranks and smaller representations. It seems that 
the sum of ranks of regular subgroups should be just equal to the rank of the group - almost up to the weak scale, when
the rest conserved symmetries are $U(1)$ (the electromagnetic charge), SU(3) (the colour charge) and $SO(1,3)$
(the spin degrees of freedom), with the sum of ranks equal to $5$.

The smallest spinor representations 
would occur for  the break of the type $(SU(2)\times SU(2))^k$. In this case all the eigenvalues of the Cartan subalgebra
would be $\pm 1/2$ and accordingly also the eigenvalues of the superpositions of $S^{ab}$ which lead to
invariant subgroups, would then be $\pm 1/2$ for doublets and $0$  for singlets. 
The group $SO(1,13)$  is not of such a type. Accordingly, instead of only $SU(2)$ (and $U(1)$'s from $SU(2)$) 
also $SU(3)$ appears (as well as $U(1)$ from $SO(6)$),  
enabling the concept of spinors
and ``anti-spinors``, as already mentioned. These is our comment on small representations of spinors in the 
Standard model. 

In the proposed approach unifying spins and charges the concept of families appears by the application 
of the operators $\tilde{S}^{ab} = -i/4 [\tilde{\gamma}^a \tilde{\gamma}^b - 
\tilde{\gamma}^b \tilde{\gamma}^a]$ on a starting family. 
The number of families at low 
energies then  depends on the interaction.

We propose a Lagrange density for spinors in $d =14$ dimensional space 
\begin{eqnarray}
\psi^+ \gamma^0 \gamma^a f^{\mu}_a\; (p_{\mu} -\frac{1}{2} {\cal S}^{bc}\; \omega_{bc\mu}\;)\; \psi.
\label{SNMBlagrange}
\end{eqnarray}
It contains the 
(gauged) gravitational field with spin connections $\omega_{bc\mu}$ and vielbeins $f^{\mu}_a$ only, 
which then manifest at low energies
as all the known gauge fields  $A^{Ai}_{\alpha}$ 
\begin{eqnarray}
\psi^+ \gamma^0 \gamma^m f^{\alpha}_m\; 
(p_{\alpha} - \sum_{{Ai}}\; \tau^{Ai} A^{Ai}_{\alpha}) \psi
\label{lagrangetau}
\end{eqnarray}
as well as the Yukawa couplings 
\begin{eqnarray}
\psi^+ \gamma^0 \gamma^h S^{h'h''} \omega_{h'h''\sigma} f^{\sigma}_h \psi,
\label{lagrangeyukawas}
\end{eqnarray}
and
\begin{eqnarray}
\psi^+ \gamma^0 \gamma^h \tilde{S}^{h'h''} \tilde{\omega}_{h'h''\sigma} f^{\sigma}_h \psi
\label{lagrangeyukawast}
\end{eqnarray}
with indices $h,h',h''$ from $\{5,6,7,8\}$). Operators of the type $S^{h h'}$ 
transform right handed $SU(2)$ singlets into left handed $SU(2)$ doublets within
one family, while  operators  of the type $\tilde{S}^{h h'}$  
cause transitions among  families, without changing properties with respect to the $S^{ab}$. 
Both types of terms manifest
accordingly at low energies as mass terms of quarks and leptons (and anti-quarks and anti-leptons). 
It is the interactions in higher dimensions which 
look like a Higgs causing the Yukawa 
couplings
in a four-dimensional part of space.

The mass matrices suggested by the approach at low energies have the dimension $4 \times 4$. 
They have, if assumed to be real, very peculiar symmetry, so that the 
number of different matrix elements is always equal to  the rank of the matrix. 
Accordingly the eigenvalues can  easily be found and also fitted to the experimental data in a way, that
they do agree with the experimental data. The approach suggests the
fourth family in agreement with the ref.\cite{okun}.

We presented\cite{holgernormaren} a possible scheme of breaking symmetry $SO(1,13)$, which manifests the
above mentioned properties of representations. 
We have so far learned from this study that the break of symmetry 
presented, seems to be the appropriate one, being able to reproduce the properties of the running
coupling constants.

The mass protection mechanism works in only even dimensional spaces 
(see \cite{norma01,holgernorma2002,holgernorma0203}),
provided that only one Weyl 
spinor (one irreducible representation) is assumed. Since we assume one Weyl spinor representation of $SO(1,13)$,
spinors in $d=14$ are massless. And as we have said, it is the interaction in $d=(1+13)$, which 
manifests in $d=(1+3)$ part of space
as mass terms. But how does it come that spinors are (almost) massless although Witten has shoun\cite{witten81}, that
whenever the part of space compactified and the symmetry accordingly breaks, the representations of both
handedness appear in the rest of space, leading to masses which are reversely proportional to the 
radius of the compactification? It is easy to see from Tables I and II that witten is write.  But one can also see,
that representations of different handedness have different charges. We found the way\cite{holgernorma2004} to
show that by choosing a Kaluza-Klein gauge charge on e chooses the handedness, making the mass protection mechanism 
working again.

It is, of course, a lot of open questions yet. But we believe that the approach of (one of) us might offer
possible answers to many of them after further studies will be done.

\section*{Acknowledgments} 
It is our pleasure to thank all the participants of the annual workshops entitled ''What comes beyond 
the Standard model'' taking place at Bled from 1999 to 2003 for fruitful discussions. In particular 
we would like to stress the essential contribution to the advance of the approach, presented in this talk,  
of Holger Bech Nielsen, whose thoughts and comments were extremely
fruitful. We acknowledge the contributions also of Dragan Lukman and Astri Kleppe.

\title*{An Example Giving Hope That Kaluza-Klein-like Theories Lead to Massless Spinors\thanks{The shortened
version of this talk was sent to Phys. Rev. Lett.}}
\author{Norma Manko\v c Bor\v stnik${}^1$  Holger Bech Nielsen${}^2$}
\institute{%
${}^1$Department of Physics, University of
Ljubljana, Jadranska 19, Ljubljana 1111, \\
and Primorska Institute for Natural Sciences and Technology,\\
C. Mare\v zganskega upora 2, Koper 6000, Slovenia\\
{${}^2$}Department of Physics, Niels Bohr Institute,
Blegdamsvej 17,\\
Copenhagen, DK-2100}

\titlerunning{An Example Giving Hope \ldots}
\authorrunning{Norma Manko\v c Bor\v stnik and Holger Bech Nielsen}
\maketitle

\begin{abstract} 
Kaluza-Klein-like theories seem to have difficulties with the existence of massless spinors after the compactification
of a part of space\cite{witten1981,witten1983}. We show  on an example of a flat torus with a
torsion - as a compactified part
of an even dimensional space - that a Kaluza-Klein charge can be defined, commuting with the operator of handedness 
for this part of space, which by marking (real) representations makes possible the choice of 
the representation of a particular handedness. Consequently 
the mass protection mechanism assures masslessness of particles in the noncompactified part of  space.
\end{abstract}

\section{Introduction}
\label{WIintroduction}

Kaluza-Klein-like theories\cite{kaluzaklein}, assuming that it is only gravitational gauge field in $d$-dimensional space,
which manifests in the four dimensional part of space as all the known gauge fields and the gravity (and unifying accordingly
all the known gauge fields into gravity, seems to be a very attractive and promising 
idea. The difficulties  
of these theories occur\cite{witten1981,witten1983}, when assuming that
the compactification  is responsible for the fact, that none but four dimensions are measurable at low energies. It looks  
namely very difficult to avoid after the compactification of a part of  space  the appearance of  
representations of both handedness (real representations) in this part of space and consequently also in the 
noncompactified part of space. Therefore, there is no  mass protection mechanism to
guarantee the masslessness of 
spinors\cite{palla,chaplineslanskymanton,wetterich,%
witten1981,witten1983} in the ``realistic`` noncompactified part of space.

The approach of one of us\cite{WInorma92,WInorma93,norma94,%
pikanorma2002,holgernormadk}, unifying spins and all the charges into only spins in more than four-dimensional spaces, is also 
a kind of Kaluza-Klein theories. If one left-handed Weyl spinor with the spin as the only internal degree of freedom in 
14-dimensional space is assumed\cite{normaixtapa2001,pikanorma2002}, then this left handed spinor representation, 
if analyzed in terms of
the quantum numbers of the subgroups $SO(1,3), SU(3), SU(2),U(1)$ of the group $SO(1,13)$, includes all the spinors and the 
''anti-spinors'' of one family: the left handed weak charged and the right handed weak chargeless quarks and leptons,
together with the right handed weak charged and the left handed weak chargeless anti-quarks
and anti-leptons in the same representation. One can easily define the handedness\cite{norma94} of a Weyl spinor in a flat 
Minkowski space
with $d=14$. Then accordingly also the handedness of each of the subspaces determining the spin ($SO(1,3)$) and the charges 
(the $SO(6)$ charge, or rather the $SU(3)$ charge and one $U(1)$ charge, and the $SO(4)$, or rather the $SU(2)$ and
another $U(1)$ charge) can be defined. All these operators of handedness 
anti-commute with the Weyl equation of motion for a free spinor.

In spaces which do not have the symmetry of $SO(1, d-1)$, however, but rather the symmetry of $SO(1,7) \times SO(6)$,
for example,
which means that the ground state solution of the Weyl equation of motion spans the manifold $M^{1+7} \times 
 S^5$ (besides the internal space), for example, the question then arises, if the handedness can still 
be defined in each of the subspaces, guaranteeing masslessness of spinors 
in this  ($SO(1,7)$) ``realistic`` part of space. The proposed theory of one of 
us\cite{WInorma92,WInorma93,norma94,pikanorma2002,holgernormadk} might lead to the 
Standard model assumptions (explaining hopefully, where the assumptions of the Standard model come from) 
only, if one can keep track of the
handedness up to the point, where the gravitational field causes the Yukawa couplings like type of interactions and
accordingly at the weak scale breaks the handedness of a subspace and leads to massive quarks and leptons.

We present in section \ref{examples1s1}  an  example  of a left handed spinor in $(1+5)$-dimen\-sional space  
with the Lorentz symmetry of $SO(1,5)$ ($M^{1+5}$), for which we can keep track of the handedness after the 
compactification of  space into the four dimensional
Minkowski space  of $SO(1,3)$ ($M^4$) and the compactified part of space, represented by a flat torus
with the torsion $S^1  \times S^ 1 $ (Theories with only torsion, while the Riemann tensor is equal to zero, are 
called the teleparallel theories\cite{milutin2002}). We prove that there exists a quantum number of handedness as well as
the two Kaluza-Klein charges for a flat torus with a torsion, which commute with each other. The 
two Kaluza-Klein charges commute and the operator of handedness anticommutes with the Weyl equations of motion 
operator, respectively, leading to the massless ground state
solutions of the Weyl equation in the ''realistic'' space with a symmetry $SO(1,3)$. We prove accordingly that
although the ``Witten's no-go theorem`` is valid, the masslessness of spinors in noncompactified part of 
space can still be guaranteed.

In section \ref{WIspinor} we present the definition of the operator of handedness for spi\-nors (if a space-time dimension 
is even, the operator of handedness is defined for any spin\cite{norma94,pikanorma2002,normaixtapa2001}), 
in subsection \ref{techniquespinors} of this section we 
present our technique for finding irreducible presentations of spinors in spaces of any dimension, which makes 
properties of spinors transparent (manifesting clearly, for example, that a
Weyl spinor representation of one handedness contains representations of both handedness with respect to
any subgroup of a starting group). We 
demonstrate this technique in subsection \ref{so15} on a case of our interest, namely on one Weyl 
spinor representation in  $d=1+5$. In subsection \ref{weylso(1,5)} we present the solution for a 
free spinor for this particular case. The main part of this paper is section \ref{kaluzaklein}, where we present 
some useful well known equations and relations for the Poincar\' e gauge fields of gravity with a torsion, 
needed in this paper, and demonstrate (as already mentioned)
on a flat space with a torsion as a compactified part of space, that the ``Witten's no-go theorem`` 
can be avoided by marking representations of the chosen handedness by a Kaluza-Klein charge. We also demonstrate
for this example the appearance of the gauge $U(1)$ field in the noncompactified part of space (subsection \ref{gaugefield}).

\section{Spinors in flat spaces}
\label{WIspinor}

In this section we define the operator of  handedness for spinors (if dimension of space is even, the definition of handedness
for any spin exists). We briefly present, following references\cite{WInorma93,pikanorma2002,WIholgernorma2002,WIholgernorma2003}, 
one Weyl spinor representation in $d$-dimensional spaces using the technique in which spinor states are products of nilpotents
and projectors, we use this technique for finding the representation of one Weyl spinor in $d=1+5$, and
we look for a solution of equations of motion for a free Weyl spinor for $d=1+5$.

Let $\gamma^a$ close the Clifford algebra
\begin{eqnarray}
\{\gamma^a,\gamma^b\}_+= 2\eta^{ab}, \quad {\rm for} \quad a,b \in\{0,1,2,3,5,...,d\}
\label{clifford}
\end{eqnarray}
and let
\begin{eqnarray}
\gamma^{a\dagger}= \eta^{aa} \gamma^a,
\label{cliffordher}
\end{eqnarray}
so that they are formally unitary $\gamma^{a\dagger} \gamma^a =I$. It follows then that $S^{ab}$ 
\begin{eqnarray}
S^{ab} = \frac{i}{4}\{\gamma^a,\gamma^b\}_-, \quad S^{ab\dagger} =\eta^{aa} \eta^{bb} S^{ab},
\label{WIsab}
\end{eqnarray}
close the Lie algebra of the Lorentz group ($\{S^{ab},S^{cd} \}_-= i(\eta^{ad} S^{bc} + \eta^{bc} S^{ad} -
\eta^{ab} S^{cd} -\eta^{cd} S^{ab} )$).

It is useful to define for spinors one of the Casimirs of the Lorentz group - the handedness $\Gamma^{(d)}_S$ 
\begin{eqnarray}
\Gamma^{(d)}{ }_S:&=& i^{d/2} \;\; \prod_a \quad \sqrt{\eta^{aa}} \;\gamma^a \quad = 2^{d/2} \quad
\prod_a \quad S^{03} S^{12} \cdots S^{d-1,d},\quad {\rm if} \quad d=2n,\nonumber\\
\Gamma^{(d)}{ }_S:&=&  i^{(d-1)/2}\;\; \prod_a \quad \sqrt{\eta^{aa} }\;\gamma^a, \quad {\rm if} \quad d=2n+1, \nonumber\\
\{\Gamma^{(d)}{ }_S&,&S^{ab}\}_- = 0, \quad {\rm for} \quad {\rm any} \quad d.
\label{handedness}
\end{eqnarray}
We understand the product of $\gamma^a$'s in ascending order with respect to index $a$: $\gamma^0 \gamma^1
\gamma^2\gamma^3\gamma^5\cdots
\gamma^d$. It follows from Eqs.(\ref{clifford},\ref{handedness}) for any choice of the signature $\eta^{aa}$ that
$\Gamma^{d\dagger}{ }_S = \Gamma^{d}{ }_S $ and $(\Gamma^{d}{ }_S)^2 =I$.
One finds that in even dimensional spaces $\Gamma^{(d)}{ }_S$  and $\gamma^a$anti-commutes, while in odd dimensional 
spaces they commute
\begin{eqnarray}
\{\Gamma^{(d)}{ }_S, \gamma^a \}_+ &=& 0, \quad {\rm if} \quad d=2n,\nonumber\\
\{\Gamma^{(d)}{ }_S, \gamma^a \}_- &=& 0, \quad {\rm if} \quad d=2n+1.
\label{handcom}
\end{eqnarray}
Accordingly the Weyl equations of motion operator for a free spinor anticommutes with $\Gamma^{(d)}{ }_S$ in even 
dimensional spaces, while in odd dimensional spaces they commute
\begin{eqnarray}
\{\Gamma^{(d)}{ }_S, \gamma^a p_a \}_+ &=& 0,\quad {\rm if} \quad d=2n,\nonumber\\
\{\Gamma^{(d)}{ }_S, \gamma^a p_a \}_- &=& 0,\quad {\rm if} \quad d=2n+1.
\label{handweyl}
\end{eqnarray}
The handedness of Eq.(\ref{handedness}) is namely the Casimir also of the Poincar\'e group, which includes
besides $M^{ab}$ also $p^a$
\begin{eqnarray}
M^{ab} = L^{ab} +S^{ab}.
\label{mab}
\end{eqnarray}
$M^{ab}$ close the same Lie algebra of the Lorentz group as $S^{ab}$ do and
\begin{eqnarray}
\{\Gamma^{(d)}{ }_S, M^{ab}\}_- =0 = \{\Gamma^{(d)}{ }_S,p^a\}.
\label{Gpmcom}
\end{eqnarray}
One also finds
\begin{eqnarray}
\{ M^{ab}, \gamma^{c} p_c \}_- &=& 0,\quad {\rm for}\;{\rm any} \quad d.
\label{mabweylcom}
\end{eqnarray}
The handedness $\Gamma^{(d)}{ }_M $
\begin{eqnarray}
\Gamma^{(d)}{ }_M:&=& \alpha \epsilon_{a_1 a_2 \cdots a_{d-1 d}} M^{a_1 a_2} M^{a_2 a_3} \cdots M^{a_{d-1}a_d}
\quad {\rm for} \quad d=2n,
\label{handednessm}
\end{eqnarray}
is the Casimir of the Lorentz group ($\{\Gamma^{(d)}{ }_M, M^{ab}\}_- =0$), but it is not the 
Casimir of the Poincar\' e group ($\{\Gamma^{(d)}{ }_M, p^a\}_- = W^a$), and $W^a$ is the Pauli-Ljubanski 
vector\cite{bojannorma}
($W^a =  \rho \,\varepsilon^{ab}{}_{a_1 a_2\ldots a_{d-3}a_{d-2}} p_b S^{a_1 a_2}\ldots
    S^{a_{d-3}a_{d-2}}$, with
$\rho = \frac{2^{n-2}}{(d-2)!}$ for spinors  and 
$\rho = \frac{1}{2^{n-1}(n-1)!^2}$ for vectors ($S^i = \pm 1, 0$)).

\subsection{One Weyl spinor representation and the technique} 
\label{techniquespinors}

We briefly repeat in this subsection the main points of the technique for generating spinor 
representations from the Clifford algebra objects, following references\cite{WIholgernorma2002,WIholgernorma2003}.
The technique indeed origins from works, presented in the papers\cite{WInorma92,WInorma93,normaixtapa2001,pikanorma2002}.
In this paper we shall pay attention on even-dimensional spaces only.

Recognizing from  the Lorentz algebra relation that two Clifford algebra objects 
$S^{ab}, S^{cd}$ with all indices different 
commute, we  select the Cartan subalgebra of the algebra of the 
Lorentz group for $d=2n$ as follows 
\begin{eqnarray}
S^{03}, S^{12}, S^{56}, \cdots, S^{d-1\; d}, \quad {\rm if } \quad d &=& 2n.
\label{choicecartan}
\end{eqnarray}
%


Following refs.\cite{WIholgernorma2002,WIholgernorma2003} we introduce the graphic representation as follows
\begin{eqnarray}
\stackrel{ab}{(k)}:&=& 
\frac{1}{2}(\gamma^a + \frac{\eta^{aa}}{ik} \gamma^b),\nonumber\\
\stackrel{ab}{[k]}:&=&
\frac{1}{2}(1+ \frac{i}{k} \gamma^a \gamma^b),
\label{WIsignature}
\end{eqnarray}
where $k$ is a sign or a sigh times $i$ obeying  $k^2 = \eta^{aa} \eta^{bb}$, a doubled eigenvalue of
$S^{ab}$ on spinor states.

It follows that
\begin{eqnarray}
S^{ab}\stackrel{ab}{(k)}=\frac{1}{2}k \stackrel{ab}{(k)},\quad
S^{ab}\stackrel{ab}{[k]}=\frac{1}{2}k \stackrel{ab}{[k]},
\label{WIgrapheigen}
\end{eqnarray}
which means that we get the same objects back multiplied by the constant $\frac{1}{2}k$. 
This also means that
$\stackrel{ab}{(k)}$ and $\stackrel{ab}{[k]}$ acting from the left hand side on anything (on a
vacuum state $|\psi_0\rangle$, for example ) are eigenvectors of $S^{ab}$.

We further find 
\begin{eqnarray}
\gamma^a \stackrel{ab}{(k)}&=&\eta^{aa}\stackrel{ab}{[-k]},\quad
\gamma^b \stackrel{ab}{(k)}= -ik \stackrel{ab}{[-k]}, \nonumber\\
\gamma^a \stackrel{ab}{[k]}&=& \stackrel{ab}{(-k)},\quad
\gamma^b \stackrel{ab}{[k]}= -ik \eta^{aa} \stackrel{ab}{(-k)}.
\label{WIgraphgammaaction}
\end{eqnarray}
It also follows that
$
S^{ac}\stackrel{ab}{(k)}\stackrel{cd}{(k)} = -\frac{i}{2} \eta^{aa} \eta^{cc} 
\stackrel{ab}{[-k]}\stackrel{cd}{[-k]}$, 
$S^{ac}\stackrel{ab}{[k]}\stackrel{cd}{[k]} = \frac{i}{2}  
\stackrel{ab}{(-k)}\stackrel{cd}{(-k)}$, 
$S^{ac}\stackrel{ab}{(k)}\stackrel{cd}{[k]} = -\frac{i}{2} \eta^{aa}  
\stackrel{ab}{[-k]}\stackrel{cd}{(-k)}$, 
$S^{ac}\stackrel{ab}{[k]}\stackrel{cd}{(k)} = \frac{i}{2} \eta^{cc}  
\stackrel{ab}{(-k)}\stackrel{cd}{[-k]}$.
It is useful to recognize that $\stackrel{ab}{(k)}$ are nilpotent operators, which are not hermitean:
$\stackrel{ab}{(k)}\stackrel{ab}{(k)}=0,$ $
\stackrel{ab}{(k)}^{\dagger}=\eta^{aa}\stackrel{ab}{(-k)}$ while $\stackrel{ab}{[k]}$ are projectors and
hermitean operators: $\stackrel{ab}{[k]}\stackrel{ab}{[k]} = \stackrel{ab}{[k]}$,
$\stackrel{ab}{[k]}^{\dagger}= \stackrel{ab}{[k]}$.

According to ref.\cite{WIholgernorma2002},  we define a vacuum state $|\psi_0>$ so that one finds
$$< \;\stackrel{ab}{(k)}^{\dagger}
 \stackrel{ab}{(k)}\; > = 1\qquad,\qquad 
< \;\stackrel{ab}{[k]}^{\dagger}
 \stackrel{ab}{[k]}\; > = 1.$$

Taking the above equations into account it is easy to find a Weyl spinor irreducible representation
for $d$-dimensional space, with $d$ even (or odd). (We advise the reader to see the 
references\cite{WIholgernorma2002,WIholgernorma2003}.) 

For $d$ even, we simply set the starting state as a product of $d/2$, let us say, only nilpotents 
$\stackrel{ab}{(k)}$, one for each $S^{ab}$ of the Cartan subalgebra  elements (Eq.(\ref{choicecartan})),  applying it 
on an (unimportant) vacuum state\cite{WIholgernorma2002}. 
Then the generators $S^{ab}$, which do not belong 
to the Cartan subalgebra, applied to the starting state from the left hand side, 
 generate all the members of one
Weyl spinor.  
\begin{eqnarray}
\stackrel{0d}{(k_{0d})} \stackrel{12}{(k_{12})} \stackrel{35}{(k_{35})}\cdots \stackrel{d-1\;d-2}{(k_{d-1\;d-2})}
\psi_0 \nonumber\\
\stackrel{0d}{[-k_{0d}]} \stackrel{12}{[-k_{12}]} \stackrel{35}{(k_{35})}\cdots \stackrel{d-1\;d-2}{(k_{d-1\;d-2})}
\psi_0 \nonumber\\
\stackrel{0d}{[-k_{0d}]} \stackrel{12}{(k_{12})} \stackrel{35}{[-k_{35}]}\cdots \stackrel{d-1\;d-2}{(k_{d-1\;d-2})}
\psi_0 \nonumber\\
\vdots \nonumber\\
\stackrel{0d}{[-k_{0d}]} \stackrel{12}{(k_{12})} \stackrel{35}{(k_{35})}\cdots \stackrel{d-1\;d-2}{[-k_{d-1\;d-2}]}
\psi_0 \nonumber\\
\stackrel{od}{(k_{0d})} \stackrel{12}{[-k_{12}]} \stackrel{35}{[-k_{35}]}\cdots \stackrel{d-1\;d-2}{(k_{d-1\;d-2})}
\psi_0 \nonumber\\
\vdots 
\label{WIgraphicd}
\end{eqnarray}
All the states of one irreducible Weyl representation have the same handedness $\Gamma^{(d)}{ }_S $, 
since $\{ \Gamma^{(d)}{ }_S, S^{ab}\}_- = 0$, 
which is easily calculated 
by multiplying from the left hand side the starting
state by $\Gamma^{(d)}{ }_S$ of Eq.(\ref{handedness}). 
We chose 
the simplest phase, setting all  phases equal to one.

We speak about left handedness when $\Gamma= -1$ and right
handedness when $\Gamma =1$ for either $d$ even or odd. 

In the reference\cite{pikanorma2002} and in the talk of one of the authors {N.M.B.}
the Weyl representation of left handedness 
in $d=1+ 13$ dimensional space is presented by 
the presented technique. Each vector of the representation is 
analyzed in terms of the quantum numbers of the subgroups $SO(1,7)$ (and accordingly further on in terms of 
$SO(1,3) \times SU(2) \times U(1) $) and
$SO(6)$ (and accordingly of $SU(3) \times U(1)$) for free spinors.  In the reference \cite{pikanorma2002} 
$ \Gamma^{(1,13)}$ (it is chosen to be $-1$), 
$\Gamma^{(1,7)}$, $\Gamma^{(6)}$, $\Gamma^{(1,3)}$ and $\Gamma^{(4)}$ are presented, as well as
all the Cartan subalgebra eigenvalues of the group (and accordingly of the subgroups $SO(1,3), SU(3), SU(2)$ and the
two $U(1)$'s.
One finds that $\Gamma^{(1,13)}$ anti commutes with the Weyl equation  ($ \{\Gamma^{(1,13)},\gamma^a p_a\}_+=0$), while 
each of the vectors of the Weyl representations has well defined handedness of all the $\Gamma^{(d)}$'s.
One Weyl left handed spinor representation contains the left handed weak charged quarks and leptons, the right handed weak
chargeless quarks and leptons, as well as the right handed weak charged anti-quarks and anti-leptons 
and the left handed weak chargeless anti-quarks and anti-leptons.

The question then arises: If coordinates are subsequently compactified, for example from $ M^d, d=1+13$ to
$M^{1+7} \times M^6$, can  one keep
track of the handedness of the subspaces to guarantee the masslessness of the spinors after the compactification?
We shall study a simpler case of a flat $M^{1+5}$ with the $SO(1,5)$ symmetry, in which  the space is 
compactified into a flat $M^{1+3}$ of the symmetry $SO(1,3)$ 
and a flat torus with torsion, and show that there exists the operator in  $S^1 \times S^1$, which indeed has the 
properties of the operator of handedness. We shall first in the next subsection \ref{so15} represent the left handed
Weyl spinor in $d=1+5$.

\subsection{Weyl representation for $SO(1,5)$}
\label{so15}

We present in this subsection the quantum numbers of the Weyl left handed spinor of $SO(1,5)$, 
using the technique\cite{WIholgernorma2002,WIholgernorma2003}. We make a choice of the Cartan subalgebra operators
$S^{03}, S^{12}, S^{5,6}$. According to Eq.(\ref{handedness}) $\Gamma^{(1,5)} = \gamma^0 \gamma^1 \gamma^2 \gamma^3 
\gamma^5 \gamma^6 = 8i S^{03} S^{12} S^{56} = \Gamma^{(1,3)} \times \Gamma^{(2)},$ with $\Gamma^{(1,3)} =
i \gamma^0 \gamma^1 \gamma^2 \gamma^3 = -4i S^{03} S^{12}$ and $\Gamma^{(2)} = -i \gamma^5 \gamma^6 = -2 S^{56}$.

Following the technique one  finds the left handed Weyl spinor by making a choice of the starting state 
$\stackrel{03}{(+i)} \stackrel{12}{(+)} \stackrel{56}{(+)}$. There are four basic states, two right handed spinors 
with respect 
to the subgroup $SO(1,3)$ with the ''$S^{56}$ charge''
equal to $+1/2$ and two left handed spinors with respect to the group $SO(1,3)$ with the ''$S^{56}$ charge'' 
equal to $-1/2$.
\begin{table} 
\centering
\begin{tabular}{|r|c||c||c|c|c||c|c|c|}
\hline
i&$$&$\psi_i$&$\Gamma^{(1,5)}$&$ \Gamma{(1,3)}$&$\Gamma^{(2)}$&
$S^{12}$&$S^{03}$&$S^{56}$\\
\hline\hline
1&$\psi_1$&$\stackrel{03}{(+i)}\stackrel{12}{(+)}|\stackrel{56}{(+)}$
&-1&+1&-1&+1/2&+i/2&+1/2\\
\hline 
2&$\psi_2$&$\stackrel{03}{[-i]}\stackrel{12}{[-]}|\stackrel{56}{(+)}$
&-1&+1&-1&-1/2&-i/2&+1/2\\
\hline\hline
3&$\psi_3$&$\stackrel{03}{[-i]}\stackrel{12}{(+)}|\stackrel{56}{[-]}$
&-1&-1&+1&+1/2&-i/2&-1/2\\
\hline 
4&$\psi_4$&$\stackrel{03}{(+i)}\stackrel{12}{[-]}|\stackrel{56}{[-]}$
&-1&-1&+1&-1/2&+i/2&-1/2\\
\hline\hline
\end{tabular}
\caption{%
Four states of the left handed Weyl representation in $SO(1,5)$ is presented.
The first two states are right handed with respect to $SO(1,3)$ ($\Gamma^{(1,3)} =1$) and the last two are left 
handed ($\Gamma^{(1,3)} =-1$). The handedness of the rest of the space ($\Gamma^{(2)}$) is opposite to the handedness
of $SO(1,3)$. The eigenvalues of the Cartan subalgebra $S^{12}, S^{03}, S^{56}$ are also presented.}
\end{table}

Table I demonstrates clearly what Witten said\cite{witten81}: A Weyl spinor of one handedness 
always contains with respect to subgroups of the starting group handedness of both types. But we also noticed that
the two representations of different  handedness differ  in the $S^{56}$ ''charge''. Can we make use of it?

\subsection{Solutions of free Weyl equations of motion}
\label{weylso(1,5)}

A free spinor in $(d=1 + 5)$-dimensional space with a momentum $p^a= (p^0,\overrightarrow{p})$, obeying the Weyl equations of
motion 
\begin{eqnarray}
(\gamma^a p_a =0)\psi,
\label{weylfree}
\end{eqnarray}
is in $M^6$ a plane wave, while the spinor part is  for a generic $p^a$ a 
superposition of all the vectors, presented on Table I 
\begin{eqnarray}
\psi(p)&=& e^{-ip^ax_a} \;{\cal N}\; \biggl\{\alpha \stackrel{03}{(+i)}\stackrel{12}{(+)}\stackrel{56}{(+)} +
\beta \stackrel{03}{[-i]}\stackrel{12}{[-]}\stackrel{56}{(+)} \biggr. \nonumber\\
&+& \frac{p^5 +i p^6}{(p^5)^2 + (p^6)^2} [ -\alpha(p^0-p^3) +\beta(p^1-ip^2)] 
\stackrel{03}{[-i]}\stackrel{12}{(+)}\stackrel{56}{[-]}  \nonumber\\
&+& \biggl.\frac{p^5 +i p^6}{(p^5)^2 + (p^6)^2} [ \alpha(p^1+ip^2) -\beta(p^0-p^3)]
\stackrel{03}{(+i)}\stackrel{12}{[-]}\stackrel{56}{[-]}\biggr\}, 
\label{weylin6}
\end{eqnarray}
with the condition $(p^0)^2 = (\overrightarrow{p})^2$ and for any $(\alpha/\beta)$  
(Eq.(\ref{weylin6}) therefore offers two independent solutions). 
If $p^5=0=p^6$, $\alpha$ and $\beta $ are related ($\beta/\alpha = (p^1 +ip^2)/(p^0 + p^3)$) and the solution can be
either left handed with respect to $SO(1,3)$ ($\Gamma^{(1,3)} =-1$) with $S^{56} = -1/2$ or right handed with respect
to $SO(1,3)$ ($\Gamma^{(1,3)} = +1$) with $S^{56} = +1/2$. Looking from the point of view of the
four-dimensional subspace, we call the eigenvalue of $S^{56}$ a charge.

In even-dimensional spaces a left handed spinor ($\Gamma^{(1,5)}=-1$) is mass protected\cite{WIholgernorma2002} 
(while in odd-dimensional spaces it is not\cite{WIholgernorma2002}) and an
interacting field (gravity) can not for $d$ even make spinors massive. In the four-dimensional subspace, however, nonzero
components of the momentum $p^a$ in higher than four dimensions (nonzero either $p^5$ or $p^6$) manifest as a mass term,
since they cause a superposition of  the left and the  right handed components of  $\Gamma^{(1,3)}$).  A spinor with nonzero
components of a momentum $p^a$ in only $d=4$, manifests in the four-dimensional subspace as a massless either left or 
right handed particle.

Let us repeat the properties of a free Weyl spinor from Eqs.(\ref{handedness},\ref{handweyl},\ref{mabweylcom},\ref{Gpmcom}),
which we shall need in the next sections, keeping in mind that we are now in even dimensional spaces and parts of spaces
\begin{eqnarray}
\{ \Gamma^{(d)}{ }_S, \gamma^a p_{a} \}_+ &=& 0, \nonumber\\
\{ M^{ab}, \gamma^c p_{c} \}_- &=& 0, \nonumber\\
\{ S^{ab}, \Gamma^{(d)}{ }_S\}_- &=& 0, \nonumber\\
\{ M^{ab}, \Gamma^{(d)}{ }_S\}_- &=& 0,\nonumber\\
\Gamma^{(d)}{ }_S \psi_{sol}= - \psi_{sol}&,& \quad {\rm for}\quad {\rm any} \quad a,b,c.
\label{weylfreeprop}
\end{eqnarray}
We also notice
\begin{eqnarray}
\{ \Gamma^{(2)}{ }_S, \gamma^h p_{h} \}_+ &=& 0,\nonumber\\
\{ M^{56}, \gamma^h p_{h} \}_- &=& 0, \nonumber\\
\{ S^{56}, \Gamma^{(2)}{ }_S\}_- &=& 0, \nonumber\\
\{ M^{56}, \Gamma^{(2)}{ }_S\}_- &=& 0,\nonumber\\
\Gamma^{(2)} \psi_{sol}&=& \pm \psi_{sol}, \nonumber\\
S^{56} \psi_{sol} &=& \pm \frac{1}{2}\psi_{sol}, \nonumber\\
\{ \Gamma^{(2)}{}_S, S^{56} \}_+ \psi_{sol}  &=& - \psi_{sol},\quad {\rm for} \quad h =5,6.
\label{weylfreeprop56}
\end{eqnarray}
$\psi_{sol}$ denotes the solution of the Weyl equations of motion in $d= (1+5)$.

\section{Weyl spinor in spaces with curvature and torsion}
\label{kaluzaklein}

We repeat some well known relations for spaces with (only) gravity,
manifesting as a gauge fields of the Poincar\ 'e group through vielbeins and spins-connections, needed 
later for our proof.

We let a spinor interact with a gravitational field through vielbeins $f^a{}_{\mu}$ and spin connections
$\omega_{ab\mu}$
\begin{eqnarray}
(\gamma^a p_{0a} =0)\psi, \quad p_{0a} = f^{\mu}{}_a p_{0\mu}, \quad p_{0\mu} = p_{\mu} - \frac{1}{2}
S^{ab} \omega_{ab \mu}.
\label{weylgravity}
\end{eqnarray}
Here $a,b,..$ denote a tangent space index, while $\alpha, \beta, \mu,\nu,..$ denote an Einstein index.

Taking into account that $\gamma^a \gamma^b = \eta^{ab}- 2i S^{ab}$,  $\{\gamma^a, S^{bc}\}_- = i
(\eta^{ac} \gamma^b - \eta^{ab} \gamma^c)$, and $e^a{}_{\mu} f^{\mu}{}_b = \delta^a{}_b,\; 
e^a{}_{\mu} f^{\nu}{}_a = \delta^{\nu}_{\mu} $, one easily finds
\begin{eqnarray}
(\gamma^a p_{0a} )^2 = p_{0}^a p_{0a} + \frac{1}{2} S^{ab} S^{cd} {\cal R}_{abcd} + 
S^{ab} {\cal T}^{\beta}{}_{ab} p_{0 \beta}, 
\label{RandT}
\end{eqnarray}
with
\begin{eqnarray}
p_{0a} &=& f_a{}^{\mu} (p_{\mu} - \frac{1}{2} S^{cd} \omega_{cd \mu}), \nonumber\\
{\cal R}_{abcd} &=& f^{\alpha}{}_a f^{\beta}{}_b (- \omega_{cd\beta,\alpha} + \omega_{cd\alpha,\beta} + 
\omega_{ce\alpha}\omega^e{}_{d\beta} - \omega_{ce\beta}\omega^e{}_{d\alpha}), \nonumber\\
{\cal T}^{\beta}{}_{ab} &=& f^{\alpha}{}_a f^{\beta}{}_{b,\alpha}- f^{\alpha}{}_b f^{\beta}{}_{a,\alpha}
+ f^{\beta}{}_c (f^{\alpha}{}_a \omega^c{}_{b\alpha} - f^{\alpha}{}_b \omega^c{}_{a \alpha}), \nonumber\\
{\cal T}^a{}_{\mu \nu} &=& e^b{}_{\mu} e^c{}_{\nu} e^a{}_{\beta}  {\cal T}^{\beta}{}_{bc} = 
e^a{}_{\mu,\nu} -  e^a{}_{\nu,\mu} + \omega^a{}_{b\mu} e^b{}_{\nu}- \omega^a{}_{b\nu} e^b{}_{\mu}.
\label{RandTdet}
\end{eqnarray}

A term  $\gamma^h p_{0h} \psi_{sol} =0,\; h \in \{ n+1,..,d \},$ where $\psi_{sol}$ is defined by 
$(\gamma^a p_{0a} =0)\psi_{sol}= (\gamma^m p_{0m} + \gamma^h p_{0h} =0)\psi_{sol}, 
{\rm with} \quad m \in \{0,1,2,3, n-1\},\;
h \in \{ n+1,..,d \},$
 manifests 
 as the n-dimensional mass term.

If all the dimensions except $n$ are compactified, then one can assure the masslessness of spinors 
in the n-dimensional part of the space, if a way of 
compactifying all but $n=2k$ dimensions can  be found,
so that the handedness in the compact part of the space is well defined.  
Then, starting in $(d=2m)$-dimensional space with a spinor of only one handedness and assuring that 
the spinor has in the compactified part of the space well defined handedness,
 then the handedness in $n$, being the product 
of the handedness in $d$ ($\Gamma^{(d)}{ }_S$) and the handedness in the compactified part of  space 
$d-n$ ($\Gamma^{(d-n)}{ }_S$), is also well defined ($\Gamma^{(n)}{ }_S =
\Gamma^{(d)}{ }_S \times \Gamma^{(d-n)}{ }_S$),
guaranteeing the masslessness of spinors in the $n$-dimensional subspace.

We shall demonstrate in subsection \ref{examples1s1} that in case when $M^{1+5}$ compactifies
to $M^{1+3}$ and a flat torus $S^1 \times S^1$ with a torsion, in $d=1+3$ massless spinors  of well defined  handedness
exist. In subsection \ref{gaugefield} we demonstrate appearance of an $U(1)$ gauge field in the 
noncompactified part of space.

We present in what follows some well known relations, needed in section \ref{examples1s1}.

Requiring that the total covariant derivative\cite{milutin2002} of a vielbein $e^a{}_{\mu}$
\begin{eqnarray}
e^a{}_{\nu ; \mu} = e^a{}_{\nu , \mu} - \omega^a{}_{b \nu} e^b{}_{\mu} - \Gamma^{\alpha}{}_{\mu \nu} e^a{}_{\alpha} 
\label{WItotaldereamu}
\end{eqnarray}
is equal to zero
\begin{eqnarray}
e^a{}_{\nu ; \mu} = 0, 
\label{totaldereamuzero}
\end{eqnarray}
 it follows  that the covariant derivative of a metric tensor $g_{\alpha \beta}= 
e^a{}_{\alpha} e_{a \beta}$  is also equal to zero $g_{\alpha \beta; \mu}= 0 $, as it should be. Then
$ \Gamma^{a}{}_{\mu \nu} = \Gamma^{\alpha}{}_{\mu \nu} e^a{}_{\alpha} =
e^a{}_{\mu , \nu} - \omega^a{}_{b \nu} e^a{}_{\mu} $, which is 
in agreement with the definition of  the torsion in Eq.(\ref{RandTdet}), provided that
\begin{eqnarray}
{\cal T}^{\alpha}{}_{\mu \nu} = \Gamma^{\alpha}{}_{\mu \nu} - \Gamma^{\alpha}{}_{\nu \mu}. 
\label{torandGamma}
\end{eqnarray}
In a flat space with a spin connection $\omega_{ab \mu} = 0$, the equation for $\Gamma^{\alpha}{}_{\nu \mu}$ simlifies to
\begin{eqnarray}
\Gamma^{\alpha}{}_{\nu \mu} = f^{\alpha}{}_{a} e^a{}_{\nu , \mu}. 
\label{totaldereamu}
\end{eqnarray}

One then finds that a covariant derivative of a torsion in a flat space (where a spin connection is equal to zero)
is equal to the ordinary derivative of a torsion
\begin{eqnarray}
{\cal T}^a{}_{\mu \nu ; \rho} &=& 
e^a{}_{\mu , \nu , \rho}- e^a{}_{\nu , \mu , \rho} = {\cal T}^a{}_{\mu \nu , \rho}.
\label{covtorsion}
\end{eqnarray}
The Bianchi identities
\begin{eqnarray}
{\cal T}^a{}_{\mu \nu , \rho} + {\cal T}^a{}_{\rho \mu , \nu } + {\cal T}^a{}_{\nu \rho , \mu} =0
\label{Tbianchi}
\end{eqnarray}
connect ordinary derivatives of the torsion components.

In a flat space with only a torsion the metric tensor $g^{\alpha \beta} = f^{\alpha}{}_a f^{\beta a}$ is
equal to the Minkowski metric $\eta^{\alpha \beta}$ and the Christoffel symbols $\{{}^{\alpha}_{\beta \gamma}\}$ 
are accordingly equal to zero.

Let us make, in a flat space with only a torsion, a choice of the following  Lagrange density for the torsion field
\begin{eqnarray}
{\cal L}_T &=& a {\cal T}^a{}_{bc} {\cal T}_a{}^{bc } + b {\cal T}^b{}_{ac} {\cal T}^a{}_{b}{}^{c} + c
{\cal T}^a{}_{ba} {\cal T}^{cb}{}_{c } \nonumber\\
&=&  a {\cal T}^{\mu}{}_{\alpha \beta} {\cal T}_{\mu}{}^{\alpha \beta }
+ b {\cal T}^{\mu}{}_{\alpha \beta} {\cal T}^{\alpha}{}_{\mu}{}^{\beta} + c
{\cal T}^{\mu}{}_{\alpha \mu} {\cal T}^{\beta \alpha}{}_{\beta},
\label{WIlagrange}
\end{eqnarray}
with arbitrary coefficients $a,b,c$. The action leads after the variation of the action
\begin{eqnarray}
S= \int \; d^d x {\cal L}_T (e^a{}_{\mu}, e^a{}_{\mu , \nu}),
\label{WIaction}
\end{eqnarray}
with  respect to  vielbeins $e^a{}_{\mu}$ to the equations of motion for the vielbeins.

\section{Spinor on $ M^{1+3} \times (S^1 \times S^1)$-torus with torsion}
\label{examples1s1}

Starting in $d=1+5$ with a left handed spinor, we  compactified $M^{1+5}$ to $M^{1+3}$ and a flat 
torus ($S^1 \times S^1$) with a torsion. In order to end up in $d=4$ with a spinor of well defined handedness, 
which would  guarantee the masslessness of a spinor,  a proposal for the Kaluza-Klein 
charges implementing the $U(1)\times U(1)$ group of the flat Riemann-space torus $S^1 \times S^1$ 
is needed. They should in $d=2$
commute with the Weyl equations of motion  with torsion, as well as anti-commute with 
the operator of handedness $\Gamma^{(2)}$. 

We propose the following two charges ${}^{''}M^{56}_{(i)}{}^{''}$, $(i)=5,6$,
\begin{eqnarray}
{}^{''}M^{56}_{(i)}{}^{''} = p^{(i)}  - \frac{m_i}{2}\Gamma^{(2)} , \quad i= \{5,6\}. 
\label{m56}
\end{eqnarray}
for which we shall prove that they have the properties 
\begin{eqnarray}
\{{}^{''}M^{56}_{(i)}{}^{''}, \gamma^h p_{0h}\}_- &=& 0, \nonumber\\
\{\Gamma^{(2)}, {}^{''}M^{56}_{(i)}{}^{''}\}_- &=& 0, \nonumber\\
{}^{''}M^{56}_{(i)}{}^{''}\psi_{sol} &=& const. \psi_{sol}, \nonumber\\ 
\{\Gamma^{(2)},{}^{''}M^{56}_{(i)}{}^{''}\}_+ \psi_{sol} &=& const. \psi_{sol}, \quad (i)=5,6, h=5,6,\nonumber\\
\{\Gamma^{(2)}, p_{0h} \gamma^h\}_+ &=& 0, \nonumber\\
\Gamma^{(2)} \psi_{sol} &=& \psi_{sol}.
\label{hand56}
\end{eqnarray}

We use the notation ${}^{''}M^{56}_{(i)}{}^{''}$ in order to remind us that in a noncompactified flat space
 the angular momentum $M^{56}$ is the constant of motion as all $M^{ab}$ are (Eq.\ref{weylfreeprop}), while 
in a flat torus $S^1 \times S^1$ the momentum $M^{56}$ is not a symmetry operator of the torus and can 
accordingly not be a candidate for a  Kaluza-Klein charge.

In a flat torus $S^1 \times S^1$ the Riemann tensor is equal to zero,
leading to:
\begin{eqnarray}
{\cal R}_{56(5)(6)} =0= \omega_{56(5),(6)} - \omega_{56(6),(5)},
\label{Rtorus1and1}
\end{eqnarray}
with $5,6$ the two tangent  indices $(a,b)$ and $(5),(6)$ the two Einstein indices $(\mu,\nu)$ of 
the two compactified dimensions.
It follows from Eq.(\ref{Rtorus1and1}) that 
$$\ (\omega_{5 6 (5)}, \omega_{5 6 (6)} ) = grad \Phi.$$ 
We can make a choice of the gauge in which
\begin{eqnarray}
(\omega_{56(5)}, \omega_{56(6)}) = (0,0).
\label{Rtorus1and10}
\end{eqnarray}
We parametrize the zweibein as follows
\begin{eqnarray}
e^a{}_{\mu} = 
\begin{pmatrix}\cos \phi   & \sin \phi \cr
- \sin \phi & \cos \phi \cr\end{pmatrix},\quad
f^{\mu}{}_{a} = 
\begin{pmatrix}\cos \phi   & -\sin \phi \cr
\sin \phi & \cos \phi \cr\end{pmatrix},
\label{f1and1}
\end{eqnarray}
with $\phi =\phi(x^{(5)},x^{(6)})$.
Then it follows from Eq.(\ref{RandTdet}) for the torsion
\begin{eqnarray}
{\cal T}^a{}_{(5)(6)} &=& e^a_{(5),(6)} - e^a_{(6),(5)},
\label{Texpla}
\end{eqnarray}
leading to
\begin{eqnarray}
{\cal T}^5{}_{(5)(6)} &=& -\sin \phi \phi_{,(6)} + \cos \phi \phi_{,(5)},\quad
{\cal T}^6{}_{(5)(6)} = \cos \phi \phi_{,(6)} + \sin \phi \phi_{,(5)}, \nonumber\\
{\cal T}^{(5)}{}_{(5)(6)} &=& \phi_{,(5)}, \quad
{\cal T}^{(6)}{}_{(5)(6)} = \phi_{,(6)}.  
\label{Texplmu}
\end{eqnarray}

In a flat space the covariant derivative of a torsion is equal to the ordinary derivative 
of a torsion (Eq.\ref{covtorsion}). The Jacobi identities then read 
${\cal T}^{\mu}{}_{\alpha \beta ,\gamma} + {\cal T}^{\mu}{}_{\gamma \alpha,\beta} 
+{\cal T}^{\mu}{}_{ \beta \gamma, \alpha} =0$.

We shall require 
that the ordinary derivative of a torsion is
zero
\begin{eqnarray}
{\cal T}^{\mu}{}_{\alpha \beta,\gamma}= 0,
\label{Tmucov}
\end{eqnarray}
from where it follows
\begin{eqnarray}
\phi_{,(5),(5)} = 0, \quad \phi_{,(5),(6)} =0,\nonumber\\
\phi_{,(6),(5)} = 0, \quad \phi_{,(6),(6)} =0. 
\label{phiexpl}
\end{eqnarray}
The solution of these equations is
\begin{eqnarray}
\phi = \alpha x^{(5)} + \beta x^{(6)} + {\rm constant},
\label{solution}
\end{eqnarray}
for a generic choice of $\{\alpha, \beta \}$.
Since $\phi$ is an angle with the periodic properties $\phi(0,x^{(6)}) = \phi(2\pi, x^{(6)}), \;
\phi(x^{(5)},0) = \phi(x^{(5)}, 2\pi)$, it must be that $\alpha $ and $\beta $ are two integers and we find
\begin{eqnarray}
e^5{}_{(5)} = \; \; \cos(m x^{(5)} +n x^{(6)}), \quad e^6{}_{(5)} = \sin(m x^{(5)} +n x^{(6)}), \nonumber\\
e^6{}_{(5)} = -\sin(m x^{(5)} +n x^{(6)}), \quad e^6{}_{(6)} = \cos(m x^{(5)} +n x^{(6)}), 
\label{zweiexpl}
\end{eqnarray}
where $m$ and $n$ are any integers.

We shall  prove later that the requirement of Eq.(\ref{Tmucov}) is in agreement with what the solutions of the
equations of motion, following from the action of Eq.\ref{WIaction}, give.

{\it The proof that the two Kaluza-Klein charges} ${}^{''}M^{56}_{(i)}{}^{''} = p^{(i)}  - \frac{m_i}{2}
\Gamma^{(2)} , \;\; i= \{5,6\} $ {\it commute with the Weyl equations of motion in a flat torus with the torsion}:

We have to prove that all the equations (\ref{hand56}) are fulfilled with ${}^{''}M^{56}_{(i)}{}^{''}$ from
Eq.(\ref{m56}).
One easily sees that the first and the third equation of Eq.(\ref{hand56}) are fulfilled, since $\Gamma^{(2)}{ }_S$
anti commutes with $\gamma^h,\; h=5,6$ and commutes with $p^h$ and itself. We check the second equation by
taking into account Eqs.(\ref{f1and1},\ref{zweiexpl})
\begin{eqnarray}
\{p^{(i)} - \frac{m_i}{2} \Gamma^{(2)}, \gamma^h f_h{}^{\mu} p_{\mu} \}_- &=& 0 \nonumber\\
= m_i i \{ \gamma^5 (-p_6) + \gamma^6 p_5 \}
 &-& \frac{m_i}{2} 2 \Gamma^{(2)}{ }_S \gamma^h p_h 
 \nonumber\\
&=& m_i \Gamma^{(2)}{ }_S \gamma^h p_h - m_i \Gamma^{(2)}{ }_S \gamma^h p_h,
\label{checkmab}
\end{eqnarray}
which shows that also the second equation is fulfilled. We assume that for the ground state solution 
\begin{eqnarray}
p^h \gamma_h \psi_{sol} =0, \quad {\rm with} \quad p_{(5)}\psi_{sol}=0=p_{(6)}\psi_{sol},
\label{checkpsi}
\end{eqnarray}
which means that the ground state $\psi_{sol}$ is independent of $p_{(5)}$ and $p_{(5)}$. 
In the internal (spinor) part, 
$\psi_{sol}$ can contain any of the four states of Table I and accordingly we conclude
\begin{eqnarray}
 \Gamma^{(2)}{}_S \psi_{sol} &=& (\pm) \psi_{sol}, \nonumber\\
 {}^{''}M^{56}_{(i)}{}^{''} \psi_{sol} &=&  - \frac{m_i}{2} 
 \Gamma^{(2)}{ }_S \psi_{sol},\nonumber\\
 \{ \Gamma^{(2)}{ }_S&,& {}^{''}M^{56}_{(i)}{}^{''} \}_+ \psi_{sol} = (1- \frac{m_i}{2}) \Gamma^{(2)}{ }_S \psi_{sol},
\label{checkpsif}
\end{eqnarray}
{\it which means that if we choose an eigenspace of the Kaluza-Klein charge} ${}^{''}M^{56}_{(i)}{}^{''} $, {\it we 
only get one eigenvalue for}  $\Gamma^{(2)}{ }_S$ {\it on that subspace}.

Let us also check  that $(\gamma^h p_h)^2 \psi_{sol}=0$, which is needed to guarantee the masslessness of spinors in the four
dimensional subspace. Taking into account Eqs.(\ref{weylgravity},\ref{RandT}), with $R_{abcd}=0, \; a,b,c,d =5,6 $,
for a flat torus,  we find that for the ground state solution, which is independent
of $x^{(5)}$ and $x^{(6)}$,  
\begin{eqnarray}
(\gamma^h p_h)^2 \psi_{sol}= (p^h p_h + 
S^{ab} {\cal T}^{\sigma}{}_{ab} p_{\sigma} ) \psi_{sol}=0, 
\label{massless}
\end{eqnarray}
contributing no mass term in the four-dimensional part of  space.

Let us at the end look for the equations of motion for the torsion field, making use of
the Lagrange density, proposed in Eqs.(\ref{WIlagrange}, \ref{WIaction}).
One easily finds
\begin{eqnarray}
{\cal L}_T = (2a +b+c) [(\phi{}_{,(5)})^2 + (\phi{}_{,(5)})^2]. 
\label{actionex}
\end{eqnarray}
Taking $\phi(x^{(5)}, x^{(6)})$ as variational fields we end up with the equations of motion
\begin{eqnarray}
2(2a +b+c) [\phi{}_{,(5),(5)} + (\phi{}_{,(5),(6)}]=0, 
\label{equation}
\end{eqnarray}
in agreement with what we obtained in Eqs.(\ref{solution},\ref{zweiexpl}).

\subsection{Appearance of $U(1)$ gauge field}
\label{gaugefield}

We demonstrate  how a gauge field $U(1)$  appears in the four-dimensional part of space after
the compactification of space $M^{1+5}$ into the flat torus $S^1 \times S^1$ with the torsion from
section \ref{examples1s1}, if spin connections are appropriately chosen. 

To simplify we take all the vielbeins and spin connections,
contributing to  gravity 
in $M^{1+3}$, equal to zero. We only let to be non zero those components of the Riemann tensor,
which contribute in $M^{1+3}$ to the $U(1)$
gauge field
\begin{eqnarray}
{\cal R}_{\sigma \rho  h k}&=&0, \quad {\rm with} \;\;h,k \in \{5,6\}, \quad \sigma,\rho  \in \{ (5), (6)\}, \nonumber\\
{\cal R}_{\alpha \beta h k}&\ne& 0, \quad {\rm with} \;\;h,k \in \{5,6\}, \quad \alpha,\beta \in 
\{ (0),\cdots (3)\}, \nonumber\\
{\cal R}_{\alpha \beta m n}&=& 0, \quad {\rm with} \;\;m,n \in \{0,\dots,3\}, \quad \alpha,\beta \in 
\{ (0),\cdots (3)\}, \nonumber\\
\nonumber\\
e^a{}_{\mu} &=& 
\begin{pmatrix}\delta^m{}_\alpha &  &0 & \cr
& 0& \cos \phi   & \sin \phi \cr
& & - \sin \phi & \cos \phi \cr\end{pmatrix}.
\label{gaugeu1def}
\end{eqnarray}
With the above choice of the Riemann tensor and the vielbein tensor we find
\begin{eqnarray}
{\cal R}_{\alpha \beta 5 6} = \omega_{56\alpha,\beta} - \omega_{56 \beta, \alpha} = :F_{\alpha \beta},\nonumber\\
{\cal R}_{abcd}\eta^{ac} \eta^{bd} =0,\nonumber\\
{\cal R}^2 = {\cal R}_{\alpha \beta 5 6} {\cal R}^{\alpha \beta 5 6} = F_{\alpha\beta} F^{\alpha \beta}. 
\label{gaugeu1der1}
\end{eqnarray}
The spin connection $A^{\alpha} :=\omega_{56}{}^{\alpha}$ appears as a gauge field $U(1)$ in $M^{(1+3)}$. Since the
compactification of two dimensions of the $d=6$-dimensional space into a flat torus with the torsion presented in 
Eq. (\ref{gaugeu1def}) makes a Weyl  spinor of one handedness, which is accordingly massless,  
in the four-dimensional part of space possible,
the Weyl equations 
of motion in the four dimensional part of space  then  read
\begin{eqnarray}
\gamma^m p_{0m} \psi =0, \quad {\rm with} \quad m \in \{0,\cdots,3\} \nonumber\\
p_{0m} = p_m - \tau A_m, \quad \tau = \frac{1}{2} S^{56}. 
\label{weylin4}
\end{eqnarray}

\section{Our conclusions}
\label{WIconclusion}

Starting with a Weyl spinor of only one handedness in a flat Riemann space $M^{1+5}$, we were able to find 
a compactified  $S^1 \times S^1$ flat Riemann space torus with a torsion (the spin connection 
$\omega_{hk \alpha} $ is accordingly equal to zero), 
for which we define the two Kaluza-Klein charges ${}^{''}M^{56}_{(i)}{}^{''}, i=5,6$. 
The two charges commute with the operator of handedness of
the compactified part of space and accordingly enable - by marking  spinors with the Kaluza-Klein charges - 
to make a choice of states of only one handedness  in also the ''physical'' ($M^{1+3}$) part of  space.
Consequently our ''physical'' spinors are massless. The Weyl equations of motion operator for the compactified space 
with the torsion ${\cal T}^{\alpha}{}_{\beta \gamma}$ - equal, for the Einstein indices 
$\alpha, \beta, \gamma = (5),(6)$, to two generic integers - commutes 
with the two Kaluza-Klein charges.

The crucial achievement of our example of compactifying space with a torsion is, that we can take the eigenvalue subspace
of one of the proposed Kaluza-Klein charges (the two Kaluza-Klein charges commute) inside the solution space
\begin{eqnarray}
 \{ \psi_{sol} | p_h \gamma^h \psi_{sol} =0\}, \quad h=5,6,
 \label{WIsol}
\end{eqnarray}
namely, for example,
\begin{eqnarray}
 \{ \psi | {}^{''}M^{56}_{(1)}{}^{''}\psi, \;\; \psi \in {solution \;\; space}\},
 \label{solcharge}
\end{eqnarray}
which accordingly contains wave functions of only one specific handedness 
$\Gamma^{(2)}{}_S$ $(=-2 S^{56})$, i.e.
$\Gamma^{(2)}{}_S \psi = \alpha \psi$ when $\gamma^a p_{0a} \psi =0$ and ${}^{''}M^{56}_{(1)}{}^{''}\psi = \beta \psi$,
with $\alpha$ determined from $\beta$.

According to the Witten's theorem\cite{witten1981,witten1983} the space of Eq.(\ref{WIsol}) must be real (reducible) 
representation
of the group $U(1)\times U(1)$ (which is demonstrated clearly in Table I) and leads accordingly  
to a real representation in also the ''physical'' $M^{1+3}$ part 
of space, which is  consequently without 
the mass protection. But from Table I it is clear that representations of both 
handedness distinguish among themselves in the $S^{56}$ charge. We proved that one can make use of this fact, if 
looking for the Kaluza-Klein charges.
Our spinor has, due to the choice of the eigenvalue subspace of the Kaluza-Klein charges, only 
one handedness in the compactified part of space and consequently also  in the ''physical'' part.
The Witten ''no-go'' theorem is in our case not violated. 

One still can make a choice of a spin connection on a way that the Weyl spinor 
interacts in the four-dimensional part of space with a gauge field, defined by the spin connection (Eq.(\ref{weylin4})). 

In the approach of one of us\cite{WInorma92,WInorma93,norma94,%
normaixtapa2001,pikanorma1998,pikanorma2002} the break of symmetries was proposed, leading from the starting 
choice of the Weyl spinor of one handedness in $d=1+13$ to particles and anti-particles of the Standard model, keeping 
track of the handedness  
of particles and anti-particles in each of the subspaces. In our simple example this is indeed achieved, by ''marking''
states with the two Kaluza-Klein charges $ {}^{''}M^{56}_{(i)}{}^{''}, \quad i=5,6,$ each of them can do the job alone,
provided that the corresponding integer $m_{(i)}$ is nonzero.

It stays to find out, whether the braking of the symmetry of the space of $SO(1,13)$ to $SO(6)\times SO(1,7)$
(which the approach of one of us proposes as one step in breaking the starting symmetry to the
symmetry $SO(1,3)\times SU(3)\times SU(2)\times U(1)$ of the Standard model) leads to Kaluza-Klein charges,
which commute with the operator of handedness and the operator of the Weyl equations of motion, as well.

One should try to prove also another ''no-go'' theorem: {\it For spaces without torsion one can never find 
a group of symmetry generators, implementing an isometry group, so that it can be used for splitting the solutions
(the real Witten's space of solutions) into  complex representations - the eigenvalue spaces of the handedness}.

\section{Acknowledgement }

It is our pleasure to thank the participants of the series of the Bled workshops, which were asking steadily one 
of the authors  (NMB)
to prove that marking spinor representations in a way to select one handedness is indeed possible in 
Kaluza-Klein-like theories. We thanks in particular C. Froggatt. NMB would like to thank M. Blagojevi\' c for
fruitful discussions on teleparallel theories.


\title*{Hierarchy Problem and a New Bound State}
\author{Colin D.~Froggatt\footnote{E-mail:
c.froggatt@physics.gla.ac.uk}${}^1$ \and 
Holger~Bech~Nielsen\footnote{E-mail:
hbech@alf.nbi.dk}${}^2$}
\institute{%
${}^1$Department of Physics and Astronomy,
Glasgow University,\\
Glasgow G12 8QQ, Scotland\\
${}^2$Niels Bohr Institute,\\
Blegdamsvej 17-21, DK 2100 Copenhagen, Denmark}

\titlerunning{Hierarchy Problem and a New Bound State}
\authorrunning{Colin D.~Froggatt and Holger~Bech~Nielsen}
\maketitle

\begin{abstract}
Instead of solving fine-tuning problems by some automatic method
or by cancelling the quadratic divergencies in the hierarchy
problem by a symmetry (such as SUSY), we rather propose to look
for a unification of the different fine-tuning problems. Our
unified fine-tuning postulate is the so-called Multiple Point
Principle, according to which there exist many vacuum states with
approximately the same energy density (i.e.~zero cosmological
constant). Our main point here is to suggest a scenario, using
only the pure Standard Model, in which an exponentially large
ratio of the electroweak scale to the Planck scale results. This
huge scale ratio occurs due to the required degeneracy of three
suggested vacuum states. The scenario is built on the hypothesis
that a bound state formed from 6 top quarks and 6 anti-top quarks,
held together mainly by Higgs particle exchange, is so strongly
bound that it can become tachyonic and condense in one of the
three suggested vacua. If we live in this vacuum, the new bound
state would be seen via its mixing with the Higgs particle. It
would have essentially the same decay branching ratios as a Higgs
particle of the same mass, but the total lifetime and production
rate would deviate from those of a genuine Higgs particle.
Possible effects on the $\rho$ parameter are discussed.

\end{abstract}

\section{Introduction}
\label{cdfhb:introduction}

There are several problems in high energy physics and cosmology of
a fine-tuning nature, such as the cosmological constant problem or
the problem of why the electroweak scale is lower than the Planck
energy scale by a huge factor of the order of $10^{17}$. In
renormalisable theories, such fine-tuning problems reappear order
by order in perturbation theory. Divergent, or rather cut-off
dependent, contributions (diagrams) have to be compensated by
wildly different bare parameters order by order. The most well-known
example is the hierarchy problem in a non-supersymmetric theory
like the pure Standard Model, which is the model we consider in this
article. New quadratic divergencies occur order by order in the square
of the Standard Model Higgs mass, requiring the bare Higgs mass squared
to be fine-tuned again and again as the calculation proceeds order by
order. If, as we shall assume, the cut-off reflects
new physics entering near the Planck scale $\Lambda_{Planck}$,
these quadratic divergencies
become about $10^{34}$ times bigger than the final mass squared of
the Higgs particle. Clearly an explanation for such a fine-tuning
by ~34 digits is needed. Supersymmetry can tame these divergencies
by having a cancellation between fermion and boson contributions,
thereby solving the technical hierarchy problem. However the problem
of the origin of the huge scale ratio still remains, in the form of
understanding why the $\mu$-term and the soft supersymmetry breaking
terms are so small compared to the fundamental mass scale
$\Lambda_{Planck}$.

In addition to the fine-tuning problems, there are circa 20
parameters in the Standard Model characterising the couplings and
masses of the fundamental particles, whose values can only be
understood in speculative models extending the Standard Model. On
the other hand, the only direct evidence for physics beyond the
Standard Model comes from neutrino oscillations and various
cosmological and astrophysical phenomena. The latter allude to
dark matter, the baryon number asymmetry  and the need for an
inflaton field or some other physics to generate inflation. In
first approximation one might ignore such indications of new
physics and consider the possibility that the Standard Model
represents physics well, order of magnitudewise, up to the Planck
scale.

In the short term, rather than a new extended model with new
fields, we have the need for an extra principle that can specify
the values of the fine-tuned parameters and give predictions for
theoretically unknown parameters. Of course we do need new fields
or particles, such as dark matter, heavy see-saw neutrinos and the
inflaton, but at present they constitute a rather weak source of
inspiration for constructing the model beyond the Standard Model.
On the other hand there is a strong call for an understanding of
the parameters, e.g.~the cosmological constant or the Higgs
particle mass, in the already well working Standard Model.

Since these problems are {\em only} fine-tuning problems, it would
{\em a priori} seem that we should look for some fine-tuning principle.
In a renormalisable theory, a fine-tuning requirement should concern
renormalised parameters rather than bare ones. This is well
illustrated by the cosmological constant problem, where simply
requiring a small value for the bare cosmological constant will not
solve the phenomenological problem. There are several contributions
(e.g.~from electroweak symmetry breaking)
to the observed value, renormalising it so to speak, which are huge
compared to the phenomenological value.


\section{A Fine-tuning Principle}
\label{finetuning}

In the spirit of renormalisable theories, it is natural to
formulate a fine-tuning postulate in terms of quantities that are
at least in principle experimentally accessible. So one is led to
consider n-point functions or scattering amplitudes, which are
functions of the 4-momenta of the external particles. However, in
order to specify the value of such a quantity for a given
configuration of particles, it would be necessary to specify all
the external momenta of the proposed configuration. One might
consider taking some integral or some average over all the
external momenta in a clever way. However, for several external
particles, it really looks rather hard to invent a fine-tuning
postulate that is simple enough to serve as a fundamental
principle to be fulfilled by Nature in choosing the coupling
constants and masses. But the situation becomes much simpler if we
think of formulating a fine-tuning principle for a zero-point
function! The zero-point function is really just the vacuum energy
density or the value of the dressed cosmological constant
$\Lambda_{cosmo}$. The cosmological constant is of course a good
idea for our purpose, in as far as the cosmological constant
problem itself would come into the fine-tuning scheme immediately
if we make the postulate that the zero-point function should
vanish. Nowadays its fitted value is not precisely zero, in as far
as about $73\%$ of the energy density in the Universe is in the
form of dark energy or a cosmological constant. However this value
of the cosmological constant is anyway very tiny compared to the
{\em a priori} expected Planck scale or even compared to the
electroweak or QCD scales.

Now an interesting question arises concerning the detailed form of
this zero cosmological constant postulate, in the case when there
are several candidate vacuum states. One would then namely ask:
should the zero cosmological constant postulate apply just to one
possible vacuum state or should we postulate that all the
candidate vacua should have their {\em a priori} different
cosmological constants set equal to zero (approximately)? It is
our main point here to answer this question by extending the zero
cosmological constant postulate to all the candidate vacua! In
fact this form of the zero cosmological constant postulate
unifies\footnote{We thank L.~Susskind for pointing out to us that
the cosmological constant being zero can be naturally incorporated
into the Multiple Point Principle} the cosmological constant
problem with our so-called Multiple Point Principle
\cite{CDHBglasgowbrioni}, which states that there exist several vacua
having approximately the same energy density.

In principle, for each proposed method for explaining why the
cosmological constant is approximately zero, we can ask whether it
works for just the vacuum that is truly realised or whether it
will make several vacuum candidates zero by the same mechanism.
For example, one would expect that the proposal of Guendelman
\cite{guendelmann}, of using an unusual measure on space-time, would
indeed easily give several vacua with zero energy density rather
than only one. However for a method like that of Tsamis and
Woodard  \cite{woodard}, in which it is the actual time
development of the Universe that brings about the effectively zero
cosmological constant, one would only expect it to work in the
actual vacuum. Similarly if one uses the anthropic principle
\cite{weinberg}, one would only expect to get zero cosmological
constant for that vacuum in which we, the human beings, live.

The main point of the present talk is to emphasize that the
Multiple Point Principle, which can be considered as a consequence
of solving the cosmological constant problem in many vacua, can be
helpful in solving other fine-tuning problems; in particular the
problem of the electroweak scale being so tiny compared to the
Planck scale.

\section{Approaching the Large Scale Ratio Problem}
\label{scaleratio}

We consider here the problem of the hierarchy between the Planck
scale and the electroweak scale. This scale ratio is so huge that
it is natural to express it as the exponential of a large number.
In fact we might look for inspiration at another scale ratio
problem for which we already have a good explanation: the ratio of
the fundamental (Planck) scale to the QCD scale. The QCD scale
$\Lambda_{QCD}$ is the energy scale at which the QCD fine
structure constant formally diverges. It is believed that the
scale ratio $\Lambda_{Planck}/\Lambda_{QCD}$ is determined by the
renormalisation group running of the QCD fine structure constant
$\alpha_s(\mu)$, with the scale ratio being essentially equal to
the exponent of the inverse of the value of the fine structure
constant $1/\alpha_s(\Lambda_{Planck})$ at the Planck scale. So we
might anticipate explaining the Planck to electroweak scale ratio
in terms of the renormalisation group and the Standard Model
running coupling constants at the fundamental scale and the
electroweak scale.

At first sight, it looks difficult to get such an explanation by
fine-tuning a running coupling -- e.g.~the top quark Yukawa
coupling -- at the electroweak scale, using our requirement of
having vacua with degenerate energy densities. The difficulty is
that, from simple dimensional arguments, the energy density or
cosmological constant tends to become dominated by the very
highest frequencies and wave numbers relevant in the quantum field
theory under consideration -- the Planck scale in our case. In
fact the energy density has the dimension of energy to the fourth
power, so that modes with Planck scale frequencies contribute
typically $(10^{17})^4$ times more than those at the electroweak
scale.  The only hope of having any sensitivity to electroweak
scale physics would, therefore, seem to be the existence of two
degenerate phases, which are identical with respect to the state
of all the modes corresponding to higher than electroweak scale
frequencies. They should, so to speak, \underline{only} deviate by
their physics at the electroweak scale and perhaps at lower scales
in energy. In such a case it could be that the energy density
difference between the two phases would only depend on the
electroweak scale physics and, thus, could more easily depend on
the running couplings taken at the electroweak scale. It is,
namely, only for the modes of this electroweak scale that the
running couplings at this scale are relevant.

So, in order to ``solve'' the large scale ratio problem using our
Multiple Point Principle, we need to have a model with two
different phases that only deviate by the physics at the
electroweak scale. So what could that now be? Different phases are
most easily obtained by having different expectation values of
some scalar field, which really means different amounts of some
Bose-Einstein condensate. A nice way to have such a condensate only 
involve physics at a certain low energy scale, the electroweak scale 
say, consists in having a condensate of bound states made out of some
Standard Model particles -- we shall actually propose top quarks
and anti-top quarks. Such bound states could now naturally have
sizes of the order of the electroweak length scale. Such a picture
would really only make intuitive sense, when the density is not
large compared to the scale given by the size of the bound states;
otherwise they would lie on top of each other and completely
disturb the binding. One might say that the physical situation for
the binding would become drastically changed, when the density in
the bound state condensate gets so high as to have huge multiple
overlap. Presumably one could naturally get a condensate with a
density which is not so far from the scale given by the size and,
thus, the electroweak scale.  In the next section we shall spell
out this idea of making a bound state condensate in more detail.
We shall then return to the large scale ratio problem in section
\ref{return} and explain how the Multiple Point Principle is used
to determine the top quark Yukawa coupling constant at the Planck
scale, in terms of the electroweak gauge coupling constants, by
postulating the existence of a third degenerate vacuum.

\section{The Bound State}
\label{boundstate}

\subsection{The Idea of a Bound State Condensate}
\label{condensate}

So we are led to consider some strongly bound states, made out of
e.g. top-quarks and using Higgs fields or other particles to bind
them, such that the energy scale of a condensate formed from them
is - by dimensional arguments - connected to the scale of the
Standard Model Higgs field vacuum expectation value or VEV (which
is of course what one usually calls the electroweak scale). For
dimensional reasons this condensate has now a density of an order
of magnitude given by this electroweak scale. Then the frequencies
or energies of the involved modes of vibration are also of this
order, in the sense that it is the modes with energies of this
order that make the difference (between two phases say).  It is
therefore also the running couplings at this scale that are the
directly relevant parameters! If we now impose some condition,
like the degeneracy of two phases resulting from this bound state
condensation dynamics, it should result in some relation or
requirement concerning the {\em running couplings at the electroweak
scale}.

Instead of simply a bound state condensing, one could {\em a priori}
also hope for some other nonlinear effect taking place in a way
involving essentially only the modes/physics at the electroweak
scale. The crux of the matter is that, at short distances compared
to the electroweak length scale, the non-perturbative effect in
question would hardly be felt. Consequently, the huge contributions
to the energy density from the short distance modes can be cancelled
out between the two phases, in imposing our Multiple Point Principle
(MPP). But the bound state idea is in a way the most natural and
simple, since bound states are already well-known to occur in many 
places in quantum physics.

\subsection{A Bound State of 6 top and 6 anti-top quarks?}
\label{12tops}

Of course, when we look for bound states in the Standard Model, we
know immediately that there is a huge number of hadronic bound
states consisting of mesons, baryons and glueballs, i.e. from QCD.
These bound states typically have the size given by the QCD
scale parameter $\Lambda_{QCD}$, which means lengths of the order 
of an inverse GeV. That is to say the strong scale rather than the
electroweak one. Nevertheless you could {\em a priori} hope that
some phase transition, involving quarks and caused by QCD, could
determine a certain quark mass by the MPP requirement of being at
the border between two phases of the vacuum. That could then in
turn lead to a fixing of the Standard Model Higgs VEV, which is
known to be responsible for all the quark masses. But, from
dimensional considerations, you would expect to get that the quark
mass needed by such an MPP mechanism would be of the order of the
strong scale. So it may work this way if the strange quark were
the one to be used, since it namely has a mass of the order of the
strong scale. But these speculated QCD-caused phase transitions
are not quite what we require. We rather seek a condensation
getting its scale from the Standard Model Higgs VEV and want to
avoid making severe use of QCD. However, at first sight, the other
gauge couplings and even the top quark Yukawa coupling (let alone
the other smaller Yukawa couplings) seem rather small for making
strongly bound states; with a binding so strong, in fact, as to
make the bound state tachyonic and to condense in the vacuum.
Indeed, if you think of bound states consisting of a couple of
particles, it is really pretty hopeless to find any case of such a
strong binding except in QCD. But now scalar particle exchange has
an important special feature. Unlike the exchange of gauge
particles, which lead to alternating signs of the interaction when
many constituents are put together, scalar particle exchange leads
to attraction in all cases: particles attract both other particles
and antiparticles and the attraction of quarks, say by Higgs
exchange, is independent of colour.

The only hope of getting very strong binding without using QCD, so
as to obtain tachyonic bound states in the Standard Model, is to
have many particles bound together and acting cooperatively -- and
then practically the use of a scalar exchange is unavoidable. So we
are driven towards looking for bound states caused dominantly by the
exchange of Higgs particles, since the Higgs particle is the only
scalar in the Standard Model and we take the attitude of
minimising the amount of new physics. Since the Yukawa couplings
of the other quarks are so small, our suggestion is to imagine
some top quarks and/or anti-top quarks binding together into an
exotic meson. It better be a boson and thus a ``meson'', since we
want it to condense.

There are, of course, bound states of say a top quark and an
anti-top quark which are mainly bound by gluon exchange, although
comparably by Higgs exchange. However these are rather loosely
bound resonances compared to the top quark mass. But, if we now add more
top or anti-top quarks to such a state, the Higgs exchange
continues to attract while the gluon exchange saturates and
gets less significant. This means that the Higgs exchange binding
potential for the whole system gets proportional to the number of
pairs of constituents, rather than to the number of constituents
itself. So, at least {\em a priori} by having sufficiently many
constituents, one might foresee the binding energy exceeding the
constituent mass of the system.

In order to get the maximal binding, one needs to put each of the
added quarks or anti-quarks into an S-wave state. Basically we can
use the same technique as in the calculation of the binding energy
of the electron to the atomic nucleus in the hydrogen atom. Once
the S-wave states are filled, we must go over to the P-wave and so
on. But the P-wave binding in a Coulomb shaped potential only
provides one quarter of the binding energy of the S-wave. In the
case of a scalar exchange, an added particle gets a binding energy
to each of the other particles already there. However, once the
S-wave states are filled, these binding energies go down by a
factor of about four in strength and it becomes less profitable
energetically to add another particle. Depending on the strength
of the coupling, it can therefore very easily turn out to be most
profitable to fill the S-wave states and then stop.

Now, when we use top quarks and anti-top quarks, one can easily
count the number of constituents, by thinking of the S-wave as
meaning that essentially all the particles are in the same state
in geometrical space. Then there are 12 different ``internal''
states into which these S-wave quarks/antiquarks can go: each
quark or anti-quark can be in two spin states and three colour
states, making up all together $2 \times 2 \times 3 = 12$ states.
Thus we can have 6 top quarks and 6 anti-top quarks in the bound
state, before it gets necessary to use the P-wave. We shall here
make the hypothesis, which to some extent we check below, that
indeed the strongly bound state which we seek is precisely this
one consisting of just 12 particles.

So we now turn to the question of whether or not this exotic 6 top
quark and 6 anti-top quark state is bound sufficiently strongly to
become tachyonic, i.e.~to get a negative mass squared. Actually,
in order to confirm our proposed MPP fine-tuning mechanism, we
need that the experimentally measured top quark Yukawa coupling
should coincide with the borderline value between a condensate of
this almost tachyonic exotic meson being formed or not being
formed. On the basis of the following crude estimate, we want to
claim that such a coincidence is indeed not excluded.

\subsection{The Binding Energy Estimate}
\label{binding_estimate}

We now make a crude estimate of the binding energy of the proposed
12 quark / anti-quark bound state. As a first step we consider the
binding energy $E_1$ of one of them to the remaining 11
constituents treated as just one collective particle, analogous to
the nucleus in the hydrogen atom. Provided that the radius of the
system turns out to be sufficiently small compared to the Compton
wavelength of the Standard Model Higgs particle, we can take this
to be given by the well-known Bohr formula for the ground state
binding energy of a one electron atom. It is simply necessary to
replace the electric charge of the electron $e$ by the top quark
Yukawa coupling $g_t/\sqrt{2}$, in the normalisation where the
running mass of the top quark is given by the formula $m_t = g_t
\, 174$ GeV, and to take the atomic number to be $Z=11$:
\begin{equation}
E_1 = -\left(\frac{11g_t^2/2}{4\pi}\right)^2 \frac{11m_t}{24}
\label{binding}
\end{equation}
Here we have used $m_t^{reduced}= 11m_t/12$ as the reduced mass of
the top quark.

In order to obtain the full binding energy for the 12 particle
system, we should multiply the above expression by 12 and divide
by 2 to avoid double-counting the pairwise binding contributions.
However this analogy with the atomic system only takes into
account the $t$-channel exchange of a Higgs particle between the
constituents. A simple estimate of the $u$-channel Higgs exchange
contribution \cite{itep} increases the binding energy by a further
factor of 16/11. So the expression for the total non-relativistic
binding energy due to Higgs particle exchange interactions
becomes:
\begin{equation}
 E_{binding} = \left(\frac{11g_t^4}{\pi^2}\right)m_t
\label{binding2}
\end{equation}

We have here neglected the attraction due to gluon exchange and
the even smaller electroweak gauge field forces. However the gluon
attraction is rather a small effect compared to the Higgs particle
exchange, in spite of the fact that the QCD coupling
$\alpha_s(M_Z) = g_s^2(M_Z)/4\pi = 0.118$. This value of the QCD
fine structure constant corresponds to an effective gluon top
anti-top coupling constant squared of:
\begin{equation}
e_{tt}^2 = \frac{4}{3}g_s^2 \simeq \frac{4}{3} 1.5 \simeq 2.0
\end{equation}
We have to compare this gluon coupling strength $e_{tt}^2 \simeq
2$ with $Zg_t^2/2 \simeq 11/2 \times 1.0$ from the Higgs particle.
This leads to an increase of the binding energy by a factor of
$(15/11)^2$ due to gluon exchange, giving our final result for the
non-relativistic binding energy:
\begin{equation}
 E_{binding} = \left(\frac{225g_t^4}{11\pi^2}\right)m_t
\label{binding3}
\end{equation}

The correction from W-exchange will be smaller than that from
gluon exchange by a multiplicative factor of about
$\left(\frac{\alpha_2(M_Z)}{\alpha_s(M_Z)} \frac{3}{4}\right)^2 \simeq
\frac{1}{25}$, and the weak hypercharge exchange is further
reduced by a factor of $\sin^4\theta_W$. Also the s-channel Higgs
exchange diagrams will give a contribution in the same direction.
There are however several effects going in the opposite direction,
such as the Higgs particle not being truly massless and that we
have over-estimated the concentration of the 11 constituents
forming the ``nucleus''. Furthermore we should consider
relativistic corrections, but we postpone a discussion of their
effects to ref.~\cite{fln}.

\subsection{Estimation of Phase Transition Coupling}
\label{coupling_estimate}

From consideration of a series of Feynman diagrams or the
Bethe-Salpeter equation for the 12 particle bound state, we would
expect that the mass squared of the bound state, $m_{bound}^2$,
should be a more analytic function of $g_t^2$ than $m_{bound}$
itself. So we now write a Taylor expansion in
$g_t^2$ for the mass {\em squared} of the bound state, crudely
estimated from our non-relativistic binding energy formula:
\begin{eqnarray}
m_{bound}^2 & = & \left(12m_t\right)^2 - 2\left(12
m_t\right)\times
E_{binding} + ...\\
& = & \left(12m_t\right)^2\left(1 -\frac{225}{66\pi^2}g_t^4 +
...\right)
\label{expansion}
\end{eqnarray}

We now assume that, to first approximation, the above formal
Taylor expansion (\ref{expansion}) can be trusted even for large
$g_t$ and with the neglect of higher order terms in the {\em mass
squared} of the bound state. Then the condition that the bound
state should become tachyonic, $m_{bound}^2 < 0$, is that the top
quark Yukawa coupling should be greater than the value given by
the vanishing of equation (\ref{expansion}):
\begin{equation}
0 =  1-\frac{225}{66\pi^2}g_t^4 + ...
\label{br0}
\end{equation}
We expect that once the bound state becomes a tachyon, we should
be in a vacuum state in which the effective field, $\phi_{bound}$,
describing the bound state has a non-zero expectation value. Thus
we expect a phase transition just when the bound state mass
squared passes zero\footnote{In fact the phase transition
(degenerate vacuum condition) could easily occur for a small
positive value of $m_{bound}$ and hence a somewhat smaller value
of $g_t^2$.}, which roughly occurs when the running top quark
Yukawa coupling at the electroweak scale, $g_t(\mu_{weak})$,
satisfies the condition (\ref{br0}) or:
\begin{equation}
g_t|_{phase \ transition} =
\left(\frac{66\pi^2}{225}\right)^{1/4} \simeq 1.3
\end{equation}

We can make an estimate of one source of uncertainty, by
considering the effect of using a leading order Taylor expansion
in $g_t^2$ for $m_{bound}$ instead of for $m_{bound}^2$. This
would have led to difference of a factor of 2 in the binding
strength and hence a correction by a factor of the fourth root of
2 in the top quark Yukawa coupling at the phase boundary; this
means a $20\%$ uncertainty in $g_t|_{phase \ transition}$. Within
an uncertainty of this order of $20\%$, we have a 1.5 standard
deviation difference between the phase transition (and thus the
MPP predicted) coupling, $g_t \simeq 1.3$, and the measured one,
$g_t \simeq 1.0$, corresponding to a physical top quark mass of
about 173 GeV. We thus see that it is quite conceivable within our
very crude calculations that, with the experimental value of the
top quark Yukawa coupling constant, the pure Standard Model could
lie on the boundary to a new phase; this phase is characterised by
a Bose-Einstein condensate of bound states of the described type,
consisting of 6 top quarks and 6 anti-top quarks!

\subsection{Mixing between the Bound State and the Higgs Particle}

Strictly speaking, if the above scenario is correct, it is not at
all obvious in which of the two vacua we live. If we live in the
phase in which the bound state condensate is present, the
interaction of the bound state particle with the Standard Model
Higgs particle can cause a bound state particle to be pulled out
of the vacuum condensate and then to function as a normal
particle. This effect will mean that the normal Higgs particle
will {\em mix} with the bound state, in a similar way as one has
mixing between the photon and the $Z^0$ gauge boson, or between
$\eta$ and $\eta'$. This means that the two observed particles
would actually be superpositions, each with some amplitude for
being the bound state and with some amplitude for being the
original Higgs particle. Both can have expectation values, or
rather the expectation value is described by some abstract vector
denoting the two different components. Also both superpositions
would be exchanged and contribute to the binding of the bound
state. Taking this two component nature of the effective Higgs
particles into account makes the discussion more complicated than
with a single Higgs particle.

Really there are three types of experimentally accessible parameters
for which we at first want to predict a relation from our bound
state model:
\begin{enumerate}
\item
The top quark mass is given in the simplest case by the top quark
Yukawa coupling times the Higgs VEV. However, in the two effective
Higgs picture (one of them being the bound state essentially just
mixed somewhat with the Higgs particle), the top quark mass
becomes of the form $h_1 v_1 + h_2 v_2$. Here $h_1$ and $h_2$ are
the Yukawa couplings of the top quark for the two superpositions,
whose VEVs are denoted by $v_1$ and $v_2$.
\item
The gauge boson masses: for example the $W$ boson mass in the
two effective Higgs picture becomes $M_{W}^2= g_2^2 (v_1^2 + v_2^2)$.
Here, for simplicity, we have taken both the fields to be doublets
with weak hypercharge $y/2= -1/2$ like the original Higgs field.
We reconsider the irreducible representation content of the bound
state field in section \ref{rhoparameter}, where we discuss the
$\rho$ parameter problem.
\item
The binding strength parameter for the bound state which
determines the vacuum phase in which the energy density is the
lowest. Even if this parameter is hard to determine
experimentally, we may at least relate it to our MPP from which it
can essentially be predicted. In the simplest case with a single
Higgs particle, the binding strength parameter is just the top
quark Yukawa coupling squared $g_t^2$ as discussed in section
\ref{coupling_estimate}. However with two effective Higgs
particles, the parameter $g_t^2$ would to first approximation be
replaced by $ h_1^2 + h_2^2$.
\end{enumerate}

We now remark that the three quantities listed above
are related by a Schwarz inequality, namely:
\begin{equation}
|h_1v_1 + h_2v_2|^2 \le (v_1^2 + v_2^2)(h_1^2 + h_2^2)
\end{equation}
(written as if we had only real numbers, but we could use complex
ones also). With this correction due to the mixing, we lose our
strict prediction of the top quark mass corresponding to two
degenerate phases, with and without a bound state condensate
respectively, unless we can estimate the mixing. In fact such an
estimate is not entirely out of question, because we know the
coupling of the Higgs to the bound state and can potentially also
estimate the density of the condensate. Qualitatively we just
predict that the resulting top quark mass will be somewhat {\em
smaller} than the estimate made in section
\ref{coupling_estimate}, but we expect it to remain of a similar
order of magnitude.

Such a mixing correction would seem to be welcome, in order to improve
the agreement of the experimental top quark Yukawa coupling with
the estimated phase transition value. We namely tend to predict
a top quark mass, which is too large by a factor of about 1.3, without
including any mixing correction. However this disagreement should not be
taken too seriously, as it is within the accuracy of our calculation.
Nonetheless, if we do live in the phase containing the bound state
condensate, the mixing correction would be good for repairing this
weak disagreement with experiment.

\section{Return to the Large Scale Ratio Problem}
\label{return}

\subsection{Three degenerate vacua in the pure Standard Model}

As discussed at Portoroz by Colin Froggatt \cite{colin}, it is
possible to determine the top quark running Yukawa coupling
$g_t(\mu)$ at the fundamental scale $\mu_{fundamental} =
\Lambda_{Planck}$ by using the Multiple Point Principle to
postulate the existence of a third degenerate vacuum, in which the
Standard Model Higgs field has a VEV of order the Planck scale
\cite{CFHBfn2}. This requires that the renormalisation group improved
effective potential for the Standard Model Higgs field should have
a second minimum near the Planck scale, where the potential should
essentially vanish. This in turn means that the Higgs
self-coupling constant $\lambda(\mu)$ and its beta function
$\beta_{\lambda}(\mu)$ should both vanish near the fundamental
scale, giving the following relationship between the top quark
Yukawa coupling $g_t(\mu_{fundamental})$ and the electroweak
$SU(2) \times U(1)$ gauge coupling constants
$g_2(\mu_{fundamental})$ and $g_1(\mu_{fundamental})$:
\begin{equation}
\label{gt4} g_t^4 = \frac{1}{48} \left(9g_2^4 + 6g_2^2g_1^2
+3g_1^4 \right)
\end{equation}
If we now input the experimental values of the gauge coupling
constants, extrapolated to the Planck scale using the Standard
Model renormalisation group equations, we obtain
$g_t(\mu_{fundamental}) \simeq 0.4$. However we note that the
numerical value of $g_t(\mu)$, determined from the expression on
the right hand side of eq.~(\ref{gt4}), is rather insensitive to
the scale, varying by approximately $10\%$ between $\mu = 246$ GeV
and $\mu = 10^{19}$ GeV.

From our assumption of the existence of three degenerate vacua in
the Standard Model, our Multiple Point Principle has provided
predictions for the values of the top quark Yukawa coupling
constant at the electroweak scale, $g_t(\mu_{weak}) \simeq 1.3$,
and at the fundamental scale, $g_t(\mu_{fundamental}) \simeq 0.4$.
So we can now calculate a Multiple Point Principle prediction for
the ratio of these scales $$\mu_{fundamental}/\mu_{weak},$$ using
the Standard Model renormalisation group equations.

\subsection{Estimation of the logarithm of the scale ratio}

We now estimate the fundamental to electroweak scale ratio by
using the leading order beta function for the Standard Model top
Yukawa coupling constant $g_t(\mu)$:
\begin{equation}
 \beta_{g_t} = \frac{dg_t}{d\ln\mu} =
 \frac{g_t}{16\pi^2}\left(\frac{9}{2}g_t^2 - 8g_3^2 -
 \frac{9}{4}g_2^2 - \frac{17}{12}g_1^2\right)
 \label{betatop}
\end{equation}
where the $SU(3) \times SU(2) \times U(1)$ gauge coupling
constants are considered as given at the fundamental scale,
$\mu_{fundamental} = \Lambda_{Planck}$. It should be noticed that,
due to the relative smallness of the fine structure constants
$\alpha_i =g_i^2/4\pi$ and particularly of
$\alpha_3(\mu_{fundamental})$, the
beta function $\beta_{g_t}$ is numerically rather small at the
Planck scale. So the logarithm of the scale ratio
$\ln\mu_{fundamental}/\mu_{weak}$ needed to generate the required
amount of renormalisation group running, between the values
$g_t(\mu_{fundamental}) \simeq 0.4$ and $g_t(\mu_{weak}) \simeq
1.3$, must be a large number. Hence the scale ratio itself must be
huge and in this way we explain why the electroweak scale
$\mu_{weak}$ is so low compared to the fundamental scale
$\mu_{fundamental}$. In practice the Multiple Point Principle only
gives the order of magnitude of the logarithm of the scale ratio,
predicting $\mu_{fundamental}/\mu_{weak} \sim 10^{16} - 10^{20}$.

We note that as the strong scale is approached, $\mu \rightarrow
\Lambda_{QCD}$, $g_3(\mu)$ and the rate of logarithmic running of
$g_t(\mu)$ becomes large. So the strong scale $\Lambda_{QCD}$
provides an upper limit to the scale ratio predicted by the
Multiple Point Principle. Indeed the predicted ratio naturally
tends to give an electroweak scale within a few orders of
magnitude from the strong scale.

\section{How to see the bound state?}
\label{how-to-see}

\subsection{Mixing with the Higgs Particle}

Such a strongly bound state as we propose, consisting of 12
constituents, will practically act as a conserved type of
particle, because energy conservation forbids its destruction by
having a few of its constituents decay. The point is that the mass
of the remaining bound state or resonance, made up of the leftover
constituents, would be larger than that of the original strongly
bound state. Considering its interaction with the relatively light
particles of the Standard Model, the bound state would therefore
still be present after the interaction. This means that the most
important effective couplings, involving an effective scalar bound
state field and Standard Model fields, would have two or four
external bound state attachments. If we further restrict ourselves
to a renormalisable effective theory, we would be left with the
bound state scalar field only having interactions involving scalar
and gauge fields. An interaction between two fermions and two
scalar fields would already make up a dimension five operator,
which is non-renormalisable.

If we live in the phase without the condensate of new bound state
particles, these considerations imply that the bound state must be
long lived; it could only decay into a channel in which all 12
constituents disappeared together.  The production cross section
for such a particle would also be expected to be very low, if it
were just crudely related to the cross section for producing 6 top
quarks and 6 anti-top quarks. However, if we live in the phase
with the condensate, there exists the possibility that the bound
state particle could disappear into the condensate, which has of
course an uncertain number of bound state particles in it. Since,
as is readily seen, there is a significant coupling of the
Standard Model Higgs particle to two bound state attachments - a
three scalar coupling vertex - we can achieve such a disappearance
very easily by means of this vertex. This disappearance results in
the bound state obtaining an effective transition mass term into
the Standard Model Higgs particle. Such a transition means that
the Higgs particle and the new bound state will - provided we are
in the condensate phase - mix with each other! This has very
important consequences for the observability of the bound state.
We shall seemingly get two Higgs particles sharing the strength of
the fundamental Higgs particle, by each being a superposition of
the latter and of the bound state.

So, provided that we presently live in the phase with the bound
state condensate, we predict that at the LHC we shall apparently see
two Higgs particles! They will each behave just like the normal
Higgs particle, except that all of its couplings will be reduced
by a mixing angle factor, common of course for all the different
decay modes of the usual Standard Model Higgs particle, but
different for the two observed Higgs particles.

\subsection{The Rho Parameter; a Problem?}
\label{rhoparameter}

If we do not live in the phase with the condensate we can
naturally not expect to observe any effects of this condensate,
but if we live in the phase with the condensate then one might
look for the effects of this condensate. An effect that at first
seems to be there, and is perhaps likely to prevent the model from
being phenomenologically viable, is that the condensate of bound
states is not invariant under the $SU(2)\times U(1)$ gauge group
for the electroweak interactions. In fact this condensate will
{\em  a priori} begin to ``help'' the Standard Model Higgs field
giving masses to the $W$ and $Z^0$ particles. Now, however, since
we imagine the bound state to exist in the background of the usual
Higgs condensate and as only being bound due to the effects of
this surrounding medium, the bound state is strongly influenced by
the $SU(2)\times U(1)$ breaking effects of these surroundings.
Thus we cannot at first consider the bound state as belonging to
any definite irreducible representation of this electroweak group.
Rather we must either describe it by a series of fields belonging
to different irreducible representations of this group or simply
describe it by a single effective field that does not have any
definite electroweak quantum numbers. But this fact means that the
condensate of bound states has to be expressed by several such
fields having non-zero expectation values. These different fields
of different irreducible representations will not give the same
mass ratio for the $W$ and the $Z^0$ bosons. Thus, provided the
bound state condensate is of such an order of magnitude that its
effect on the gauge boson masses is not negligible, it will in 
general generate a $\rho$ parameter in disagreement with experiment.

So far our calculations have not supported the hope that, by some
mathematical accident, the $\rho$ parameter comes out to be
essentially equal to unity. Rather it seems that, in order for our
model to be consistent with the remarkably good agreement of the
Standard Model predictions with experiment, we require one of the
following situations to occur:

1) We do not live in the phase with the condensate but rather in
the one without the condensate.

2) The contribution of the condensate expectation value to the
gauge boson masses is simply very much smaller than that of the
genuine Standard Model Higgs field.

3) For irreducible representations other than the singlet and the
doublet with weak hypercharge $y/2 =1/2$, some self-interaction
or renormalisation group effect has made the irreducible
representation content of the bound state very small or vanishing.

As we shall see below, there is some weak evidence that our model
favours the idea that we actually live in the phase {\em with} the 
bound state condensate and even with an appreciable expectation value
compared to that of the genuine Higgs condensate. So it would seem
to fit our model best, if we could get the third of the above
possibilities to work and thereby avoid causing problems for the value of
the $\rho$ parameter.

\section{In which phase do \underline{we} live?}

In a model like ours, where there are many vacuum states, one must
identify which of those states is the vacuum around {\em us}. We
definitely live in a phase with a remarkably small Higgs field VEV
compared to the ``fundamental'' scale or natural unit for Higgs
field VEVs, which we take to be the Planck scale in our model.
Among the three Standard Model vacua discussed above, there are
thus two possibilities corresponding to the phases with the low
value of the Standard Model Higgs VEV. So what remains to be
decided is whether or not there is a condensate of the bound
states in the vacuum in which we live.

In section \ref{how-to-see} we discussed possible observational
effects related to the bound state, which could discriminate
between the two phases. Here we shall investigate which vacuum
phase is likely to emerge from the Big Bang and then assume that
it survives to the present epoch.

There is, however, no {\em a priori} reason to believe in the
absence of vacuum phase transitions since the first minutes after
the Big Bang. They might even have occurred in the era when stars
and galaxies were already present, but then one could imagine that
there should be astrophysical signatures revealing such transitions.
Indeed one might even wonder if the claims for a time variation of
the fine structure constant, indicated by some spectral investigations,
could be a consequence of such phase transitions. But it must be
admitted that the domain walls between phases would have such a
huge energy per unit surface area that they might be expected to
disturb all of cosmology as we understand it. So it seems likely that
there were no later phase transitions and that we do live in the
phase that emerged after the first minutes of the Big Bang.
If the other vacuum phases are to occur anywhere or
anytime at all, it must then be in the future.

We now turn to the question: what phase is likely to have come out
of the Big Bang? Of course the phase that emerges depends very
sensitively on the vacuum energy density. The higher energy
density vacua are expected to decay into the one with the lowest
energy density, provided though that sufficient thermal energy is
present to surmount any energy barriers between the vacua. Hence
the question of which vacuum emerges will be settled at the epoch
when the temperature is still just high enough that the phase
border can be passed, i.e.~when it is still possible to produce
the walls between the phases by thermal fluctuations.

At that epoch it is the Helmholz free energy density $f$ rather
than the true energy density $u$ that matters. The difference is
the term $-sT$, where $T$ is the temperature and $s$ is the
entropy density. Assuming the true energy density is exactly the
same in the two phases, the emergent phase should be the one
having the highest entropy density $s$ at the temperature in
question. That in turn should be the phase with highest number of
light species. Now, in the Standard Model, the known fermions and
gauge bosons, $W$ and $Z^0$, get their masses from the Higgs VEV
in the vacuum in question. So the emergent phase should 
be the one with the lowest Higgs VEV, when these 
particles have the smallest masses, giving in turn the larger
entropy, and then the lower free energy density. Now the presence
of the many bound states in the condensate tends to reduce the
Higgs field VEV. So it is indeed the phase {\em with} the
condensate, which is expected to come out from the early Universe.
Our tentative conclusion is thus that we should live in the phase
with the bound state condensate, provided of course that we are
correct in assuming that a new phase did not take over at a later
epoch.

Phenomenologically this phase with the bound state condensate present
today is the more interesting possibility, in as far as it leads to
the mixing of the Higgs particle and the bound state. It thereby
gives us the possibility of seeing this bound state much more easily,
namely as another ``Higgs'' particle. However, in this case we must
face up to the challenge of calculating the $\rho$ parameter.

\section{Conclusion}

In this talk, we have put forward a scenario for how the huge
hierarchy in energy scale comes about between a supposed
fundamental scale, taken as the Planck scale, and the electroweak
scale, meaning the scale of the $W$ and $Z^0$ particles and the
Higgs particle etc. This consists of introducing a fine-tuning
postulate -- the Multiple Point Principle -- according to which
there are many different vacua, in each of which the cosmological
constant or energy density is very small. In this way our main
fundamental assumption is that the cosmological constant problem
is solved, in some way or other, {\em several times}. The
remarkable result of the present article is that, as well as
fine-tuning the cosmological constants, this principle can lead to
a solution of a separate mystery, namely of why the electroweak
scale of energy is so low compared to the Planck scale. This
problem, which is separate from but closely related to the
technical hierarchy problem, gets solved in our scenario to the
degree that we even obtain a crude value for the logarithm of the
large scale ratio. We even get a suggestive explanation for why,
compared to its logarithmic distance from the Planck scale, the
electroweak scale is relatively close to the strong scale,
$\Lambda_{QCD}$. We, of course, have to input the large ratio 
of the Planck to QCD scales, in the form of the value of the 
QCD coupling constant at the fundamental scale.

In our scenario the pure Standard Model is assumed to be valid up
close to the Planck scale, apart from a possible minor
modification at the neutrino see-saw scale. We then postulate that
there are just three vacuum states all having, to first
approximation, zero energy density. In addition to specifying
information about the bare cosmological constant, this postulate
leads to two more restrictions between the parameters of the 
Standard Model. They are, in principle, complicated relations
between all the coupling constants and masses and it is
non-trivial to evaluate their consequences.  However we took the
values of the gauge coupling constants, which are anyway less
crucial, from experiment and these two relations then gave values
for the top quark Yukawa coupling constant at the
electroweak scale and the ``fundamental'' scale respectively:
$g_t(\mu_{weak}) \simeq 1.3$ and $g_t(\mu_{fundamental}) \simeq
0.4$.

The main point then is that we need an appreciable running of the
top quark Yukawa coupling, in order to make the two different
values compatible. That is to say we need a huge scale ratio,
since the running is rather slow due to the smallness of the
Standard Model coupling constants in general from the
renormalisation group point of view. This is our suggested
explanation for the mysterious huge hierarchy found empirically
between the Planck and electroweak scales. Indeed it even leads to
an approximately correct value for the logarithm of the huge scale
ratio!

It is crucial for our scenario that there should exist the
possibility of a phase with a certain bound state condensing in
the vacuum. The existence of such a bound state is {\em a priori}
a purely calculational problem, in which no fundamentally new
physics comes in. We suggest that this bound state should be
composed of 6 top quarks and 6 anti-top quarks held together by
Higgs exchange and, maybe to some extent, also by the exchange of
the bound state itself -- in a bootstrap-like way. If we live in
the vacuum without a bound state condensate, it would be difficult
to obtain direct experimental evidence for the bound state.
However if we live in the vacuum with a bound state condensate, 
which actually seems to be the
most likely situation in our scenario, it should be possible to
see the effects of this condensate. There should then be a
significant mixing between the bound state and the Higgs particle.
This implies the existence of two physical particles, sharing the
coupling strength and having the same decay branching ratios as
the conventional Standard Model Higgs particle. The resulting
effective 2 Higgs doublet model deviates from supersymmetry 
inspired models, by both ``Higgs"
particles having the same ratio of the couplings to the $-1/3$
charged quarks and the $2/3$ charged quarks. This distinguishing
feature puts a high premium on being able to detect
the charm anti-charm quark decay modes as well as the bottom
anti-bottom quark decays of Higgs particles at the LHC. It should also be
possible to calculate the contribution of the bound state to the
$\rho$ parameter, but this seems to be rather difficult in
practice.

At present the strongest evidence in favour of our scenario is
that the experimental top quark Yukawa coupling constant is,
within the crude accuracy of our calculations, in agreement with
the value at which the phase transition between the two vacua
should take place. If this agreement should persist with a more
accurate calculation of the phase transition coupling, it would
provide strong evidence in support of our scenario

There is, of course, a need for some physical mechanism underlying
our model, which could be responsible for the needed fine-tuning.
It seems likely that some kind of non-locality, through space-time
foam or otherwise, is needed \cite{CDHBglasgowbrioni}

\newcommand{\QSagt}{\hbox{\lower2pt\hbox{$>\atop\raise3pt\hbox{$\sim$}$}}} 
\newcommand{\QSalt}{\hbox{\lower2pt\hbox{$<\atop\raise3pt\hbox{$\sim$}$}}} 
\newcommand{\QSgtrsim}{\QSagt} 
\newcommand{\QSlesssim}{\QSalt}

\title*{What Comes Next? (Beyond the Standard Model)}
\author{Qaisar Shafi}
\institute{%
Bartol Research Institute, University of Delaware, \\Newark, DE 19716, USA
}

\titlerunning{What Comes Next? (Beyond the Standard Model)}
\authorrunning{Qaisar Shafi}
\maketitle

\begin{abstract}
Experimental evidence supporting the presence of new
 physics beyond the standard model has been steadily mounting, especially in recent years. 
I discuss a number of such topics including supersymmetric GUTs 
with intimate connection to inflation and leptogenesis (and with crucial 
input from neutrino oscillations), extra dimensions, warped  geometry, 
cosmological constant problem, and $D$-brane inflation. Supersymmetry and 
extra dimensions can be expected to continue to play an important role in the search 
for a more fundamental theory.
\end{abstract}

\section{Introduction}

Despite its remarkable successes on many fronts
 the standard model (SM) is finally showing some cracks because it cannot
 explain some important experimental observations. For example:

\begin{itemize} 
 
\item It is unable to explain the atmospheric neutrino anomaly 
which is most easily understood via $\nu_{\mu} - \nu_{\tau}$ 
oscillations entailing essentially maximal mixing and  $\delta m^2 \sim 10^{-3} 
\,\mbox{eV}^2 $\cite{r1}.  Within the SM framework plus gravity, 
$\delta {m^2}$ is expected to be $\QSalt 10^{-10}\, \mbox{eV}^2$,  based on dimension
 five operators.

From a particle physics viewpoint this provides the most convincing evidence 
for physics beyond the SM. In this talk I shall 
consider two extensions that can explain the atmospheric and solar 
neutrino anomalies in terms of neutrino oscillations. One of them is well known 
and invokes a SM singlet neutrino\cite{r2}, while the second extension 
utilizes a fifth dimension that is warped\cite{r3} but does not introduce 
any new fields!

\item The SM fails to provide a satisfactory 
non-baryonic dark matter candidate. From big bang nucleosynthesis, and more 
recently  $\delta T/T$ measurements,  it seems clear that baryons make up just
 a few percent of the critical energy density of the universe. There is 
growing experimental evidence for dark matter on the order of 30-35\% of 
the critical density\cite{r4}, and a satisfactory model should provide 
some reasonable candidate(s). Plausible ones include the LSP, axion and 
perhaps even long lived superheavy particles. All three require new physics
 beyond the SM:  

\begin{itemize}         

\item Despite several attempts it appears to be the case that         
the SM cannot provide a satisfactory explanation for the         
observed baryon asymmetry in the universe. We will consider an         
attractive scenario in which inflation and leptogenesis are         
intimately related. Speaking of inflation\cite{r5},    
 it does not         seem possible to realize it within 
the SM framework. Note that         a satisfactory inflationary scenario 
should\cite{r6} resolve the         flatness and horizon problems; provide
 a satisfactory source of         density fluctuations; explain the 
observed baryon asymmetry in         the universe; 

\end{itemize} 
\end{itemize}                

There are, in addition, several theoretical 
motivations for physics beyond the SM: 

\begin{itemize}        

\item The SM has a plethora of undetermined parameters. The        list includes the 
three gauge couplings, the weak mixing angle        $\theta_W$, the CKM 
and MNS mixing angles and phases, fermion        masses, etc. Some progress
 can be achieved by resorting to grand        unification;

 \item The gauge hierarchy problem is a major technical flaw in the      
 SM. Remedies that will be briefly discussed include supersymmetry       
and warped geometry;

     \item Last but not least, it has so 
far not been possible to      provide a consistent unification of the SM 
and gravity. 

\end{itemize}

 I will discuss some extensions of the SM that 
are motivated by one or more of the points listed above.

\section{The MSSM route}

Ignoring $\nu$ oscillations, MSSM presumably is the most
 compelling 'minimal' extension of SM. In this approach, the gauge 
hierarchy problem is at least partially resolved (the supersymmetric $\mu$ 
term poses a new problem, namely how its coefficient happens to be of order
 $M_W$ rather than $M_P=2.4\times 10^{18}\,\mbox{GeV}$). The model also 
offers an attractive dark matter candidate known as the LSP. This particle
 arises because of the existence of unbroken $Z_2$ matter parity which 
is imposed in MSSM and $SU(5)$ by hand. The $Z_2$ symmetry can be nicely 
embedded\cite{r7} in the center $Z_4$ of $SO(10)$, and remains unbroken 
provided one utilizes tensor representations for the symmetry breaking of $SO
(10)$ to MSSM. Another attractive feature of MSSM has to do with gauge 
coupling unification\cite{r8}. The three gauge couplings nicely unify at a 
scale $\sim 2\times 10^{16}$ GeV, given an MSSM spectrum of particles in 
the TeV mass range and below. The MSSM scenario offers a particularly rich
 phenomenology which will be tested at the LHC and hopefully other 
accelerators. Among the shortcomings of MSSM one could list: The appearance 
of new flavor and CP violating processes that must be carefully monitored;
 The issue of proton stability (or rather instability!). Although $Z_2
$ matter parity in MSSM prevents rapid proton decay, dimension five 
superpotential couplings such as $QQQL$ are still permitted and lead to decay 
channels such as  $p \rightarrow K + \nu$. The new lower bounds from 
SuperK on these decay channels\cite{r1} require that the dimensionless 
coefficients associated with the dimension five operators be  $\QSalt  10^{-8}
$. Clearly, this is not a very satisfactory state of affairs; The 
mechanism responsible for SUSY breaking has no explantion within the MSSM 
setting. Proposed scenarios based on supergravity (SUGRA), gauge mediation an
d extra dimensions have been extensively discussed.

\section{From MSSM to SUSY GUTS}

Grand unification is an `obvious' and certainly 
elegant extension of the SM. Some well known candidates are $SU(5)$\cite{r9}
 $SO(10)$\cite{r10} and $E_6$\cite{r11}, and the subgroups  $SU(4)_c 
\times SU(2)_L \times SU(2)_R$\cite{r12} of  $SO(10)$ and $SU(3)_c \times 
SU(3)_L \times SU(3)_R$ of $E_6$.  The hallmarks of grand unification 
include:  

Electric charge quantization which, it turns out, is intimately 
related to the existence of topologically stable magnetic monopoles. The 
discovery of such mono\-poles may help distinguish between the various GUTS,
 even in the absence of proton decay or other signatures. For instance,
 if the monopoles carry two quanta of Dirac magnetic charge\cite{r13}, 
the underlying theory should be $SO(10)$ breaking via its 4-2-2 subgroup 
above. If the lightest monopoles carry three Dirac quanta, then it would 
signal the presence of $E_6$ breaking via its 3-3-3 subgroup\cite{r14};

The existence of $B$ and $L$ violating processes in GUTS which 
makes it rather easy to generate the observed baryon asymmetry;

In the $SO(10)$ and $E_6$ schemes there exist SM singlet fields which lead to the 
appearance of non-zero masses for the known neutrinos via the see saw 
mechanism\cite{r2};

Gauge coupling unification which seems to be in good
 agreement with the measurements, especially in the supersymmetric 
framework;

Good (but sometimes bad!) asymptotic relations, such as $m_b 
= m_\tau$, and $m_\mu = 3 m_s$;

Proton decay, which alas has so 
far not been observed. Indeed, it is perhaps reasonable to state that the 
minimal versions of supersymmetric $SU(5)$ and $SO(10)$ are excluded 
because they predict proton decay via dimension five operators with a 
lifetime of order $10^{28} - 10^{32} \mbox{yrs}$, in conflict with the SuperK 
lower bound of $10^{33}$ yrs.

I now wish to emphasize the predictive power of SUSY  $SO(10)/SU(4)_c \times SU(2)_L \times SU(2)_R$  
by considering the following four topics: Top Quark Mass, Higgs Boson Mass, 
Inflation and Leptogenesis.

\begin{enumerate}

\item To see how the top mass was predicted in $SO(10)$\cite{r15},  let us consider the third 
family superpotential coupling $16_3 \times 16_3 \times 10$,  where the 
10-plet contains the MSSM electroweak doublets $H_u$ and $H_d$.

This yields the asymptotic Yukawa relations $h_t = h_b = h_{\tau}$ and 
tan $\beta \sim m_t /m_b~(>>1)$.  The requirement that $m_b (m_b) \sim 3 m_{\tau
}$ then requires that $m_t (m_t) \sim 175 \,\mbox{GeV}$,  which is in 
very good agreement with the experimental numbers. Radiative corrections can
 have important implications and are discussed in \cite{r16}.

\item 
Regarding the SM higgs mass, let us consider the ``Weinberg Salam'' limit
 of MSSM such that only a single higgs survives below the susy breaking 
scale. Since $\tan\beta \gg 1$, the tree level mass of the surviving (SM) 
higgs is essentially equal to $M_Z$. After radiative corrections, one
 obtains for the higgs mass a value of about  115 - 125 GeV\cite{r17}.

\item To see how inflation can be elegantly realized\cite{r18,r19}  in the 
framework of SUSY GUTs, let us consider the gauge symmetry breaking $G 
\rightarrow H$,  such that (global) SUSY is unbroken.  Introduce a suitable 
$U(1)_R$ symmetry which permits the following unique superpotential:
 
 \begin{equation} W = \kappa S ( \phi \bar \phi - M^2) \,  \end{equation}  

where $\phi,\bar\phi$ are conjugate fields whose scalar components acquire vevs 
that break $G\rightarrow H$,  while SUSY is unbroken. The singlet 
superfield S provides the inflaton. Note that both $W$ and $S$ carry one unit of
 $R$ charge. In the absence of SUSY breaking the vev of $S$ is zero, but 
the latter is shifted by an amount proportional to the gravitino mass 
$m_{3/2}$, once SUSY breaking a la supergravity is introduced. An 
inflationary scenario in the early universe is realized by assuming that the fields 
were displaced sufficiently far from their present day minima.  Thus,
 for $\langle S\rangle \gg M$,  $\phi,\bar\phi \rightarrow 0$, so that $G
$ was restored in the early universe, and the tree level scalar potential is given 
by  
\begin{equation} V_{\mathrm{tree}} = \kappa^2 M^4 \,. 
\end{equation} 
\noindent With SUSY thus broken, there are radiative corrections from the
  $\phi-\bar\phi$ supermultiplets that provide logarithmic corrections 
to the potential which lead to inflation. 
In one loop approximation the inf\mbox{}lationary effective potential is given by \cite{r18}
\begin{eqnarray} \label{loop}
V_{LOOP} &= & \kappa^{2}M^{4}\bigg[1 +\frac{\kappa^{2}\mathcal{N}}{32\pi^{2}} \Big(
2\ln\frac{\kappa^{2}|S|^{2}}{\Lambda^{2}}+(z+1)^{2}\ln(1+z^{-1})+ \nonumber \\ 
& &\hspace{4.5cm} +(z-1)^{2}\ln(1-z^{-1})\Big) \bigg]\,,
\end{eqnarray}
\noindent where $z=x^{2}=|S|^{2}/M^{2}$, $\mathcal{N}$ is the dimensionality of the representations to which
$\phi$, $\overline{\phi}$ belong, and $\Lambda$ is a renormalization mass scale. From Eq. (\ref{loop}) the
quadrupole anisotropy is found to be \cite{r18,lss}:
\begin{equation} \label{quad}
\left(\frac{\delta T}{T}\right)_{Q}\approx\frac{8\pi}{\sqrt{\mathcal{N}}}\left(\frac{N_{Q}}{45}\right)^{\frac{1}{2}}
\left(\frac{M}{M_{P}}\right)^{2}x_{Q}^{-1}y_{Q}^{-1}f (x^{2}_{Q})^{-1}\,,
\end{equation}
\noindent with
\begin{equation}
f(z)=\left(z+1\right)\ln\left(1+z^{-1}\right)+\left(z-1\right)\ln\left(1-z^{-1}\right)\,,
\end{equation}
\begin{equation}
y_{Q}^{2}=\int_{1}^{x^{2}_{Q}}\frac{\textrm{d} z}{z f(z)}\quad ,y_Q\ge 0\,.
\end{equation}
\noindent Here, the subscript $Q$ denotes the epoch when the present horizon scale crossed outside the inf\mbox{}lationary horizon
and $N_{Q}$ is the number of e-foldings it underwent during inf\mbox{}lation. From Eq. (\ref{loop}) one also obtains
\begin{equation} \label{m_kap}
\kappa\approx\frac{8\pi^{3/2}}{\sqrt{\mathcal{N}N_{Q}}}\,y_{Q}\,\frac{M}{M_{P}}\,.
\end{equation}
\noindent For relevant values of the parameters ($\kappa\ll1$), the slow roll conditions are violated only `infinitesimally' close
to the critical point at $x=1$ ($|S|=M$) \cite{r19}. So inf\mbox{}lation continues practically until this point is reached,
where the `waterfall' occurs.

Several comments are in order:

\begin{itemize}
\item For $x_{Q}\gg1$ (but $|S_{Q}|\ll M_{P}$), $y_{Q}\to x_{Q}$ and $x_{Q}\,y_{Q}\,f(x^{2}_{Q})\to1^{-}$.

\item Comparason of Eq. (\ref{quad}) with the COBE result $(\delta T/T)_{Q}\simeq6.3\times10^{-6}$ \cite{cobe} shows that
the gauge symmetry breaking scale $M$ is naturally of order $10^{16}$ GeV.

\item Suppose we take $G=SO(10)$, with $\phi(\overline{\phi})$ belonging to the $\textbf{16}(\overline{\textbf{16}})$
representation, so that $G$ is spontaneously broken to $SU(5)$ at scale $M$. Taking $\mathcal{N}=16$, and
 $x_{Q}\,y_{Q}\,f(x^{2}_{Q})\to1^{-}$, $M$ is determined to be $10^{16}$ GeV, which essentially coincides with the
SUSY GUT scale. The dependence of $M$ on $\kappa$ is displayed in Fig:\,\ref{m_kappa}. Note that a five dimensional supersymmetric
$SO(10)$ model in which inf\mbox{}lation is associated with this symmetry breaking was presented in \cite{kyae}. 

\begin{figure}[htb]
\psfrag{N=16}{\scriptsize{$\mathcal{N}=16$}}
\psfrag{N=2}{\scriptsize{$\mathcal{N}=2$}}
\includegraphics[angle=0, width=12cm]{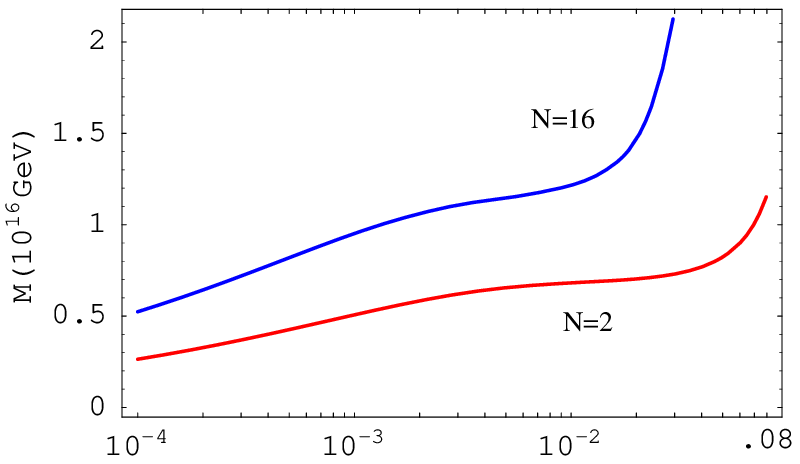}
\vspace{-.8cm}
\begin{center}
{\large \qquad $\kappa$}
\end{center}
\caption{The gauge symmetry breaking scale $M$ as a function of the coupling constant $\kappa$.
$\mathcal{N}=16\ (2)$ corresponds to the breaking $SO(10)\to SU(5)$ and $SU(2)_L\times SU(2)_R\times U(1)_{B-L}\to
SU(2)\times U(1)$ respectively.}
\label{m_kappa}
\end{figure}

\item Another realistic example is given by
$G=SU(3)_c\times SU(2)_{L}\times SU(2)_{R}\times U(1)_{B-L}$, corresponding to $\mathcal{N}=2$, and the
scale $M$ is then associated with the breaking of $SU(2)_{R}\times U(1)_{B-L}\to U(1)_{Y}$ \cite{lss,dls}.

\item The scalar spectral index $n_s$ is given by \cite{ll}
\begin{equation}
n_s\cong1-6\epsilon+2\eta,\qquad\epsilon\equiv \frac{m^2_P}{2}\left(\frac{V'}{V}\right)^2,\qquad\eta\equiv\frac{m^2_P V''}{V},
\end{equation}
\noindent where $m_P$ is the reduced Planck mass $M_P/\sqrt{8\pi}$; hereafter we take $m_P=1$. The primes denote derivatives with 
respect to the normalized real scalar field $\sigma\equiv\sqrt{2}|S|$. For $x_{Q}\gg1$ (but $\sigma_{Q}\ll1$), $n_s$
approaches \cite{r18}
\begin{equation}
n_{s}\simeq1+2\eta\simeq 1-\frac{1}{N_{Q}}\simeq0.98
\end{equation}
where $N_{Q}\approx60$ denotes the number of e-foldings.\footnote{
$N_{Q}\simeq56.5+(1/3)\ln(T_r/10^9\,\textrm{GeV})+(2/3)\ln(\mu/10^{15}\,\textrm{GeV})$ \cite{r19}, 
where $T_r$ is the reheating temperature
and $\mu$ is the false vacuum energy density.} The dependence of $n_{s}$ on $\kappa$ is displayed in Fig:\,\ref{n_kappa} 
(the behavior of
$n_s$ for large $\kappa$ is inf\mbox{}luenced by the SUGRA correction, as discussed below).

\begin{figure}[htb]
\psfrag{N=16}{\scriptsize{$\mathcal{N}=16$}}
\psfrag{N=2}{\scriptsize{$\mathcal{N}=2$}}
\includegraphics[angle=0, width=12cm]{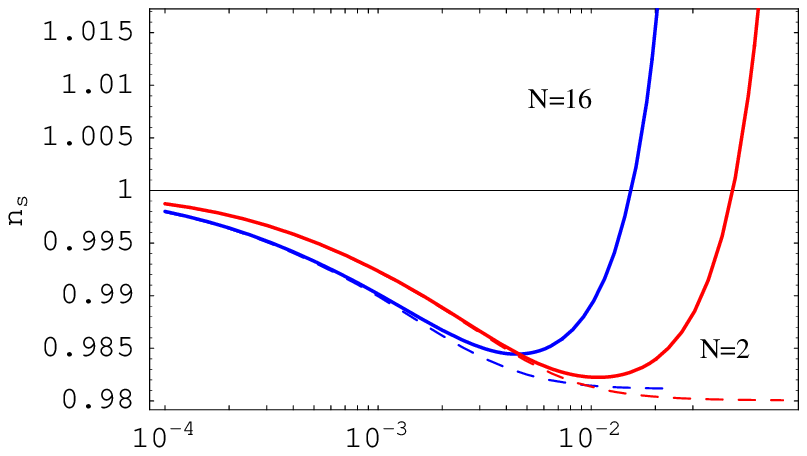}
\vspace{-.8cm}
\begin{center}
{\large \qquad $\kappa$}
\end{center}
\caption{The spectral index $n_s$ at $k=0.05 \ \textrm{Mpc}^{-1}$ as a function of the coupling constant $\kappa$
(dashed line--without SUGRA correction, solid line--with SUGRA correction).}
\label{n_kappa}
\end{figure}

\item The minimum number of e-foldings ($\approx60$) required to solve the horizon and flatness problems can be achieved
even for $x_{Q}$ very close to unity, provided that $\kappa$ is taken to be sufficiently small. This follows from Eq. (\ref{m_kap}).
An important constraint on $\kappa$ can arise from considerations of the reheat temperature $T_{r}$ after inf\mbox{}lation,
taking into account the gravitino problem. The latter requires that $T_{r}\QSlesssim10^{10}$ GeV \cite{gravitino}, unless
some mechanism is available to subsequently dilute the gravitinos. 

The inf\mbox{}laton mass is $\sqrt{2}\kappa M$ (recall that both $S$, and $\phi$, $\overline{\phi}$ oscillate about their
minima after inf\mbox{}lation is over, and they have the same mass), and so to prevent inf\mbox{}laton decay via gauge interactions
which would cause $T_r$ to be too high ($\sim M\gg10^{10}$ GeV), the coupling $\kappa$ should not exceed unity.
A more stringent constraint on $\kappa\ (\leqslant0.1)$ appears when SUGRA corrections are included.

\item For $G=SO(10)$ or $SU(3)_c\times SU(2)_{L}\times SU(2)_{R}\times U(1)_{B-L}$, 
the inf\mbox{}laton produces right handed neutrinos \cite{ls,lss,lsv,Pati:2002pe} whose subsequent out of
equilibrium decay leads to the observed baryon asymmetry via leptogenesis \cite{lepto,ls}.

\item For sufficiently large values of $\kappa$, SUGRA corrections become important, and more often than not, these tend to derail
an otherwise succesful infla\-tionary scenario by giving rise to scalar mass$^2$ terms of order $H^2$, where $H$ denotes
the Hubble constant. Remarkably, it turns out that for a canonical SUGRA potential (with minimal K\"ahler potential
$|S|^2+|\phi|^2+|\overline{\phi}|^2$), the problematic mass$^2$ term cancels out for the superpotential $W_1$ in
Eq.\,(1) \cite{copeland}. This may be considered an attractive feature of the inf\mbox{}lationary scenario. Note
that this property persists even when non-renormalizable terms that are permitted by the $U(1)_R$ symmetry are included
in the superpotential.

\end{itemize}

\noindent The SUGRA scalar potential is given by
\begin{equation} \label{sugra}
V=e^{K/m^2_p}\left[\left|\frac{\partial W}{\partial z_i}+\frac{z^*_i W}{m^2_p}\right|^2-3\frac{|W|^2}{m^2_p}\right]\,,
\end{equation}
\noindent where the sum extends over all fields $z_i$, and $K=\sum_i |z_i|^2$ is the canonical K\"ahler potential.
From Eq. (\ref{sugra}), the SUGRA correction to the potential 
is \cite{copeland,QSpanagio,QSlinde,QSkawasaki}
\begin{equation}
V_{SUGRA}=\kappa^{2}M^{4}\left[\frac{1}{8}\sigma^{4}+\ldots\right]\,,
\end{equation}
\noindent where $\sigma=\sqrt{2}|S|$ is a normalized real scalar field, and we have set the reduced Planck mass $m_P=1$. 
The effective inf\mbox{}lationary potential $V_1$ can be written to a good approximation as the sum of the radiative and SUGRA corrections.
For $1\gg\sigma\gg \sqrt{2}M$,
\begin{equation} \label{v_approx}
V_1\approx\kappa^{2}M^{4}\left[1+\frac{\kappa^{2}\mathcal{N}}{32\pi^{2}}2\ln\frac{\kappa^{2}\sigma^{2}}{2\Lambda^{2}}+
\frac{1}{8}\sigma^{4}\right]\,,
\end{equation}
\noindent and comparing the derivatives of the radiative and SUGRA corrections one sees that the radiative term dominates for
$\sigma^2\QSlesssim\kappa\sqrt{\mathcal{N}}/2\pi$. From $3H\dot{\sigma}=-V'$, $\sigma^2_Q\simeq\kappa^2 \mathcal{N}N_Q/4\pi^2$ for the
one-loop effective potential, so that SUGRA effects are negligible only for $\kappa\ll2\pi/\sqrt{\mathcal{N}}N_{Q}\simeq
0.1/\sqrt{\mathcal{N}}$. (For $\mathcal{N}=1$, this essentially agrees with \cite{QSlinde}).

\noindent From Eq. (\ref{v_approx}), the scalar spectral index is given by
\begin{equation}
n_s\simeq1+2\eta\simeq1+2\left(3\sigma^2-\frac{\kappa^2\mathcal{N}}{8\pi^2\sigma^2}\right)\,,
\end{equation}
\noindent and it exceeds unity for $\sigma^2\QSgtrsim\kappa\sqrt{\mathcal{N}}/2\sqrt{3}\pi$. For $x_Q\gg1$,
\begin{equation} \label{nq}
N_Q=\int_{\sigma_{end}}^{\sigma_Q}\frac{V}{V'}\textrm{d}\sigma
\approx\frac{\pi}{2\sigma^2_Q}\frac{\kappa}{\kappa_c}\tan\left(
\frac{\pi}{2}\frac{\kappa}{\kappa_c}\right)\,,
\end{equation}
\noindent where $\kappa_c=\pi^2/\sqrt{\mathcal{N}}N_Q\simeq0.16/\sqrt{\mathcal{N}}$.  
Using Eq. (\ref{nq}), one finds that the spectral index exceeds unity for $\kappa\simeq2\pi/\sqrt{3\mathcal{N}}N_Q\simeq0.06/\sqrt{\mathcal{N}}$.

The quadrupole anisotropy is found from Eq. (\ref{v_approx}) to be
\begin{equation}
\left(\frac{\delta T}{T}\right)_{Q}=\frac{1}{4\pi\sqrt{45}}\frac{V^{3/2}_1}{V'_1}\approx
\frac{1}{2\pi\sqrt{45}}\frac{\kappa\,M^2}{\sigma^3_Q}\,.
\end{equation}
\noindent In the absence of the SUGRA correction, the gauge symmetry breaking scale $M$ calculated from the observed quadrupole 
anisotropy approaches
the value $\mathcal{N}^{1/4}\cdot6\times10^{15}$ GeV for $x_Q\gg1$ (from Eq. (\ref{quad}), with $x_{Q}\,y_{Q}\,f(x^{2}_{Q})\to1^{-}$).
The presence of the SUGRA term leads to larger values of $\sigma_Q$ and hence larger values of $M$ for 
$\kappa\QSgtrsim0.06/\sqrt{\mathcal{N}}$. The dependence of $M$ on $\kappa$ including the full one-loop
potential (Eq. (\ref{loop})) and the leading SUGRA correction is presented in Fig:\,\ref{m_kappa}.

\begin{figure}[htb]
\psfrag{N=16}{\scriptsize{$\mathcal{N}=16$}}
\psfrag{N=2}{\scriptsize{$\mathcal{N}=2$}}
\includegraphics[angle=0, width=12cm]{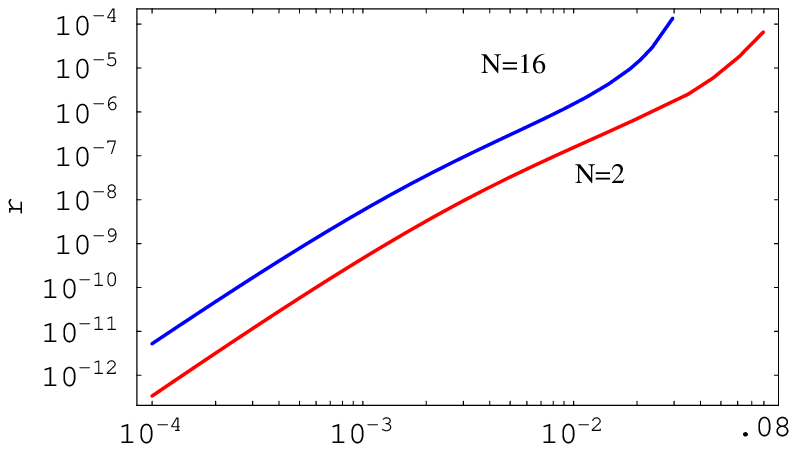}
\vspace{-.8cm}
\begin{center}
{\large \qquad $\kappa$}
\end{center}
\caption{The tensor to scalar ratio $r$ as a function of the coupling constant $\kappa$.}
\label{fig:r_kappa}
\end{figure}

The scalar spectral index in this class of models is close to unity for small $\kappa$, has a minimum 
at $\simeq0.98$ for
$\kappa\simeq0.02/\sqrt{\mathcal{N}}$, and exceeds unity for $\kappa\QSgtrsim0.06/\sqrt{\mathcal{N}}$ (Fig:\,\ref{n_kappa}).
The experimental data seems not to favor $n_s$ values in excess of unity on smaller scales (say $k\sim0.05$ Mpc$^{-1}$),
which leads us to restrict ourselves to $\kappa\QSlesssim0.06/\sqrt{\mathcal{N}}$. 
Thus, even though the symmetry breaking scale $M$ is of order
$10^{16}$ GeV (Fig.\,\ref{m_kappa}), the vacuum energy density during inf\mbox{}lation is smaller than $M^4_{GUT}$. Indeed,
the tensor to scalar ratio $r\QSlesssim10^{-4}$ (Fig:\,\ref{fig:r_kappa}). Finally, the quantity $\textrm{d}n_{s}/\textrm{d}\ln k$ is negligible
for small $\kappa$ and $\sim10^{-3}$ as the spectral index crosses unity \cite{QSkawasaki} (Fig:\,\ref{n_k}). 
The WMAP team has reported a value for 
$\textrm{d}n_{s}/\textrm{d}\ln k=-0.042^{+0.021}_{-0.020}$ \cite{wmap}, 
but the statistical significance of this conclusion has been questioned by the authors of \cite{seljak}. Clearly, more data is
necessary to resolve this important issue. Modifications of the models discussed here has been proposed in \cite{QSkawasaki} to
generate a much more significant variation of $n_s$ with $k$.

\begin{figure}[htb]
\includegraphics[angle=0, width=12cm]{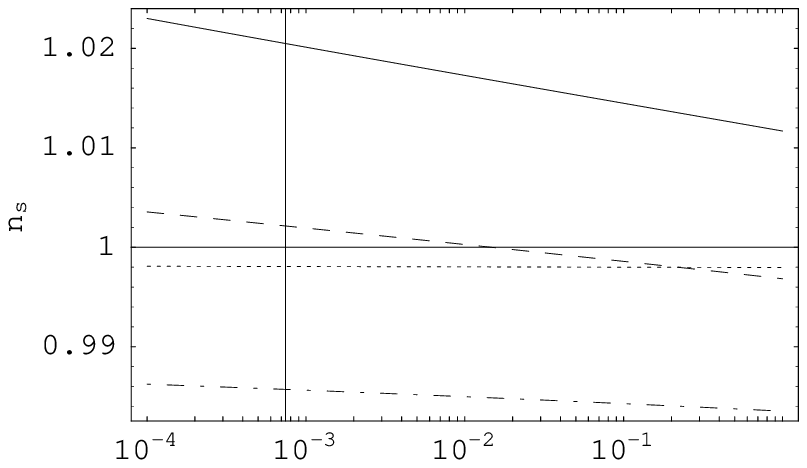}
\vspace{-.8cm}
\begin{center}
{\large \qquad $k/\textrm{Mpc}^{-1}$}\end{center}
\caption{The spectral index $n_s$ as a function of the wavenumber $k$ $(\mathcal{N}=16)$:
$\kappa=0.004$ (dot-dashed), $1\times10^{-4}$ (dotted), $0.015$ (dashed), $0.02$ (solid).}
\label{n_k}
\end{figure}

Let's now consider leptogenesis via inflaton decay for a particular model. 
An interesting example is $G= SU(4)_c \times SU(2)_L \times SU(2)_R$ and  $H = SU(3)_c \times SU(2)_L 
\times U(1)$, with $\phi = (4,1,2)$ and $\bar\phi = (\bar 4, 1, 2)$\cite{r20}.  
At the end of inflation the scalar fields $\phi$, $\bar\phi$ 
and $S$ oscillate about their respective minima. Taking account of the 
following superpotential couplings that Provide masses for the right handed
 neutrinos,  

\begin{equation} W_1 = (\gamma_i/ M_P) \phi \bar \phi F_I^c F_I^c
 \,, \end{equation} 

 where $F_I^c$ belong to $(\bar 4,1,2)$ of $G$, one sees 
that the three fields $\phi, \bar\phi$ (and $S$) decay into right handed 
neutrinos  (sneutrinos) [$M_P = 2.4 \times 10^{18}$ GeV is the reduced 
Planck mass].  The decay width is given by  

\begin{equation} \Gamma \sim (m_{\mathrm{inf}}/8\pi) \left(M_i/M_P\right)^2, \end{equation}  

where $M_i$ denotes the mass of the heaviest right handed neutrino that the inflaton can decay 
into. Assuming an MSSM spectrum below the GUT scale, the reheat 
temperature is estimated to be 

 \begin{equation} T_r \sim 1/3(\Gamma M_P )^{1/2} \sim (1/
10) (50/ N_Q)^{1/2} M_i \,. \end{equation} 

The gravitino constraint requires that
  $T_r \QSalt 10^9 - 10^{10}\, \mbox{GeV}$\cite{gravitino}.  
Thus, we have $M_i \QSalt 10^{
10}\, \mbox{GeV}$.

In order to decide on the decay product of the 
inflaton, let us take a look at the atmospheric and solar neutrino oscillations. 
We begin with the atmospheric neutrino ($\nu_\mu- \nu_\tau$) 
oscillations, and assume that the light neutrinos have hierarchical masses, with $
m_3 >>m_2 >> m_1$. Then,  

\begin{equation}                   \Delta m_{\mathrm{atm}}^2 = m_3^2 = (m_{D3}^2/ M_3) , \end{equation}  

where $m_{D3}$ ($= m_
t$) denotes the third family Dirac mass and equals the asymptotic top 
quark mass because of $SU(4)_c$ . It turns out  $m_t \sim 110 \,\mbox{GeV}$ 
[ ],  so that $M_3 \sim \mbox{few} \times 10^{14}\, \mbox{GeV}$.  Thus, the
 inflaton will not decay into the third family right handed neutrino. 
Similar arguments\cite{r22}  taking account of solar neutrino oscillations 
show that the inflaton also cannot decay into the second heaviest right 
handed neutrino, whose mass is estimated to be of order $10^{12}- 10^{13} 
\,\mbox{GeV}$.  Thus, we are led to the conclusion that the inflaton decays 
into the lightest (first family) right handed neutrino which has mass 
$\QSalt 10^{10}\, \mbox{GeV}$. As an example we consider $M = 8 \times 10^{
15}\, \mbox{GeV}$,  $\kappa \sim 10^{-3}$ and $m_{\mathrm{inf}}$ (= $2^{1/2
} \kappa M$)  $\sim 10^{13}\, \mbox{GeV}$.  For the inflationary scenario
 this yields a scalar spectral index $n=0.985$. Furthermore, the vacuum
 energy during inflation is $\sim 10^{-6} M_{\mathrm{GUT}}^4$,  so that 
the gravitational wave contribution to the quadrupole anisotropy is negligible ($ \QSalt 1\%$).

To summarize, the inflationary scenario we have 
discussed has the unique property that it predicts a mass scale linked 
with inflation that is tantalizingly close to the SUSY GUT scale, 
automatically leads to leptogenesis via inflaton decay, and predicts an essentially
 scale invariant ($n = \mbox{.98 -. 99}$) spectrum for the density 
fluctuations. Furthermore, the 50-60 e-foldings required to solve the horizon
 and flatness problems occur when the inflaton field value is quite close
 to the GUT scale (to within a factor of order unity or so). Thus, 
Planck scale corrections can be safely ignored. With a minimal Kaehler 
potential, the SUGRA corrections also do not disrupt the `flatness', which is 
a non-trivial result. Last but not least, let us note that the $SU(4)_c 
\times SU(2)_L \times SU(2)_R$ breaking produces monopoles and how the 
primordial monopole problem is avoided has been extensively discussed  in 
\cite{r20}.

\end{enumerate} 

\section{Extra Dimensions (Large, Small and, of course, Intermediate)} 
 This subject has created much recent excitement for a variety of 
reasons. Before discussing them, let us recall some earlier motivations: 
 \begin{itemize}

\item extra dimensions may lead to `deeper' 
unification than grand unification . For instance, through inclusion of gravity,
 and unification of `matter', gauge and `higgs' sectors.

\item Superstring theories for their consistency require 10 spacetime dimensions. 
 \item  A particularly promising approach is based on the presumed 
existence of $M$-theory, whose low energy limit, it is speculated, is 11-$D$ 
supergravity.

\end{itemize}

Alas, the predictive power of these 
higher dimensional theories is quite limited. Indeed, it is far from clear
 how this top-down approach leads to the SM (or MSSM) at low energies. 
For instance, compactification of the ten dimensional $E_8 \times E_8$ 
heterotic string theory on a suitable six dimensional Calabi-Yau manifold 
leads to three chiral families and $N = 1$ SUSY which is quite remarkable.
 Unfortunately, the detailed phenomenology fails to work out. It is hard,
 if not impossible, to keep the proton sufficiently stable, there is 
proliferation of undetermined parameters associated with the Internal (CY) space, 
unification of the three gauge couplings is not easy because the
 low energy spectrum does not coincide with that from MSSM, etc. Put 
briefly, the top-down approach has not been particularly successful as far as
 phenomenology is concerned. It may greatly help matters if a bridge can 
be developed between the low energy world of the SM and MSSM and physics 
at the highest energies, such that the presence of the extra dimensions 
is somehow clearly exploited. A number of recent attempts try to show, for
 instance, how the presence of one or more extra dimension can be 
exploited to yield possible resolution of the gauge hierarchy problem, the 
doublet-triplet problem, the apparent stability of the proton, and so on. 
Three of these approaches will be briefly discussed:

\subsection{Large extra dimension(s)}

It is assumed that the fundamental (quantum gravity) 
scale $M_f$ is more or less comparable to the electroweak scale, and 
this can be made consistent within a $4+N$ dimensional setting through a 
relation of the type\cite{r24}  

\begin{equation} M_P^2 \simeq M_f^{2 + N} V_N\,,
 \end{equation} %

\noindent where $V_N$ denotes the characteristic volume of the $N$ 
dimensional compact space, and $M_P$, as usual, is the reduced Planck scale. 
Clearly, $V_N$ must be orders of magnitude larger than $M_f^{-N}$, for 
$M_P\gg M_f$ . For $M_f \sim \mbox{TeV}$, say, $R \sim 10^{(30/N - 17)}
\mbox{cm}$.  Since the gravitational force is currently being tested in the
 submillimeter region, the large extra dimension scenario is still 
plausible for $N=2$ from this point of view. Because higher dimensional 
gravity is infrared soft, rare processes such as $K\rightarrow \pi + 
\mbox{graviton}$  have branching ratios that are proportional to  
$(m_K / M_f)^{N+2} \sim 10^{-12}$, for $N=2$ and $M_f \sim \mbox{TeV}$. 
($m_K$ denotes the $K$ meson mass). With $M_f$ close to the elctroweak scale, it was
 hoped that the guage hierarchy problem would simply go away. There are
 at least two difficulties with this idea;   

\begin{itemize} 

\item A new hierarchy has been introduced , $M_f^{-N}/V_N \ll 1$. 
How does this arise ?

\item Even though laboratory constraints on $M_f$ are still very
 mild  ($\QSagt 1 - .6$ TeV) as $N$ varies from 2 to 5,  far more 
stringent lower bounds on $M_f$ arise from astrophysical considerations based on
 supernova and neutron stars\cite{r25}. Thus, for $N=2$ (3), $M_f$ may 
be at least as high as 1700 (60) TeV. This makes the approach look 
somewhat less appealing.

\end{itemize}

\subsection{SUSY GUTS and (Flat) Extra Dimension(s)} 

The presence of a fifth dimension with the 
topology of $S^1/Z_2 \times Z_2$ has been extensively utilized to construct
 supersymmetric models that: 

\begin{itemize} 

\item Achieve doublet-triplet splitting (without many tears!)\cite{r26};  

\item eliminate dimension five nucleon decay;  incorporate fermion mass hierarchies and 
fermion mixings without introducing small Yukawa couplings; 

\item implement the idea of 'neutrino' democracy which yields large neutrino mixings (with
out making them bimaximal)\cite{r27};  

\item realize inflation and leptogenesis in realistic $5D$ SUSY GUTS\cite{r22} \end{itemize}

With the 
rather stringent lower limits from SuperK on the proton lifetime, it is 
important that a five dimensional framework allows us to either completely
 eliminate or sufficiently suppress dimension five $p$ decay, and perform
 a clean doublet-triplet splitting without introducing a large , highly 
non-minimal Higgs sector required in four dimensional theories to realize 
the same goals.

As far as fermion masses and mixings are concerned, 
the five           dimensional approach is not necessarily more predictive,
 but it        does provide a new and geometric way to think about the
 hierarchies.    The fermion wave functions carry a non-trivial dependence 
on the        fifth dimension, which can be exploited to yield the 
effective four     dimensional Yukawa (determined by their overlap with the
 higgs          doublets) More interestingly, perhaps, a fifth dimension plays
 an       essential role in realizing the democratic neutrino ansatz in 
SUSY      GUTS\cite{r27}.

\subsection{Warped Geometry} 

The idea
 that one or  more of the extra dimensions may be warped can be
 motivated as follows:  

 \begin{itemize}

\item Provides possible resolution of the gauge hierarchy problem, namely how $M_W \ll M_P$\cite{r3}. Note
 that this can be achieved without invoking SUSY; 

 \item The wave functions of bulk fields are far more constrained compared to the case of 
'flat' geometry. In principle, this could lead to more predictive power;

\item The observed atmospheric and solar neutrino oscillations can be
 accomodated within the SM framework without invoking additional fields
 such as singlet (right handed) neutrinos. For a discussion of seesaw mechanism in warped
geometry, see\cite{Huber:2003sf};

\item The approach may be
 consistent with grand unification, which is not the case when 
dealing, say, with large extra dimensions;

\item There exists the intriguing 
possibility that `warping' could be exploited to understand the 'tiny' vacuum 
energy density  $\rho_v \sim 10^{-120} M_P^4$\cite{r28};

\item There 
are experimental predictions which can be experimentally tested. For 
instance, the $KK$ excitations of bulk SM fields are expected to have masses 
in the several-TeV range\cite{r29} which,  hopefully, can be tested at 
the LHC. Since there are no large extra dimensions, it is warping which 
makes these excitations experimentally accessible (without warping the 
masses would be of order $M_P$ and therefore inaccessible for ever!)

\end{itemize}

\subsubsection{Warped Geometry and Neutrino oscillations}
Warped geometry allows the possibility of accommodating the observed atmospheric and solar neutrino oscillations 
within the SM framework without invoking additional fields such as singlet (right handed) neutrinos. 
Especially the bimaximal mixing scenario are
incorporated in models with a small but warped extra dimension.
In one scenario Dirac neutrino masses arise from a coupling to
right-handed neutrinos in the bulk. Alternatively, dimension five
interactions may induce a Majorana mass for the neutrinos. In
both cases the mixing angle $U_{e3}$ is expected be close to
its present bound. While the direct detection of KK states is still
waiting, rare processes, such as proton decay,
neutrinoless double beta decay and flavor violation can provide
important hints to the higher dimensional theory.

\vspace{.5cm}
\noindent \textbf{The Dirac Case}

It was
suggested in ref.~\cite{GN} that in warped models small Dirac neutrino
masses can arise from a coupling to sterile (right-handed) bulk neutrinos.
In order to generate masses in the sub-eV range, the sterile
neutrinos have to be localized close to the Planck-brane, while
the SM model neutrinos were confined to the
TeV-brane. We generalized this scenario to incorporate
bulk SM neutrinos \cite{HS3}.

After electroweak symmetry breaking, the Dirac neutrino masses
are generated from
the Yukawa-type coupling between the SM neutrinos, contained in the
SU(2) doublets $L_i$, and the three sterile neutrinos $\psi_i$
\begin{equation}
{\cal L}=h^{(5)}_{ij}\bar{L}_i\psi_jH+{\rm h.c.~} +~\dots
\end{equation}
where $h^{(5)}_{ij}$ are the 5D Yukawa couplings. The
4D Dirac masses are given by
\begin{equation} \label{3.2}
m_{ij}=\int_{-\pi R}^{\pi R}\frac{dy}{2\pi R}h^{(5)}_{ij}
e^{-4\sigma}H(y) f_{0iL}(y)f_{0j\psi}(y),
\end{equation}
where the Higgs field is confined to the TeV-brane,
i.e.~$H(y)=H_0\delta(y-\pi R)$, to maintain the solution
of the hierarchy problem.

The fermion wave functions and
consequently the induced  4D Dirac masses crucially depend
on the 5D mass parameters of the
left- and right-handed fermions, $c_L$ and $c_{\psi}$ respectively,
which enter Eq.~(\ref{3.2}). As the 5D Dirac mass, i.e.~$c$
parameter of the fermion increases, it gets localized closer
towards the Planck-brane. Its overlap with the Higgs profile
at the TeV-brane is consequently reduced, which is reflected in
a smaller 4D fermion mass from electroweak symmetry breaking.
In ref.~\cite{HS2} we have shown that this geometrical picture
nicely generates the charged
lepton and quark mass hierarchies, as well as quark mixings,
by employing $c$-parameters and 5D Yukawa couplings
of order unity.

From the KK reduction of the left-handed neutrino field $\nu_L$
we obtain a left-handed zero mode $\nu_L^{(0)}$, corresponding to
the SM neutrino,  and an infinite
tower of left- and right-handed KK excited states $\nu_L^{(i)}$ and
 $\nu_R^{(i)}$, where we omit flavor indices. The sterile neutrinos
decompose into the right-handed zero mode $\psi_R^{(0)}$ and the
KK excited states $\psi_L^{(i)}$ and $\psi_R^{(i)}$. After electroweak
symmetry breaking the mass matrix takes the form
\begin{equation} \label{nu_mass}
M_{\nu}= (\nu_L^{(0)},\nu_L^{(1)}, \psi_L^{(1)},\dots) \left(\begin{array}{cccc}
m^{(0,0)} & 0 &m^{(0,1)} & \cdots \\[.1cm]
m^{(1,0)} & m_{L,1} & m^{(1,1)}& \cdots \\[.1cm]
0 & 0 & m_{\psi,1} & \cdots \\
\vdots & \vdots & \vdots & \ddots
\end{array}\right)\left(\begin{array}{c}\psi_R^{(0)} \\[.1cm]   \nu_R^{(1)} \\[.1cm]
\psi_R^{(1)} \\ \vdots \end{array}\right)
\end{equation}
where we again suppress flavor indices, i.e.~every entry represents
a $3\times 3$ matrix in flavor space. The masses $m^{(i,j)}$
are obtained by inserting the relevant wave functions
into Eq.~(\ref{3.2}), and $m_{L,i}$ and $m_{\psi,i}$ denote the KK masses
of the excited neutrino states. The zeros in (\ref{nu_mass}) follow
from the $Z_2$ orbifold properties of the wave functions.

The squares of the physical neutrino masses are the eigenvalues of the
hermitian mass matrix $M_{\nu}M_{\nu}^{\dagger}$. The unitary matrix $U_{\nu}$,
such that $U^{\dagger}_{\nu}M_{\nu}M_{\nu}^{\dagger}U_{\nu}$ is diagonal,
relates the left-handed mass eigenstates $N_L^{\rm phys}$ to the
interaction eigenstates $N_L=(\nu_L^{(0)},\nu_L^{(i)},\psi_L^{(i)})$ via
$N_L=U_{\nu}N_L^{\rm phys}$. The observed neutrinos $\nu_L^{\rm phys}$
correspond to the three lightest states in $N_L^{\rm phys}$. The
right-handed mass eigenstates are obtained from a unitary matrix
$V_{\psi}$ which diagonalizes  $M_{\nu}^{\dagger}M_{\nu}$.
The physical neutrino mixings
\begin{equation}
U=U_l^{\dagger}U_{\nu}
\end{equation}
also depend on the rotations of the left-handed charged
leptons $U_l$. If the Yukawa couplings of the charged leptons are diagonal, the
mixing angles of the physical neutrinos can be directly read off
from $U$.
Like in the CKM matrix, there are three mixing angles, and a single complex
phase $\delta$ which induces CP violation in the lepton sector.

The atmospheric neutrino data imply \cite{SK}
\begin{equation} \label{atm}
1\cdot10^{-3}{\rm eV}^2 <\Delta m^2_{\rm atm}<5\cdot10^{-3}{\rm eV}^2,
\quad \sin^22\theta_{\rm atm}>0.85.
\end{equation}
There are several solutions to the solar neutrino anomaly \cite{BGP}.
Most favored is the LMA solution
\begin{equation} \label{sol}
2.5\cdot10^{-5}{\rm eV}^2 <\Delta m^2_{\rm sol}<4\cdot10^{-4}{\rm eV}^2,
\quad 0.3 <\sin^22\theta_{\rm sol}< 0.93.
\end{equation}
The SMA solution gives only a poor fit to the data. The CHOOZ reactor
experiment constrains $|U_{e3}|^2\equiv s_2^2$ to be at most a few
percent \cite{CHOOZ}. Nothing is known about the CP violating phase $\delta$.

Various constraints on the scenario with bulk gauge and fermion fields
have been discussed in the literature \cite{HS,r29,HPR}.
With bulk gauge fields for instance, the SM relationship between the gauge couplings
and masses of the Z and W bosons gets modified. The electroweak
precision data then requires the lowest KK excitation of the gauge bosons
to be heavier than about 10 TeV \cite{HS,r29}.
This bound becomes especially important if the fermions are localized
towards the Planck-brane $(c>1/2)$. As discussed in ref.~\cite{HS2},
this applies to all SM fermions, with the exception of the top-quark, if the
fermion mass hierarchy arises from their different locations in the
extra dimension. In this case the bounds induced by the contribution
of KK  excitations of the SM gauge bosons to  the electroweak
precision observables are weak. They only require the KK masses
to be above about 1 TeV \cite{GP}.
In the following we will therefore assume that
the mass of the first KK gauge boson is $m_1^{(G)}=10$ TeV.
The corresponding masses of the KK fermions are then in the range
10 to 16 TeV, for $0<c<1.5$. The mass of the lightest KK graviton
is 16 TeV.

In order to induce large mixings, the neutrinos have to be located at
similar positions in the extra dimension. Therefore, we take
$c_L^{(1)}=c_L^{(2)}=c_L^{(3)}\equiv c_L$. The charged lepton
masses are, for instance,  reproduced by $c_L=0.567$,
$c_E^{(1)}=0.787$, $c_E^{(2)}=0.614$ and $c_E^{(3)}=0.50$,
where for simplicity we have assumed diagonal Yukawa couplings
for the charged leptons. The neutrino mixings are then solely governed by
the neutrino mass matrix $M_{\nu}$. We take $k=M_5=M_{\rm Pl}$.
From $m_1^{(G)}=10$ TeV we determine the brane separation
$kR=10.83$.  A parameter set which implements the
large angle MSW solution is
\begin{eqnarray} \label{LMSW}
c_{\psi}^{(1)}=1.43,~ c_{\psi}^{(2)}=1.36, ~c_{\psi}^{(3)}=1.30,
\nonumber\\[.2cm]
\frac{h_{ij}^{(5)}}{g_2^{(5)}}=\left(\begin{array}{ccc}
-2.0 & 1.5 & -0.5 \\
-1.8 & -1.1 & 1.9 \\
0.5 & 1.9 & 1.7
\end{array}\right).
\end{eqnarray}
We obtain the light neutrino masses $m_{\nu1}=1.0\cdot 10^{-3}$ eV,
$m_{\nu2}=1.0\cdot 10^{-2}$ eV and $m_{\nu3}=7.1\cdot 10^{-2}$ eV.
For the neutrino oscillation parameters we find $\Delta m^2_{\rm atm}=
4.9\cdot10^{-3}$ eV$^2$, $\sin^22\theta_{\rm atm}=0.99$,
$\Delta m^2_{\rm sol}=1.0\cdot10^{-4}$ eV$^2$ and
$\sin^22\theta_{\rm sol}=0.90$. Also, $|U_{e3}|^2=0.036$
is close to the experimental bound which is typical for the
solutions we find.

The SM neutrinos mix not only with each other but also
with the left-handed KK states of the sterile neutrinos
$\psi_L^{(i)}$ \cite{GN}. This effect diminishes the effective weak
charge of the light neutrinos. Thus, the effective number
of neutrinos contributing to the width of the Z boson is reduced
to $n_{\rm eff}=3-\delta n$, where $\delta n$ is obtained from
summing the relevant squared entries of $U_{\nu}$. Measurements
of the Z width induce the constraint $\delta n QSlsim 0.005$.
For the parameter set (\ref{LMSW}) we find the result $\delta n = 2\cdot 10^{-5}$,
well below the experimental sensitivity.

%
%
%
%
%
%
\vspace{.5cm}
\noindent \textbf{The Majorana case}

In the warped SM the suppression scale for non-renormalizable
operators can be anywhere between a few TeV and the Planck scale,
depending on the localization of the fermions \cite{GP,HS2}.
Majorana masses for the left-handed neutrinos are generated by the
dimension-five operator \cite{GP,HS2,HS4},
\begin{equation} \label{nu3}
\int d^4x\int dy \sqrt{-g}\frac{l_{ij}}{M_5^2}H^2\Psi_{iL}C\Psi_{jL}
\equiv \int d^4x ~M^{(\nu)}_{ ij}\Psi_{iL}^{(0)}C\Psi_{jL}^{(0)},
\end{equation}
where $l_{ij}$ are dimensionless couplings constants and
$C$ is the charge conjugation operator. The neutrino mass matrix
reads
\begin{equation} \label{nu}
M^{(\nu)}_{ ij}=\int_{-\pi R}^{\pi R}\frac{dy}{2\pi R}\frac{l_{ij}}{M_5^2}e^{-4\sigma(y)}
H^2(y)f_{0iL}(y) f_{0jL}(y).
\end{equation}
By localizing $f_{0L}$ close to the Planck-brane
($c_L\gg1/2$), $M^{(\nu)}$ can be made very small. However,
at the same time, the masses of the charged leptons, residing
in SU(2) multiplets with the neutrinos, can become very small
as well \cite{GP,HS2}. Since the tau is the heaviest lepton,
it has to reside closest to the TeV-brane, and induces the
most stringent constraint on $l$.
In ref.~\cite{HS4} we
considered the somewhat favorable case $\lambda_{\tau}^{(5)}=10g^{(5)}_2$
and $k/M_5=0.01$, obtaining  $m_{\nu,\tau}=l\cdot 25$ eV. A neutrino
mass of 50 meV can then be generated by $l=0.002$, which
is only a moderately small number.

Following the spirit of ``anarchy'', we parametrize the coefficients
in (\ref{nu}) by
\begin{equation}
l_{ij}\equiv \Lambda\cdot \tilde l_{ij}.
\end{equation}
The supposedly
fundamental theory responsible for the effective interaction
(\ref{nu3}) is represented through the order unity coefficients $\tilde l_{ij}$.
In order to generate large mixings we locate the left-handed
doublets at the same position in the extra dimension,
\begin{equation}
c_{e,L}=c_{\mu,L}=c_{\tau,L}=0.72.
\end{equation}
To accommodate the tau, muon and electron masses we are led to
$c_{\tau,R}=1/2$, $c_{\mu,R}=0.64$ and $c_{e,R}=0.85$.
Taking the order unity coefficients to be homogeneously
distributed $1/2<|\tilde l_{ij}|<2$ with random phases from 0 to $2\pi$, we find
the most favorable value for the overall scale to be
$\Lambda=1.9\cdot10^{-3}$. We randomly generate
parameter sets for $\tilde l_{ij}$, calculate the neutrino
mass matrix from (\ref{nu}) and compute the neutrino
masses and mixings. The left-handed charged lepton rotations
are similar to the neutrino rotations because both fields sit
together  in the extra dimension. We include
them by generating random Yukawa couplings for the charged leptons
as well.

Focusing on the LMA solution of the solar neutrino anomaly,
which turns out to be clearly favored, we find the following
picture \cite{HS4}. From our neutrino parameter sets about 70 \% reproduce
$\Delta m^2_{\rm atm}$ (\ref{atm}). Imposing in addition the
constraint from $\Delta m^2_{\rm sol}$ (\ref{sol}), we are
left with about 28 \% of the parameter sets. The solar
and atmospheric mixings angles bring this number down
to about 6 \%, which is still a considerable fraction
given the number of constraints. The most stringent
constraint turns out to come from the CHOOZ experiment.
Adding the requirement $|U_{e3}|^2<0.05$, the fraction
of viable parameter sets shrinks to about 0.7 \%.
This result is closely related to an average $\langle |U_{e3}|^2\rangle =0.22$,
which is considerably above the experimental bound.
Averaging only over the sets which pass all experimental constraints,
we find $\langle |U_{e3}|^2\rangle =0.022$.
This means that  $|U_{e3}|^2$ is probably close to the experimental bound
and can likely be tested at future neutrino experiments,
such as MINOS, ICARUS and OPERA which are
 sensitive down to $|U_{e3}|^2\sim0.01$ \cite{LONGB}.
Since we typically start with large phases in the neutrino and
charged lepton mass matrices, the complex phase $\delta$ in the mixing
matrix $U$ is found to be most likely of order unity.
\begin{table}[t] \centering
\begin{tabular}{|c||c|c|c|} \hline
&  $\Delta m^2_{\rm atm, sol}$ &$+\sin^22\theta_{\rm atm, sol}$ & $+|U_{e3}|^2<0.05$
\\ \hline
LMA & 44.4 (28.1) & 5.8 (5.9) & 5.0 (0.7)   \\ \hline
SMA & 1.3 (0.04) & 0.3 ($<0.001$) & 0.3 ($<0.001$)  \\ \hline
LOW & 0.008 ($<0.001$) & 0.002 ($<0.001$)  &  0.002 ($<0.001$)  \\ \hline
\end{tabular}
\caption{Probability in percent that a randomly generated set of coefficients
$1/2<|\tilde l_{ij}|<2$ satisfies the constraints from $\Delta m^2_{\rm atm, sol}$
(first column) and $\sin^22\theta_{\rm atm, sol}$ (second column) and
 $|U_{e3}|^2<0.05$  (third column). The results are given for the case
$c_{e,L}>c_{\mu,L}=c_{\tau,L}$ ($c_{e,L}=c_{\mu,L}=c_{\tau,L}$).
 }
\label{t_L}
\end{table}
The SMA, LOW and VAC solutions to the solar neutrino anomaly
can be realized only with severe fine-tuning.
Our results  are summarized in table \ref{t_L}.

So far the constraint from the CHOOZ experiment is the most
stringent one. Its inclusion reduces the probability that a parameter
set realizes the LMA solution from about 6 \% to
less than one percent. With all entries in the neutrino mass matrix being
of similar magnitude, the ensuing mixing angles are typically
large. The fit to the neutrino data  improves considerably if the
electron neutrino is somewhat separated from the other two
neutrino species. Shifting the electron neutrino closer to the
Planck-brane induces small elements in the neutrino mass
matrix. As a result, small values of  $\Delta m^2_{\rm sol}$
and $|U_{e3}|^2$ become more probable. The neutrino mass
matrix (\ref{nu})
acquires the following structure
\begin{equation} \label{eps}
M^{(\nu)}\sim\left(\begin{array}{ccc}
\epsilon^2 & \epsilon & \epsilon  \\[.1cm]
\epsilon & 1 & 1 \\[.1cm]
\epsilon & 1 & 1
\end{array}\right),
\end{equation}
where $\epsilon\approx f_{0,e}^{(\nu)}(\pi R)/f_{0,\mu}^{(\nu)}(\pi R)
\approx \exp(-(c_{e,L}-c_{\mu,L})\pi kR)$.

We keep the muon and tau neutrinos at the previous locations
$c_{\mu,L}=c_{\tau,L}=0.72$. A separation of these two
would make the fit to the atmospheric neutrino data
more difficult since the corresponding mixing is reduced.
For the electron neutrino
the most favorable choice turns out to be $c_{e,L}=0.79$ leading to
$\epsilon=0.15$.
The positions
of the right-handed leptons are found to be $c_{e,R}=0.79$,
$c_{\mu,R}=0.62$ and $c_{\tau,R}=0.49$.

Let us again focus on the LMA solution.
Assuming $1/2<|\tilde l_{ij}|<2$ for the order unity coefficients,
we find the best value  $\Lambda=2.6\cdot10^{-3}$. The correct values
of the $\Delta m^2$'s are fitted by 44 \% of the parameter sets.
Taking into account also the constraints from the solar and
atmospheric mixing angles, this fraction shrinks to 5.8 \%.
This reduction is mostly due to the solar mixing angle which is
suppressed if $\epsilon$ is small.
Most important, however, the CHOOZ constraint $|U_{e3}|^2<0.05$
is now satisfied almost automatically, and we are finally left with 5.0 \%
of the parameter sets. Compared to the case of $\epsilon=1$, the
probability for a parameter set to satisfy all observational
constraints is enhanced by almost an order of magnitude.
The distribution of $|U_{e3}|^2$ is now peaked at
small values. Averaging only over parameter sets
which satisfy all constraints, we find a moderately small value
$\langle |U_{e3}|^2\rangle=0.010$. As a consequence,
the next generation neutrino experiments still has a chance
to detect a non-vanishing $|U_{e3}|^2$ \cite{LONGB}.

Taking even smaller values of $\epsilon$, it is possible to
implement the SMA solution to the solar neutrino problem.
The LOW and VAC solutions are very difficult to accommodate
because of their small solar  $\Delta m^2$. Thus, the LMA solution
is by far the most favored scenario. A collection of our results
is given table \ref{t_L}.

\vspace{.5cm}
\noindent \textbf{Rare processes}

The most direct evidence
for a Majorana mass of neutrinos would be the discovery of
neutrinoless double beta decay ($0\nu\beta\beta$).
The non-observation of this process in the Heidelberg-Moscow
experiment implies an upper bound on
$|M^{(\nu)}_{11}|\equiv m_{ee}<0.35$ eV \cite{HMC}.  There are
plans to bring this limit down to about 0.01 eV.
If all neutrinos are at the same position in the extra dimension,
we expect $m_{ee}\sim0.02$ eV, which is far below the current
experimental sensitivity but within reach of future experiments.
Once the electron neutrino is localized closer towards the Planck-brane
$0\nu\beta\beta$ becomes drastically suppressed proportional to
$\epsilon^2$. For the clearly favored LMA scenario we find
$m_{ee}\sim0.001$ eV, which is too small to be detected in the
near future. For the SMA, LOW and VAC solutions $m_{ee}$ is
even smaller. It is therefore questionable if our model induces
$0\nu\beta\beta$ at a detectable level. In any case the Majorana
masses we find are far below the recently claimed evidence
of about 0.4 eV \cite{KDHK}.

The low cut-off scale dramatically amplifies the impact of
non-renormalizable operators on rare processes, such as
proton decay and flavor violation. With bulk fields localized
towards the Planck-brane the corresponding suppression
scales can be significantly enhanced \cite{GP,HS2}.
However, there are limits because the SM fermions need
to have sufficient overlap with the Higgs field at the TeV-brane
to acquire their observed masses. In the case of proton
decay, typical dimension-six operators still have to be suppressed
by small couplings of order $10^{-8}$ to be compatible with
observations. One may be tempted to completely eliminate
these small numbers by imposing additional symmetries.
In ref.~\cite{HS3} we used lepton number conservation
to make the proton stable and to forbid the potentially
large Majorana neutrino masses of Eq.~(\ref{nu3}).
In the case of Majorana neutrinos proton stability
can be ensured by lepton parity. In both cases baryon
number violation may still be present, and processes
like  neutron anti-neutron oscillations and double
nucleon decay could be close to their
experimental bounds.

In the scenario of Dirac neutrino masses the rate of
$\mu\rightarrow e\gamma$ transitions is considerably
enhanced by the presence of heavy sterile neutrino states.
If the SM neutrinos are confined to the TeV-brane,
its large branching ratio pushes the KK scale up to 25 TeV
and thus imposes the most stringent constraint on the model
\cite{K00}.
However, the rate for $\mu\rightarrow e\gamma$ is very
sensitive to the mixing between light and heavy neutrino states.
With bulk neutrinos the mixing with heavy states is considerably
reduced. In the case of the large angle MSW solution (\ref{LMSW})
we obtain Br$(\mu\rightarrow e\gamma)\approx 10^{-15}$ \cite{HS3}.

\subsubsection{Origin of the Non-Zero Vacuum Energy Density}
Warping could be exploited to possibly identify the origin of the 
non-zero vacuum energy density reported by recent observations\cite{r4}. Let 
us assume that the observed $\rho_v$  is associated with a 
'false' vacuum where we happen to live, and which is separated by a 
potential barrier from the 'true' vacuum with zero energy density. Our main goal 
then is to identify the source of $\rho_v$.

Although SUSY will be 
important, especially when we discuss radiative corrections and quintessence\cite{r4}, the idea is 
best illustrated by considering a bulk real scalar field in a warped 
background:  

\begin{equation} ds^2 = e^{-2\sigma(y)}\eta_{\mu \nu}dx^\mu dx^\nu + 
dy^2 \,, \end{equation}  

\noindent where $y$ denotes the fifth coordinate, and  $\sigma(y) = k|y|$ . The 5D scalar $\mbox{mass}^2$ consists, 
in general, of both bulk and brane contributions 

\begin{equation} M^2 (y) = b^2 k^2 + a^2 k \delta (y- \pi R) + \tilde a ^2 k \delta(y)\,, \end{equation} 

\noindent where $k$ denotes the curvature of the warped space $S^1/Z_2$, and  $a, \tilde a$, 
and $b$ are dimensionless parameters.  In the massless case, the zero mode 
wave function is constant along the fifth ($y$) direction and, as expected,
 the $KK$ states have TeV scale masses and are localized close to the TeV
 brane. If one swithches on a bulk and/or a Planck brane mass term then, 
roughly speaking, the zero mode acquires a mass comparable 
($\sim \mbox{TeV}$) to the first $KK$ state.  The result is dramatically different, 
however, if instead a Planck size mass term is introduced on the TeV brane. The effective 
zero mode gets warped by two powers of $\Omega$  
($ = \exp(\pi k R)$, $k R \sim 10$, where $R$  denotes the size of the fifth dimension): 

\begin{equation} M_0 = a k \Omega^{-2} \,. \end{equation}  

\noindent In particular, for a TeV-brane mass of order $M_P$,  

\begin{equation} M_0 \sim \mbox{TeV}^2 / M_P \sim 
10^{-30} M_P \sim 10^{-3}\,\mbox{eV} \sim  \rho_v^{1/4} \,. \end{equation}  

\noindent This is certainly intriguing but the important thing now is to try to estab
lish a link between $m_0$ and $\rho_v$. Consider the following contributi
on to the 5D action:  

\begin{eqnarray} \Delta S_5 & = & \int d^4x 
d^4y\sqrt{-g}\delta(y - \pi R) \nonumber\\ &&\mbox{}\times\left[\frac{1}{
2}a^2 k |\Phi|^2 +  \frac{1}{3}E_5 k^{-\frac{1}{2}}|\Phi|^3 +  \frac{1}
{4}\lambda_5 k^{-2}|\Phi|^4 \right] \,. \end{eqnarray} 

\noindent Integrating over $y$, the 4D effective potential for the zero mode is given by
 
\begin{equation} V_4 \sim \frac{1}{2}m^2\phi^2 + \frac{1}{3}E\phi^3 + \frac{1}{4}\lambda\phi^4 \end{equation} 

\noindent where 

 \begin{eqnarray} m^2 & = & a^2 k^
2\Omega^{-4} \nonumber\\ E & = & E_5 k\Omega^{-4} \nonumber\\ \lambda
 & = & \lambda_5 \Omega^{-4} \,. \end{eqnarray} 

\noindent Note that all three are warped by four powers of $\Omega^{-1}$ (a bulk quartic term does not
 undergo warping). Suppose that $m^2 < 0$, in which case $\langle 
\phi\rangle \sim k\Omega^{-2}$,  and we could arrange an energy splitting of the desired magnitude
magnitude 

\begin{equation} \Delta V \sim m^2 \langle \phi\rangle^2 \sim k^4 \Omega^
{-8} \sim  (10^{-3}\mbox{eV})^4 \,. \end{equation} 

 Although we ignored SUSY
 so far, its presence should help ensure that radiative corrections do 
not spoil the scenario. Moreover, SUSY is crucial it seems if one wishes to
 extend the discussion to include the idea of quintessence,
a slow rolling scalar field with energy density adjusted to explain the observed $
\rho_v$.

In the presence of SUSY\cite{r28}, the zero mode scalar wave 
function has the form 

 \begin{equation} f_0(y) \sim e^{-\frac{a^2}{2}\sigma(y)
} \end{equation}

\noindent which yields the following dependence on the warped factor:
 
 \begin{eqnarray} \label{couplings_susy} m^2 & \sim & k^2 \Omega^{-4 - a^2} \nonumber\\ 
E & \sim & E_5 k \Omega^{-4 - \frac{3}{2}a^2} \nonumber\\ \lambda & \sim 
& \lambda_5 k \Omega^{-4 - 2a^2}\,. \end{eqnarray}

 For $a^2 > 0$, the zero mode is localized close to the Planck brane , and one could 
contemplate a quintessence -like solution of the vacuum energy problem. For 
instance, $m^2 \phi^2 \sim m^2 M_P^2$ requires  $m^2 \sim \Omega^{-8} M_P^2
$.  This means $a^2 \sim 4$, which corresponds to $m^2 \sim 10^{-33}\mbox
{eV}$(!) The BIG unanswered questions are: Can this be realized in a 
realistic scheme? What about radiative corrections? Can SUSY and warped 
geometry together protect the tiny mass scales $\sim10^{-3}\, \mbox{eV}$ (cosmological constant)
and $\sim10^{-33}\, \mbox{eV}$ (quintessence)?

Let's consider the case with a vanishing bulk mass which can readily be embedded in the supersymmetric
framework by setting $a=0$ in Eq.~(\ref{couplings_susy}).
Then supersymmetry helps to tame radiative corrections.
As in four dimensions the soft mass of the scalar zero mode, $m^2$,
will receive a 1-loop quantum correction
$\delta m^2\sim(1/16\pi^2)m^2\lambda\ln(\Lambda)$
by the exchange of the
scalar zero mode and its superpartner. Here $\Lambda$
denotes the momentum cut-off and $\lambda$ stems from the
self coupling localized at Planck-brane. As discussed above, in the case
$a=0$ we have $m^2\sim k^2\Omega^{-4}$ and $\lambda\sim 1$.
In the warped model there are additional radiative corrections
to $m^2$ by the exchange of KK states of the scalar field.
The KK states are localized towards the TeV-brane. Therefore
they acquire a larger soft mass of order $k^2\Omega^{-2}$.
The quartic coupling between two zero modes and two KK states
of the scalar field is of order $\Omega^{-2}$. Thus loops
with  KK states of the scalar are of the same order as
zero mode loop. It remains to be seen if performing the
sum over the KK contributions, which are individually small,
leads to a destabilization of the soft mass of the zero mode.

Since we assume the scalar field to be a gauge singlet, further
quantum corrections can only come from gravity. Exchange of
a zero mode gravitino in the loop gives rise to
$\delta m^2\sim(1/16\pi^2)m^2_{3/2}(\Lambda/M_{\rm Pl})^2$,
where $m_{3/2}$ is the gravitino mass. Since for the zero mode
of the gravitino $m_{3/2}\sim k\Omega^{-3}$, we obtain a
tiny correction even for a Planck-size cut-off. For the KK
gravitinos the situation is different. Being localized towards the
TeV-brane, their supersymmetry breaking mass is TeV-size $(k\Omega^{-1})$
\cite{GP2} and their dimension-six interaction with the scalar
zero mode is suppressed by $1/(\Omega M_{\rm Pl}^2)$. The
corresponding correction to the scalar mass is then
$\delta m^2\sim(1/16\pi^2)k^2\Omega^{-3}(\Lambda/M_{\rm Pl})^2$.
Since the dimension-six operator has its support at the TeV-brane,
it seems plausible that the KK gravitino loop is cut off at the TeV-scale.
In this case its contribution to the scalar mass
$\delta m^2\sim(1/16\pi^2)k^2\Omega^{-5}$ is safely suppressed.
However, given the limitations and uncertainties which are inherent
in our treatment of radiative corrections, the issue of the quantum
stability of our approach to the dark energy problem is not yet
satisfactorily settled.

\subsection{D-branes and Inflation}

Let us consider inflation 
in the context of D-branes, which are soliton Like 
configurations in type II A, II B and type I superstring theories\cite{r30}.  
For instance, in type II A and II B string theories, the elementary 
excitations are closed strings. The D-p branes in these Theories are 
p- dimensional objects whose dynamics is described by the theory of open strings who
se ends are constrained to lie on the D-p brane, In contrast to ordinary 
solitons, D-p branes are unusual in the sense that their mass (tension) 
is proportional to $1/g$, and their non-perturbative effects behave as 
$e^{-1/g}$, where $g$ denotes the string coupling. A D-p brane also carries 
a non-zero RR (Ramond- Ramond)  charge. A suitable stack of D-p branes 
can be put together to yield four dimensional gauge theories in the low 
energy limit. We will not pursue this any further and focus instead on their
 use in implementing inflationary scenarios\cite{r31}  that do not 
suffer from the formidable challenges encountered in supersymmetric/ 
supergravity models, as mentioned earlier. To begin with, consider the 
interaction energy between two static, identical D-branes . The interaction proceed
s via the exchange of graviton, dilaton and antisymmetric tensor fields. 
For separation r in the transverse plane, the interaction energy per unit
 brane volume is given by

\begin{equation} 2\kappa_{(10)}^2\left[\rho^2_{(p)}
 - T^2_{(p)}\right] \Delta^E_{(9-p)}(r)\,, \end{equation}

\noindent where $\kappa_{10
}^2 = 16 \pi G_N^{(10)}$($G_N^{(10)}$ is the ten dimensional Newton' 
constant), $\rho_{(p)}$ is the brane charge density, $T_{(p)}$ is the brane
 tension, and $\Delta^E_{(9-p)}(r)$ denotes the scalar propagator in $9-p
$ transerve dimensions. It turns out that $T_{(p)} = \rho_p$, so that the
 RR repulsion exactly cancels the gravitational and dilaton attraction.
 For a brane- antibrane pair, however, the net force is attractive, with
 the interaction energy per unit volume given by

\begin{equation} E(r) = 4 
\pi(4\pi^2 \alpha^\prime)^{3-p} \Delta^E_{(9-p)}(r)\,, \end{equation}

\noindent where we utilized the relation

\begin{equation} T^2_{(p)} = \rho_{(p)}^2 = \frac{\
pi}{\kappa_{(10)}^2} (4\pi^2 \alpha^\prime)^{3-p}. \end{equation}

For D-3 branes in particular, we find for the potential energy

\begin{equation} V(r) =
 2 T_{(3)} (1- 1/16\pi^2 r^4 T_{(3)}). \end{equation}

Introducing a canonically normalized 'scalar; field  $\phi = T_{(3)}^{1/2}r$, the potential 
becomes

\begin{equation} \label{potential} V(\phi) \simeq M^4\left(1 - \frac{T_
{(3)}}{16\pi^2\phi^4}\right). \end{equation}

Let us now see how this can 
lead to four dimensional inflation by first outlining some assumptions:
  
 \begin{itemize}

 \item the size of the extra dimension is considered 
frozen during inflation, which means that that the radion $\mbox{mass}^2 
\gg 2 T/ M_P^2$ (= $H^2$, $H =$ Hubble constant during inflation); 
 The inter-brane motion caused by the attractive force will drive four 
dimensional inflation.

\item The effective four dimensional Hubble 
size $H^{-1}$ should be larger than the (common) size $R_c$ of the extra 
dimensions. This enables us to treat the universe as four dimensional at 
distances $\sim H^{-1}$ . 

\end{itemize}

Under these assumptions, 
the salient features of the inflationary scenario that follow from the 
potential (\ref{potential}) can be worked out. Three predictions are 
especially significant: The scalar spectral index n of density fluctuations 
is approximately 0.97- 0.98 ; The microwave temperature anisotropy is 
proportional to ($M_c/M_P$),  which leads to $M_c =$ compactification scale
 $\sim  10^{12}$ GeV;  The Hubble constant during inflation is of order $
10^9$ GeV,  so that the gravitational wave contribution to the observed
 anisotrpy is negligible ($\ll 1\%$) .

 An important conclusion we 
draw from these considerations is that the scale of the extra dimensions 
which play a role in inflation is neither 'large' ($\QSagt \mbox{TeV}^{-1}$ 
) nor small  ($\sim M_P^{-1}$), but rather intermediate ($\sim 10^{-12}
\,\mbox{GeV}^{-1}$).

 \section{Scorecard}

\begin{itemize}

\item Solar and Atmospheric Neutrino Oscillations: If neutrinos are Majorana 
particles with non-zero masses, we could either invoke the standard 
see-saw mechanism which utilizes heavy SM singlet neutrinos, or utilize the 
SM dimension five operators in a warped background. While the existence 
of heavy right handed (singlet) neutrinos would be hard anytime soon to confirm 
experimentally, the $KK$ excitations of the SM fields with some luck may be discovered
 at the LHC\cite{r29};

\item Dark Matter: Although the LSP is a particularly 
attractive candidate, another good candidate is the axion; in some models 
superheavy quasi-stable particles can be identified which may contribute 
to or even comprise all of the dark matter;

\item Inflation: 
Attractive GUT models have been found\cite{r18} in which the SUSY GUT scale is 
directly related to the inflationary scenario. It was recently shown
\cite{r22} how this can be extended to five dimensions , with the fifth dimension
 compactified on $S^1/Z_2 \times Z_2$. As seen above, inflation also can 
be implemented within the D- brane setting, where the characteristic size
 of the extra dimension is of order  $10^{-12}\mbox{GeV}^{-1}$.

\item Baryon-Lepton Asymmetry: By the author's definition an inflationary 
scenario is incomplete without an explanation of the origin of the observed
 baryon asymmetry. Those scenarios in which the inflaton decay produces 
the asymmetry via leptogenesis are particularly compelling.

\item Dark Energy: The existence of one (or possibly more) warped dimension 
could lead to the warping of energy scales which could provide at least a 
partial explanation of the origin of the observed vacuum energy density. 
 \end{itemize}

Experimental challenges for this decade include:  
 \begin{itemize} \item Finding the SM (or the lightest MSSM) higgs 
boson; 

\item Confirmation of neutrino oscillations, measurement of 
$\theta_{13}$, and ongoing sear\-ches for neutrinoless double beta decay
to verify the Majorana nature of the $\nu$;  
 
\item SUSY and extended higgs sector searches at the LHC, Fermilab and TESLA(?) 

\item Direct and indirect detection of dark matter 
(LSP, axion, $KK$ matter); 

\item Searches for proton decay, magnetic
 monopoles and stable fractionally charged color singlet states in large 
underground detectors; 

\item Searches for other rare processes 
such as $n-\bar n$ oscillations,  $\mu \rightarrow e + \gamma$, $\tau 
\rightarrow \mu + \gamma$, etc. 

\item  Searches for the TeV scale $KK$ 
excitations of $W$ and $Z$ at the LHC, expected from setting the SM in a 
warped background.

\end{itemize}

Finally, MAP , PLANCK, SNAP and
 other astrophysics/cosmology based projects should provide new impetus for
 an even closer collaboration between particle physics and cosmology. 

\section{Conclusions}

It has become increasingly clear in recent
 years that there is life beyond the Standard Model. In a bottom-up 
approach one needs to provide masses for the known neutrinos that appear to be
 cosmologically insignificant ($\ll \mbox{eV}$), unless neutrino mass 
degeneracy is at  work. Since neutrinos cannot fully solve the dark 
matter problem, an LSP or an axion or some other cold dark matter candidate 
must be invoked. The recent dramatic measurements of temperature 
anisotropies of the cosmic microwave background radiation on various angular scales
 demand new physics (such as inflation) that is not incorporated in the 
SM. Supersymmetric GUTS offer a particularly compelling extension of the SM. 
There is good evidence for gauge coupling unification, and in an $SO(10)$ 
setting with the MSSM parameter $\tan\beta \gg 1$, respectable 
predictions for the top mass and the SM higgs mass have been made. A major 
stumbling block for these theories has to do with proton decay and, more 
specifically, dimension five nucleon decay operators. (Recall that a long time 
ago proton decay also sealed the fate of non- SUSY $SU(5)$, despite 
attempts to save it). It is therefore quite exciting to note that extra 
dimensions may help to stabilize the proton by banishing dimension five nucleon 
decay. Furthermore, they also seem to provide a nice mechanism for 
resolving the doublet- splitting problem. More recently, it has been 
shown\cite{r22} how inflation and baryo-leptogenesis is possible in five dimensional
 models, with compactification of the fifth dimension realized on 
$S^1/Z_2 \times Z_2$.  It seems to me that these models deserve further 
attention and it would be interesting to try to derive them from a more 
fundamental theory. 

\vspace{0.5cm}
\noindent {\bf Acknowledgements}

I offer my sincere thanks to Norma Mankoc Borstnik, Colin Frogatt, Anne Guehl and Holger Nielsen
for organizing a particularly stimulating meeting at such a beautiful location.


\title*{Loops Versus Strings\thanks{IFT-UAM/CSIC-03-23}}
\author{Enrique \'Alvarez}
\institute{%
Instituto de F\'{\i}sica Te\'orica UAM/CSIC, C-XVI,
and  Departamento de F\'{\i}sica Te\'orica, C-XI,\\
Universidad Aut\'onoma de Madrid, 
E-28049-Madrid, Spain}

\titlerunning{Loops Versus Strings}
\authorrunning{Enrique \'Alvarez}
\maketitle

\begin{abstract}
Two popular attempts to understand the quantum physics of gravitation
are critically assessed. The talk on which this paper is based 
was intended for a general particle-physics audience.    
\end{abstract}

\section{General Questions on Quantum Gravity}
It is not clear at all what is the problem in quantum gravity (cf. \cite{Alvarez} or 
\cite{Alvarez-Gaume} for 
general reviews, written in the same spirit as the present one).
The answers to the following questions are not known, and I believe it can do no harm
to think about them before embarking in a more technical discussion.
\par
Actually, some people  proposed that gravity should not be quantized, owing to its
special properties as determining the background on which all other fields propagate.
There is a whole line of thought on the possibility that gravity is not a fundamental theory,
and this is certainly an alternative  one has to bear in mind. Indeed, even the {\em holographic
principle} of G. 't Hooft, to be discussed later, can be interpreted in this sense.
\par
Granting that, the next question is whether 
it does  make any sense to consider gravitons
  propagating in some background; that is, whether there is some useful
approximation in which there is a particle physics approach to the physics
of gravitons as quanta of the gravitational field. A related question is whether 
semiclassical gravity, i.e., the approximation
in which the source of the classical Einstein equations is replaced by the
expectation value of the energy momentum tensor of some quantum theory has some physical 
(\cite{Duff}) validity
in some limit.
\par
At any rate, even if it is possible at all, the at first sight easy problem of graviton interactions
in an otherwise flat background has withstood analysis of several generations of physicists.
The reason is that the coupling constant has mass dimension $-1$, so that the structure
of the perturbative counterterms involve higher and higher orders in the curvature invariants
(powers of the Riemann tensor in all possible independent contractions), schematically,
\begin{equation}
S=\frac{1}{2\kappa_R^2}\int R+\int R^2 + \kappa_R^2 \int R^4+\ldots
\end{equation}
Nobody knows how to make sense of this approach, except in one case, to be mentioned later on.
\par
It could be possible, {\em sensu stricto} to stop here. But if we believe that 
quantum gravity should  give answers to such questions as to the fate of the initial 
cosmological singularity, its is almost unavoidable to speak of the wave function of the universe.
This brings its own set of problems, such as to whether it is possible to do quantum mechanics
 without classical observers or whether  the wave function of the Universe
 has a probablilistic interpretation. Paraphrasing C. Isham \cite{Isham}, one would not known when to 
qualify a probabilistic prediction on the whole Universe as a successful one.
\par
The aim of the present paper is to discuss in some detail established results on the field.
In some strong sense, the review could be finished at once, because there are none. 
There are, nevertheless, some interesting attempts, which look promising from certain points of view.
Perhaps the two approaches that have attracted more attention have been the loop approach,
on the one hand and strings on the other. We shall try to critically assess prospects in both.
 Interesting related papers are \cite{Horowitz}\cite{Smolin}.

\section{The issue of background independence}
One of the main differences between both attacks to the quantum gravity problem
 is the issue of background
 independence, by which it is understood that no particular background should enter
into the definition of the theory itself. Any other approach is purportedly at
variance with diffeomorphism invariance.
\par
Work in particle physics in the second half of last century led to some
understanding of ordinary gauge theories. Can we draw some lessons from there?
\par
Gauge theories can be formulated in the {\em bakground field approach}, as
introduced by B. de Witt and others (cf. \cite{dewitt}). In this approach,
the quantum field theory depends on a background field, but not on any one in particular,
and the theory enjoys background gauge invariance.
\par
Is it enough to have quantum gravity formulated in such a way? This was, incidentally, the way
G.  Hooft and M. Veltman did the first complete one-loop calculation (\cite{thv}).
\par
It can be argued that the only 
vacuum expectation value consistent with diffeomorphisms invariance is
\begin{equation}
<0|g_{\alpha\beta}|0>=0
\end{equation}
in which case the answer to the above question ought to be in the negative, because this is a singular
background and curvature invariants do not make sense.
It all boils down as to whether the ground state of the theory is diffeomorphisms invariant or not.
There is an example, namely three-dimensional gravity in which invariant quantization can be performed
\cite{Witten3}. In this case at least, the ensuing theory is almost topological. 
\par
In all attempts of a canonical quantization of the gravitational field, 
one always ends up with an (constraint) equation
corresponding physically to the fact that the total hamiltonian of a parametrization invariant theory
should vanish. When expressed in the Schr\"odinger picture, this equation is often 
dubbed the {\em Wheeler-de Witt equation}. This equation is plagued by
operator ordering and all other sorts of ambiguities.
It is curious to notice that in ordinary quantum field theory there also exists a Schr\"odinger
representation, which came recently to be controlled well enough as to be able to perform lattice 
computations (\cite{Luscher}).
\par
Gauge theories can be expressed in terms of gauge invariant operators, such as
Wilson loops . They obey a complicated
set of equations, the loop equations, which close in the large $N$ limit
as has been shown by Makeenko and Migdal (\cite{MM}). These equations can be properly regularized,
e.g. in the lattice. Their explicit solution
is one of the outstanding challenges in theoretical physics. Although many conjectures
have been advanced in this direction, no definitive result is available.
\section{ Loops}
The whole philosophy of this approach is canonical, i.e., an analysis of the evolution of
variables defined classically through a foliation of spacetime by a family of spacelike three-surfaces
$\Sigma_t$. The standard choice in this case (cf. for example \cite{Alvarez}) is the 
three-dimensional metric, $g_{ij}$, and its canonical conjugate, related to the extrinsic curvature.
Due to the fact that the system is reparametrization invariant, the total hamiltonian vanishes,
and this hamiltonian constraint is usually called the Wheeler- de Witt equation.
\par
Here, as in any canonical approach the way one chooses the canonical variables is fundamental.
\par
Ashtekar's clever insight started from the definition of an original set of 
variables (\cite{Ashtekar}) stemming from the Einstein-Hilbert 
lagrangian written in the form \footnote{Boundary terms have to be considered as well. We refer to 
the references for
details.}
 \begin{equation}
S=\int e^a\wedge e^b\wedge R^{cd}\epsilon_{abcd}  
\end{equation}
where $e^a$ are the one-forms associated to the tetrad,
\begin{equation}
e^a\equiv e^a_{\mu}dx^{\mu}.
\end{equation}
Tetrads are defined up to a local Lorentz transformation
\begin{equation}
(e^a)^{\prime}\equiv L^a\,_b(x)e^b
\end{equation}
The associated $SO(1,3)$ connection one-form $\omega^a\,_b$ is usually called the 
{\em spin connection}. Its field strength is the  curvature expressed as a two form:
\begin{equation}
R^a\,_b\equiv d\omega^a\,_b+\omega^a\,_c\wedge \omega^c\,_b.
\end{equation}
Ashtekar's variables are actually based on the $SU(2)$ self-dual connection
 \begin{equation}
 A=\omega - i * \omega
\end{equation}
Its field strength is
\begin{equation}
F\equiv d A + A\wedge A
\end{equation}
The dynamical variables are then $(A_i, E^i\equiv F^{0i})$. The main virtue of these variables is that
constraints are then linearized.
One of them is exactly analogous to            
Gauss'law: 
\begin{equation}
D_i E^i=0.
\end{equation}
There is another one related to three-dimensional diffeomorphisms invariance,
\begin{equation}
tr\, F_{ij}E^i=0
\end{equation}
 and, finally, there is the Hamiltonian constraint,
\begin{equation} 
 tr F_{ij}E^i E^j=0
\end{equation}
\par
On a purely mathematical basis, there is no doubt that Astekhar's variables are of a great ingenuity.
As a physical tool to describe the metric of space, they are not real in general. This forces
a reality condition to be imposed, which is akward. For this reason it is usually prefered
to use the Barbero-Immirzi (\cite{Barbero}\cite{Immirzi}) 
formalism in which the connexion depends on a free parameter, $\gamma$,
\begin{equation}
A_a^i=\omega_a^i +\gamma K_a^i
\end{equation}
$\omega$ being the spin connection and $K$ the extrinsic curvature. When $\gamma=i$ Astekhar's
formalism is recovered; for other values of $\gamma$ the explicit form of the constraints 
is more complicated. Thiemann (\cite{Thiemann}) has proposed a form for the Hamiltonian constraint
which seems promising, although it is not clear whether the quantum constraint 
algebra is isomorphic to the classical algebra (cf.\cite{Rovellir}). A comprehensive reference
is \cite{Thiemannr}
\par
Some states which satisfy the Astekhar constraints are given by the
 loop representation, which can be introduced from the construct (depending both on
the gauge field $A$ and on a parametrized loop $\gamma$)
\begin{equation}
 W (\gamma , A)\equiv tr\, P e^{\oint_{\gamma}A}
\end{equation}
and a functional transform mapping functionals of the gauge field $\psi(A)$ into functionals
of loops, $\psi(\gamma)$:
\begin{equation}
 \psi(\gamma)\equiv \int {\cal D}A\, W(\gamma,A)\psi(A)
\end{equation}
When one divides by diffeomorphisms, it is found that
functions of knot classes (diffeomorphisms classes of smooth, non self-intersecting loops)
 satisfy all the constraints. 
 \par
Some particular states sought to reproduce smooth spaces at coarse graining are
the {\em Weaves}. It is not clear to what extent they also approach the conjugate variables (
that is, the extrinsic curvature) as well.
\par
In the presence of a cosmological constant the hamiltonian constraint reads:
\begin{equation}
\epsilon_{ijk}E^{ai}E^{bj}(F^k_{ab}+\frac{\lambda}{3}\epsilon_{abc}E^{ck})=0
\end{equation}
A particular class of solutions of the constraint \cite{Smolinc} are self-dual solutions of the form
\begin{equation}
F^i_{ab}=-\frac{\lambda}{3}\epsilon_{abc}E^{ci}
\end{equation}
Kodama (\cite{Kodama} has shown that the Chern-Simons state 
 \begin{equation}
 \psi_{CS}(A)\equiv e^{\frac{3}{2\lambda} S_{CS}(A)}
\end{equation} 
is a solution of the hamiltonian constraint. He even suggested that the {\em sign} of the coarse
grained, classical  cosmological constant was always positive, irrespectively of the sign of the 
quantum 
parameter $\lambda$, but it is not clear whether this result is general enough.
 There is some concern that this state as such is not normalizable with the usual norm. It 
has been argued that
this is only natural, because the physical relevant norm must be very different from the na\"{\i}ve one
(cf. \cite{Smolin}) and indeed normalizability of the Kodama state has been suggested as a criterion
for the correctness of the physical scalar product.
\par
Loop states in general (suitable symmetrized) can be represented as 
spin network (\cite{RS}) states: colored lines (carrying some $SU(2)$ representation) 
meeting at nodes where intertwining $SU(2)$ operators act. A beautiful graphical
representation of the group theory has been succesfully adapted for this purpose.
There is a clear relationship between this representation and the Turaev-Viro \cite{Turaev} 
invariants
Many of these ideas have been foresighted by Penrose (cf. \cite{penrose}).
\par
There is also a 
path integral representation, known as  {\em spin foam} (cf.\cite{Baez}), a topological theory 
of colored surfaces representing the evolution of a spin network.
These are closely related to topological BF theories, and indeed, independent generalizations 
have been proposed.
Spin foams can also be considered as an independent approach to the quantization of the gravitational 
field.(\cite{Barrett})
\par
In addition to its specific problems, this approach shares with all canonical approaches
to covariant systems
the problem of time. It is not clear its definition, at least in the absence of matter.
Dynamics remains somewhat mysterious; the hamiltonian constraint does not say in what sense
(with respect to what) the three-dimensional dynamics evolve.

\subsection{Big results}
One of the main successes of the loop approach is that the
area (as well as the volume) operator is discrete. This allows, assuming that 
a black hole has been formed (which is a process that no one knows how to represent
in this setting), to explain the formula for the black hole entropy . The result
is expressed in terms of the Barbero-Immirzi parameter (\cite{RSS}). The physical meaning
 of this dependence is not well understood.
\par
 
It has been pointed out \cite{Bekenstein} that there  is a potential drawback
in all theories in which the area (or mass) spectrum is discrete with eigenvalues $A_n$ 
if the level spacing between eigenvalues $\delta A_n$ is uniform because of the predicted thermal character
of Hawking's radiation. The explicit computations in the present setting, however, lead to
an space between (dimensionless) eigenvalues
\begin{equation}
 \delta\, A_n\sim e^{-\sqrt{A_n}}, 
 \end{equation} 
which seemingly avoids this set of problems.
  \par 
It has also been pointed out that \cite{Freidel} not only the spin foam, but almost
all other theories of gravity can be expressed as topological BF theories with constraints.
While this is undoubtely an intesting and potentially useful remark, it is
important to remember that the difference between the linear sigma model (a free field theory)
and the nonlinear sigma models is just a matter of constraints. This is enough to produce a mass gap
and asymptotic freedom in appropiate circumstances.

\section{Strings}
It should be clear by now that we probably still do not know
what is exactly the problem to which string theories are the answer. At any rate, the starting 
point is that all elementary particles are viewed as quantized excitations of a one dimensional object,
the string, which can be either open (free ends) or closed (a loop). Excellent books are 
avaliable, such as \cite{Greens}\cite{Polchinskis}.
\par
String theories enjoy a convoluted history. Their origin can be traced to the
Veneziano model  of strong interactions. A crucial step was the reinterpretation 
by Scherk and Schwarz 
(\cite{Scherk}) of the massless spin two state in the closed sector (previously 
thought to be related to the 
Pomeron) as the  
graviton and consequently of the whole string theory as
a potential 
theory of quantum gravity, and potential unified theories of all interactions. Now the wheel has
made a complete turn, and we are perhaps back
through the Maldacena conjecture (\cite{ea-Maldacena}) to a closer relationship than previously thought 
with ordinary gauge theories.
\par
From a certain point of view, their dymamics is determined by a two-dimen\-sional non-linear sigma
model, which geometrically is a theory of imbeddings of a two-dimensional surface (the world
sheet of the string) to a (usually ten-dimensional) target space:
\begin{equation}
x^{\mu}(\xi): \Sigma_2\rightarrow M_n
\end{equation}
There are two types of interactions to consider.
Sigma model interactions (in a given two-dimensional surface) 
are defined as an expansion in powers of momentum, where a new
dimensionful parameter, $\alpha^{\prime}\equiv l_s^2$ sets the scale. This scale
is {\em a priori} believed to be of the order of the Planck length. The first terms in the action
always include a coupling to the massless backgrounds: the spacetime metric, the two-index Maxwell
like field known as the Kalb-Ramond or $b$-field, and the dilaton. To be specific,
\begin{equation}
 S= \frac{1}{l_s^2}\int_{\Sigma_2}g_{\mu\nu}(x(\xi))\partial_a x^{\mu}(\xi)\partial_b 
x^{\nu}(\xi)\gamma^{ab}(\xi)
+\ldots 
\end{equation} 
There are also string interactions, (changing the two-dimensional surface) 
proportional to the string coupling constant, $g_s$, whose
variations are related to the logarithmic variations of the dilaton field.
Open strings (which have gluons in their spectrum) {\em always} contain closed strings 
(which have gravitons in their spectrum) as intermediate
states in higher string order ($g_s$)  corrections. This interplay open/closed is one of the 
most fascinating aspects of the whole string theory.
\par
It has been discovered by Friedan (cf. \cite{Friedan}) that in order for the quantum theory
to be consistent with all classical symmetries (diffeomorphisms and conformal invariance), the 
beta function of
the generalized couplings \footnote{There are corrections coming from both dilaton and Kalb-Ramond
fields. The quoted result is the first term in an expansion in derivatives, with expansion parameter
 $\alpha^{\prime}\equiv l_s^2$.} must vanish:
\begin{equation}
 \beta (g_{\mu\nu})=R_{\mu\nu}=0 
\end{equation} 
  This result remains until now as one of the most important ones in string theory, hinting at a 
deep relationship between Einstein's equations and the renormalization group.
   
Polyakov (\cite{Polyakov}) introduced the so called {\em non-critical strings} which have
in general a two-dimensional cosmological constant (forbidden otherwise by Weyl invariance).
The dynamics of the conformal mode (often called Liouville in this context) is, however, 
poorly understood.

\subsection{General setup}

Fundamental strings live in  D=10 spacetime dimensions, and so a Kaluza-Klein mecanism of sorts
must be at work in order to explain why we only see four non-compact dimensions at low energies.
Strings have in general tachyons in their spectrum, and the only way to construct seemingly
consistent string theories (cf. \cite{Gliozzi}) 
is to project out those states, which leads to supersymmetry. This means in turn that all low energy
predictions heavily depend on the supersymmetry breaking mechanisms. 
\par
  
 String perturbation theory is probably well defined although a full proof is not available.
\par
Several stringy symmetries are believed to be exact: T-duality, relating large and small compactification
volumes, and $S$-duality,
relating the strong coupling regime with  the weak coupling one.
Besides, extended configurations ({\em D branes}); topological defects in which open strings can
end are known to be important \cite{Polchinski}. They couple to Maxwell-like fields which are p-forms
called Ramond-Ramond (RR) fields.
These dualities \cite{Hull} relate  all five string theories (namely, Heterotic $E(8)\times E(8)$ 
Heterotic $SO(32)$, Tipe $I$, $IIA$ and $IIB$) and it is conjectured that there is an 
unified eleven -dimensional theory, dubbed $M$-theory
of which ${\cal N}=1$ supergravity in $d=11$ dimensions is the low energy limit.

\subsection{Big results}
Perhaps the main result is that graviton physics in flat space is well defined for the first time, and
this is no minor accomplishment.
\par
Besides, there is evidence that at least some geometric singularities are harmless
in the sense that strings do not feel them.
Topology change amplitudes do not vanish in string theory.
\par
The other Big Result \cite{Strominger} is that one can
correctly count states
 of extremal black holes
 as a function of charges. This is at the same time astonishing and disappointing.
It clearly depends strongly on the objets being BPS states (that is, on supersymmetry),
and the result has not been extended to non-supersymmetric configurations.
On the other hand, as we have said, it {\em exactly} reproduces the entropy as a 
function of a sometimes large number of charges, without any adjustable parameter.
     
\subsection{The Maldacena conjecture}
 Maldacena \cite{ea-Maldacena} proposed as a conjecture that $IIB$ string theories in a background
$AdS_5\times S_5$ with  radius $l\sim l_s (g_s N)^{1/4}$ and N units of RR flux 
is equivalent to a four dimensional ordinary gauge theory in flat four-dimensional Minkowski
space, namely ${\cal N}=4$ super Yang-Mills with gauge group $SU(N)$ and coupling constant
$g=g_s^{1/2}$.
 \par
Although there is much supersymmetry in the problem and the kinematics largely determine correlators,
(in particular, the symmetry group $SO(2,4)\times SO(6)$ is realized as an isometry group on the
gravity side and as an $R$-symmetry group as well as conformal invariance on the gauge theory side)
this is not fully so \footnote{The only correlators that are completely determined
through symmetry are the two and three-point functions.}
and the conjecture has passed many tests in the semiclassical approximation
to string theory.
\par
 This is the first time that a precise holographic description of spacetime in terms of a (boundary) 
gauge theory is proposed and, as such it is of enormous potential interest. It has been conjectured 
by 't Hooft \cite{'tHooft} and further developed by Susskind \cite{Susskind} that there should be
much fewer degrees of freedom in quantum gravity than previously thought. The conjecture claims that
it should be enough with one degree of freedom per unit Planck surface in the two-timensional boundary
of the three-dimensional volume under study. The reason for that stems from an analysis of the 
Bekenstein-Hawking \cite{Bekenstein}\cite{Hawking} 
entropy associated to a black hole, given in terms of the
two-dimensional area $A$ \footnote{The area of the horizon for a Schwarzschild black hole is given by:
\begin{equation}
A=\frac{8\pi G^2}{c^4}M^2
\end{equation}
}
of the horizon by
\begin{equation}
S=\frac{ c^3}{4 G\hbar}A.
\end{equation}

This is  a deep result indeed, still not fully understood.
\par
It is true on the other hand that the Maldacena conjecture has only been checked for the time being 
in some corners of 
parameter space, namely when strings can be approximated by supergravity in the appropiate background.
The way it works \cite{ea-Witten} is that the supergravity action corresponding to fields with prescribed
boundary values is related to gauge theory correlators of certain gauge invariant operators 
corresponding to the particular field studied:
\begin{equation}
e^{- S_{sugra}[\Phi_i]}\bigg|_{\Phi_i|_{\partial AdS}=\phi_i}= 
<e^{\int {\cal O}_i\phi_i}>_{CFT}
\end{equation}

 \section{Observational prospects}

In the long term, advances in the field, as in any other branch of physics will be determined 
by experiment. The prospects here are quite dim. It has been advertised \cite{Alfaro} that
as a consequence of loop quantum gravity 
\footnote{In string theory, with a  string scale of the order of the Planck mass, there is effectively
a minimal length, namely the self dual radius but then
 the corrections are of the type \cite{Myers}
\begin{equation}
E^2 = (\vec{p})^2 + m^2 + m^2\sum_{n=1} c_n (\frac{m}{m_P})^n
\end{equation}
much more difficult to observe experimentally}
anomalous dispersion relations of the form
\begin{equation}
E^2 = (\vec{p})^2 + m^2 + E^2\sum_{n=1} c_n (\frac{E}{m_P})^n
\end{equation}
could explain some strange facts on the cosmic ray spectrum. Although this is an interesting 
suggestion (cf.\cite{Coleman}) it is not specific to loop quantum gravity; noncommutative models 
make similar predictions as indeed does any theory with a fundamental scale. In spite of some optimism, it is not easy to perform specific 
experiments which
could discriminate between different quantum gravity alternatives. This should not by any means
be taken as an indication that the experiments themselves are not interesting. Nothing could be most
exciting that to discover deviations from the suposedly exact symmetries of Nature, and it is 
amazing that present observations already seemingly exclude some alternatives \cite{Myers}.
\par
On the string side, perhaps some effects related to specific stringy states, such as
the winding states could be experimentally verified ( cf. for example some suggestions in 
\cite{Alvarezz}). It has also been proposed that the string scale
could be lowered, from the Planck scale down to the TeV regime \cite{Antoniadis}.
It is difficult to really pinpoint what is exactly stringy about those models, and in particular,
all string predictions are difficult to disentangle from supersymmetric 
model predictions and rely heavily
on the mechanisms of supersymmetry breaking.

\section{Summary: the state of the art in quantum gravity}
In the loop approach one is working
 with  nice candidates for a quantum
  theory. The theories are interesting, probably related to topological field theories (\cite{Blau})
and background independence as well as diffeomorphism invariance are clearly implemented.
On the other hand, it is not clear that their low energy limit  is related to 
Einstein gravity.
    
\par 
 Strings start from a perturbative approach more familiar to a particle physicist.
However, they carry all the burden of supersymmetry and Kaluza-Klein. It has proved to be very
difficult to study nontrivial non-supersymmetric dynamics.
\par
Finally, and this applies to  all approaches, the holographic ideas seem intriguing; there are
many indications of a deep relationship between gravity and gauge theories.
\par
We would like to conclude by insisting on the fact that although there is not much we know for 
sure on quantum effects on the gravitational field, even
the few things we know are a big feat, given the difficulty to do physics 
without experiments.

\section*{Acknowledgments}
I wish like to thank Norma Mankoc-Borstnik and Holger Bech Nielsen for the oportunity to
participate in a wonderful meeting in the pleasant atmosphere of Portoroz.
I have benefited from discussions with J. Alfaro, J. Barrett, C. G\'omez, W. Kummer  and  E. Verdaguer.
This work ~~has been partially supported by the
European Commission (HPRN-CT-200-00148) and CICYT (Spain).

               
\newcommand{\Lizzieqn}[1]{(\ref{#1})}
\def\FLbra#1{\left\langle #1\right|}
\def\FLket#1{\left| #1\right\rangle}
\def\FLhs#1#2{\left\langle #1\right|\left. #2\right\rangle}
\newcommand{\FLdm}[2]{\FLket{#1}\FLbra{#2}}

\title*{Fuzzy Two-dimensional Spaces}
\author{Fedele Lizzi}
\institute{%
Dipartimento di Scienze Fisiche\\
Universit\`{a} di Napoli {\sl Federico II}\\ 
and {\it INFN, Sezione di Napoli}\\
Monte S.~Angelo, Via Cintia, 80126 Napoli, Italy\\
lizzi@na.infn.it}

\titlerunning{Fuzzy Two-dimensional Spaces}
\authorrunning{Fedele Lizzi}
\maketitle

\begin{abstract}
We give a short introduction to three fuzzy spaces: the torus, the
sphere and the disc.
\end{abstract}

\section{Introduction}
\setcounter{equation}{0} This conference was ambitiously called
\emph{What Comes Beyond the Standard Model} without a question
mark. We had several lively discussions, probably there were are
as many opinions on what comes beyond the SM as there were
participants. My personal point of view on the matter is:
\emph{Beyond the standard model there is the structure of
spacetime}. The standard view assigns to spacetime a large number
of symmetries, which is a ``good thing'', but also, so to speak, a
large number of points. Too many points in fact! The short
distance structure is what gives rise to infinities in a field
theory, and create problems in the quantization of gravity. Here I
will describe how ``fuzzy'' approximations\footnote{For a general
introduction to fuzzy spaces see~\cite{Madorebook}} can be related
to a spacetime with few degrees of freedom, but with the
symmetries which we usually associate to the continuum. None of
the examples presented here is operatively proposed as a
description of spacetime, to start with they are all two
dimensional, and they are still too rigid as theories. But I think
they point to another direction in which the use of noncommutative
geometry~\cite{ncgbooks} may be a viable possibility for the
description of spacetime at Planck's distances. Other structures
of spacetime with a smaller number of  degrees of freedom inspired
by noncommutative geometry have been studied for example
in~\cite{nonfuzzy}.

Fuzzy spaces are finite approximations based on noncommutative
geometry. Spaces are described by a matrix approximation to the
functions defined on them. This means that their commutative
algebra defined on an ordinary space will be approximate by a
noncommutative algebra. Often, but not necessarily, in the limit
in which the rank of the matrices goes to infinity, the
noncommutativity parameter goes to zero. In this paper, mainly for
space limitations, I will consider two-dimensional examples, and
describe their geometry concentrating on the eigenvalues of the
Laplacian. While this does not give an unambiguous
characterization of a space (one cannot hear the shape of
drum~\cite{Kac}) it still provides considerable information.

The first and foremost fuzzy space is Madore's fuzzy
sphere~\cite{Madorefuzzysphere}, which I describe in
section~\ref{se:sphere}, after the ``zeroth fuzzy space'', the
fuzzy torus which can be traced back to H.~Weyl~\cite{Weyl}. In
section~\ref{se:disc} I present the recent fuzzy approximation to
the disc which I developed with P.~Vitale and A.~Zampini.

\section{The Fuzzy Torus}
\setcounter{equation}{0}

Consider an ordinary torus described by the coordinates
$x_1,x_2\in [0,1]$, a function can be expressed as a series
\begin{equation}
f(x_1,x_2)=\sum_{n_1,n_2=-\infty}^\infty f_{n_1n_2} e^{in_1x_1}
e^{in_2x_2} \label{torus}
\end{equation}
The coefficients $f_{n_1n_2}$ decrease as $n_i\to\infty$ in a way
which defines various classes of functions (continuous, smooth
etc.). The composition law is the usual commutative convolution:
\begin{equation}
(fg)_{n_1n_2}=\sum_{m_1,m_2=-\infty}^\infty f_{m_1 m_2}
g_{n_1-m_1\ n_2 -m_2} \label{convolution}
\end{equation}
The torus has a $U(1)\times U(1)$ symmetry which acts on $f$
multiplying its fourier coefficients by a phase:
\begin{equation}
f(x_1+\alpha_1,x_1+\alpha_1)=\sum_{n_1,n_2=-\infty}^\infty
f_{n_1n_2}e^{n_1\alpha_1+n_2\alpha_2}\, e^{in_1x_1} e^{in_2x_2}
\end{equation}
The Laplacian is $\partial_{x_1}^2+\partial_{x_2}^2$ and its spectrum is
given by all integers of the form $m_1^2+m_2^2$.

Truncating the series~\Lizzieqn{torus} while keeping the (also
truncated) convolution~\Lizzieqn{convolution} gives a nonassociative
algebra. Another possibility is to consider a \emph{fuzzy}
approximation, that is we consider a finite expansion
\begin{equation}
f(x_1,x_2)=\sum_{n_1,n_2=-N}^N f_{n_1n_2} U_1^{n_1} U_2^{n_2}
\label{fuzzytorus}
\end{equation}
with
\begin{equation}
U_1U_2=e^{\frac{2\pi}{2N+1}}U_2U_1 \label{nctorus}
\end{equation}
This relation, which characterize the noncommutative
torus~\cite{Rieffel}, has become very popular lately since its
appearance in string theory~\cite{CDS}. The algebra defined
by~\Lizzieqn{nctorus} has a finite representation of rank $2N+1$:
\begin{eqnarray} {U}_1&=&\left({\begin{array}{lllll} 1& & & & \\
&e^{\frac{2\pi i}{2N+1}}& & & \\ & &e^{2\, \frac{2\pi i}{2N+1}}& &
\\& & &\ddots& \\ & & & &
e^{(N-1)\, \frac{2\pi i}{N}}
\end{array}}\right)\nonumber\\
{U}_2&=&\left({\begin{array}{lllll} 0&1& & &0\\ &0&1& & \\
& &\ddots&\ddots&
\\ & & &\ddots&1\\ 1& & & &0\end{array}}\right) \ .
\end{eqnarray}
Now the harmonics retained are finite, the space is \emph{finite}
dimensional and the products is consistent, at the price of
loosing the commutativity of the algebra, which is recovered as
$n\to\infty$. One can define two outer derivatives,
$\partial_iU_j=\delta_{ij}U_i$ which defines a Laplacian whose
spectrum is of course finite, and coincides with the one of the
ordinary torus for the first $2N+1$ eigenvalues. The $U(1)\times
U(1)$ symmetry acts exactly in the same way, and is again a
symmetry of the Laplacian.

From the mathematical point of view the way the approximation
works is delicate. The algebra of the torus, as well as the one of
the noncommutative torus, is not the inductive limit of finite
dimensional algebras\footnote{It is however the inductive limit of
algebra of matrices of functions defined on two
circles~\cite{ElliottEvans}.} , but the limit for the
noncommutative can be understood in a weak
sense~\cite{LandiLizziSzaboLargeN}.

\section{The Fuzzy Sphere \label{se:sphere}}
\setcounter{equation}{0}

A sphere of radius $r$ is the subspace of ${\mathbb R}^3$ with
coordinates $x_i$ defined by the constraint
\begin{equation}
\sum_i x_i^2=r^2 \label{sphereconstraint}
\end{equation}
A function on the sphere can be expanded as
\begin{equation}
f(x)=f_0+f_ix_i+f_{ij}x_ix_j+\ldots=\sum_{l=0}^\infty f_{i_1\ldots
i_l}x_{i_1}\ldots x_{i_l} \label{fuzzyexpx}
\end{equation}
Equation~\Lizzieqn{sphereconstraint} imposes constraints on the
independence of the $f$'s, they have to be totally symmetric and
all partial traces (that is sum over pair of indices) must vanish.
Since the spherical harmonics $Y_{lm}(\theta,\varphi)$ are a basis
complex homogeneous polynomials of degree $l$, the
expansion~\Lizzieqn{fuzzyexpx} is but a rearrangement of the standard
expansion of functions on the sphere in spherical harmonics:
\begin{equation}
f(\theta,\varphi)=\sum_{l=0}^\infty\sum_{m=-l}^lf_{lm}Y_{lm}(\theta,\varphi)
\label{fuzzyexpy}
\end{equation}
This makes the counting of independent coefficients at a given
level $l$ easier: there are $2l+1$ of them. Again a truncation of
the expansion would give a nonassociative algebra.

The idea behind the fuzzy sphere~\cite{Madorefuzzysphere} is the
substitution of the $x_i$ with a quantity proportional to the
generators of $SU(2)$ in the $2N+1$ dimensional representation
($N$ integer or half-integer):
\begin{equation}
x_i=\frac{r}{\sqrt{N(N+1)}}\, L_i\ \label{defxifuzz}
\end{equation}
The constraint~\Lizzieqn{sphereconstraint} is just the usual Casimir
relation, but now the $x$'s do not commute anymore:
\begin{equation}
[x_i,x_j]=i\frac{r}{\sqrt{N(N+1)}}\varepsilon_{ijk}x_k
\label{fuzzyalg}
\end{equation}
The algebra of functions on the sphere has become a noncommutative
geometry. We have discretized the sphere, not by substituting the
continuous manifold with a finite set of points, but, with a
procedure similar to the fuzzy torus, with a change of the
algebra. The group $SU(2)$ is a symmetry of the sphere and acts in
a natural way of the algebra.


As in the fuzzy torus the algebra of the fuzzy sphere is finite,
this time of rank $N(N+1)$. Recalling the
expansion~\Lizzieqn{fuzzyexpy}, consider the matrix expression of the
spherical harmonics for $l\leq N$ as elements of the fuzzy sphere,
thus defining ``fuzzy harmonics''. This can easily be done
recalling from the standard theory of angular momentum that
\begin{equation} Y_{ll}\propto x_+^l\end{equation} and that \begin{equation} x_\pm Y_{lm}=\sqrt{l(l+1)- m(m
\pm 1)}Y_{l\,m\pm 1}\end{equation} Since in the fuzzy sphere the $x$'s are
$(2N+1)\times (2N+1)$ matrices, so will be the fuzzy harmonics.
The product and commutator of two fuzzy harmonics can be
calculated~\cite{Harmon}:
\begin{eqnarray}
Y^{(N)}_{lm}Y^{(N)}_{l'm'}&=&\sum_{l"=0}^\infty\sum_{m"=-l}^l
\sqrt{\frac{(2N+1)(2l+1)(2l'+1)(2l"+1)}{4\pi}}
\nonumber\\&&(-1)^{N+l+l'+l"+m"}{\small \left(
\begin{array}{ccc} l&l'&l"\\m&m'&-m" \end{array}\right)
\left\{
\begin{array}{ccc}
l&l'&l"\\N&N&N \end{array}\right\} } Y^{(N)}_{l"m"}\nonumber\\
\label{prodY}
\end{eqnarray}
where the quantities in round and curly brackets are Wigner's
3-$j$ and 6-$j$ symbols respectively, in the normalization
of~\cite{Varsalovich}. Using properties of these it is possible to
prove that relation~\Lizzieqn{prodY} is associative, in the limit
$N\to\infty$ it converges to the usual product of two ordinary
spherical harmonics, and the commutator of two fuzzy harmonics
goes as $N^{-1}$.

There are three natural derivations on the fuzzy sphere, which
close the $SU(2)$ algebra:
\begin{equation}
{\cal L}_i f:=[L_i,f]:=L^R_if+L^L_if
\end{equation}
where the $L_i$ have been defined in~\Lizzieqn{defxifuzz} and the
$L(R)$ index indicate that the operator is acting from the
left(right). In the continuum these operators go to
$-i\varepsilon_{ijk}x_j\nabla_k$ acting on function on the sphere.
it is intrinsic in the construction that the Laplacian will have
the usual spectrum of the one on the sphere (integers of the form
$l(l+1)$ truncated at $l\leq N$.

\section{The Fuzzy Disc \label{se:disc}}
\setcounter{equation}{0}

The previous examples were of compact spaces, we now present the
case of a space with a boundary: the
disc~\cite{fuzzydisc}\footnote{A similar model has been introduce
in~\cite{BalGuptaKurkcuoglu}.}. Start from functions of
coordinates $x$ and $y$.  Then ``quantize'' the plane with a
nontrivial commutator $ [\hat x,\hat y]=i\frac\theta 2$. It is
convenient to consider the plane as a complex space with $z=x+iy$.
The quantized versions of $z$ and $\bar z$ are the usual
annihilation and creation operators, $a=\hat x +i \hat y$ and
$a^\dagger=\hat x -i \hat y$ with the unusual relation
\begin{equation}
[a,a^\dagger]=\theta \ . \label{aadcomm}
\end{equation}
The parameter $\theta$, which has the dimensions of a square
length, has no physical meaning, like the distance between sites
in a lattice approximation.

We have associated operators to functions, but there are
ambiguities in this association. To give a unambiguous procedure
we introduce a map which associates an operator on an Hilbert
space to a function on the plane. This is the Weyl
map~\cite{Weyl}. Given a function on the plane we define
$\Omega_\theta$ which to the function $\varphi(z,\bar z)$
associates the operator $\hat \varphi$ defined as follows. Given
the function $\varphi(\bar z,z)$ consider its Taylor expansion:
\begin{equation}
\varphi(\bar z,z)=\varphi^{\mathrm Tay}_{mn}\bar z^m z^n  \ , \label{taylorphi}
\end{equation}
to this function we associate the operator
\begin{equation}
\Omega_\theta(\varphi):=\hat\varphi=\varphi^{\mathrm Tay}_{mn}{a^\dagger}^m a^n  \ ,
\label{taylorexp}
\end{equation}
thus we have ``quantized'' the plane using a normal ordering
prescription\footnote{Actually this map is not the one introduced
by Weyl but a variation of it.}. The map $\Omega_\theta$ is
invertible. Its inverse can be efficiently expressed defining the
\emph{coherent} states:
\begin{equation}
a\FLket{z}=z\FLket{z}  \ ,
\end{equation}
then it results
\begin{equation}
\Omega^{-1}_\theta(\hat\varphi)=\varphi(\bar z,z)=\FLbra{z}\hat
\varphi\FLket{z}  \ . \label{Omegam1}
\end{equation}
There is another useful basis on which it is possible to represent
the operators. Consider the number operator
\begin{equation}
{\rm N}=a^\dagger a  \ , \label{defnumb}
\end{equation}
and its eigenvectors which we indicate\footnote{There is a
possible notational confusion, since we never consider coherent
states with integer values we refrain from introducing a different
symbol for the eigenvectors of ${\rm N}$.} by $\FLket{n}$:
\begin{equation}
{\rm N}\FLket{n}=n\theta\FLket{n}  \ .
\end{equation}
We can then express the operators with a density matrix notation:
\begin{equation}
\hat\varphi=\sum_{m,n=0}^\infty\varphi_{mn}\FLdm{m}{n}  \ .
\label{dmexp}
\end{equation}
The elements of the density matrix basis have a very simple
multiplication rule:
\begin{equation}
\FLket{m}\FLhs{n}{p}\FLbra{q}=\delta_{np}\FLdm{m}{q}  \ . \label{densmult}
\end{equation}
The connection between the expansions~\Lizzieqn{taylorexp}
and~\Lizzieqn{dmexp} is given by:
\begin{eqnarray}
a&=&\sum_{n=0}^\infty \sqrt{(n+1)\,\theta}\FLdm{n}{n+1}\nonumber\\
a^\dagger&=&\sum_{n=0}^\infty \sqrt{(n+1)\,\theta}\FLdm{n+1}{n} \ .
\end{eqnarray}
Applying the map \Lizzieqn{Omegam1} to the operator $\hat\varphi$ in
the number basis we obtain for the function $\varphi$ a new
expression, analogous to the Taylor expansion \Lizzieqn{taylorphi} in
terms of the coefficient $\varphi_{nm}$, that is
\begin{equation}
\varphi(\bar
z,z)=e^{-\frac{|z|^2}{\theta}}\sum_{m,n=0}^\infty\varphi_{mn}
\frac{\bar z^m z^n}{\sqrt{n!m!\theta^{m+n}}}  \ , \label{newtay}
\end{equation}
The maps $\Omega$ and $\Omega^{-1}$ yield a procedure of going
back and forth from functions to operators. Moreover,  the product
of operators being noncommutative, a noncommutative $*$ product
between functions is implicitly defined as
\begin{equation}
\left(\varphi*\varphi'\right)(\bar z,
z)=\Omega^{-1}\left(\Omega(\varphi)\, \Omega(\varphi')\right)  \ .
\label{starint}
\end{equation}
This product (a variation of the Gr\"onewold-Moyal product) was
first introduced by Voros \cite{Voros}. We will indicate the
algebra of functions on the plane with this product as ${\cal
A}_\theta$. In the density matrix basis, the $*$ product
simplifies to an infinite row by column matrix multiplication. It
is easy to see that, when $\theta\to0$, the $*$ product goes to
the ordinary commutative product.

The main point of the previous discussion is that the map $\Omega$
and its inverse provide a manner to associate to each function an
infinite dimensional matrix. And again we pay the price of
noncommutativity.

\begin{figure}[htbp]
\centering
\includegraphics[width=8cm]{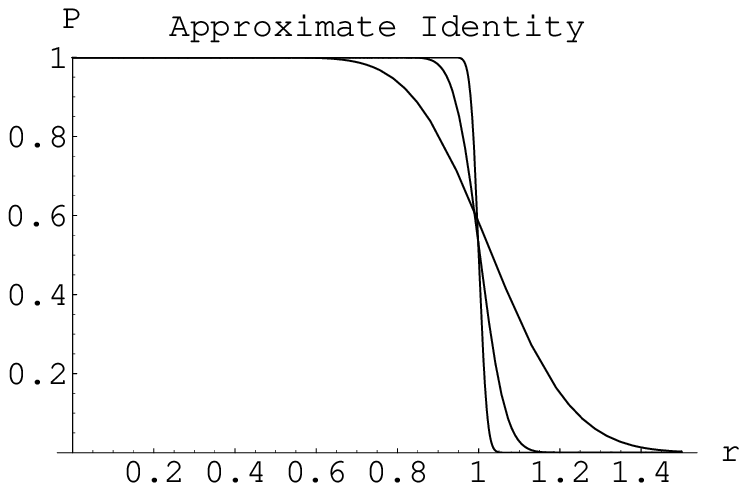}
\caption{%
Profile of the spherically symmetric function $P^N_\theta$
for the choice $R^2=N\theta=1$ and $N=10, 10^2, 10^3$. As $N$
increases the step becomes sharper.}
\label{identity}
\end{figure}

Define now the subalgebras (of functions whose
expansion~\Lizzieqn{newtay} terminates when either $n$ or $m$ is larger
than a given integer $N$. It is immediate to see that these
functions form a subalgebra isomorphic to $N\times N$ matrices. It
can be obtained easily from the full algebra of functions, via a
projection using the idempotent function introduced in a similar
context in~\cite{PinzulStern}:
\begin{equation}
P^N_\theta=\sum_{n=0}^N \FLhs{z}{n}\FLhs{n}{z}=\sum_{n=0}^N\frac{
r^{2n}}{n!\theta^n}e^{-\frac{r^2}{\theta}}  \ ,
\end{equation}
where we have used the polar decomposition of $z=r e^{i\phi}$. The
finite sum may be performed, yielding
\begin{equation}
P^N_\theta=\frac{\Gamma(N+1, {r^2\over \theta})}{\Gamma(N+1)}  \ ,
\end{equation}
where $\Gamma(n,x)$ denotes  the  incomplete gamma function. In
this notation it is then clear that, in the limit $N\to \infty$
and $\theta\to 0$ with
$
R^2\equiv N\theta
$
fixed, the sum converges to $1$ if $r^2/\theta<N$ (namely $r<R$),
and converges to 0 otherwise. It has cylindrical symmetry since
$\phi$ does not appear. For $N$ finite the function vanishes
exponentially for $r$ larger than $R$. Already for $N= 10^3$ it is
well approximated (see figure~\ref{identity}) by a step function.

The function $P^N_\theta$ is a projector of the $*$-algebra of
functions on the plane: $ P^N_\theta*P^N_\theta=P^N_\theta$,  and
the subalgebra ${\cal A}^N_\theta$ is defined as$ {\cal
A}_\theta^N=P^N_\theta*{\cal A}_\theta*P^N_\theta$. Truncating at
a finite $N$ the expansion provides an infrared cutoff. The cutoff
is fuzzy in the sense that functions in the subalgebra are still
defined outside the cutoff, but are exponentially damped.

The starting point to define the matrix equivalent of the
derivations is:
\begin{eqnarray}
\partial_z\varphi&=&\frac{1}{\theta}\FLbra{z}
[a^\dagger,\Omega(\varphi)]\FLket{z}\nonumber\\
\partial_{\bar
z}\varphi&=&\frac{1}{\theta}\FLbra{z}[a,\Omega(\varphi)]\FLket{z}  \ .
\end{eqnarray}
The above expression is exact. Acting on an element of the
subalgebra ${\cal A}_\theta^N$ the derivative takes the functions
out of the algebra, a phenomenon not uncommon in noncommutative
geometry. However
\begin{equation}
\partial_z({\cal A}_\theta^N)\subset {\cal A}_\theta^{N+1}
\end{equation}
analogously for $\partial_{\bar z}$, so that the derivatives of
functions of the subalgebra can still be considered finite
matrices. Notice that $ \partial_z P^N_\theta*\varphi*P^N_\theta\neq
P^N_\theta*(\partial_z\varphi)*P^N_\theta$, the equality obviously
holding in the limit. We will use $\partial_z P^N_\theta*\varphi
P^N_\theta$ to define the Laplacian below.

The fact that we keep $a$ and $a^\dagger$ operators on the full
space (hence they are still infinite matrices) is crucial for the
identification of the algebra of $N\times N$ matrices with the
approximation to the disc. If we were just to truncate the
matrices $a$ and $a^\dagger$, their commutator would not be
proportional to the identity.

Rotations are well defined, in fact the generator of rotations on
the fuzzy disc is nothing but the number operator ${\rm N}$
introduced in~\Lizzieqn{defnumb}. It commutes with $\hat P_\theta^N$,
just as the generator of rotations on the ordinary disc, the
angular momentum operator $\partial_\phi$, commutes with $P_\theta^N$.
Just as the fuzzy sphere maintains the invariance group of the
sphere, the fuzzy disc retains the fundamental symmetry of the
disc. Note that the eigenvalue equations
\begin{equation}
\frac 1\theta [a,\varphi]=\lambda \varphi\ \ , \ \ \frac 1\theta
[a^\dagger,\varphi]=\lambda \varphi
\end{equation}
have no solution in the space of $N\times N$ matrices, just as in
the commutative case translations on the disc have no
eigenvectors. Nevertheless the fuzzy Laplacian
\begin{equation}
\hat\nabla^2\hat\varphi:=\frac{4}{\theta^2}[a,[a^\dagger,\varphi]]
=\frac{4}{\theta^2}[a^\dagger,[a,\varphi]] \label{lapladef}
\end{equation}
can be defined. In particular consider the following matrix model:
\begin{equation}
S=\frac 1{2\pi\theta}\int d^2z\; \varphi^* *(\nabla^2\varphi)
=\tr\Phi_{mn}^\dagger \left(\nabla^2\right)_{mnpq}\Phi_{pq}  \ ,
\label{action}\end{equation} where $\left(\nabla^2\right)_{mnpq}$ is
implictly defined by~\Lizzieqn{lapladef}. It is an operator acting on a
finite space of dimension $(N+1)^2$, and its eigenvalues can be
calculated and compared with the exact commutative case.

\begin{figure}[htbp]
\centering
\includegraphics[width=6cm]{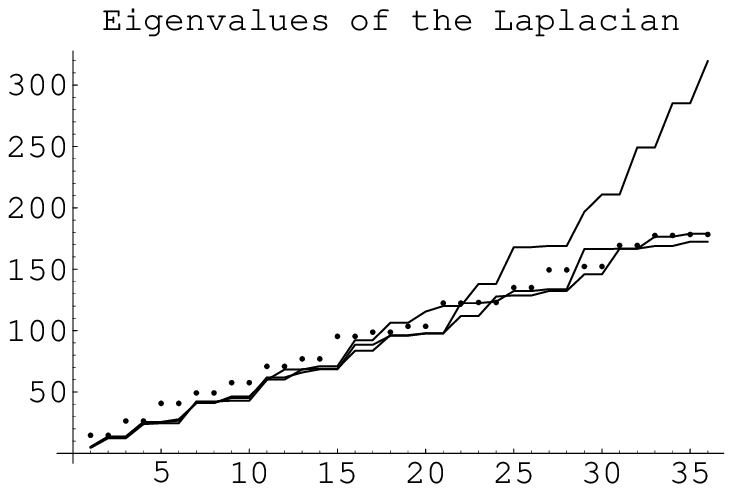}
\includegraphics[width=6cm]{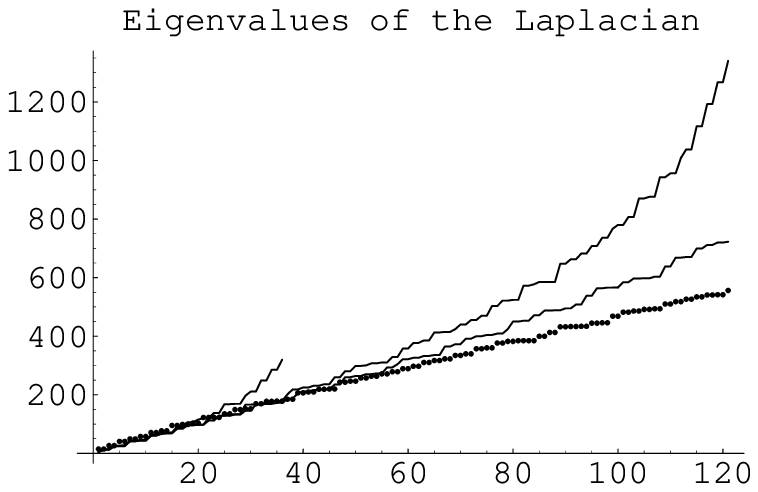}
\caption{The first eigenvalues of the
Laplacian on the disc (dots) and the fuzzy Laplacian (solid lines)
for $N=5, 10,15$. The lines corresponding to the three cases can
be distinguished because the agreement with the exact case
improves as $N$ grows. In the figure on the right the curve which
interrupts is the one corresponding to $N=5$, for which there are
only 36 eigenvalues. }
\label{figlapl}
\end{figure}

In figure~\ref{figlapl} we show a comparison between the
eigenvalues for the exact and approximate Laplacians for three
values of $N$.
The agreement is good, a fuzzy drum sounds pretty much like a
regular drum. Discrepancies start occurring  for the $4 N^{th}$
eigenvalues, this is to be expected because $4 N / R^2=4/\theta$
is the energy cutoff of the theory\footnote{We use units for which
$\hbar=c=1$}. Most of the eigenvalues (but not all) of the fuzzy
Laplacian are doubly degenerate, but the unmatched ones become
sparser as $N$ increases. Also the approximate Green's function
obtained from the action~\Lizzieqn{action} shows an excellent agreement
with the exact solution~\cite{fuzzydisc}

\section{Conclusions}
We have seen a few examples of finite dimensional (matrix)
algebras which approximate, in a definite sense, continuous
geometries. The examples exhibited are quite peculiar, and I am
not proposing them operatively as model of spacetime. Nevertheless
they are useful hints of a structure of spacetime where symmetries
are implemented on a space with a drastic reduction of the
fundamental degrees of freedom.

\section*{Acknowledgments}
I thank Colin Froggatt, Norma Mankoc-Borstnik, and Holger-Bech
Nielsen for organizing a most stimulating conference in a
beautiful place, and especially for inviting me! I also thank
P.~Vitale and A.~Zampini for a very enjoyable collaboration. This
work has been supported in part by the {\sl Progetto di Ricerca di
Interesse Nazionale {\em SInteSi}}.



\begin{thebibliography}{99}
\bibitem{Sakharov}
A.D. Sakharov, Pis'ma Zh. Eksp.
Teor. Fiz. {\bf 5,} 32 (1967)
[JETP Lett. {\bf 5,} 24 (1967)]
\bibitem{FY}
M. Fukugita and Y. Yanagida, Phys. Lett. {\bf B174,} 45 (1986).
\bibitem{FGY}
P.H. Frampton, S.L. Glashow and T. Yanagida.
Phys. Lett. {\bf B548,} 119 (2002).
{\tt hep-ph/0208157}
\bibitem{Y}
T. Yanagida,
in {\it Proceedings of the Workshop
on Unified Theories and Baryon
Number of the Universe}
Editors:O.Sawada and A. Sugamoto,
(Tsukuba, Japan, 1979),
KEK Report
KEK-79-18, page 95.\\
S.L. Glashow, in Quarks and Leptons; Cargese, July 1979.
Editors, M. Levy {\it et al.}  Plenum (1980) p.707.\\
M. Gell-Mann, P. Ramond and R. Slansky,
in {\it Supergravity}
Editors: D.Z. Freedman and P. Van Nieuwenhuizen,
(North-Holland, Amsterdam, 1979).
\bibitem{buchpascos}
W. Buchm\"{u}ller, in
Proceedings of the Eighth Symposium
on Particles, Strings and Cosmology.
Editors: P.H. Frampton and Y.J. Ng.
Rinton Press, New Jersey (2001), page 97.
\bibitem{FGM}
P.H. Frampton, S.L. Glashow and D. Marfatia,
Phys. Lett.  {\bf B536,} 79 (2002). {\tt hep-ph/0201008}
\bibitem{Buch}
See {\it e.g.}
W. Buchm\"{u}ller and M. Plumacher,
Phys. Reps. {\bf 320,} 329 (1999), and
references therein.
\bibitem{branco}
G.C. Branco, T. Morozumi, B.M. Nobre
and M.N. Rebelo.
Nucl. Phys. {\bf B617,} 475 (2001)
{\tt hep-ph/0107164}\\
M. Rebelo, Phys. Rev. {\bf D67,} 013008 (2003). {\tt hep-ph/0207236}.
\end{thebibliography}

\begin{thebibliography}{99}

\bibitem{afshordi}
N.~Afshordi, Y.~S.~Loh and M.~A.~Strauss,
arXiv:astro-ph/0308260.




\bibitem{ACFM}A. Albrecht,  D. Coulson, P.G. Ferreira and
	J. Magueijo, {\em Phys. Rev. Lett.} {\bf 76}, 1413 (1996).

\bibitem{avelino}
P.~P.~Avelino {\it et al.},
Phys.\ Rev.\ D {\bf 64} (2001) 103505
[arXiv:astro-ph/0102144].


\bibitem{barger}
V.~Barger, H.~S.~Lee and D.~Marfatia,
Phys.\ Lett.\ B {\bf 565} (2003) 33
[arXiv:hep-ph/0302150].

\bibitem{bartolo} N.~Bartolo, S.~Matarrese and A.~Riotto,
Phys.\ Rev.\ D {\bf 64} (2001) 123504
[arXiv:astro-ph/0107502].

\bibitem{freese}
M.~Bastero-Gil, K.~Freese and L.~Mersini-Houghton,
index,''
arXiv:hep-ph/0306289.


\bibitem{bms}
R.~Bean, A.~Melchiorri and J.~Silk,
Phys.\ Rev.\ D {\bf 68} (2003) 083501
[arXiv:astro-ph/0306357].

\bibitem{bo}
R.~Bean and O.~Dore,
arXiv:astro-ph/0307100.


\bibitem{bennett}
C.~L.~Bennett {\it et al.},
Astrophys.\ J.\ Suppl.\  {\bf 148} (2003) 1
[arXiv:astro-ph/0302207].

\bibitem{benoit}
A.~Benoit {\it et al.}  [Archeops Collaboration],
Astron.\ Astrophys.\  {\bf 399} (2003) L19
[arXiv:astro-ph/0210305].

\bibitem{review4} J.~R.~Bond,
Class.\ Quant.\ Grav.\  {\bf 15} (1998) 2573.

\bibitem{boughn}
S.~Boughn and R.~Crittenden,
arXiv:astro-ph/0305001.


\bibitem{bridle}
S.~L.~Bridle, A.~M.~Lewis, J.~Weller and G.~Efstathiou,
Mon.\ Not.\ Roy.\ Astron.\ Soc.\  {\bf 342} (2003) L72
[arXiv:astro-ph/0302306].


\bibitem{AMburles}
S.~Burles, K.~M.~Nollett and M.~S.~Turner,
Astrophys.\ J.\  {\bf 552}, L1 (2001)
[arXiv:astro-ph/0010171].

\bibitem{caldwell}
R.~R.~Caldwell and M.~Doran,
arXiv:astro-ph/0305334.


\bibitem{cen}
R.~Cen,
Astrophys.\ J.\  {\bf 591} (2003) L5
[arXiv:astro-ph/0303236].


\bibitem{ciardi}
B.~Ciardi, A.~Ferrara and S.~D.~M.~White,
Mon.\ Not.\ Roy.\ Astron.\ Soc.\  {\bf 344} (2003) L7
[arXiv:astro-ph/0302451].

\bibitem{contaldi}
C.~R.~Contaldi, M.~Peloso, L.~Kofman and A.~Linde,
JCAP {\bf 0307} (2003) 002
[arXiv:astro-ph/0303636].

\bibitem{AMcrotty}
P.~Crotty, J.~Garcia-Bellido, J.~Lesgourgues and A.~Riazuelo,
Phys.\ Rev.\ Lett.\  {\bf 91} (2003) 171301
[arXiv:astro-ph/0306286].

\bibitem{julien}
P.~Crotty, J.~Lesgourgues and S.~Pastor,
Phys.\ Rev.\ D {\bf 67} (2003) 123005
[arXiv:astro-ph/0302337].


\bibitem{cyburt}
R.~H.~Cyburt, B.~D.~Fields and K.~A.~Olive,
Phys.\ Lett.\ B {\bf 567} (2003) 227
[arXiv:astro-ph/0302431].


\bibitem{AMcopeland} E.~J.~Copeland, I.~J.~Grivell and A.~R.~Liddle,
arXiv:astro-ph/9712028.

\bibitem{dode}
S.~Dodelson,
AIP Conf.\ Proc.\  {\bf 689} (2003) 184
[arXiv:hep-ph/0309057].

\bibitem{dore}
O.~Dore, G.~P.~Holder and A.~Loeb,
arXiv:astro-ph/0309281.

\bibitem{knox} S.~Dodelson and L.~Knox,
Phys.\ Rev.\ Lett.\  {\bf 84}, 3523 (2000)
[arXiv:astro-ph/9909454].

\bibitem{doste}S.~Dodelson and E.~Stewart,
arXiv:astro-ph/0109354.


\bibitem{review5} R.~Durrer, arXiv:astro-ph/0109522.


\bibitem{efstathiou2}
G.~Efstathiou,
arXiv:astro-ph/0310207.

\bibitem{freedman} W. Freedman {\it et al.}, 
Astrophysical Journal, 553, 2001, 47.

\bibitem{gordon} C.~Gordon, D.~Wands, B.~A.~Bassett and R.~Maartens,
Phys.\ Rev.\ D {\bf 63} (2001) 023506
[arXiv:astro-ph/0009131].


\bibitem{grainge}
K.~Grainge {\it et al.},
Mon.\ Not.\ Roy.\ Astron.\ Soc.\  {\bf 341} (2003) L23
[arXiv:astro-ph/0212495].

\bibitem{kkmr}
W.~H.~Kinney, E.~W.~Kolb, A.~Melchiorri and A.~Riotto,
arXiv:hep-ph/0305130.

\bibitem{kogut}
A.~Kogut {\it et al.},
Astrophys.\ J.\ Suppl.\  {\bf 148} (2003) 161
[arXiv:astro-ph/0302213].

\bibitem{komatsu}
E.~Komatsu {\it et al.},
Astrophys.\ J.\ Suppl.\  {\bf 148} (2003) 119
[arXiv:astro-ph/0302223].


\bibitem{kosowsky} A.~Kosowsky and M.~S.~Turner,
Phys.\ Rev.\ D {\bf 52} (1995) 1739
[arXiv:astro-ph/9504071].


\bibitem{halverson} N.~W.~Halverson {\it et al.},
arXiv:astro-ph/0104489.

\bibitem{hannestad}
S.~Hannestad,
JCAP {\bf 0305} (2003) 004
[arXiv:astro-ph/0303076].


\bibitem{review2} W.~Hu, D.~Scott, N.~Sugiyama and M.~J.~White,
Phys.\ Rev.\ D {\bf 52}, 5498 (1995)
[arXiv:astro-ph/9505043].

\bibitem{review} W.~Hu, N.~Sugiyama and J.~Silk,
Nature {\bf 386}, 37 (1997)
[arXiv:astro-ph/9604166].

\bibitem{dasipol}
J.~Kovac, E.~M.~Leitch, P.~C., J.~E.~Carlstrom, H.~N.~W. and W.~L.~Holzapfel,
Nature {\bf 420} (2002) 772
[arXiv:astro-ph/0209478].

\bibitem{acbar}
C.~l.~Kuo {\it et al.}  [ACBAR collaboration],
arXiv:astro-ph/0212289.

\bibitem{langlois} D.~Langlois and A.~Riazuelo,
Phys.\ Rev.\ D {\bf 62} (2000) 043504.


\bibitem{leach}
S.~M.~Leach and A.~R.~Liddle,
arXiv:astro-ph/0306305.


\bibitem{lee} A.~T.~Lee {\it et al.},
Astrophys.\ J.\  {\bf 561} (2001) L1
[arXiv:astro-ph/0104459].

\bibitem{luminet}
J.~P.~Luminet, J.~Weeks, A.~Riazuelo, R.~Lehoucq and J.~P.~Uzan,
Nature {\bf 425} (2003) 593
[arXiv:astro-ph/0310253].


\bibitem{martins}
C.~J.~A.~Martins, A.~Melchiorri, G.~Rocha, R.~Trotta, P.~P.~Avelino and P.~Viana,
arXiv:astro-ph/0302295.


\bibitem{mauskopf} P.~D.~Mauskopf {\it et al.}  [Boomerang Collaboration],
Astrophys.\ J.\  {\bf 536}, L59 (2000)
[arXiv:astro-ph/9911444].


\bibitem{mmot}
A.~Melchiorri, L.~Mersini, C.~J.~Odman and M.~Trodden,
Phys.\ Rev.\ D {\bf 68} (2003) 043509
[arXiv:astro-ph/0211522].

\bibitem{odman}
A.~Melchiorri and C.~Odman,
Phys.\ Rev.\ D {\bf 67} (2003) 081302
[arXiv:astro-ph/0302361].

\bibitem{melksilk} A.~Melchiorri and J.~Silk,
arXiv:astro-ph/0203200.

\bibitem{melchiorri} A.~Melchiorri {\it et al.}  [Boomerang Collaboration],
Astrophys.\ J.\  {\bf 536} (2000) L63
[arXiv:astro-ph/9911445].

\bibitem{miller} A.~D.~Miller {\it et al.},
Astrophys.\ J.\  {\bf 524}, L1 (1999)
[arXiv:astro-ph/9906421].

\bibitem{netterfield} C.~B.~Netterfield {\it et al.}  
[Boomerang Collaboration], arXiv:astro-ph/0104460.


\bibitem{nolta}
M.~R.~Nolta {\it et al.},
arXiv:astro-ph/0305097.


\bibitem{costa}
A.~de Oliveira-Costa, M.~Tegmark, M.~Zaldarriaga and A.~Hamilton,
arXiv:astro-ph/0307282.



\bibitem{page}
L.~Page {\it et al.},
arXiv:astro-ph/0302220.


\bibitem{pearson}
T.~J.~Pearson {\it et al.},
Astrophys.\ J.\  {\bf 591} (2003) 556
[arXiv:astro-ph/0205388].

\bibitem{Peeb1970} P.J.E. Peebles, and Yu, J.T. 1970,
Ap.J. 162, 815  


\bibitem{peiris}
H.~V.~Peiris {\it et al.},
Astrophys.\ J.\ Suppl.\  {\bf 148} (2003) 213
[arXiv:astro-ph/0302225].

\bibitem{pierpaoli}
E.~Pierpaoli,
Mon.\ Not.\ Roy.\ Astron.\ Soc.\  {\bf 342} (2003) L63
[arXiv:astro-ph/0302465].


\bibitem{scranton}
R.~Scranton {\it et al.}  [SDSS Collaboration],
arXiv:astro-ph/0307335.


\bibitem{anze}
A.~Slosar {\it et al.},
Mon.\ Not.\ Roy.\ Astron.\ Soc.\  {\bf 341} (2003) L29
[arXiv:astro-ph/0212497].

\bibitem{spergel}
D.~N.~Spergel {\it et al.},
Astrophys.\ J.\ Suppl.\  {\bf 148} (2003) 175
[arXiv:astro-ph/0302209].



\bibitem{SZ70}Sunyaev, R.A. \& Zeldovich,
Ya.B., 1970,  Astrophysics and Space Science 7, 3 


\bibitem{tegb97} M.~Tegmark,
Astrophys.\ J.\  {\bf 514}, L69 (1999)
[arXiv:astro-ph/9809201].


\bibitem{torbet} E.~Torbet {\it et al.},
Astrophys.\ J.\  {\bf 521}, L79 (1999)
[arXiv:astro-ph/9905100].


\bibitem{jussi}
J.~Valiviita and V.~Muhonen,
Phys.\ Rev.\ Lett.\  {\bf 91} (2003) 131302
[arXiv:astro-ph/0304175].

\bibitem{verde}
L.~Verde {\it et al.},
Astrophys.\ J.\ Suppl.\  {\bf 148} (2003) 195
[arXiv:astro-ph/0302218].


\bibitem{wang2}
X.~Wang, M.~Tegmark, B.~Jain and M.~Zaldarriaga,
arXiv:astro-ph/0212417.

\bibitem{wlewis}
J.~Weller and A.~M.~Lewis,
arXiv:astro-ph/0307104.

\bibitem{review3} M.~J.~White, D.~Scott and J.~Silk,
Ann.\ Rev.\ Astron.\ Astrophys.\  {\bf 32} (1994) 319.

\bibitem{wilson} M.~L.~Wilson and J.~Silk,
Astrophys.\ J.\  {\bf 243} (1981) 14.

\end{thebibliography}

\begin{thebibliography}{99}
\bibitem{Maldacena}
J. Maldacena, Adv. Theor. Math. Phys. {\bf 2,} 231 (1998).
{\tt hep-th/9711200}.\\
S.S. Gubser, I.R. Klebanov and A.M. Polyakov, Phys. Lett.
{\bf B428,} 105 (1998). {\tt hep-th/9802109}.\\
E. Witten, Adv. Theor. Math. Phys. {\bf 2,} 253 (1998).
{\tt hep-th/9802150}.
\bibitem{Frampton}
P.H. Frampton, Phys. Rev. {\bf D60,} 041901 (1999). {\tt hep-th/9812117}.
\bibitem{FS}
P.H. Frampton and W.F. Shively, Phys. Lett. {\bf B454,} 49 (1999).
{\tt hep-th/9902168}.
\bibitem{FV}
P.H. Frampton and C. Vafa. {\tt hep-th/9903226.}
\bibitem{KS}
S. Kachru and E. Silverstein, Phys. Rev. Lett. {\bf 80,} 4855 (1998).
{\tt hep-th/9802183}.
\bibitem{DGG}
A. De R\'{u}jula, H. Georgi and S.L. Glashow.
{\it Fifth Workshop on Grand Unification}.
Editors: P.H. Frampton, H. Fried and K.Kang.
World Scientific (1984) page 88.
\bibitem{ADFFL}
U. Amaldi, W. De Boer, P.H. Frampton, H. F\"{u}rstenau and J.T. Liu.
Phys. Lett. {\bf B281,} 374 (1992).
\bibitem{F2}
P.H. Frampton, Phys. Rev. {\bf D60,} 085004 (1999). {\tt hep-th/9905042}.
\bibitem{PS}
J.C. Pati and A. Salam,
Phys. Rev. {\bf D8,} 1240 (1973); {\it ibid} {\bf D10,} 275 (1974).
\bibitem{GG}
H. Georgi and S.L. Glashow, Phys. Rev. Lett. {\bf 32,} 438 (1974).
\bibitem{GQW}
H. Georgi, H.R. Quinn and S. Weinberg, Phys. Rev. Lett.
{\bf 33,} 451 (1974).
\bibitem{FG}
P.H. Frampton and S.L. Glashow, Phys. Lett. {\bf B131,} 340 (1983).
\bibitem{ADF}
U. Amaldi, W. de Boer and H. F\"{u}rstenau, Phys. Lett. {\bf B260,} 447 (1991).
\bibitem{Murayama}
H. Murayama and A. Pierce, Phys. Rev. {\bf D65,}
055009 (2002).
\bibitem{PP}
F. Pisano and V. Pleitez, Phys. Rev. {\bf D46,} 410 (1992).
\bibitem{PF}
P.H. Frampton,
Phys. Rev. Lett. {\bf 69,} 2889 (1992).
\bibitem{DK}
S. Dimopoulos and D. E. Kaplan, Phys. Lett. {\bf B531,} 127 (2002).
\bibitem{bershadsky1}
M. Bershadsky, Z. Kakushadze and C. Vafa,
Nucl. Phys. {\bf B523,} 59 (1998).
\bibitem{bershadsky2}
M. Bershadsky and A. Johansen, Nucl. Phys. {\bf B536,} 141 (1998).
\bibitem{FMink}
P.H. Frampton and P. Minkowski, {\tt hep-th/0208024}.
\bibitem{Witten}
E. Witten, JHEP {\bf 9812:012} (1998).
\bibitem{PDG}
Particle Data Group. {\it Review of Particle Physics.}
Phys. Rev. {\bf D66,} 010001 (2002).
\bibitem{Csaki}
C. Csaki, J. Erlich, G.D. Kribs and J. Terning,
Phys. Rev. {\bf D66,} 075008 (2002).
\bibitem{FRT}
P.H. Frampton, R.M. Rohm and T. Takahashi. Phys. Lett. {\bf B570,} 67 (2003).
{\tt hep-ph/0302074}.
\bibitem{Peskin}
M. Peskin and T. Takeuchi, Phys. Rev. Lett. {\bf 65,} 964 (1990);
Phys. Rev. {\bf D46,} 381 (1992).
\bibitem{DRW}
S. Dimopoulos, S. Raby and F. Wilczek, Phys. Rev. {\bf D24,} 1681 (1981);
Phys. Lett. {\bf B112,} 133 (1982).
\bibitem{DG}
S. Dimopoulos and H. Georgi, Nucl. Phys. {\bf B193,} 150 (1981).
\bibitem{FGMY}
P.H. Frampton and S.L. Glashow, Phys. Lett. {\bf B461,} 95 (1999).
P.H. Frampton, S.L. Glashow and D. Marfatia, Phys. Lett. {\bf B536,} 79 (2002).
P.H. Frampton, S.L. Glashow and T. Yanagida,  Phys. Lett. {\bf B548,} 119 (2002).

\end{thebibliography}

\begin{thebibliography}{99}

\bibitem{W1} E.Witten, {\it Noncommutative Geometry and String Field
Theory},
Nucl.Phys. {\textbf {B268}} (1986) 253.

\bibitem{Sen} A.Sen {\it Descent Relations among Bosonic D--Branes},
Int.J.Mod.Phys. {\textbf {A14}} (1999) 4061, [hep-th/{9902105}].
{\it Tachyon Condensation on the Brane Antibrane System}
JHEP {\textbf {9808}} (1998) 012, [hep-th/{9805170}].
{\it BPS D--branes on Non--supersymmetric Cycles}, JHEP {\textbf {9812}}
(1998) 021, [hep-th/{9812031}].


\bibitem{BMP1} L.Bonora, C.Maccaferri, P.Prester {\it Dressed Sliver
Solutions in Vacuum String Field Theory}, to appear

\bibitem{BMP2} L.Bonora, C.Maccaferri, P.Prester {\it Dressed Sliver:
the spectrum}, to appear


\bibitem{leclair1} A.Leclair, M.E.Peskin, C.R.Preitschopf, {\it String Field
Theory on the Conformal Plane. (I) Kinematical Principles},
Nucl.Phys. {\textbf {B317}} (1989) 411.

\bibitem{RSZ1} L.Rastelli, A.Sen and B.Zwiebach, {\it String field
theory around the tachyon vacuum},
Adv.\ Theor.\ Math.\ Phys.\  {\textbf 5} (2002) 353
[hep-th/{0012251}].

\bibitem{HKw} H.Hata and T.Kawano, {\it Open string states around
a classical solution in vacuum string field theory},
JHEP {\textbf {0111}} (2001) 038 [hep-th/{0108150}].

\bibitem{GRSZ1} D.Gaiotto, L.Rastelli, A.Sen and B.Zwiebach,
{\it Ghost Structure and Closed Strings in Vacuum String Field
Theory}, [hep-th/{0111129}].

\bibitem{GJ1} D.J.Gross and A.Jevicki, {\it Operator Formulation
of Interacting String Field Theory}, Nucl.Phys. {\textbf {B283}} (1987) 1.

\bibitem{RSZ2} L.Rastelli, A.Sen and B.Zwiebach, {\it Classical
solutions in string field theory around the tachyon vacuum},
Adv.\ Theor.\ Math.\ Phys.\  {\textbf 5} (2002) 393 [hep-th/{0102112}].

\bibitem{KP} V.A.Kostelecky and R.Potting, {\it Analytical construction
of a nonperturbative vacuum for the open bosonic string},
Phys.\ Rev.\ D {\textbf {63}} (2001) 046007
[hep-th/{0008252}].

\bibitem{Oku2} K.Okuyama, {\it Ghost Kinetic Operator of Vacuum
String Field Theory}, JHEP {\textbf {0201}} (2002) 027 [hep-th/{0201015}].


\end{thebibliography}

\begin{thebibliography}{199}
\bibitem{norma92} N. S. Manko\v c Bor\v stnik,``Spin connection as a superpartner of a vielbein''
{\it Phys. Lett.} {\bf B 292}, 25-29 (1992).
%
\bibitem{norma93} N. S. Manko\v c Bor\v stnik, 
``Spinor and vector representations in four dimensional
Grassmann space'', {\it J. Math. Phys.} {\bf 34}, 3731-3745 (1993).
%
\bibitem{normasuper94} N. Manko\v c Bor\v stnik, 
''Poincar\' e algebra in ordinary and Grassmann space and supersymmetry'',
{\it J. Math. Phys.} {\bf 36}, 1593-1601(1994), 
%
\bibitem{norma95} N. S. Manko\v c Bor\v stnik, ''Unification of spins and charegs in Grassmann space?'',
{\it Modern Phys. Lett. A} {\bf 10}, 587-595 (1995),
%
\bibitem{holgernorma00} N. S. Manko\v c Bor\v stnik, H. B. Nielsen, 
''Dirac-K\" ahler approach conneceted to quantum mechanics in Grassmann space'',
{\it Phys. Rev.} {\bf 62} (04010-14) (2000),
%
\bibitem{norma97} N. S. Manko\v c Bor\v stnik and S. Fajfer, 
''Spins and charges, the algebra and subalgebra of the group SO(1,14) and Grassmann space,
{\it N. Cimento} {\bf 112B}, 1637-1665(1997).
%
\bibitem{pikanormaproceedings1} A. Bor\v stnik, N. S. Manko\v c Bor\v stnik, 
``Are Spins and Charges Unified? How Can One
Otherwise Understand Connection Between Handedness (Spin) and
Weak Charge?'',
{\it Proceedings to the International Workshop on
``What Comes Beyond the Standard Model, Bled,
Slovenia, 29 June-9 July 1998},
Ed. by N. Manko\v c Bor\v stnik,
H. B. Nielsen, C. Froggatt, DMFA Zalo\v zni\v stvo 1999, p.52-57,
hep-ph/9905357.
%
\bibitem{pikanormaproceedings2} A. Bor\v stnik, N. S. Manko\v c Bor\v stnik, 
``Weyl spinor of SO(1,13), families of spinors of the Standard model and their masses'',
{\it Proceedings to the International Workshop on
``What Comes Beyond the Standard Model'', Bled 2000,2001,2002 Volume 2},
Ed. by N. Manko\v c Bor\v stnik,
H. B. Nielsen, C. Froggatt, Dragan Lukman, DMFA Zalo\v zni\v stvo 2002, p.27-51,
hep-ph/0301029, and the paper in preparation.
%
\bibitem{norma01} N. S. Manko\v c Bor\v stnik,
``Unification of spins and charges'',
{\it Int. J. Theor. Phys.} {\bf 40}, 315-337  (2001) and references therein.
%
\bibitem{bojannorma2001} B. Gornik, and N. S. Manko\v c Bor\v stnik, 
``Linear equations of motion for massless particles of any spin in any
even-dimensional spaces'', hep-th/0102067, hep-th/ 0102008 (2001).
%
\bibitem{holgernorma0203} N. S. Manko\v c Bor\v stnik, H. B. Nielsen, ``The internal space is making the
choice of the signature of space-time'', in preparation.
%
\bibitem{holgernorma2003} N. S. Manko\v c Bor\v stnik, H. B. Nielsen, ``How to generate families of'',
		        {\it J. of Math. Phys.} {\bf 44} (2003) 4817-4827, hep-th/0303224.
%
\bibitem{holgernorma2002} N. S. Manko\v c Bor\v stnik, H. B. Nielsen, ``How to generate spinor 
representations in any dimension in terms of projection operators, 
accepted in J. of Math. Phys. {\bf 43}, 5782-5803, hep-th/0111257.
%
\bibitem{holgernormawhy} N. S. Manko\v c Bor\v stnik, H. B. Nielsen, ``Why Nature has made a choice of 
one time and three space coordinates?, hep-ph/0108269, J.Phys.A:Math.Gen. {\bf 35} 2002,in print. 
%
\bibitem{holgernormaren} N. S. Manko\v c Bor\v stnik, H. B. Nielsen, ``Coupling constant 
unification in spin-charge unifying model agreeing with proton decay measurement``, in preparation.
%
\bibitem{georgi} H. Georgi {\it Lie algebras in particle physics } (The Benjamin/cummings, London 1982). 
%
\bibitem{astrinorma2003} N. S. Manko\v c Bor\v stnik, H. B. Nielsen, `` How to generate families of spinors'',
{\it J. of Math. Phys.} {\bf 44} (2003) 4817-4827, hep-th/0303224.
%
\bibitem{okun} V.A. Novikov, L.B. Okun,A.N. Royanov, M.I. Vysotsky, ``Extra generations and
discrepancies of electroweak precision data'', hep-ph/0111028.
%
\bibitem{astridragannorma} A. Kleppe, D. Lukman, N.S. Manko\v c Bor\v stnik, ''About families of quarks and
leptons'', in this Proceedings.
%
\bibitem{fritsch} H. Fritsch, Phys. Lett.{\bf 73 B},317 (1978).
%
\bibitem{witten81} E. Witten, ''Search for realistic Kaluza-Klein theory'',
Nucl. Phys. {\bf B 186} (1981) 412; ''Fermion quantum numbers in Kaluza-Klein theory'',
Princeton Technical Rep. PRINT-83-1056, Oct 1983.
%
\bibitem{holgernorma2004} N. S. Manko\v c Bor\v stnik, H. B. Nielsen, ''Example of Kaluza-Klein-like
theories leading to  massless spinors after compactification'', sent to Phys. Rev. Lett., hep-th/0311037.

\end{thebibliography}

\begin{thebibliography}{99}

%
\bibitem{witten1981} E. Witten, ''Search for realistic Kaluza-Klein theory'', {\it Nucl. Phys.}
{\bf B 186} (1981) 412-428.
%
\bibitem{witten1983} E.Witten, ''Fermion quantum numbers in Kaluza-Klein theorie'',  
Princeton Technical Rep. PRINT -83-1056, October 1983.
%
\bibitem{kaluzaklein} T. Kaluza, ''On the unification problem in Physics'', {\it Sitzungsber. d. Berl. Acad.}
 (1918) 204, O. Klein, ''Quantum theory and five-dimensional relativity'', {\it Zeit. Phys.} {\bf 37}(1926) 895. 

%
\bibitem{palla} L. Palla, Proceedings on 1978 Tokyo Conference on high energy physics, p. 629.
%
\bibitem{chaplineslanskymanton} N.S. Manton, {\it Nucl. Phys. } {\bf B 193} (1981) 391; G. Chapline and N. S. Manton,
{\it Nucl. Phys.} {\bf B 184} (1981) 391; G. Chapline and R. Slanski, {\it Nucl. Phys.} {\bf B 209} (1982) 461.
%
\bibitem{wetterich} G. Watterich, {\it Nucl. Phys.} {\bf B 187} (1981) 343.
%
\bibitem{WInorma92} N. S. Manko\v c Bor\v stnik,
''Spin connection as a superpartner of a vielbein'',
{\it Phys. Lett.} {\bf B 292} (1992) 25-29.
%
\bibitem{WInorma93} N. S. Manko\v c Bor\v stnik, 
''Spinor and vector representations in four dimensional
Grassmann space'', {\it J. Math. Phys.} {\bf 34}, (1993)3731-3745.
%
\bibitem{norma94} N. S. Manko\v c Bor\v stnik,
              ``Spinors, vectors and scalars in Grassmann space and canonical
              quantization for fermions and bosons``,
              International Journal of Modern Physics { \bf A 9 } (1994), 1731-
              1745.
%
\bibitem{normaixtapa2001} N. S. Manko\v c Bor\v stnik,
''Unification of spins and charges'',
{\it Int. J. Theor. Phys.} {\bf 40} (2001) 315-337 and references therein.
%
\bibitem{pikanorma1998}  Anamarija Bor\v stnik, Norma Manko\v c Bor\v stnik, 
              ``Are spins and charges unified? How can one otherwise 
              understand connection bewtween handedness(spin) and 
              weak charge?``,
              Proceedings to the Intenational workshop    
              What comes beyond the standard model, Bled, Slovenia, 
              29 June - 9 July 1998, Ed.by N. Manko\v c Bor\v stnik, 
              H. Bech Nielsen, C. Froggatt, DMFA-Zalo\v zni\v stvo 1999,
              p. 52-57, hep-ph/9905357.
%
\bibitem{pikanorma2002} A. Bor\v stnik Bra\v ci\v c, N. S. Manko\v c Bor\v stnik, 
''Weyl spinor of SO(1,13), families of spinors of the Standard model and their masses'',
			    Proceedings to the $5^{th}$ International
				Workshop ''What Comes Beyond the Standard Model'', 13 -23 of July, 2002,
				VolumeII, Ed. N. S. Manko\v c Bor\v stnik, H. B. Nielsen, C. Froggatt, D. Lukman, DMFA 
				Zalo\v zni\v stvo, Ljubljana December 2002, hep-ph/0301029 (2002)27-51.
%
\bibitem{holgernormadk} Norma Manko\v c Bor\v stnik, Holger Bech Nielsen,
              Dirac-K\" ahler approach connected to quantum mechanics in
              Grassmann space,  Phys. Rev. D 62,044010, (2000)1-14, 
			  hep-th/9911032.
%
\bibitem{WIholgernorma2002} N. S. Manko\v c Bor\v stnik, Holger B. Nielsen,
''How to generate
spinor representations in any dimension in terms of projection operators'', 
			   J. of Math. Phys. {\bf 43} (2002), (5782-5803), hep-th/0111257.               
%
\bibitem{WIholgernorma2003} N. S. Manko\v c Bor\v stnik, Holger B. Nielsen,
''How to generate families of 
spinors'', {\it J. of Math. Phys.} {\bf 44} (2003), (4817-4827), hep-th/0303224.
%
\bibitem{milutin2002} M. Blagojevi\' c, {\it Gravitation and gauge symmetries} 
(Institute of Physics Pub., London 2002, ISGN 0750307676).
%
%
\bibitem{bojannorma} B. Gornik and N. S. Manko\v c Bor\v stnik, 
``Linear equations of motion for massless particles of any spin in any
even-dimensional spaces'' hep-th/0102067, hep-th/ 0102008, to appear in Hadr. J. 
{\bf 26} June 2003, No.3 and 4.
%


\end{thebibliography}

\begin{thebibliography}{99}
\bibitem{CDHBglasgowbrioni}
D.L.~Bennett, C.D.~Froggatt and H.B.~Nielsen, Proceedings of the
27th International Conference on High Energy Physics, p. 557,  ed.
P. Bussey and I. Knowles (IOP Publishing Ltd, 1995); Perspectives
in Particle Physics '94, p. 255, ed. D. Klabu\u{c}ar, I. Picek and
D. Tadi\'{c} (World Scientific, 1995) [arXiv:hep-ph/9504294].
\bibitem{guendelmann}
E.I.~Guendelman, arXiv:gr-qc/0303048.
\bibitem{woodard}
N.C.~Tsamis and R.P.~Woodard, Ann. Phys. {\bf 238}, 1 (1995).
\bibitem{weinberg}
S.~Weinberg, Sources and detection of dark matter and dark energy
in the universe: Proceedings, ed. D.~Cline {Springer Verlag, 2001}
[arXiv:astro-ph/0005265].
\bibitem{itep}
C.D.~Froggatt and H.B.~Nielsen, arXiv:hep-ph/0308144.
\bibitem{fln}
C.D.~Froggatt, L.V.~Laperashvili and H.B.~Nielsen, Hierarchy
Problem and Multiple Point Principle (in preparation).
\bibitem{colin}
C.D.~Froggatt, These Proceedings.
\bibitem{CFHBfn2}
C.D.~Froggatt and H.B.~Nielsen, Phys. Lett. B {\bf 368}, 96 (1996)
[arXiv:hep-ph/9511371];\\
C.D.~Froggatt, H.B.~Nielsen and Y.~Takanishi, Phys. Rev. D {\bf
64}, 113014 (2001) [arXiv:hep-ph/0104161].
\end{thebibliography}

\begin{thebibliography}{99}
\bibitem{r1} SuperK contribution to these proceedings.

\bibitem{r2} G. Altarelli, these proceedings.

\bibitem{r3} Q. Shafi and S. Huber, Phys. Lett. {\bf B 544}, 295 (2002)

\bibitem{r4} A. Masiero, these proceedings.

\bibitem{r5} A. Guth, Phys. Rev. {\bf D 23}, 347 (1981).

\bibitem{r6} Q. Shafi and A. Vilenkin, Phys. Rev. Lett. {\bf 52}, 691 (1984).

\bibitem{r7} T. W. B. Kibble, G. Lazarides and Q. Shafi,  Phys. Lett. {\bf B 113}, 237 (1982).

\bibitem{r8} H. Georgi, H. Quinn and S. 
Weinberg,    Phys. Rev. Lett. {\bf 33}, 451 (1974);   S. Dimopoulos, S.
 Raby and F. Wilczek, Phys. Rev. {\bf D 24}, 1681 (1991).

\bibitem{r9} H. Georgi and S. L. Glashow,  Phys. Rev. Lett. {\bf 28}, 1494 (1972).

\bibitem{r10} H. Georgi, in Particles \& Fields,  edited by C. E. 
Carlson (AIP, New York) 1975;     H. Fritzsch and P. Minkowski, Annals 
Phys. {\bf 93}, 193, (1975).  

\bibitem{r11} F. Fursey, P. Ramond and P.
 Sikivie,    Phys. Lett. {\bf B 60}, 177 (1976);      Y. Achiman and B.
 Stech, Phys. Lett. {\bf B 77}, 389 (1978);      Q. Shafi, Phys. Lett. 
{\bf B 79}, 291 (1978).

\bibitem{r12} J. C. Pati and A. Salam, Phys. 
Rev. {\bf D 10}, 275, (1974).

\bibitem{r13} G. Lazarides, M. Magg and 
Q. Shafi,    Phys. Lett. {\bf B 97}, 87 (1980).

\bibitem{r14} Q. Shafi, Proc. of Monopole '83, Ann Arbor, Michigan (1983).       G. Lazarides,
 Q. Shafi and N. Tomaras,        Phys. Rev. {\bf D 39}, 1239 (1989).

  
 \bibitem{r15} B. Ananthanarayan, G. Lazarides and Q. Shafi,  Phys. Rev. {\bf D 44}, 1613 (1991).

\bibitem{r16} L.J. Hall , R. Rattazzi and 
U. Sarid,    Phys. Rev. {\bf D 50}, 7048 (1994);   
S. Raby, hep-ph/0211025, and references therein.

\bibitem{r17} Q. Shafi and B. Ananthanarayan,    Proc. of ICTP Summer School (1991).

\bibitem{r18} G. Dvali, Q. Shafi and R. K. Schaefer,    Phys. Rev. Lett. {\bf 73}, 1886 (1994).\\
For a recent discussion and some new results, see V. N. Senoguz and Q. Shafi, hep-ph/0305089.

\bibitem{r19} G. Lazarides, hep-ph/0111328.

\bibitem{lss}G. Lazarides, R. K. Schaefer and Q. Shafi, Phys. Rev. \textbf{D56}, 1324 (1997) \\
\indent [hep-ph/9608256].

\bibitem{cobe} G. F. Smoot \emph{et.al.}, Astrophys. J. Lett. \textbf{396}, L1(1996); \\ \indent C. L. Bennett
\emph{et.al.}, Astrophys. J. Lett. \textbf{464}, 1 (1996); \\ \indent 
E. F. Bunn, A. R. Liddle, and M. White, Phys. Rev. \textbf{D54}, 5917
(1996) \\ \indent [astro-ph/9607038].

\bibitem{kyae} B. Kyae and Q. Shafi, hep-ph/0212331.

\bibitem{dls}G. Dvali, G. Lazarides, and Q. Shafi, Phys. Lett. \textbf{B424}, 259 (1998) \\
\indent [hep-ph/9710314].

\newpage

\bibitem{ll}A. R. Liddle, and D. H. Lyth, Phys. Lett. \textbf{B291}, 391 (1992) [astro-ph/9208007];\\ \indent
Phys. Rep. \textbf{231}, 1 (1993) [astro-ph/9303019]

\bibitem{gravitino} M. Yu. Khlopov, and A. D. Linde, Phys. Lett. {\bf B138},
265 (1984);\\J. Ellis, J. E. Kim, and D. Nanopoulos,
Phys. Lett. {\bf B145}, 181 (1984).

\bibitem{ls} G. Lazarides, and Q. Shafi, Phys. Lett. {\bf B258}, 305 (1991).

\bibitem{lsv} G. Lazarides, Q. Shafi, and N. D. Vlachos, Phys. Lett {\bf B427}, 53 (1998) \\
\indent [hep-ph/9706385]; \\ \indent G.~Lazarides, and Q.~Shafi,
Phys.\ Rev.\ {\bf D58}, 071702 (1998)
[hep-ph/9803397]. 

\bibitem{Pati:2002pe}
J.~C.~Pati,
hep-ph/0209160; \\ \indent T. Asaka, hep-ph/0304124.

\bibitem{lepto} M. Fukugita, and T. Yanagida,
Phys. Lett. {\bf B174}, 45 (1986). \\ \indent For a recent discussion of thermal leptogenesis and additional references see
W. Buchmuller, P. diBari, and M. Plumacher, hep-ph/0302092. 
Note that in the models we are discussing thermal leptogenesis is possible
if we allow the reheat temperature to be close to $10^{10}$ GeV.

\bibitem{copeland} E. J. Copeland, A. R. Liddle, D. H. Lyth, E. D. Stewart, and D. Wands, 
\\ \indent Phys. Rev. \textbf{D49}, 6410 (1994).

\bibitem{QSpanagio}C. Panagiotakopoulos, Phys. Rev. \textbf{D55}, 7335 (1997) [hep-ph/9702433];
\\ \indent W.~Buchmuller, L.~Covi, and D.~Delepine,
Phys.\ Lett.\ {\bf B491}, 183 (2000) \\ \indent [hep-ph/0006168]. 

\bibitem{QSlinde} A. Linde, and A. Riotto, Phys. Rev. \textbf{D56}, 1841 (1997) [hep-ph/9703209].

\bibitem{QSkawasaki}
M.~Kawasaki, M.~Yamaguchi, and J.~Yokoyama,
hep-ph/0304161.

\bibitem{wmap}D. N. Spergel \emph{et.al.}, astro-ph/0302209; H. V. Peiris \emph{et.al.}, astro-ph/0302225.

\bibitem{seljak} U. Seljak, P. McDonald, and A. Makarov, astro-ph/0302571.

\bibitem{r20} R. Jeannerot, S. Khalil, G. Lazarides and Q. Shafi,    JHEP {\bf 0010}, 012 
(2000) [hep-ph/0002151].

\bibitem
{r22} B. Kyae and Q. Shafi, hep-ph/0211059.     

\bibitem{r24} N. Arkani-Hamed, S. Dimopoulos 
and G. Dvali,    Phys. Rev. {\bf D 59}, 086004 (1998).   I. Antoniadis 
et.al. hep-ph/9804398.

\bibitem{r25} H. Hannestad and G.G. Raffelt,  
  Phys. Rev. Lett. {\bf 88}, 071301 (2002).

\bibitem{r26} Y. Kawamura, Prog. Ther. Phys. {\bf 105}, 999 (2001) 

\bibitem{r27} Q. Shafi and
 Z. Tavartkiladze, hep-ph/0208162; ibid 0210181.

\bibitem{Huber:2003sf}
S.~J.~Huber and Q.~Shafi,
arXiv:hep-ph/0309252.

\bibitem{r28} S. Huber and Q. Shafi, hep-ph/0207232 .

\bibitem{r29} S. Huber, C. A. Lee and Q. Shafi,    Phys. Lett. {\bf B 531}, 112 (2002)

\bibitem{LED}Contributions to these proceedings.
\bibitem{QS-RS} L.~Randall and R.~Sundrum, {\em Phys. Rev. Lett.}
             {\bf 83} (1999) 3370; M.~Gogberashvili, hep-ph/9812296.




\bibitem{GN} Y.~Grossman and M.~Neubert,
                     {\em Phys. Lett.} {\bf B474} (2000) 361.

\bibitem{GP} T.~Gherghetta and A.~Pomarol,
               {\em Nucl. Phys.} {\bf B586} (2000) 141.

\bibitem{HS2} S.J.~Huber and Q.~Shafi, {\em Phys. Lett.} {\bf B498} (2001) 256.

\bibitem{HS3} S.J.~Huber and Q.~Shafi, {\em Phys. Lett.} {\bf B512} (2001) 365.



\bibitem{BGP}
                   V.~Barger, D.~Marfatia, K.~Whisnant and B.P.~Wood, hep-ph/0204253;
                   J.N.~Bahcall, M.C.~Gonzalez-Garcia, C.~Pena-Garay, hep-ph/0204314.

\bibitem{SK}T.~Toshito [SuperKamiokande Collaboration],  hep-ex/0105023.

\bibitem{CHOOZ} M.~Apollonio {\em et al.} [CHOOZ Collaboration],
                          {\em Phys. Lett.} {\bf B466} (1999) 415.

\bibitem{HS}S.J.~Huber and Q.~Shafi, {\em Phys. Rev.} {\bf D63} (2001) 045010.


\bibitem{HPR} J.L.~Hewett, F.J.~Petriello and T.G.~Rizzo, {\em JHEP} {\bf 0209}
                     (2002) 030.

\bibitem{HS4}S.J.~Huber and Qaisar Shafi {\em Phys. Lett.} {\bf B544} (2002) 295.

\bibitem{LONGB} M.~Komatsu, P.~Migliozzi and  F.~Terranova, hep-ph/0210043.

\bibitem{HMC}The Heidelberg-Moscow collaboration, H.V.~Klapdor-Kleingrothaus
{\em el al.} in the proceedings of {\em DARK 2000}, Heidelberg, Germany, 10-16 July 2000,
520-533 [hep-ph/0103062].

\bibitem{KDHK}H.V.~Klapdor-Kleingrothaus, A.~Dietz, H.L.~Harney and I.V.~Krivosheina.,
               {\em Mod. Phys. Lett.} {\bf A16} (2001) 2409 [hep-ph/0201231].


\bibitem{K00}R.~Kitano, {\em Phys. Lett.} {\bf B481} (2000) 39 [hep-ph/0002279].
\bibitem{GP2}T.~Gherghetta and A.~Pomarol, {\em Nucl.~Phys.} {\bf B602} (2001) 3  

\bibitem{r30} J. Polchinski ,    String Theory, Vols. 1 and 2, Cambride University Press.

\bibitem{r31} G. Dvali, Q. Shafi and S. Solganik, hep-th/0105203 
and references      therein.  For earlier work on higher dimensional cosmology see,
for instance, Q. Shafi and C. Wetterich, {\em Phys. Lett.} {\bf B129} (1983) 387;
{\em Nucl. Phys.} {\bf B289} (1987) 787.  

\end{thebibliography}

\begin{thebibliography}{99}

\bibitem{Alfaro}
J.~Alfaro and G.~Palma,
``Loop quantum gravity corrections and cosmic rays decays,''
Phys.\ Rev.\ D {\bf 65} (2002) 103516
[arXiv:hep-th/0111176].
\\
``Loop quantum gravity and ultra high energy cosmic rays,''
Phys.\ Rev.\ D {\bf 67} (2003) 083003
[arXiv:hep-th/0208193].



\bibitem{Alvarez}
E.~Alvarez,
``Quantum Gravity: A Pedagogical Introduction To Some Recent Results,''
Rev.\ Mod.\ Phys.\  {\bf 61} (1989) 561.

\bibitem{Alvarezz}
E.~Alvarez, L.~Alvarez-Gaume and Y.~Lozano,
``An introduction to T duality in string theory,''
Nucl.\ Phys.\ Proc.\ Suppl.\  {\bf 41} (1995) 1
[arXiv:hep-th/9410237].


\bibitem{Alvarez-Gaume}
L.~Alvarez-Gaume and M.~A.~Vazquez-Mozo,
``Topics in string theory and quantum gravity,''
arXiv:hep-th/9212006.


\bibitem{Antoniadis}
I.~Antoniadis, N.~Arkani-Hamed, S.~Dimopoulos and G.~R.~Dvali,
``New dimensions at a millimeter to a Fermi and superstrings at a TeV,''
Phys.\ Lett.\ B {\bf 436} (1998) 257
[arXiv:hep-ph/9804398].


\bibitem{Ashtekar}
A.~Ashtekar,
``New Hamiltonian Formulation Of General Relativity,''
Phys.\ Rev.\ D {\bf 36} (1987) 1587.



\bibitem{Astekhars}
A. Ashtekar, ``Quantum Geometry And Black Hole Entropy'',
 gr-qc/0005126 [=Adv.Theor.Math.Phys. 4 (2000) 1]\\
``Quantum Geometry Of Isolated Horizons And Black Hole Entropy'',
 Phys.Rev. D57 (1998) 1009 [=gr-qc/9705059]

\bibitem{Baez}
J.Baez, ``An Introduction To Spin Foam Models Of Quantum Gravity And Bf Theory'', gr-qc/0010050


\bibitem{Barbero}
J.~F.~Barbero,
``Real Ashtekar variables for Lorentzian signature space times,''
Phys.\ Rev.\ D {\bf 51} (1995) 5507
[arXiv:gr-qc/9410014].


\bibitem{Barrett}
J.~W.~Barrett and L.~Crane,
``A Lorentzian signature model for quantum general relativity,''
Class.\ Quant.\ Grav.\  {\bf 17} (2000) 3101
[arXiv:gr-qc/9904025].



\bibitem{Bekenstein}
J. Bekenstein, {\em Black  Holes And Entropy}
Phys.Rev. D7 (1973) 2333\\
(with V. Mukhanov), ``Spectroscopy Of The Quantum Black Hole'', Commun.Math.Phys. 125 (1989) 417


\bibitem{Blau}
M. Blau, ``Topological Gauge Theories Of Antisymmetric Tensor Fields'',
 hep-th/9901069 [=Adv.Theor.Math.Phys. 3 (1999) 1289]

\bibitem{Coleman}
S.~R.~Coleman and S.~L.~Glashow,
``High-energy tests of Lorentz invariance,''
Phys.\ Rev.\ D {\bf 59} (1999) 116008
[arXiv:hep-ph/9812418].


\bibitem{Duff}
M.~J.~Duff,
``Inconsistency Of Quantum Field Theory In Curved Space-Time,''
ICTP/79-80/38
{\it Talk presented at 2nd Oxford Conf. on Quantum Gravity, Oxford, Eng., Apr 1980}




\bibitem{Freidel}
L.~Freidel, K.~Krasnov and R.~Puzio,
``BF description of higher-dimensional gravity theories,''
Adv.\ Theor.\ Math.\ Phys.\  {\bf 3} (1999) 1289
[arXiv:hep-th/9901069].



\bibitem{Friedan}
D.~Friedan,
``Nonlinear Models In Two Epsilon Dimensions,''
Phys.\ Rev.\ Lett.\  {\bf 45} (1980) 1057.





\bibitem{Gliozzi}
F.~Gliozzi, J.~Scherk and D.~I.~Olive,
``Supersymmetry, Supergravity Theories And The Dual Spinor Model,''
Nucl.\ Phys.\ B {\bf 122} (1977) 253.



\bibitem{Green}
M.~B.~Green and J.~H.~Schwarz,
``Anomaly Cancellation In Supersymmetric D=10 Gauge Theory And Superstring Theory,''
Phys.\ Lett.\ B {\bf 149} (1984) 117.

\bibitem{Greens}
M.~B.~Green, J.~H.~Schwarz and E.~Witten,
``Superstring Theory. Vol. 1: Introduction,''
\\
``Superstring Theory. Vol. 2: Loop Amplitudes, Anomalies And Phenomenology,''





\bibitem{Hawking}
S. Hawking, {\em  Particle Creation By Black Holes}
 Commun.Math.Phys. 43 (1975) 199

\bibitem{Horowitz}
G.~T.~Horowitz,
``Quantum gravity at the turn of the millennium,''
arXiv:gr-qc/0011089.

 
\bibitem{Horowitz1}
G. Horowitz,''Exactly Soluble Diffeomorphism Invariant Theories'',
 Annals Phys. 205 (1991) 130


\bibitem{Hull}
C.~M.~Hull and P.~K.~Townsend,
``Unity of superstring dualities,''
Nucl.\ Phys.\ B {\bf 438} (1995) 109
[arXiv:hep-th/9410167].



\bibitem{Immirzi}
G.~Immirzi,
``Quantum gravity and Regge calculus,''
Nucl.\ Phys.\ Proc.\ Suppl.\  {\bf 57} (1997) 65
[arXiv:gr-qc/9701052].

\bibitem{Isham}
C.~J.~Isham,
``Structural issues in quantum gravity,''
arXiv:gr-qc/9510063.



\bibitem{Kodama}
H.~Kodama,
``Holomorphic Wave Function Of The Universe,''
Phys.\ Rev.\ D {\bf 42} (1990) 2548.



\bibitem{Luscher}
M.~Luscher, R.~Narayanan, P.~Weisz and U.~Wolff,
``The Schrodinger functional: A Renormalizable probe for nonAbelian gauge theories,''
Nucl.\ Phys.\ B {\bf 384} (1992) 168
[arXiv:hep-lat/9207009].



\bibitem{MM}
Y.~M.~Makeenko and A.~A.~Migdal,
``Exact Equation For The Loop Average In Multicolor QCD,''
Phys.\ Lett.\ B {\bf 88} (1979) 135
[Erratum-ibid.\ B {\bf 89} (1980) 437].





\bibitem{ea-Maldacena}J. Maldacena, 
                   {\em The large N limit of superconformal field
                    theories and supergravity},
                   Adv.Theor.Math.Phys.2:231-252,1998, 
                    {\tt hep-th/9711200}. 

   

\bibitem{Myers}
R.~C.~Myers and M.~Pospelov,
``Experimental challenges for quantum gravity,''
arXiv:hep-ph/0301124.







\bibitem{penrose}
R. Penrose, {\em Angular momentum: an approach to combinatorial spacetime},in 
{\em Quantum theory and beyond}, T. Bastin ed.(Cambridge University Press, 1971)

\bibitem{Polchinski}
J.~Polchinski,
``Dirichlet-Branes and Ramond-Ramond Charges,''
Phys.\ Rev.\ Lett.\  {\bf 75} (1995) 4724
[arXiv:hep-th/9510017].

\bibitem{Polchinskis}
J.~Polchinski,
``String Theory. Vol. 1: An Introduction To The Bosonic String,''
\\
``String Theory. Vol. 2: Superstring Theory And Beyond,''



\bibitem{Polyakov}
A.~M.~Polyakov,
``Quantum Geometry Of Bosonic Strings,''
Phys.\ Lett.\ B {\bf 103} (1981) 207.




\bibitem{Rovellir}
C.~Rovelli,
``Loop quantum gravity,''
Living Rev.\ Rel.\  {\bf 1} (1998) 1
[arXiv:gr-qc/9710008].



\bibitem{Rovelli}
C. Rovelli 
{\em  Loop Quantum Gravity} gr-qc/9606088 [=Phys.Lett. B380 (1996) 257]\\
``The Immirzi Parameter In Quantum General Relativity'',
 gr-qc/9505012 [=Phys.Lett. B360 (1995) 7]


\bibitem{RS}
C.~Rovelli and L.~Smolin,
``Spin networks and quantum gravity,''
Phys.\ Rev.\ D {\bf 52} (1995) 5743
[arXiv:gr-qc/9505006].

\bibitem{RSS}
C.~Rovelli and L.~Smolin,
``Discreteness of area and volume in quantum gravity,''
Nucl.\ Phys.\ B {\bf 442} (1995) 593
[Erratum-ibid.\ B {\bf 456} (1995) 753]
[arXiv:gr-qc/9411005].


\bibitem{Scherk}
J.~Scherk and J.~H.~Schwarz,
``Dual Models For Nonhadrons,''
Nucl.\ Phys.\ B {\bf 81} (1974) 118.









\bibitem{Smolin}
L.~Smolin,
``How far are we from the quantum theory of gravity?,''
arXiv:hep-th/0303185.

\bibitem{Smolinc}
L.~Smolin,
``Quantum gravity with a positive cosmological constant,''
arXiv:hep-th/0209079.


\bibitem{Strominger}
A.~Strominger and C.~Vafa,
``Microscopic Origin of the Bekenstein-Hawking Entropy,''
Phys.\ Lett.\ B {\bf 379} (1996) 99
[arXiv:hep-th/9601029].


\bibitem{Susskind}
L.~Susskind,
``The World as a hologram,''
J.\ Math.\ Phys.\  {\bf 36} (1995) 6377
[arXiv:hep-th/9409089].

\bibitem{'tHooft}
G.~'t Hooft,
``Dimensional Reduction In Quantum Gravity,''
arXiv:gr-qc/9310026.



\bibitem{thv}
G.~'t Hooft and M.~J.~Veltman,
``One Loop Divergencies In The Theory Of Gravitation,''
Annales Poincare Phys.\ Theor.\ A {\bf 20} (1974) 69.

\bibitem{Thiemannr}
T.~Thiemann,
``Introduction to modern canonical quantum general relativity,''
arXiv:gr-qc/0110034.




\bibitem{Thiemann}
T.~Thiemann,
``Anomaly-free formulation of non-perturbative, four-dimensional  Lorentzian quantum gravity,''
Phys.\ Lett.\ B {\bf 380} (1996) 257
[arXiv:gr-qc/9606088].\\
``Gauge field theory coherent states (GCS). I: General properties,''
Class.\ Quant.\ Grav.\  {\bf 18} (2001) 2025
[arXiv:hep-th/0005233].




\bibitem{Turaev}
V.~G.~Turaev and O.~Y.~Viro,
``State Sum Invariants Of 3 Manifolds And Quantum 6j Symbols,''
Topology {\bf 31} (1992) 865.












               




\bibitem{dewitt}B. S. DeWitt,
                          ``Quantum Theory of Gravity. II. The Manifestly Covariant Theory'',
                          Phys. Rev. 162, 1195-1239 (1967).








\bibitem{ea-Witten} E.~Witten,
                 {\sl Anti de Sitter Space and Holography},
                 {\tt hep-th/9802150};\\
 {\sl Anti-de Sitter space,
                   thermal phase transition and confinement in gauge
                   theories}
                 {\tt hep-th/9803131}.
                
\bibitem{Witten3}
E.~Witten,
``(2+1)-Dimensional Gravity As An Exactly Soluble System,''
Nucl.\ Phys.\ B {\bf 311} (1988) 46.
\end{thebibliography}

\begin{thebibliography}{99}


\bibitem{Madorebook} J. Madore, {\it An introduction to Noncommutative Differential
Geometry and its Physical Applications} Lect. not. series Lond.
Math. Soc. {\bf 206}, Cambridge University Press (1995).

\bibitem{ncgbooks}
A. Connes, {\it Noncommutative Geometry}, Academic Press,
 (1994); G.~Landi,
{\it An introduction to noncommutative spaces and their geometry},
Springer (1997) arXiv:hep-th/9701078;\\
J.~M.~Gracia-Bondia, J.~C.~Varilly and H.~Figueroa, {\it Elements
Of Noncommutative Geometry}, Birkhauser (2000).


\bibitem{nonfuzzy}
M.~Requardt, {\it Cellular networks as models for Planck-scale
physics}, J.\ Phys.\ A {\bf 31} (1998) 7997
arXiv:hep-th/9806135;\\
G.~Landi and F.~Lizzi, {\it Projective systems of noncommutative
lattice as a pregeometric substratum,} arXiv:math-ph/9810011;\\
L.~C.~de Albuquerque, J.~L.~deLyra and P.~Teotonio-Sobrinho,
{Fluctuating dimension in a discrete model for quantum gravity
based on  the spectral action}, Phys.\ Rev.\ Lett.\  {\bf 91}
(2003) 081301 arXiv:hep-th/0305082.

\bibitem{Kac} M.~Kac, {\it Can One Hear the Shape of a Drum?},
Amer.\ Math.\ Monthly {\bf 73} (1966) 1.

\bibitem{Madorefuzzysphere} J. Madore,  {\it The fuzzy sphere},
Class. Quant. Grav. {\bf 9} (1992) 69.

\bibitem{Weyl} H.~Weyl, {\it The theory of groups and Quantum Mechanics},
Dover (1931).

\bibitem{Rieffel} M.A.~Rieffel, {\it $C^*$-Algebras Associated
with Irrational Rotations}, Pacific J. Math. 93 (1981) 415.

\bibitem{CDS} A.~Connes, M.~R.~Douglas and A.~Schwarz,
{\it Noncommutative geometry and matrix theory: Compactification
on tori}, JHEP {\bf 9802} (1998) 003 arXiv:hep-th/9711162;\\
G.~Landi, F.~Lizzi and R.~J.~Szabo, {\it String geometry and the
noncommutative torus}, Commun.\ Math.\ Phys.\  {\bf 206} (1999)
603 arXiv:hep-th/9806099.


\bibitem{fuzzydisc} F.~Lizzi, P.~Vitale and A.~Zampini,
{\it The fuzzy disc}, JHEP {\bf 0308} (2003) 057
arXiv:hep-th/0306247;
{\it From the fuzzy disc to edge currents in Chern-Simons theory},
arXiv:hep-th/0309128.

\bibitem{Harmon}J.~Hoppe, {\it Quantum Theory of a massless relativistic surface
and a two-dimensional bound state problem}, M.I.T.\ Ph.~D. Thesis,
1982;\\
C.~S.~Chu, J.~Madore and H.~Steinacker, {\it Scaling limits of the
fuzzy sphere at one loop},
JHEP {\bf 0108} (2001) 038 arXiv:hep-th/0106205;\\
S.~Iso, Y.~Kimura, K.~Tanaka and K.~Wakatsuki, {\it Noncommutative
gauge theory on fuzzy sphere from matrix model}, Nucl.\ Phys.\ B
{\bf 604} (2001) 121 arXiv:hep-th/0101102.

\bibitem{Varsalovich} D.A.~Varsalovich, A.N.~Moskalev and
V.K.~Kernsky, {\it Quantum Theory of Angular Momentum}, World
Scientific (1988).

\bibitem{BalGuptaKurkcuoglu} A.P.~Balachandran, K,~Gupta and
S.~K\"urk\c{c}\"{u}o\v{g}lu, {\it Edge Currents in Noncommutative
Chern--Simons Theory from a new Matrix Model}, hep-th/0306255.

\bibitem{ElliottEvans} G.A.~Elliott and D.E.~Evans, Ann. Math. {\bf 138}, 477
(1993);\\
G.~Landi, F.~Lizzi and R.~J.~Szabo, {\it A new matrix model for
noncommutative field theory}, arXiv:hep-th/0309031.
G.~Landi, F.~Lizzi and R.~J.~Szabo, {\it Matrix Quantum Mechanics
and Soliton Regularization of Noncommutative Field Theory}, to
appear.


\bibitem{LandiLizziSzaboLargeN} G. Landi, F. Lizzi and R. J. Szabo, {\it From
large N matrices to the noncommutative torus}, Comm. Math. Phys.
{\bf 217} (2001) 181. arXiv: hep-th/9912130.


\bibitem{Voros} H. Gr\"onewold, {\it On principles of quantum mechanics}, Physica {12}
(1946) 405; J. E. Moyal,  Quantum mechanics as a statistical
theory, Proc.Cambridge Phil.Soc. {\bf 45} (1949) 99; A. Voros,
{\it Wentzel-Kramers-Brillouin method in the Bargmann
representation}, Phys. Rev. {\bf A 40} (1989) 6814.

\bibitem{PinzulStern}  A. Pinzul and  A. Stern,
{\it Absence of the holographic principle in noncommutative
Chern-Simons  theory}, JHEP {\bf 0111}, 023 (2001)
arXiv:hep-th/0107179; {\it Edge States from Defects on the
Noncommutative Plane},arXiv:hep-th/0307234.

\end{thebibliography}
\end{document}